\documentclass[10pt,a4paper,twoside,openright]{report}
\pdfoutput=1

\usepackage[top=20mm,bottom=20mm,left=15mm,right=15mm]{geometry} 
\usepackage[T1]{fontenc}
\usepackage[latin1]{inputenc}                                    

\usepackage{fancyhdr}                                            
\usepackage{amsmath}                                             
\usepackage{amsfonts}                                            
\usepackage{amssymb}                                             
\usepackage{bm}
\usepackage[frenchb]{babel}

\usepackage[pdftex]{graphicx}                                    
\usepackage{multicol}
\usepackage{multirow}
\usepackage{color}
\usepackage{subfig}
\usepackage{array}
\usepackage{lmodern}
\usepackage[pdftex,colorlinks,bookmarks,pdfpagemode=UseOutlines,pdfstartview=FitBH]{hyperref}
\usepackage{algorithm,algorithmic} 

\def\maketitle{
\null
\thispagestyle{empty}
\vfill
\begin{center}
\leavevmode
\normalfont

{\bf \LARGE Régularisation et optimisation \par

\vskip 0.5cm

pour l'imagerie sismique des fondations de pylônes} \par

\vskip 6cm

{\large Denis Vautrin, Matthieu Voorons, Jérôme Idier, Yves Goussard} \par

\vskip 1cm

\texttt{denis.vautrin@irccyn.ec-nantes.fr}, \texttt{jerome.idier@irccyn.ec-nantes.fr}

Institut de Recherche en Communications et Cybernétique de Nantes (IRCCyN)

ECN - BP 92101 - 1 rue de la Noë - 44321 Nantes

\vskip 1cm

\texttt{matthieu.voorons@gmail.com}, \texttt{yves.goussard@polymtl.ca}

Ecole Polytechnique de Montréal

C.P. 6979, succ. Centre-ville - Montréal (Québec) H2S 2S2

\vskip 3cm

\end{center}
\vfill
\null
\cleardoublepage
}

\floatname{algorithm}{Algorithme}

\newcommand{\argmin}[1]{\underset{#1}{\operatorname{arg\,min}}\;} 
\def\indentit{\mbox{l\hspace{-0.50em}1}}  
\def\Diag{\mathrm{Diag}}   
\def\red{\mathrm{red}}   
\def\capt{\mathrm{capt}}   
\def\ZE{\mathrm{ZE}}   

\usepackage{soul}

\input{alphabet}
\def\XS{\xspace}

\def\rem#1{}                    

\def\aprio{\textit{a priori}\XS}

\def\cad{c'est-\`a-dire\XS}

\def\etal{\textit{et al.}\XS}

\def\matlab{\textit{Matlab}\XS}

\def\vav{vis-\`a-vis\XS}

\begin{document}

\maketitle

\pagestyle{plain}

~
\vfill
\begin{center}
\begin{minipage}{0.8\linewidth}
{\centering \bf Résumé :\\}

\medskip

Ce rapport de recherche résume l'avancement de travaux menés conjointement par l'IRCCyN et l'Ecole Polytechnique de Montréal concernant la résolution du problème inverse pour l'imagerie sismique des fondations de pylônes électriques. Nous abordons plusieurs méthodes de type \og cartographie \fg{}. Nous nous intéressons plus particulièrement à des méthodes basées sur une formulation bilinéaire du problème direct d'une part (CSI, gradient modifié, etc.) et à des méthodes basées sur une formulation dite \og primale \fg{} d'autre part. Les performances de ces méthodes sont évaluées sur des données synthétiques.

Ces travaux ont été partiellement financés par RTE -- CNER, qui est à l'initiative du projet, et ont été effectués avec la collaboration de EDF R\&D.
\end{minipage}
\end{center} \par

\vskip 6cm

\vfill

\tableofcontents

\chapter*{Introduction \markboth{Introduction}{}}
\addcontentsline{toc}{chapter}{Introduction}
\pagestyle{fancy}

Le problème d'imagerie des fondations de pylônes consiste à déterminer la forme d'objets enfouis dans le sol. Pour cela, on génère des ondes sismiques à l'aide d'une source placée en surface et on mesure une partie du champ de vitesse résultant à l'aide d'une série de capteurs également placés en surface. Plusieurs tirs sont effectués en modifiant la position de la source d'un tir à l'autre.

Dans un premier temps, nos deux équipes (l'IRCCyN à Nantes et l'Ecole Polytechnique de Montréal) ont abordé ce problème en travaillant de façon commune sur une première méthode d'inversion (la méthode CSI). Ce travail a fait l'objet du précédent rapport d'avancement \cite{Vautrin09}. Nous nous sommes ensuite partagé les tâches de manière à travailler sur deux familles de méthodes différentes. Le tableau ci-dessous détaille les points abordés par chacune des équipes. Afin de pouvoir comparer les performances des différentes méthodes abordées, nous les avons testées sur les mêmes jeux de données synthétiques.

\begin{center} \begin{tabular}{r|l} \hline

	& {\bf - Travail commun entre Nantes et Montréal -} \\
	& \quad $\cdot$ Prise en main de l'algorithme de résolution du problème direct \\
	& \quad $\cdot$ Recherche bibliographique sur la méthode CSI \\
	& \quad $\cdot$ Adaptation de la méthode CSI à notre problème \\
{\sl Jusqu'en juillet 2009} \quad
	& \quad $\cdot$ Début de l'implémentation de la méthode \\ \hline

	& {\bf - Partage des tâches entre Nantes et Montréal -} \\
	& \quad A l'Ecole Polytechnique de Montréal : \\
	& \quad $\cdot$ Poursuite du travail concernant la méthode CSI \\
	& \quad $\cdot$ Proposition de plusieurs variantes \\
	& \quad A l'IRCCyN : \\
{\sl Jusqu'en septembre 2009} \quad
	& \quad $\cdot$ Recherche bibliographique sur les méthodes adaptées à la formulation "primale" \\ \hline

	& \quad A l'Ecole Polytechnique de Montréal : \\
	& \quad $\cdot$ Travail sur la formulation du problème direct \\
	& \quad $\cdot$ Implémentation des variantes de la méthode CSI et premiers résultats \\
	& \quad $\cdot$ Travail sur le problème de conditionnement \\
	& \quad A l'IRCCyN : \\
	& \quad $\cdot$ Travail sur l'optimisation de la formulation primale \\
	& \quad $\cdot$ Implémentation des méthodes de type gradient et premiers résultats \\
{\sl Jusqu'en décembre 2009} \quad
	& \quad $\cdot$ Utilisation du changement de variables et premiers résultats \\ \hline

\end{tabular} \end{center}

Nous avons conservé les hypothèses formulées dans le rapport d'avancement précédent \cite{Vautrin09}. On rappelle notamment les points suivants :
\begin{itemize}
\item l'approche abordée est de type "cartographique". Nous cherchons donc à reconstruire des cartes du sous-sol sans utiliser d'information \textit{a priori} sur la forme de l'objet recherché ;
\item nous considérons que la distribution de deux caractéristiques mécaniques (les vitesses des ondes de pression et cisaillement) est suffisante pour décrire un milieu ;
\item la méthode de résolution du problème direct, qui permet de construire un jeu de données synthétiques pour un milieu donné, est basée sur les différences finies.
\end{itemize}

Dans la première partie de ce rapport, nous reviendrons sur la formulation du problème direct. Une étude approfondie nous a permis de mettre en évidence une structure particulière de la matrice dite "d'impédance" faisant intervenir des opérateurs de filtrage.

La seconde partie concerne les différentes méthodes d'inversion. Nous reviendrons dans un premier temps sur la méthode CSI puis nous présenterons les autres méthodes d'inversion que nous avons abordées. Les premières sont analogues à la méthodes CSI et utilisent une formulation faisant intervenir des variables auxiliaires. Les autres se basent sur la formulation dite "primale", il s'agit de méthodes n'agissant que sur les variables d'intérêt (pas de recours à des variables auxiliaires).

\chapter{Le problème direct}

\section{Les équations de propagation}

Le problème direct consiste à construire des données synthétiques en modélisant la propagation des ondes dans un milieu donné. On utilise pour cela les équations de propagation en milieu élastique. Ces équations sont présentées dans le domaine fréquentiel (on note $\omega$ la pulsation de la fonction excitatrice et $N_f$ le nombre de fréquences considérées) et sont discrétisées à l'aide des différences finies, ce qui permet d'écrire le problème sous la forme d'un système d'équations linéaire (voir \cite{Kerzale09} et \cite{Vautrin09} pour plus de détails). Afin d'alléger les notations, nous considérerons dans un premier temps qu'il n'y a qu'une seule position de tir et que nous travaillons en monofréquentiel.

\begin{figure}[htbp!]
\begin{center}
\includegraphics[scale=1]{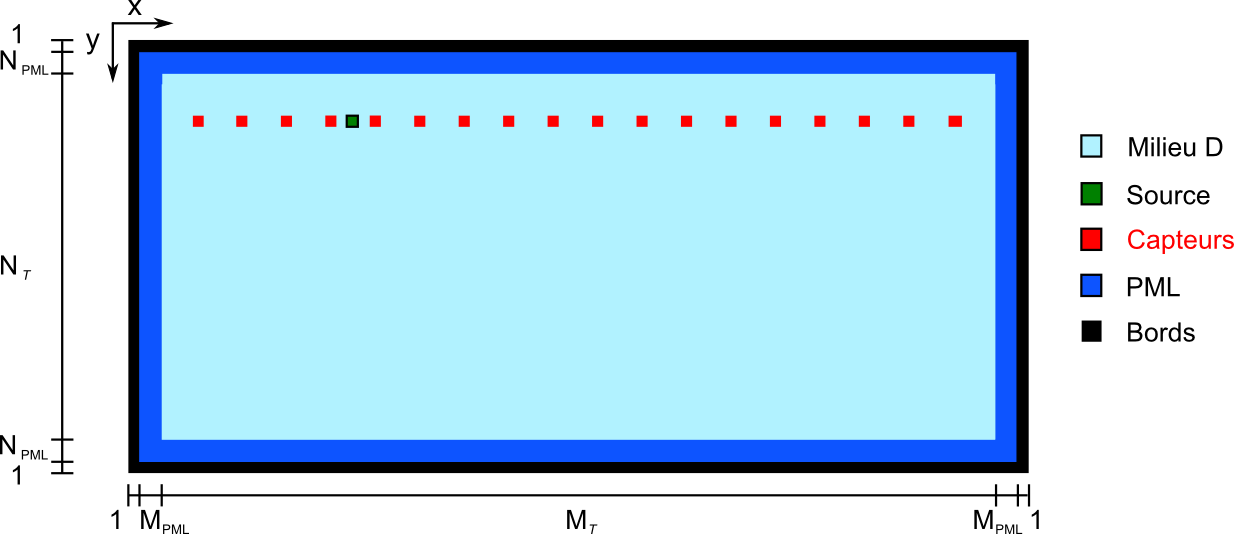}
\caption{Description du modèle proposé}
\label{FigMilieu}
\end{center}
\end{figure}

Nous schématisons sur la Figure \ref{FigMilieu} le domaine d'étude. Le milieu de propagation D de dimensions finies est entouré d'une zone "PML", ce qui permet de simuler la propagation des ondes dans un milieu aux dimiension infinies (les PML atténuent les ondes de manière à éviter une réflexion sur les bords).

A l'intérieur du milieu $D$ et dans la zone PML (voir Figure \ref{FigMilieu}), les équations de propagation dans le domaine fréquentiel s'écrivent :

\begin{equation} \label{eqDomaine}
\begin{split}
- i\omega f_x(r) = \omega^2 V_x(r)
   & + \alpha_x(r,\omega)\partial_{x}(\mathit{v}_p(r)^2 \alpha_x(r,\omega) \partial_{x}V_x(r))   \\
   & + \alpha_x(r,\omega)\partial_{x}((\mathit{v}_p(r)^2-2\mathit{v}_s(r)^2) \alpha_y(r,\omega) \partial_{y}V_y(r))   \\
   & + \alpha_y(r,\omega)\partial_{y}(\mathit{v}_s(r)^2 \alpha_x(r,\omega) \partial_{x}V_y(r))   \\
   & + \alpha_y(r,\omega)\partial_{y}(\mathit{v}_s(r)^2 \alpha_y(r,\omega) \partial_{y}V_x(r))   \\
- i\omega f_y(r) = \omega^2 V_y(r)
   & + \alpha_x(r,\omega)\partial_{x}(\mathit{v}_s(r)^2 \alpha_x(r,\omega) \partial_{x}V_y(r)   \\
   & + \alpha_x(r,\omega)\partial_{x}(\mathit{v}_s(r)^2 \alpha_y(r,\omega) \partial_{y}V_x(r))   \\
   & + \alpha_y(r,\omega)\partial_{y}(((\mathit{v}_p(r)^2 - 2\mathit{v}_s(r)^2) \alpha_x(r,\omega) \partial_{x}V_x(r))   \\
   & + \alpha_y(r,\omega)\partial_{y}((\mathit{v}_p(r)^2 \alpha_y(r,\omega) \partial_{y}V_y(r))
\end{split}
\end{equation}
où $f_x(r)$ et $f_y(r)$ sont les composantes de la fonction source au point $r$ (rapport de la force sur la masse volumique), $V_x(r)$ et $V_y(r)$ sont les composantes de la vitesse au point $r$, $\mathit{v}_p(r)$ et $\mathit{v}_s(r)$ sont les caractéristiques recherchées (respectivement les vitesses des ondes P et S) et $\alpha_x(r,\omega)$ et $\alpha_y(r,\omega)$ sont des coefficients introduits par les PML qui sont égaux à $1$ à l'intérieur du domaine.

Sur les bords du domaine, on a les contraintes suivantes :

\begin{equation} \label{eqContraintesDomaine}
V_x(r) = 0 \quad \text{ et } \quad V_y(r) = 0
\end{equation}

\section{Discrétisation des équations de propagation}

\subsection{Le stencil de Saenger} \label{Part_StencilSaenger}

Ces équations de propagation sont discrétisées en utilisant le stencil proposé par Saenger \cite{Saenger00} et en introduisant des opérateurs de différences finies adaptés. On utilise deux grilles de points de résolution $\Delta x$ suivant l'axe $\vec{x}$ et de $\Delta y$ suivant l'axe $\vec{y}$ disposées en quinconce et définies sur l'ensemble du milieu.

\begin{figure}[htbp!]
\begin{center}
\includegraphics[scale=0.8]{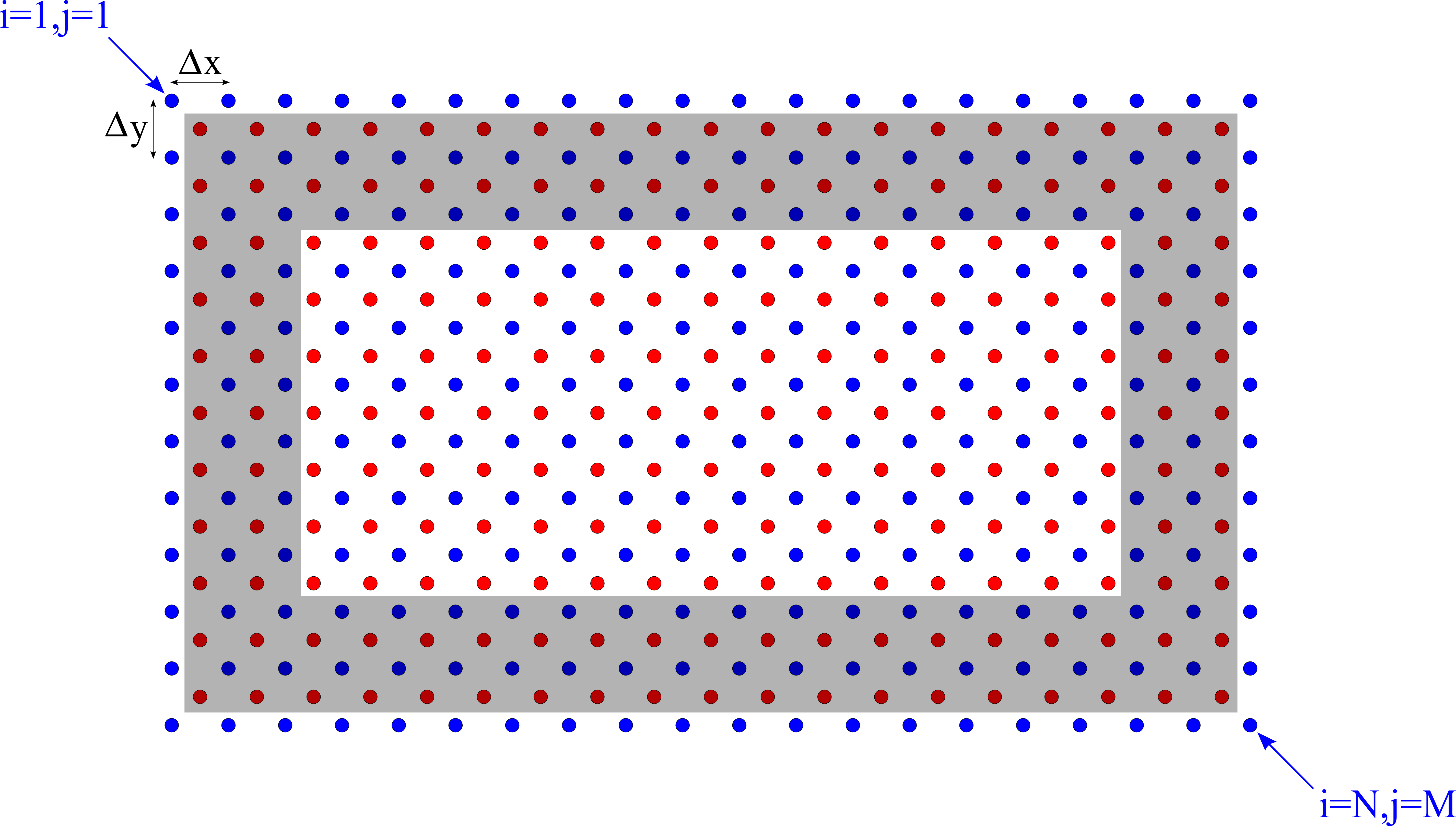}
\caption{Grilles utilisées pour la discrétisation : la grille bleue est associée aux champs de forces ($f_x$ et $f_y$) et de vitesses ($V_x$ et $V_y$) et la rouge aux champs des caractéristiques ($\mathit{v}_p$ et $\mathit{v}_s$). Les coefficients associés aux PML ($\alpha_x$ et $\alpha_y$) sont définis sur les deux grilles, ils sont différents de 1 uniquement dans la zone PML (surface grisée).}
\label{FigureGrilleSaenger}
\end{center}
\end{figure}

La première grille (représentée par les points bleus sur la figure \ref{FigureGrilleSaenger}) est liée aux champs de force $f_x$ et $f_y$ (source) et de vitesse $V_x$ et $V_y$. Les points de cette grille, de coordonnées $x_i = i \Delta x$ et $y_j = j \Delta y$ avec $i$ et $j$ entiers, sont désignés par les indices $i$ et $j$.

La seconde grille (représentée par les points rouges sur la figure \ref{FigureGrilleSaenger}) est liée aux caractéristiques du milieu $\mathit{v}_p$ et $\mathit{v}_s$. Les points de cette grille, de coordonnées $x_i = (i \pm \frac{1}{2}) \Delta x$ et $y_j = (j \pm \frac{1}{2}) \Delta y$ avec $i$ et $j$ entiers, sont désignés par les indices $i \pm \frac{1}{2}$ et $j \pm \frac{1}{2}$.

Les coefficients $\alpha_x(r,\omega)$ et $\alpha_y(r,\omega)$ sont quant à eux définis sur les deux grilles.

\bigskip

{\bf Remarque :} Si la première grille est constituée de $M$ lignes et de $N$ colonnes (si l'on reprend les notations de la Figure \ref{FigMilieu}, on a $M = M_T + 2M_{PML} +2$ et $N = N_T + 2N_{PML} +2$), la seconde grille est constituée de $M-1$ lignes et de $N-1$ colonnes.

\bigskip

\subsection{Introduction des opérateurs de différences finies}

Nous allons reprendre les équations différentielles de propagation (Equations \ref{eqDomaine} et \ref{eqContraintesDomaine}) et y introduire quatre opérateurs de différences finies adaptés au stencil de Saenger correspondant aux dérivées partielles spatiales $\partial_{x}$ et $\partial_{y}$ :

\begin{itemize}

\item un opérateur $\mathcal{G}^x$ correspondant à la dérivée partielle selon $\vec{x}$ d'une grandeur associée à la première grille :

\begin{minipage}[c]{0.75\linewidth} \begin{align}
\mathcal{G}^x(X)_{i+\frac{1}{2},j+\frac{1}{2}} = \dfrac{1}{2\Delta x} \left[ (X)_{i+1,j+1} + (X)_{i+1,j} - (X)_{i,j+1} - (X)_{i,j} \right]
\label{Eq_ExpressionGx}
\end{align} \end{minipage}
\hfill
\begin{minipage}[c]{0.2\linewidth}
\centering
\includegraphics[scale=2]{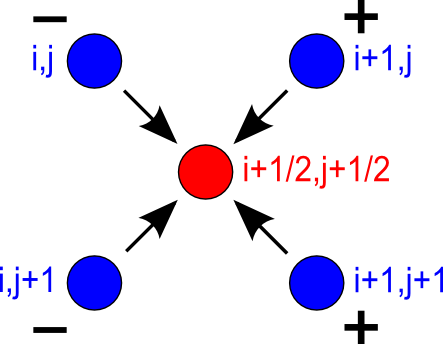}
\end{minipage}

\item un opérateur $\mathcal{G}^y$ correspondant à la dérivée partielle selon $\vec{y}$ d'une grandeur associée à la première grille :

\begin{minipage}[c]{0.75\linewidth} \begin{align}
\mathcal{G}^y(X)_{i+\frac{1}{2},j+\frac{1}{2}} = \dfrac{1}{2\Delta y} \left[ (X)_{i+1,j+1} - (X)_{i+1,j} + (X)_{i,j+1} - (X)_{i,j} \right]
\label{Eq_ExpressionGy}
\end{align} \end{minipage}
\hfill
\begin{minipage}[c]{0.2\linewidth}
\centering
\includegraphics[scale=2]{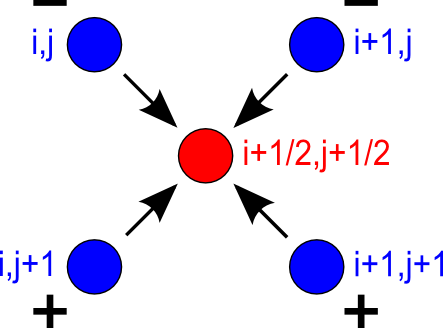}
\end{minipage}

\item un opérateur $\mathcal{H}^x$ correspondant à la dérivée partielle selon $\vec{x}$ d'une grandeur associée à la seconde grille :

\begin{minipage}[c]{0.75\linewidth} \begin{align}
\mathcal{H}^x(X)_{i,j} = \dfrac{1}{2\Delta x} \left[ (X)_{i+\frac{1}{2},j+\frac{1}{2}} + (X)_{i+\frac{1}{2},j-\frac{1}{2}} - (X)_{i-\frac{1}{2},j+\frac{1}{2}} - (X)_{i-\frac{1}{2},j-\frac{1}{2}} \right]
\label{Eq_ExpressionHx}
\end{align} \end{minipage}
\hfill
\begin{minipage}[c]{0.2\linewidth}
\centering
\includegraphics[scale=2]{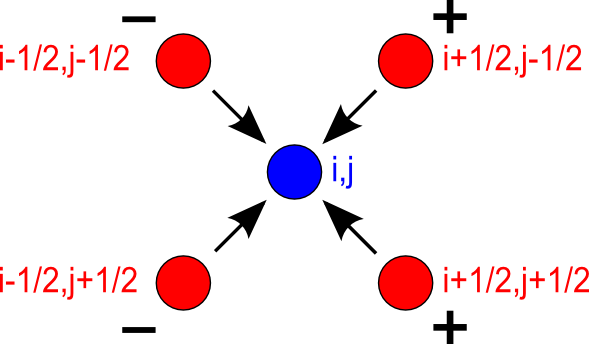}
\end{minipage}

\item un opérateur $\mathcal{H}^y$ correspondant à la dérivée partielle selon $\vec{y}$ d'une grandeur associée à la seconde grille :

\begin{minipage}[c]{0.75\linewidth} \begin{align}
\mathcal{H}^y(X)_{i,j} = \dfrac{1}{2\Delta y} \left[ (X)_{i+\frac{1}{2},j+\frac{1}{2}} - (X)_{i+\frac{1}{2},j-\frac{1}{2}} + (X)_{i-\frac{1}{2},j+\frac{1}{2}} - (X)_{i-\frac{1}{2},j-\frac{1}{2}} \right]
\label{Eq_ExpressionHy}
\end{align} \end{minipage}
\hfill
\begin{minipage}[c]{0.2\linewidth}
\centering
\includegraphics[scale=2]{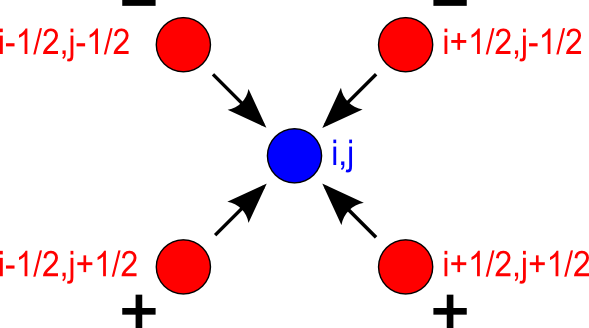}
\end{minipage}

\end{itemize}

\bigskip

{\bf Remarque :} L'utilisation de l'un de ces quatre opérateurs induit le passage d'une grille à l'autre.

\bigskip

A l'intérieur du milieu $D$ et dans la zone PML, on obtient alors les équations suivantes :

\begin{equation} \label{Eq_MilieuAprDiscr}
\begin{split}
(F_x)_{i,j} = \omega^2 (V_x)_{i,j}
              &  + (\alpha_x \mathcal{H}^x(\mathit{v}_p^2 \alpha_x \mathcal{G}^x(V_x)))_{i,j}   \\
              &  + (\alpha_x \mathcal{H}^x((\mathit{v}_p^2-2\mathit{v}_s^2) \alpha_y \mathcal{G}^y(V_y)))_{i,j}   \\
              &  + (\alpha_y \mathcal{H}^y(\mathit{v}_s^2 \alpha_x \mathcal{G}^x(V_y)))_{i,j}   \\
              &  + (\alpha_y \mathcal{H}^y(\mathit{v}_s^2 \alpha_y \mathcal{G}^y(V_x)))_{i,j}   \\
(F_y)_{i,j} = \omega^2 (V_y)_{i,j}
              &  + (\alpha_x \mathcal{H}^x(\mathit{v}_s^2 \alpha_x \mathcal{G}^x(V_y)))_{i,j}   \\
              &  + (\alpha_x \mathcal{H}^x(\mathit{v}_s^2 \alpha_y \mathcal{G}^y(V_x)))_{i,j}   \\
              &  + (\alpha_y \mathcal{H}^y((\mathit{v}_p^2-2\mathit{v}_s^2) \alpha_x \mathcal{G}^x(V_x)))_{i,j}   \\
              &  + (\alpha_y \mathcal{H}^y(\mathit{v}_p^2 \alpha_y \mathcal{G}^y(V_y)))_{i,j}
\end{split}
\end{equation}

Sur les bords du domaine, on impose :

\begin{equation} \label{Eq_BordsAprDiscr}
(V_x)_{i,j} = 0 \quad \text{ et } \quad (V_y)_{i,j} = 0
\end{equation}

\section{Ecriture sous la forme d'un système linéaire d'équations} \label{Part_SystLinPbDirect}

À partir de l'ensemble des équations discrétisées (Eq. \ref{Eq_MilieuAprDiscr} et \ref{Eq_BordsAprDiscr}), on forme un système linéaire d'équations reliant les forces aux vitesses que l'on écrit de la façon suivante :
\begin{equation} \mathit{F} \ = \ \mathbf{A} \ . \ \mathit{V} \end{equation}
ou, en distinguant les composantes horizontales et verticales des champs :
\begin{equation} \begin{bmatrix}   F_x   \\   F_y   \end{bmatrix}
=	\begin{bmatrix}   \mathbf{A}_{xx}   &   \mathbf{A}_{xy}   \\   \mathbf{A}_{yx}   &   \mathbf{A}_{yy}   \end{bmatrix}
	\begin{bmatrix}   V_x   \\   V_y   \end{bmatrix} \end{equation}

Si la première grille est constituée de $M$ lignes et de $N$ colonnes, les vecteurs $\mathit{F}$ et $\mathit{V}$ sont de longueur $2MN$ et la matrice $\mathbf{A}$, appelée "matrice d'impédance", est une matrice bande de taille $2MN \times 2MN$. Ses coefficients sont déduits de l'ensemble des équations discrétisées ; ils dépendent donc des caractéristiques du milieu de propagation (champs $\mathit{v}_p$ et $\mathit{v}_s$) ainsi que des coefficients liés aux PML ($\alpha_x(r,\omega)$ et $\alpha_y(r,\omega)$). On donne ci-dessous les expressions des quatre sous-matrices qui la constituent :

\begin{equation}
\mathbf{A}_{xx} = \left[ \mathbf{A}_{\omega,x}	+ \Diag\{ \alpha_1^x \} \mathbf{H^x} \Diag\{ \mathit{v}_p^2 \} \Diag\{ \alpha_2^x \} \mathbf{G^x}
						+ \Diag\{ \alpha_1^y \} \mathbf{H^y} \Diag\{ \mathit{v}_s^2 \} \Diag\{ \alpha_2^y \} \mathbf{G^y} \right]
\end{equation}
\begin{equation}
\mathbf{A}_{xy} = \left[ \Diag\{ \alpha_1^x \} \mathbf{H^x} \Diag\{ \mathit{v}_p^2-2\mathit{v}_s^2 \} \Diag\{ \alpha_2^y \} \mathbf{G^y}
						+ \Diag\{ \alpha_1^y \} \mathbf{H^y} \Diag\{ \mathit{v}_s^2 \} \Diag\{ \alpha_2^x \} \mathbf{G^x} \right]
\end{equation}
\begin{equation}
\mathbf{A}_{yx} = \left[ \Diag\{ \alpha_1^x \} \mathbf{H^x} \Diag\{ \mathit{v}_s^2 \} \Diag\{ \alpha_2^y \} \mathbf{G^y}
						+ \Diag\{ \alpha_1^y \} \mathbf{H^y} \Diag\{ \mathit{v}_p^2-2\mathit{v}_s^2 \} \Diag\{ \alpha_2^x \} \mathbf{G^x} \right]
\end{equation}
\begin{equation}
\mathbf{A}_{yy} = \left[ \mathbf{A}_{\omega,y}	+ \Diag\{ \alpha_1^x \} \mathbf{H^x} \Diag\{ \mathit{v}_s^2 \} \Diag\{ \alpha_2^x \} \mathbf{G^x}
						+ \Diag\{ \alpha_1^y \} \mathbf{H^y} \Diag\{ \mathit{v}_p^2 \} \Diag\{ \alpha_2^y \} \mathbf{G^y} \right]
\end{equation}
où $\Diag \{ w \}$ ($w$ étant un vecteur) désigne une matrice diagonale dont la diagonale est le vecteur $w$.

Les matrices $\mathbf{A}_{\omega,x}$ et $\mathbf{A}_{\omega,y}$ sont identiques. Ce sont des matrices diagonales telles que :
\begin{equation}
\mathbf{A}_{\omega,x}(j+(M-1)i ; j+(M-1)i) = \mathbf{A}_{\omega,y}(j+(M-1)i ; j+(M-1)i) =
		\begin{cases}
					1						& \quad \text{si $(i,j)$ appartient aux bords}			\\
					\omega^2			& \quad \text{sinon}
		\end{cases}
\end{equation}

Les matrices $\mathbf{G^x}$, $\mathbf{G^y}$, $\mathbf{H^x}$ et $\mathbf{H^y}$ sont des matrices rectangulaires associées aux opérateurs de différences finies.
\begin{itemize}
\item $\mathbf{G^x}$ et $\mathbf{G^y}$ sont de taille $(M-1)(N-1) \times MN$. Chaque ligne de ces deux matrices contient les coefficients de l'opérateur de différences finies associé ($\pm \frac{1}{2\Delta x}$ et $\pm \frac{1}{2\Delta y}$).
\item $\mathbf{H^x}$ et $\mathbf{H^y}$ sont de taille $MN \times (M-1)(N-1)$. Certaines de leurs lignes ne contiennent que des zéros (lignes associés aux points du bord du domaine) ; les autres contiennent chacune les coefficients de l'opérateur de différences finies associé ($\pm \frac{1}{2\Delta x}$ et $\pm \frac{1}{2\Delta y}$).
\end{itemize}

Afin de prendre en compte les différentes pulsations qui constituent la fonction excitatrice et les différentes positions de tir de la source ($N_k$ tirs), nous écrirons maintenant le système linéaire reliant le champ des forces au champ des vitesses de la façon suivante :
\begin{equation} \label{Eq_SystLinPbDirect}
\mathit{F}_{\omega,k} = \mathbf{A}_{\omega,p,s} . \mathit{V}_{\omega,k}
\end{equation}
où $\omega$ correspond la pulsation de la fonction excitatrice considérée ($\omega = \omega_1, \dots, \omega_{N_f}$) et $k$ désigne à la position de la source ($k = 1, \dots, N_k$).

La résolution du problème direct consiste à résoudre le système linéaire donné par l'équation \ref{Eq_SystLinPbDirect} pour chaque fréquence et chaque position de la source et à retenir les composantes du champ de vitesse mesurées par les capteurs.

\section{Décomposition de la matrice d'impédance sous forme d'une somme} \label{PartStructMatrImpedance}

La matrice d'impédance $\mathbf{A}_{\omega,p,s}$ du système linéaire \ref{Eq_SystLinPbDirect} peut être décomposée en la somme de trois termes :
\begin{equation} \mathbf{A}_{\omega,p,s} = \mathbf{A}_\omega + \mathbf{A}^p + \mathbf{A}^s \end{equation}
où :
\begin{itemize}
\item $\mathbf{A}^p$ contient les éléments de $\mathbf{A}_{\omega,p,s}$ qui dépendent des $(\mathit{v}_p)_{i,j}$ ;
\item $\mathbf{A}^s$ contient les éléments de $\mathbf{A}_{\omega,p,s}$ qui dépendent des $(\mathit{v}_s)_{i,j}$ ;
\item $\mathbf{A}_\omega$ contient les éléments qui ne dépend ni des $(\mathit{v}_p)_{i,j}$, ni des $(\mathit{v}_s)_{i,j}$ (cette matrice ne dépend que de la fréquence $\omega$).
\end{itemize}

En reprenant les expressions données dans la partie précédente, on obtient les expressions suivantes :

\begin{equation}
\mathbf{A^p}	=
\begin{bmatrix}	\Diag\{ \alpha_1^x \} \mathbf{H^x}			\\
						\Diag\{ \alpha_1^y \} \mathbf{H^y}			\\
\end{bmatrix}
\Diag\{ \mathit{v}_p^2 \}
\begin{bmatrix}	\Diag\{ \alpha_2^x \} \mathbf{G^x}			&			\Diag\{ \alpha_2^y \} \mathbf{G^y}			\\
\end{bmatrix}	\\
\end{equation}

\begin{equation*}
\mathbf{A^s}	=
\begin{bmatrix}	\Diag\{ \alpha_1^y \} \mathbf{H^y}			\\
						\Diag\{ \alpha_1^x \} \mathbf{H^x}
\end{bmatrix}
\Diag\{ \mathit{v}_s^2 \}
\begin{bmatrix}	\Diag\{ \alpha_2^y \} \mathbf{G^y}			&			\Diag\{ \alpha_2^x \} \mathbf{G^x}
\end{bmatrix}
\end{equation*}
\begin{equation}
					+
\begin{bmatrix}	\Diag\{ \alpha_1^x \} \mathbf{H^x}			\\
						- \Diag\{ \alpha_1^y \} \mathbf{H^y}
\end{bmatrix}
\Diag\{ \mathit{v}_s^2 \}
\begin{bmatrix}	\Diag\{ \alpha_2^x \} \mathbf{G^x}			&			- \Diag\{ \alpha_2^y \} \mathbf{G^y}
\end{bmatrix}
\end{equation}
\begin{equation*}
					+
\begin{bmatrix}	- \Diag\{ \alpha_1^x \} \mathbf{H^x}		\\
						- \Diag\{ \alpha_1^y \} \mathbf{H^y}
\end{bmatrix}
\Diag\{ \mathit{v}_s^2 \}
\begin{bmatrix}	\Diag\{ \alpha_2^x \} \mathbf{G^x}			&			\Diag\{ \alpha_2^y \} \mathbf{G^y}
\end{bmatrix}
\end{equation*}

La matrice $\mathbf{A}_\omega$ est diagonale ; on a :
\begin{equation}
\mathbf{A_\omega} =
\begin{bmatrix}	A_{\omega,x}			&		0							\\
						0							&		A_{\omega,y}			\\
\end{bmatrix}
\end{equation}

Cette décomposition sera utilisée lors de la construction des équations utilisées pour résoudre le problème inverse.

\begin{figure}[htbp!]
\begin{center}
\includegraphics[scale=1]{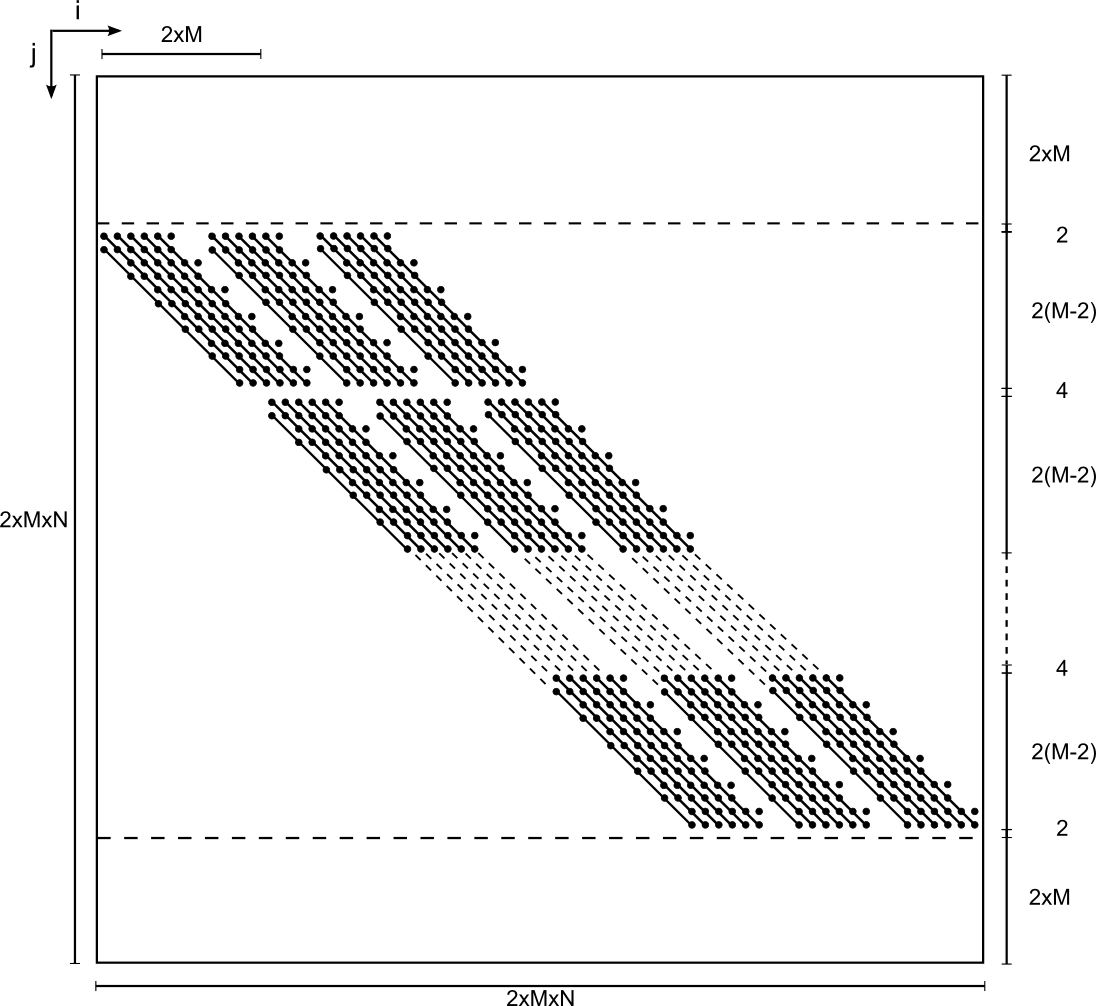}
\caption{Structure des matrices $\mathbf{A}^p$ et $\mathbf{A}^s$ si l'on construit $\mathit{F}_{\omega,k}$ et $\mathit{V}_{\omega,k}$ en alternant composantes horizontales et composantes verticales. Chaque ligne comporte 0 ou 18 coefficients non nuls.}
\label{FigMatrice}
\end{center}
\end{figure}

$\mathbf{A}^p$ et $\mathbf{A}^s$ sont des matrices bandes. Si l'on construit les vecteurs $\mathit{F}_{\omega,k}$ et $\mathit{V}_{\omega,k}$ en alternant composantes horizontales et composantes verticales, la strucure des matrices $\mathbf{A}^p$ et $\mathbf{A}^s$ est telle que représentée sur la Figure \ref{FigMatrice}. On remarque que chaque ligne de ces matrices comporte 0 ou 18 coefficients non nuls. En effet, à l'intérieur du domaine $D$ et dans la zone PML, chaque composante du vecteur $\mathit{F}_{\omega,k}$ en un point du milieu s'écrit en fonction des deux composantes de la vitesse au même point et à ses huit voisins. Il s'agit en fait d'un voisinage du deuxième ordre puisque l'expression d'une composante de $\mathit{F}_{\omega,k}$ en un point passe par l'utilisation des opérateurs $\mathbf{G^x}$, $\mathbf{G^y}$, $\mathbf{H^x}$ et $\mathbf{H^y}$ qui induisent le passage d'une grille à l'autre par une combinaison linéaire en quatre points voisins (cf équations \ref{Eq_ExpressionGx} à \ref{Eq_ExpressionHy}). Sur les bords, la relation entre $\mathit{F}_{\omega,k}$ et $\mathit{V}_{\omega,k}$ est donnée par la matrice $\mathbf{A}_\omega$.

\bigskip

{\bf Remarque :} Les coefficients des matrices $\mathbf{A}^p$ et $\mathbf{A}^s$ associés à la zone PML dépendent des $\alpha_{x}(\omega)$ et $\alpha_{y}(\omega)$. Une partie de ces deux matrices est donc dépendante de la fréquence considérée $\omega$.

\chapter{Introduction sur les méthodes de résolution du problème inverse} \label{Chap_IntroPbInv}

Nous abordons maintenant le problème inverse. L'objectif est de retrouver la distribution des grandeurs caractéristiques d'un milieu en utilisant les données mesurées par des capteurs et la fonction source. Pour réduire la sous détermination du problème, nous utilisons les données issues de différentes positions de tir. Nous nous limiterons à la recherche de deux caractéristiques ($\mathit{v}_{p}$ et $\mathit{v}_{s}$) et nous considèrerons que les données mesurées correspondent à la vitesse verticale $V_y$ au niveau des capteurs situés en surface.

Nous proposons plusieurs méthodes d'inversion qui ont toutes en commun le fait de minimiser un critère de façon itérative. On peut regrouper ces différentes méthodes en deux familles :
\begin{itemize}
\item la première comprend les méthodes utilisant une formulation faisant intervenir des variables auxiliaires. L'avantage de ces formulations est qu'elles permettent de se ramener à la résolution de plusieurs sous-problèmes plus simples (minimisation d'un critère quadratique par exemple). Cependant, ce type de formulation nécessite l'optimisation d'un plus grand nombre de variables (variables d'intérêt et variables auxiliaires).
\item la seconde est associée aux méthodes n'utilisant pas de variable auxiliaire. Ces méthodes ont l'avantage de n'agir que sur les variables d'intérêt et donc de travailler dans un espace de dimension plus réduite. Cependant, la formulation utilisée est non linéaire ce qui rend le processus d'optimisation plus complexe.
\end{itemize}

Pour la plupart des méthodes proposées, la recherche des caractéristiques physiques du milieu consistera à retrouver leurs variations par rapport à un milieu de référence. Le milieu de référence est choisi par l'utilisateur ; il s'agit, par exemple, d'un milieu composé uniquement du matériau qui entoure l'objet diffractant (dans notre étude, les caractéristiques du milieu de référence sont celles de la terre).

Les formulations établies ici découlent de l'équation matricielle associée au problème direct et de la propriété de décomposition de la matrice d'impédance sous forme de somme (voir Partie \ref{PartStructMatrImpedance} page \pageref{PartStructMatrImpedance}).

\section{Construction d'une première formulation bilinéaire} \label{part_FormulationBilineaire1}

\subsection{Expression des vitesses en fonction des caractéristiques du milieu recherché}

Plaçons nous tout d'abord dans le milieu de référence qui est choisi par l'utilisateur. Pour désigner ses caractéristiques, nous utiliserons les notations $\mathit{v}_{p,0}$ et $\mathit{v}_{s,0}$ . Pour une fréquence et une position de la source données, le champ de vitesse qui se propage dans ce milieu de référence est dit "incident" ; on le note $\mathit{V}_{\omega,k}^0$. On a la relation :
\begin{equation}
\begin{aligned}
\mathit{F}_{\omega,k}  &  =  (\mathbf{A}_{\omega,p,s})_0.\mathit{V}_{\omega,k}^0 \\
              &  =  [\mathbf{A}_\omega + \mathbf{A}^p_0 + \mathbf{A}^s_0].\mathit{V}_{\omega,k}^0
\end{aligned}
\label{EqMilieuRef}
\end{equation}
où la matrice d'impédance $(\mathbf{A}_{\omega,p,s})_0$ est liée aux caractéristiques du milieu de référence $\mathit{v}_{p,0}$ et $\mathit{v}_{s,0}$ (cette matrice est donc connue) ; elle peut être décomposée en la somme de trois termes comme expliqué dans la Partie \ref{PartStructMatrImpedance}.

Considérons ensuite le milieu recherché, c'est-à-dire le milieu en présence de l'objet diffractant. Il est caractérisé par les champs $\mathit{v}_{p}$ et $\mathit{v}_{s}$. Dans ce milieu et pour une fréquence et une position de la source données, le champ de vitesse est dit "total" ; on le note $\mathit{V}_{\omega,k}$.
\begin{equation}
\begin{aligned}
\mathit{F}_{\omega,k}  &  =  \mathbf{A}_{\omega,p,s}.\mathit{V}_{\omega,k} \\
              &  = [\mathbf{A}_\omega + \mathbf{A}^p + \mathbf{A}^s].\mathit{V}_{\omega,k}
\end{aligned}
\label{EqMilieuDif}
\end{equation}

En soustrayant \eqref{EqMilieuRef} à \eqref{EqMilieuDif}, on obtient :
\begin{equation}
\mathit{V}_{\omega,k} = \mathit{V}_{\omega,k}^0 - (\mathbf{A}_{\omega,p,s})_0^{-1}([\mathbf{A}^p - \mathbf{A}^p_0].\mathit{V}_{\omega,k} + [\mathbf{A}^s - \mathbf{A}^s_0].\mathit{V}_{\omega,k})
\label{EqVtot1}
\end{equation}

Nous avons donc établi une relation entre le champ des vitesses $\mathit{V}_{\omega,k}$ et les caractéristiques du milieu recherché qui interviennent dans les matrices $\mathbf{A}^p$ et $\mathbf{A}^s$.

\subsection{Introduction des contrastes et des sources de contraste}\label{partDef_contrastes}

Dans cette première formulation, nous nous intéressons aux variations des caractéristiques du milieu recherché par rapport au milieu de référence. Nous chercherons plus particulièrement à retrouver la distribution spatiale de deux contrastes :
\begin{itemize}
\item $(\mathit{\chi}_p)_{i,j} = (\mathit{v}_p^2 - \mathit{v}_{p,0}^2)_{i,j}$ est le contraste en vitesse des ondes P ;
\item $(\mathit{\chi}_s)_{i,j} = (\mathit{v}_s^2 - \mathit{v}_{s,0}^2)_{i,j}$ est le contraste en vitesse des ondes S.
\end{itemize}
Ils caractérisent le milieu en présence de l'objet diffractant au même titre que les champs $\mathit{v}_p$ et $\mathit{v}_s$ puisque le milieu de référence est connu. $\mathit{\chi}_p$ et $\mathit{\chi}_s$ sont des vecteurs de longueur $(M-1)(N-1)$ (voir Partie \ref{Part_StencilSaenger}).

Les deux contrastes recherchés interviennent dans l'équation \ref{EqVtot1} via les matrices $\mathbf{A}^p - \mathbf{A}^p_0$ et $\mathbf{A}^s - \mathbf{A}^s_0$. Cela nous amène à définir deux matrices de contraste :
\begin{itemize}
\item $\mathbf{X}_p = \mathbf{A}^p - \mathbf{A}^p_0$ est la matrice de contraste en vitesse des ondes P ; chacun de ses éléments est une combinaison linéaire des $(\mathit{\chi}_p)_{i,j}$ ;
\item $\mathbf{X}_s = \mathbf{A}^s - \mathbf{A}^s_0$ est la matrice de contraste en vitesse des ondes S ; chacun de ses éléments est une combinaison linéaire des $(\mathit{\chi}_s)_{i,j}$.
\end{itemize}
$\mathbf{X}_p$ et $\mathbf{X}_s$ sont des matrices de taille $2MN \times 2MN$. En reprenant les expressions données dans la partie \ref{PartStructMatrImpedance}, on obtient les expressions des matrices de contraste en fonction des contrastes :
\begin{equation}
\begin{aligned}
\mathbf{X^p} & = \begin{bmatrix}
                   \Diag\{ \alpha_1^x \} \mathbf{H^x} \\
                   \Diag\{ \alpha_1^y \} \mathbf{H^y} \\
                 \end{bmatrix}
                 \Diag\{ \chi_p \}
                 \begin{bmatrix}
                   \Diag\{ \alpha_2^x \} \mathbf{G^x} & \Diag\{ \alpha_2^y \} \mathbf{G^y} \\
                 \end{bmatrix}
\end{aligned}
\end{equation}

\begin{equation}
\begin{aligned}
\mathbf{X^s} & = \begin{bmatrix}
                   \Diag\{ \alpha_1^y \} \mathbf{H^y}\\
                   \Diag\{ \alpha_1^x \} \mathbf{H^x}
                 \end{bmatrix}
                 \Diag\{ \chi_s \}
                 \begin{bmatrix}
                   \Diag\{ \alpha_2^y \} \mathbf{G^y} & \Diag\{ \alpha_2^x \} \mathbf{G^x}
                 \end{bmatrix}\\
             & + \begin{bmatrix}
                   \Diag\{ \alpha_1^x \} \mathbf{H^x} \\
                   - \Diag\{ \alpha_1^y \} \mathbf{H^y}
                 \end{bmatrix}
                 \Diag\{ \chi_s \}
                 \begin{bmatrix}
                   \Diag\{ \alpha_2^x \} \mathbf{G^x} & - \Diag\{ \alpha_2^y \} \mathbf{G^y}
                 \end{bmatrix} \\
             & + \begin{bmatrix}
                   - \Diag\{ \alpha_1^x \} \mathbf{H^x} \\
                   - \Diag\{ \alpha_1^y \} \mathbf{H^y}
                 \end{bmatrix}
                 \Diag\{ \chi_s \}
                 \begin{bmatrix}
                   \Diag\{ \alpha_2^x \} \mathbf{G^x} & \Diag\{ \alpha_2^y \} \mathbf{G^y}
                 \end{bmatrix}
\end{aligned}
\end{equation}

On introduit finalement les sources de contraste : $\mathit{W}_{\omega,k} = (\mathbf{X}_p + \mathbf{X}_s) \mathit{V}_{\omega,k}$. Il s'agit des variables auxiliaires introduites pour construire une première formulation bilinéaire. Ce sont des vecteurs de taille $2MN$.

L'équation (\ref{EqVtot1}) peut alors s'écrire de la façon suivante :
\begin{equation}
\mathit{V}_{\omega,k} = \mathit{V}_{\omega,k}^0 - (\mathbf{A}_{\omega,p,s})_0^{-1} \mathit{W}_{\omega,k}
\label{EqVtot2}
\end{equation}

\subsection{Construction des équations de données}\label{partDef_eqDonnees}

Les équations de données correspondent à l'expression des données mesurées au niveau des capteurs en fonction des variables auxiliaires, pour chaque fréquence et chaque position de la source. Pour établir ces équations, on utilise l'expression de $\mathit{V}_{\omega,k}$ précédente (Eq. \ref{EqVtot2}) en ne retenant que les composantes mesurées par les capteurs :
\begin{equation} (\mathit{V}_{\omega,k})_{\capt} = (\mathit{V}_{\omega,k}^0)_{\capt} - \mathbf{E}_1 (\mathbf{A}_{\omega,p,s})_0^{-1} \mathit{W}_{\omega,k} \end{equation}
où :
\begin{itemize}
\item $\mathbf{E}_1$ est une matrice d'échantillonnage de taille $N_c \times 2MN$. Elle correspond à l'ensemble des $N_c$ lignes de la matrice identité associées aux composantes mesurées par les capteurs.
\item $(\mathit{V}_{\omega,k})_{\capt} = \mathbf{E}_1 \mathit{V}_{\omega,k}$ (composantes du champ de vitesse total mesurées par les capteurs)
\item $(\mathit{V}_{\omega,k}^0)_{\capt} = \mathbf{E}_1 \mathit{V}_{\omega,k}^0$ (composantes du champ de vitesse incident mesurées par les capteurs)
\end{itemize}

\subsection{Construction des équations de couplage}

Les équations de couplage correspondent à l'expression des variables auxiliaires (les sources de contraste $\mathit{W}_{\omega,k}$) en fonction des variables d'intérêt (les contrastes $\chi_p$ et $\chi_s$). Pour établir ces équations, on reprend l'équation \eqref{EqVtot2} et on multiplie les termes de droite et de gauche par la matrice $\mathbf{X}_p + \mathbf{X}_s$. On détermine ainsi une équation de couplage pour chaque fréquence et chaque position de la source :
\begin{equation} \mathit{W}_{\omega,k} = (\mathbf{X}_p + \mathbf{X}_s) (\mathit{V}_{\omega,k}^0 - (\mathbf{A}_{\omega,p,s})_0^{-1} \mathit{W}_{\omega,k}) \end{equation}

\subsection{Réduction du problème à une zone d'étude}\label{partReductionZE}

Une réduction de la taille du problème est possible si l'on s'intéresse uniquement à un sous-domaine du milieu $D$ (domaine situé à l'intérieur des PML, voir Figure \ref{FigMilieu} page \pageref{FigMilieu}). On définit une zone appelée "zone d'étude" à l'intérieur de laquelle on autorise les paramètres caractéristiques ($\mathit{v}_p$ et $\mathit{v}_s$) à prendre des valeurs différentes de celles du milieu de référence. Autrement dit, la zone d'étude est la région à l'intérieur de laquelle les contrastes $\mathit{\chi}_p$ et $\mathit{\chi}_s$ sont susceptibles de prendre des valeurs non nulles ; à l'extérieur de cette zone, les contrastes restent nuls.

Il en résulte que plusieurs lignes et colonnes des matrices $\mathbf{X}_p$ et $\mathbf{X}_s$ sont nulles. On peut donc se ramener à deux matrices de taille réduite de taille $N_0 \times N_0$, où $N_0$ désigne le nombre de lignes et de colonnes non nulles de $\mathbf{X}_p$ et $\mathbf{X}_s$ :
\begin{equation} (\mathbf{X}_p)_{\red} = \mathbf{E}_2 \mathbf{X}_p \mathbf{E}_2^t \quad \text{(et donc } \mathbf{X}_p = \mathbf{E}_2^t (\mathbf{X}_p)_{\red} \mathbf{E}_2 \text{)} \end{equation}
\begin{equation} (\mathbf{X}_s)_{\red} = \mathbf{E}_2 \mathbf{X}_s \mathbf{E}_2^t \quad \text{(et donc } \mathbf{X}_s = \mathbf{E}_2^t (\mathbf{X}_s)_{\red} \mathbf{E}_2 \text{)} \end{equation}
où la matrice d'échantillonnage $\mathbf{E}_2$ est utilisée pour éliminer les lignes et les colonnes nulles des matrices de contraste. Elle est de taille $N_0 \times 2MN$ et correspond à l'ensemble des $N_0$ lignes de la matrice identité associées aux lignes non nulles des matrices $\mathbf{X}_p$ et $\mathbf{X}_s$.

Étant donné que $\mathit{W}_{\omega,k} = (\mathbf{X}_p + \mathbf{X}_s) \mathit{V}_{\omega,k}$, certains coefficients de $\mathit{W}_{\omega,k}$ sont nuls. On peut donc éliminer les coefficients nuls et se ramener à des vecteurs de source de contraste de taille réduite :
\begin{equation} (\mathit{W}_{\omega,k})_{\red} = \mathbf{E}_2 \mathit{W}_{\omega,k} \quad \text{(et donc } \mathit{W}_{\omega,k} = \mathbf{E}_2^t (\mathit{W}_{\omega,k})_{\red} \text{)} \end{equation}

On peut alors réécrire les équations de données et de couplage en faisant intervenir les matrices de contraste et les vecteurs de source de contraste de taille réduite :
\begin{eqnarray}
(\mathit{V}_{\omega,k})_{\capt} & = & (\mathit{V}_{\omega,k}^0)_{\capt} - \mathbf{E}_1 (\mathbf{A}_{\omega,p,s})_0^{-1} \mathbf{E}_2^t (\mathit{W}_{\omega,k})_{\red} \\
(\mathit{W}_{\omega,k})_{\red} & = & \left( (\mathbf{X}_p)_{\red} + (\mathbf{X}_s)_{\red} \right) \left( \mathbf{E}_2 \mathit{V}_{\omega,k}^0 - \mathbf{E}_2 (\mathbf{A}_{\omega,p,s})_0^{-1} \mathbf{E}_2^t (\mathit{W}_{\omega,k})_{\red} \right)
\end{eqnarray}

La réduction de la taille du problème a également un impact sur l'expression des matrices de contraste en fonction des contrastes ($(\chi_p)_\ZE$ et $(\chi_s)_\ZE$ désignent les composantes des vecteurs de contraste appartenant à la zone d'étude) :
\begin{equation} (\mathbf{X}_p)_{\red}
 = \begin{bmatrix} \mathbf{H^x} \\ \mathbf{H^y} \end{bmatrix} \Diag\{ (\chi_p)_\ZE \}	\begin{bmatrix} \mathbf{G^x} & \mathbf{G^y} \end{bmatrix} \end{equation}
\begin{equation} (\mathbf{X}_s)_{\red}
 = \begin{bmatrix} \mathbf{H^y} \\ \mathbf{H^x} \end{bmatrix} \Diag\{ (\chi_s)_\ZE \}	\begin{bmatrix} \mathbf{G^y} & \mathbf{G^x} \end{bmatrix}
 + \begin{bmatrix} \mathbf{H^x} \\ - \mathbf{H^y} \end{bmatrix} \Diag\{ (\chi_s)_\ZE \}	\begin{bmatrix} \mathbf{G^x} & - \mathbf{G^y} \end{bmatrix}
 + \begin{bmatrix} - \mathbf{H^x} \\ - \mathbf{H^y} \end{bmatrix} \Diag\{ (\chi_s)_\ZE \}	\begin{bmatrix} \mathbf{G^x} & \mathbf{G^y} \end{bmatrix} \end{equation}
où les matrices $\mathbf{G^x}$, $\mathbf{G^y}$, $\mathbf{H^x}$ et $\mathbf{H^y}$ sont de taille réduite.

{\bf Remarque :} Les expressions ci-dessus ne font plus apparaître les coefficients $\alpha_1^x$, $\alpha_1^y$, $\alpha_2^x$ et $\alpha_2^y$ car la zone d'étude se situe à l'intérieur du milieu D (pas d'intersection avec la zone PML) ; ces coefficients sont donc égaux à 1.

\subsection{Simplification des écritures} \label{Part_SimplifEcr}

Afin de simplifier les écritures, et en tenant compte des remarques précédentes concernant la réduction des matrices et des vecteurs, nous écrivons désormais les équations de couplage et de données de la façon suivante :
\begin{eqnarray}
(\mathit{V}_{\omega,k})_{\capt} & = & (\mathit{V}_{\omega,k}^0)_{\capt} - \mathbf{B}^d_\omega \mathit{W}_{\omega,k}   \label{EqDonnees} \\
\mathit{W}_{\omega,k} & = & (\mathbf{X}_p + \mathbf{X}_s) (\mathit{V}_{\omega,k}^0 - \mathbf{B}^c_\omega \mathit{W}_{\omega,k})   \label{EqCouplage}
\end{eqnarray}
où $\mathbf{B}^c_\omega = \mathbf{E}_2 (\mathbf{A}_{\omega,p,s})_0^{-1} \mathbf{E}_2^t$ et $\mathbf{B}^d_\omega = \mathbf{E}_1 (\mathbf{A}_{\omega,p,s})_0^{-1} \mathbf{E}_2^t$.

Nous avons donc construit une formulation bilinéaire où les équations de données donnent l'expression des données synthétiques ($(\mathit{V}_{\omega,k})_{\capt}$) en fonction d'un jeu de variables auxiliaires ($\mathit{W}_{\omega,k}$) et où les équations de couplage font le lien entre les variables auxiliaires et les variables d'intérêt ($\mathit{\chi}_p$ et $\mathit{\chi}_s$).

Nous simplifions également l'expression des matrices de contraste en fonction des contrastes :
\begin{eqnarray}
\mathbf{X}_p & = & \mathbf{H^p} \Diag \{ \chi_p \} \mathbf{G^p} \\
\mathbf{X}_s & = & \sum_{i=1}^3 \mathbf{H_i^s} \Diag \{ \chi_s \} \mathbf{G_i^s}
\end{eqnarray}

\section{Construction de la formulation primale} \label{part_FormulationPrimale}

La seconde famille de méthodes abordée correspond aux méthodes utilisant la formulation dite "primale", c'est-à-dire l'expression des données synthétiques en fonction des variables d'intérêt sans faire intervenir les variables auxiliaires. Cette relation peut se déduire facilement de l'expression bilinéaire proposée précédemment.

Isolons tout d'abord le terme $\mathit{W}_{\omega,k}$ dans l'équation de couplage (Eq. \ref{EqCouplage}), pour une fréquence $\omega$ et une position $k$ de la source données :
\begin{equation} \mathit{W}_{\omega,k} = [I + (\mathbf{X}_p + \mathbf{X}_s) \mathbf{B}^c_\omega]^{-1} (\mathbf{X}_p + \mathbf{X}_s) \mathit{V}_{\omega,k}^0 \end{equation}

Reportons cette expression dans l'équation de données (Eq. \ref{EqDonnees}) :
\begin{equation} \label{Eq_FormulationPrimale}
(\mathit{V}_{\omega,k})_{\capt} =  (\mathit{V}_{\omega,k}^0)_{\capt} - \mathbf{B}^d_\omega [I + (\mathbf{X}_p + \mathbf{X}_s) \mathbf{B}^c_\omega]^{-1} (\mathbf{X}_p + \mathbf{X}_s) \mathit{V}_{\omega,k}^0
\end{equation}

Nous obtenons ainsi l'expression des composantes de la vitesse mesurées par les capteurs $(\mathit{V}_{\omega,k})_{\capt}$ en fonction des contrastes recherchés ($\mathit{\chi}_p$ et $\mathit{\chi}_s$).

\section{Procédures de minimisation utilisées} \label{Part_IntroAlgoMinimisation}

Pour les différentes méthodes d'inversion envisagées, l'objectif est de minimiser un critère $\mathcal{C}$ que l'on exprime en fonction des contrastes $\chi_p$ et $\chi_s$ ainsi que plusieurs variables auxiliaires pour certaines des méthodes abordées. Nous utilisons pour cela une méthode locale de type gradient qui consiste en une succession de minimisations en une dimension : à chaque itération, on définit une direction de recherche dans l'espace de représentation puis on détermine un pas de progression efficace le long de cette direction de recherche.

\subsection{La définition de la direction de recherche}

\begin{description}
\item[Dans le cas où le critère à minimiser est quadratique,] (cas rencontré pour certaines méthodes utilisant des variables auxiliaires) nous utilisons le gradient conjugué linéaire. Pour cette méthode, on définit un ensemble de directions conjuguées par combinaison linéaire entre le gradient au point courant et la direction de recherche choisie à l'itération précédente. Cela guarantit la convergence du critère en au plus $n$ itérations, $n$ désignant la taille du problème traité.
\item[Dans le cas où le critère à minimiser n'est pas quadratique,] nous utilisons une des méthodes suivantes :
\begin{itemize}
\item l'algorithme du gradient conjugué non linéaire : cette méthode découle de la méthode du gradient conjugué linéaire et consiste donc à définir une direction de recherche par combinaison linéaire entre le gradient au point courant et la direction de recherche choisie à l'itération précédente.
\item l'algorithme L-BFGS : cette méthode peut être vue comme une généralisation de la méthode du gradient conjugué non linéaire. Pour un entier $m$ choisi par l'utilisateur, elle consiste à effectuer une combinaison linéaire entre le gradient au point courant et les directions de recherche choisies lors des $m$ itérations précédentes.
\end{itemize}
\end{description}
Dans les deux cas, nous écartons d'emblée la méthode de plus forte pente qui s'avère généralement moins efficace.

\subsection{Le choix d'un pas de progression}\label{part_PasProgression}

Pour les différents algorithmes envisagés, il est nécessaire de définir un pas de progression $\alpha$ à chaque itération après la définition de la direction de recherche $d$. Pour cela, on considère la fonction $\Phi$ dont les variations sont celles du critère le long de la direction considérée : $\Phi(\alpha) = \mathcal{C}(\chi + \alpha d)$. A priori, le pas retenu doit correspondre à un minimum local : on doit rechercher $\alpha$ tel que $\Phi^{\prime} (\alpha) = 0$.

Pour l'algorithme du gradient conjugué linéaire (cas de la minimisation d'un critère quadratique), le minimiseur du critère est unique et s'obtient de manière analytique. Lorsque le critère n'est pas quadratique, le minimiseur s'obtient de manière itérative. Dans ce cas, il est préférable de retenir un pas vérifiant les conditions dites de Wolfe \cite{Nocedal99}. L'intérêt est double : d'une part, cela permet de s'approcher d'un minimum de $\Phi$ pour un nombre d'évaluations du critère et du gradient raisonnable. D'autre part, on s'assure d'obtenir un pas de progression suffisamment proche d'un minimum pour que les méthodes de type gradient proposées donnent une direction de descente et que l'algorithme converge globalement. Cette procédure de minimisation est résumée dans l'algorithme \ref{AlgoMinimisationGeneral}.

\begin{algorithm}
\begin{algorithmic}
\STATE Initialisation
\REPEAT
   \STATE Calcul du critère et du gradient en $\chi_k$
   \STATE Définition d'une direction de recherche $\mathbf{d}_k$
   \REPEAT
      \STATE Choix d'un pas de progression $\alpha$
      \STATE Calcul du critère et du gradient en $\chi + \alpha \mathbf{d}$
   \UNTIL Vérification des conditions de Wolfe
   \STATE $\chi_{k+1} \leftarrow \chi_k + \alpha_k \mathbf{d}_k$
\UNTIL Convergence
\end{algorithmic}
\caption{Algorithme itératif - Procédure générale de minimisation}
\label{AlgoMinimisationGeneral}
\end{algorithm}

On distingue deux conditions de Wolfe :
\begin{description}
\item[La première condition de Wolfe ou condition d'Armijo :]
\begin{equation} \mathcal{C}(\chi + \alpha \mathbf{d}) \leq \mathcal{C}(\chi) + c_1 \alpha \nabla \mathcal{C}(\chi)^T \mathbf{d} \end{equation}
\item[La seconde condition de Wolfe :]
\begin{equation} \nabla \mathcal{C}(\chi + \alpha \mathbf{d})^T \mathbf{d} \geq c_2 \alpha \nabla \mathcal{C}(\chi)^T \mathbf{d} \end{equation}
\item[ou la seconde condition de Wolfe forte :]
\begin{equation} \vert \nabla \mathcal{C}(\chi + \alpha \mathbf{d})^T \mathbf{d} \vert \leq \vert c_2 \alpha \nabla \mathcal{C}(\chi)^T \mathbf{d} \vert \end{equation}
\end{description}
Les coefficients $c_1$ et $c_2$ intervenant dans les inégalités précédentes doivent être choisis tels que : $0 < c_1 < c_2 < 1$.

Pour déterminer un pas de progression, nous utiliserons l'algorithme proposé par Moré et Thuente \cite{More94} qui a l'avantage de déterminer un pas satisfaisant les conditions fortes de Wolfe pour un nombre limité de calculs du critère et du gradient.

L'ensemble des méthodes présentées ici ont l'avantage d'être bien adaptées aux problèmes de grande taille puisque l'on définit une nouvelle direction de descente pour un coût de calcul faible (on la calcule par simple combinaison linéaire des gradients aux itérations précédentes et au point courant, on ne passe pas par le calcul du Hessien du critère). C'est la raison pour laquelle certaines méthodes de type Newton-Kantorovitch ne seront pas utilisées ici : elles nécessitent l'inversion d'une matrice (inversion du Hessien du critère ou d'une forme approchée) et seraient alors très coûteuses en calcul. De plus, un petit nombre de vecteurs suffit à définir la nouvelle direction de recherche à chaque itération, ce qui réduit l'espace mémoire nécessaire pour stocker les variables utilisées.

\section{Utilisation d'autres variables} \label{Part_IntroChgtVariable}

Les équations établies précédemment font intervenir les contrastes $\chi_p$ et $\chi_s$. Cependant, parmi les différentes méthodes que nous aborderons, certaines montrerons un problème de sensibilité du critère vis-à-vis des variations de $\chi_p$ et $\chi_s$ autour de valeurs élevées, ce qui a tendance à ralentir la convergence des algorithmes. Nous serons alors amenés à utiliser d'autres variables notées $\sigma_p$ et $\sigma_s$ afin d'améliorer la sensibilité du critère. Elles sont choisies de sorte que de faibles variations de $\sigma_p$ et $\sigma_s$ induisent de fortes variations de $\chi_p$ et $\chi_s$ pour des valeurs de contraste élevées.

\begin{figure}[!ht]
\begin{center}
\includegraphics[scale=0.6]{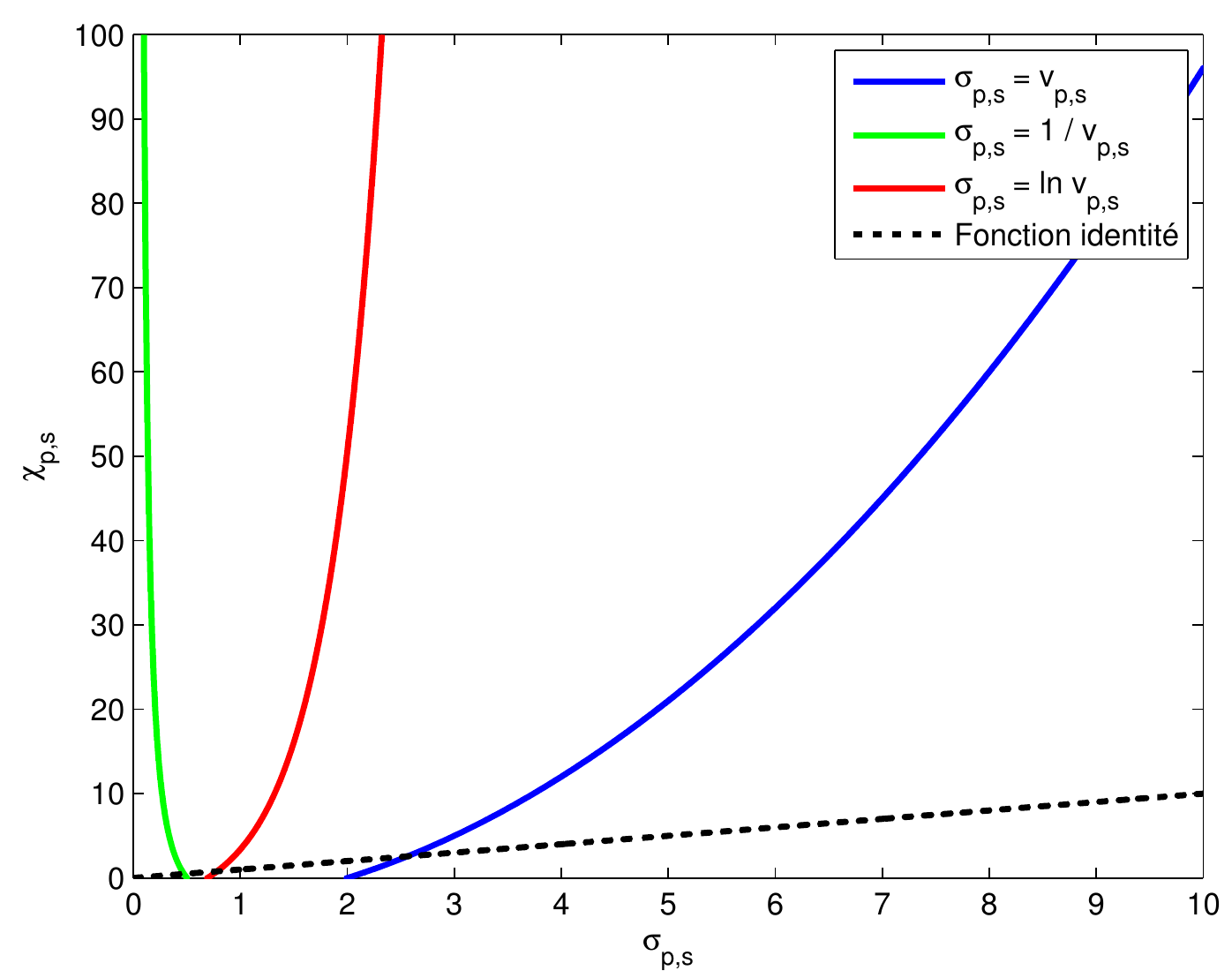}
\caption{Allure des fonctions associées aux différents changements de variable (dans cet exemple, $\mathit{v}_{p,0} = \mathit{v}_{s,0} = 2$)}
\label{Fig_TracesChangementsVariables}
\end{center}
\end{figure}

Nous proposons plusieurs changements de variable. On les liste dans le tableau suivant en donnant leurs relations par rapport à $\chi_p$ et $\chi_s$ ainsi que les valeurs caractéristiques de la terre et du béton qui leurs sont associées. On donne également l'allure des fonctions associées aux différents changements de variables proposés sur la Figure \ref{Fig_TracesChangementsVariables}.

\begin{center}
\begin{tabular}{|c|c|m{3cm}|m{3cm}|} \hline
\multirow{2}{*}{\bf Variables proposées}
   & {\bf Expressions de $\chi_p$ et $\chi_s$} & \multicolumn{2}{c|}{\bf Valeurs caractéristiques} \\ \cline{3-4}
   & {\bf en fonction de $\sigma_p$ et $\sigma_s$} & {\bf Terre} & {\bf Béton} \\ \hline
$\sigma_p = \mathit{v}_p$
   & $\chi_p = \sigma_p^2 - \mathit{v}_{p,0}^2$
   & $\sigma_{p,\text{Terre}} = 300$ & $\sigma_{p,\text{Béton}} = 4000$ \\
$\sigma_s = \mathit{v}_s$
   & $\chi_s = \sigma_s^2 - \mathit{v}_{s,0}^2$
   & $\sigma_{s,\text{Terre}} = 150$ & $\sigma_{s,\text{Béton}} = 2200$ \\ \hline
$\sigma_p = 1/\mathit{v}_p$
   & $\chi_p = (1/\sigma_p)^2 - \mathit{v}_{p,0}^2$
   & $\sigma_{p,\text{Terre}} = 3,3.10^{-3}$ & $\sigma_{p,\text{Béton}} = 2,5.10^{-4}$ \\
$\sigma_s = 1/\mathit{v}_s$
   & $\chi_s = (1/\sigma_s)^2 - \mathit{v}_{s,0}^2$
   & $\sigma_{s,\text{Terre}} = 6,7.10^{-3}$ & $\sigma_{s,\text{Béton}} = 4,5.10^{-4}$ \\ \hline
$\sigma_p = \ln{\mathit{v}_p}$
   & $\chi_p = \exp{2 \sigma_p} - \mathit{v}_{p,0}^2$
   & $\sigma_{p,\text{Terre}} = 5,7$ & $\sigma_{p,\text{Béton}} = 8,3$ \\
$\sigma_s = \ln{\mathit{v}_s}$
   & $\chi_s = \exp{2 \sigma_s} - \mathit{v}_{s,0}^2$
   & $\sigma_{s,\text{Terre}} = 5,0$ & $\sigma_{s,\text{Béton}} = 7,7$ \\ \hline
\end{tabular}
\end{center}

\section{Données utilisées pour tester les différentes méthodes} \label{Part_MilieuxTests}

Afin de comparer les performances des méthodes abordées, nous avons effectué des tests sur un jeu de données synthétiques générées à l'aide de l'algorithme de résolution du problème direct. Nous avons travaillé sur un milieu de taille réduite puis sur un milieu de taille intermédiaire. Nous ne travaillerons pas sur des milieux de taille réelle (dimensions plus grandes et résolution plus fine) car la place mémoire requise serait trop importante.

Nous présentons ci-dessous les deux milieux utilisés. Sur chaque figure, on représente :
\begin{itemize}
\item en bleu clair la partie du milieu D qui reste invariante (le contraste reste nul, les caractéristiques restent celles du milieu de référence) ;
\item en rouge la zone d'étude (zone dans laquelle les caractéristiques évoluent), l'élément en béton est représenté en rouge plus foncé ;
\item en bleu foncé la zone PML ;
\item en jaune la position des capteurs (on mesure la composante verticale de la vitesse) ;
\item en vert les différentes positions de la source.
\end{itemize}

Pour les différentes méthodes d'inversion proposées, seules les caractéristiques des pixels appartenant à la zone d'étude évoluent. Par conséquent, nous présenterons les résultats obtenus en n'affichant que le contenu de la zone d'étude.

\subsection{Milieu de petite taille} \label{Part_DonneesPetites}

\begin{figure}[htbp!]
\begin{center}
\includegraphics[scale=0.5]{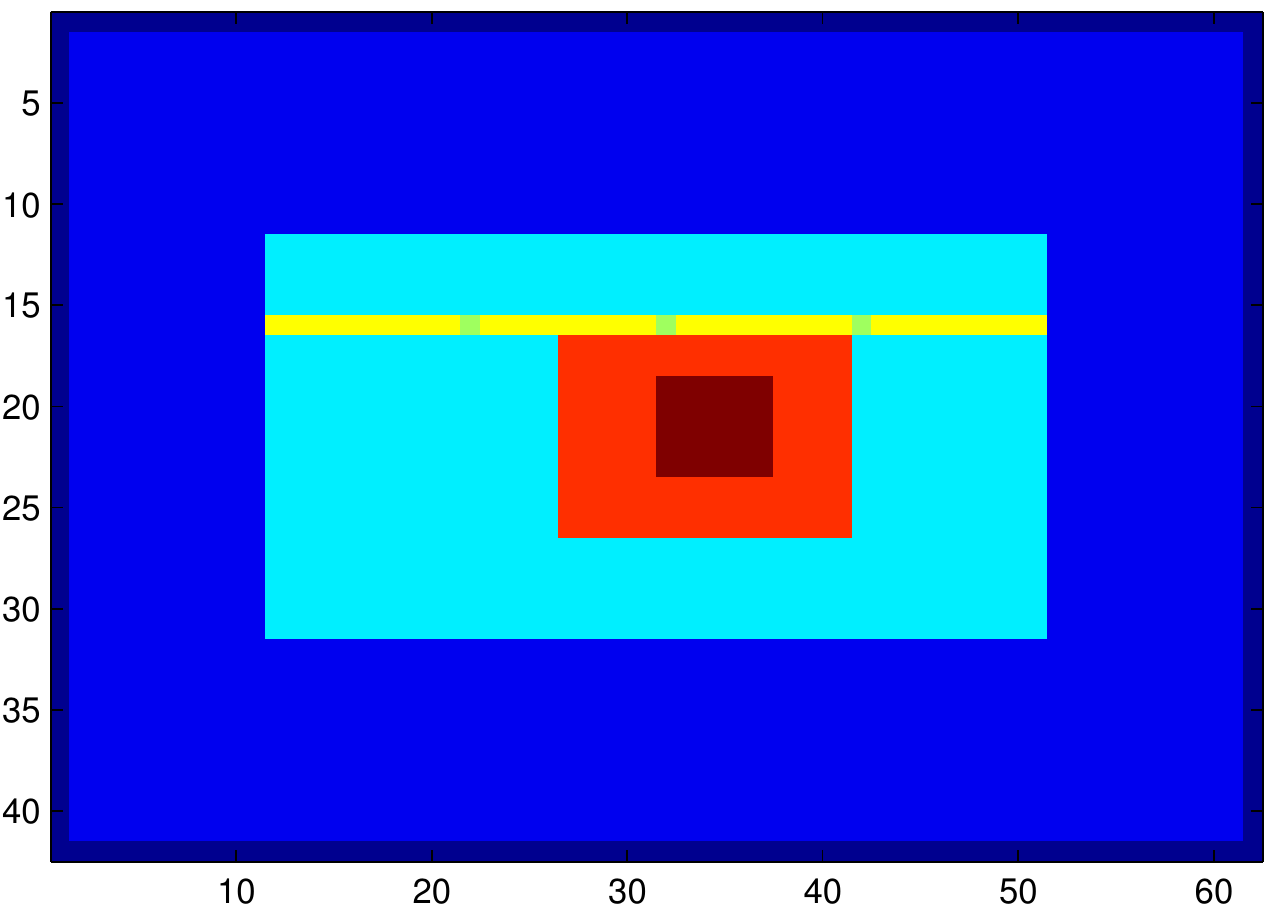}
\caption{Milieu utilisé pour les premiers tests avec les capteurs (jaune), les positions successives de la source (vert), la zone PML (bleu foncé), la zone d'étude contenant le bloc de béton (rouge)}
\label{Fig_PetitMilieuTest}
\end{center}
\end{figure}

Le milieu de petite taille entouré de la zone PML est représenté sur la Figure \ref{Fig_PetitMilieuTest}.

On donne ci-dessous ses caractéristiques :
\begin{itemize}
\item la taille du milieu est de 1 m de profondeur et 2 m de largeur ;
\item le signal source est un ricker centré à 200 Hz ;
\item on retient 15 fréquences réparties de façon uniforme entre 46,7 Hz et 700 Hz ;
\item la source est positionnée à 0,2 mètre en profondeur et est placée successivement à 0,5 m, 1 m et 1,5 m sur l'axe horizontal ;
\item les capteurs sont positionnés à 0,2 mètre en profondeur et sont espacés de 5 cm sur toute la longueur du milieu ;
\item le milieu recherché comprend un bloc de béton (20 cm de hauteur, 25 cm de largeur) entouré de terre (pas d'air) ;
\item la résolution en x et en y est de 0,05 m ;
\item il n'y a pas d'atténuation des ondes ;
\item la taille de la zone PML est de 50 cm.
\end{itemize}

\subsection{Milieu de taille intermédiaire}

\begin{figure}[htbp!]
\begin{center}
\includegraphics[scale=0.5]{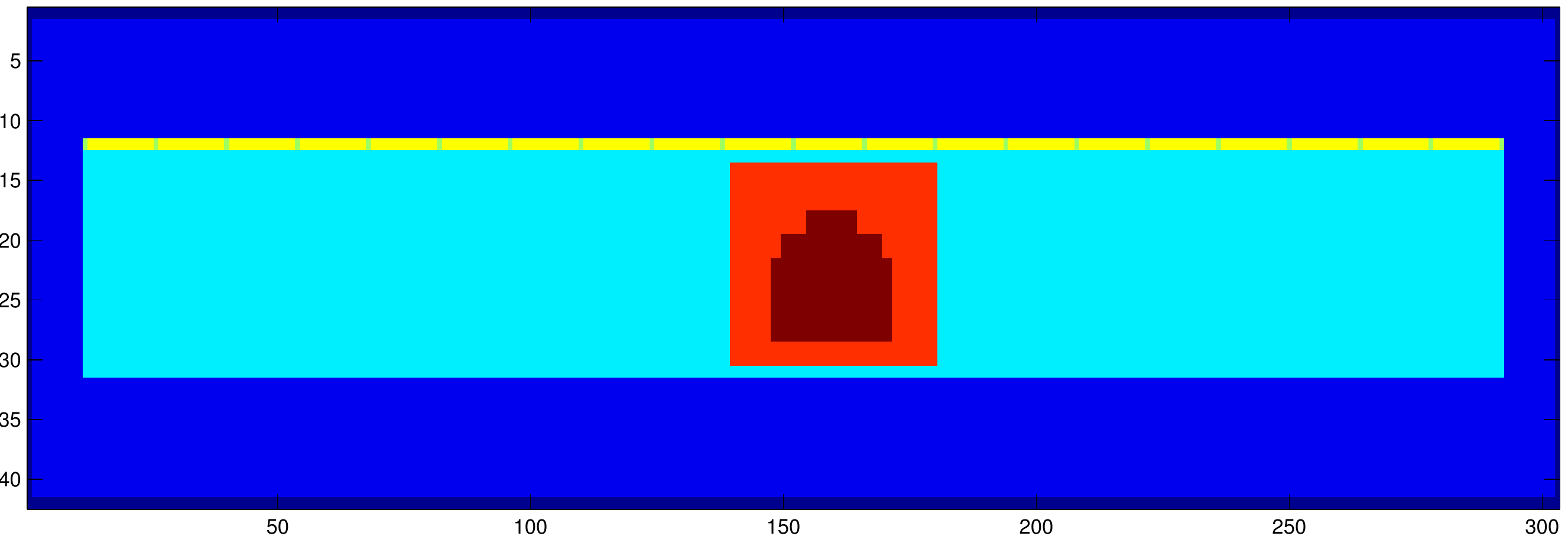}
\caption{Milieu de plus grande taille avec les capteurs (jaune), les positions successives de la source (vert), la zone PML (bleu foncé), la zone d'étude contenant le bloc de béton (rouge)}
\label{Fig_GrandMilieuTest}
\end{center}
\end{figure}

On représente le milieu de taille intermédiaire entouré de la zone PML sur la Figure \ref{Fig_GrandMilieuTest}.

On donne ci-dessous ses caractéristiques :
\begin{itemize}
\item la taille du milieu est de 1 m de profondeur et 14 m de largeur ;
\item le signal source est un Ricker centré à 200 Hz ;
\item on retient 15 fréquences réparties de façon uniforme entre 46,7 Hz et 700 Hz ;
\item la source est positionnée à 0 mètre en profondeur et est placée de 0 à 14 m par pas de 0,7 m sur l'axe horizontal ;
\item les capteurs sont positionnés à 0 mètre en profondeur et sont espacés de 5 cm sur toute la longueur du milieu ;
\item le milieu recherché comprend est une superposition de dalles de béton (hauteur totale : 0,5 m, largeur totale : 1,15 m) entourées de terre (pas d'air) ;
\item la résolution en x et en y est de 0,05 m ;
\item il n'y a pas d'atténuation des ondes ;
\item la taille de la zone PML est de 50 cm.
\end{itemize}


\chapter{Méthodes d'inversion fondées sur une formulation bilinéaire}

Toutes les méthodes basées sur une formulation bilinéaire
suivent une démarche commune pour déterminer la
solution du problème d'inversion. Nous allons détailler, dans une
première partie, les spécificités de ce type d'inversion, puis nous
présenterons chaque méthode ainsi que les résultats obtenus.

\section{Idée générale de l'inversion basée sur la formulation bilinéaire}\label{partIdeeGenerale}

Le modèle direct permet, à partir des équations de propagation et
d'une fonction source connue, de calculer les composantes du champ de
vitesse en tout point d'un milieu souterrain dont on connaît les
caractéristiques physiques. Dans le problème d'inversion, les
caractéristiques physiques du milieu sont inconnues, de même que les
composantes du champ de vitesse à l'intérieur du domaine. Seules sont
connues les composantes verticales du champ de vitesse que l'on mesure
en surface du milieu considéré. Dans le cas des formulations
bilinéaires, on estime non seulement les variables d'intérêt liées aux
caractéristiques du milieu, mais aussi celles liées au champ de
vitesse, en tentant de compenser l'augmentation du nombre de
paramètres à estimer par une plus grande simplicité de la formulation
algébrique du problème. Ceci se traduit par des procédures
d'estimation plus faciles à mettre en {\oe}uvre et plus efficaces
algorithmiquement. De plus, nous avons le choix entre travailler avec
des variables représentant les grandeurs physiques, ou utiliser le
contraste de ces grandeurs par rapport à un milieu de référence. Ce
choix permet de définir un certain nombre de variantes de la méthode
d'inversion bilinéaire et a une influence directe sur les
caractéristiques numériques des méthodes d'estimation. Dans ce qui
suit, nous donnons une formulation générale du problème, et nous
tentons de mettre en évidence les éléments qui ont l'impact le plus
significatif sur les caractéristiques de la méthode d'inversion.

\subsection{Variables à estimer}

Comme indiqué
  précédemment, la formulation bilinéaire met en jeu deux ensembles
de variables $\mathit{x}$ et $\mathit{z}$ que l'on définit comme suit
:
\begin{description}
 \item[Les variables d'intérêt] $\mathit{x}_p$ et $\mathit{x}_s$ qui
   ne dépendent que de la position et qui paramétrisent les caractéristiques
   du milieu que sont les vitesses des ondes P et S
 \item[Les variables auxiliaires] $\mathit{z}_{\omega,k}$ qui dépendent de la
   fréquence et de la position de la source et qui paramétrisent le champ de vitesse $\mathit{V}_{\omega,k}$.
\end{description}
Comme nous l'avons dit précédemment, les variables $x_p$, $x_s$ et $z_{\omega,k}$ peuvent être choisies de plusieurs
  manières, ce qui influe sur les caractéristiques des méthodes
  d'estimation correspondantes. De plus, il est possible d'introduire
  des changements de variables scalaires sur les quantités $x_p$, $x_s$ et
  $z_{\omega,k}$. L'effet de ces choix est décrit au
  paragraphe~\protect\ref{sec:choix-variables}.

\subsection{Système d'équations}
Quelle que soit la nature
  exacte des $x_p$, $x_s$ et $z_{\omega,k}$, la relation entre le milieu inconnu et les
  mesures $y_{\omega, k}$ est modélisée à l'aide d'un système de deux
équations qui prennent la forme générale suivante :
\begin{description}
\item[Une équation d'observation]
  qui relie le vecteur des mesures $\mathit{y}_{\omega,k}$ aux variables auxiliaires à travers la fonction $\mathcal{R}(\mathit{z}_{\omega,k})$ :
\begin{align*}
  \mathit{y}_{\omega,k} = \mathbf{E}_1 \mathcal{R}(\mathit{z}_{\omega,k}) + b_{\omega, k}
\end{align*}
où $\mathbf{E}_{1}$ est la matrice d'échantillonnage aux capteurs définie dans la section~\ref{partDef_eqDonnees}, $b_{\omega, k}$ désigne un
  bruit qui représente les erreurs de mesure et de modélisation, et $\mathcal{R}(\mathit{z}_{\omega,k})$ est une fonction des variables auxiliaires dont l'expression sera détaillée ultérieurement.
\item[Une équation de couplage] qui relie les variables auxiliaires
  aux variables d'intérêt et qui peut être considérée comme une
  contrainte sur le champ total de vitesse. Cette
  équation prend la forme générique suivante :
\begin{align*}
  \mathcal{H}(\mathit{z}_{\omega,k})=\mathbf{K}(\mathit{x}_p,
  \mathit{x}_s) \mathcal{R}(\mathit{z}_{\omega,k})
\end{align*}
\end{description}
où $\mathcal{H}(\mathit{z}_{\omega,k})$ est une fonction de $\mathit{z}_{\omega,k}$ dont la forme sera précisée ultérieurement.

  La principale propriété du système d'équations ci-dessus est son
  caractère \emph{bilinéaire}, ce qui signifie que chacune des
  équations est linéaire (ou plus précisément affine) par rapport aux variables $x_p$, $x_s$,
  \emph{et} par rapport à $z$. En particulier, de par la linéarité de l'équation d'observation
  par rapport à $z$, on a nécessairement :
\begin{align}
  \mathcal{R}(\mathit{z}_{\omega,k}) = \mathbf{M}_d\mathit{z}_{\omega,k} +
  \mathit{u}_d \label{eq_Def_R}
\end{align}

  où $\mathbf{M}_d$ et $\mathit{u}_d$ désignent respectivement une
  matrice et un vecteur constants.

De même, l'équation de couplage est
  linéaire par rapport à $z$, ce qui implique :
\begin{align}
\mathcal{H}(\mathit{z}_{\omega,k}) =
\mathbf{M}_c\mathit{z}_{\omega,k} + \mathit{u}_c \label{eq_Def_H}
\end{align}
avec $\mathbf{M}_c$ une matrice constante et $\mathit{u}_c$ un
  vecteur constant. Par ailleurs, la linéarité de l'équation de
  couplage par rapport à $x$ permet d'affirmer que les éléments de la
  matrice \Kv sont des combinaisons linéaires des $\mathit{x}_p$ et
  $\mathit{x}_s$. De plus, on a vu dans la section~\ref{PartStructMatrImpedance} que la matrice d'impédance peut se
décomposer en une somme de trois matrices indépendantes. De la même
façon, la matrice \Kv peut se décomposer en trois matrices
$\mathbf{K}^p$, $\mathbf{K}^s$ et $\mathbf{K}^\omega$ (cette dernière
pouvant être nulle suivant la méthode employée). $\mathbf{K}^p$ et
$\mathbf{K}^s$ sont respectivement des combinaisons linéaires des
$\mathit{x}_p$ pour la première et des $\mathit{x}_s$ pour la
deuxième. On a alors les
deux égalités suivantes :
\begin{equation}
\begin{aligned}
\mathbf{K}^p \mathcal{R}(z_{\omega,k}) & = \mathbf{\Delta}_{\omega,k}^p \mathit{x}_p \\
\mathbf{K}^s \mathcal{R}(z_{\omega,k}) & = \mathbf{\Delta}_{\omega,k}^s \mathit{x}_s
\end{aligned}
\label{eq_EgaliteBilineaire}
\end{equation}
où $\mathbf{\Delta}_{\omega,k}^p$ et $\mathbf{\Delta}_{\omega,k}^s$
sont des matrices qui dépendent linéairement de $z$. Ces égalités
permettent d'exprimer explicitement la linéarité des équations par rapport à chacun des jeux de variables.

En utilisant la décomposition des matrices $\mathbf{K}^p$ et
$\mathbf{K}^s$ donnée dans la section~\ref{PartStructMatrImpedance},
on établit les expressions explicites des
matrices $\mathbf{\Delta}_{\omega,k}^p$ et
$\mathbf{\Delta}_{\omega,k}^s$ :
\begin{equation}
\begin{aligned}
\mathbf{\Delta}_{\omega,k}^p	& =
\begin{bmatrix}
\Diag\{ \alpha_1^x \} \mathbf{H^x}			\\
\Diag\{ \alpha_1^y \} \mathbf{H^y}			\\
\end{bmatrix}
\Diag\{
\begin{bmatrix}
\Diag\{ \alpha_2^x \} \mathbf{G^x}			&			\Diag\{ \alpha_2^y \} \mathbf{G^y}			\\
\end{bmatrix} \mathcal{R}(z_{\omega,k})\}	\\
\mathbf{\Delta}_{\omega,k}^s	&=
\begin{bmatrix}	\Diag\{ \alpha_1^y \} \mathbf{H^y}			\\
						\Diag\{ \alpha_1^x \} \mathbf{H^x}
\end{bmatrix}
\Diag\{
\begin{bmatrix}	\Diag\{ \alpha_2^y \} \mathbf{G^y}			&			\Diag\{ \alpha_2^x \} \mathbf{G^x}
\end{bmatrix}\mathcal{R}(z_{\omega,k}) \}\\
 & + \begin{bmatrix}
\Diag\{ \alpha_1^x \} \mathbf{H^x}			\\
\Diag\{ \alpha_1^y \} \mathbf{H^y}
\end{bmatrix}
\Diag\{
\begin{bmatrix}	\Diag\{ \alpha_2^x \} \mathbf{G^x}			&			- \Diag\{ \alpha_2^y \} \mathbf{G^y}
\end{bmatrix}\mathcal{R}(z_{\omega,k}) \}\\
 & + \begin{bmatrix}
-\Diag\{ \alpha_1^x \} \mathbf{H^x}		\\
- \Diag\{ \alpha_1^y \} \mathbf{H^y}
\end{bmatrix}
\Diag\{
\begin{bmatrix}	\Diag\{ \alpha_2^x \} \mathbf{G^x}			&			\Diag\{ \alpha_2^y \} \mathbf{G^y}
\end{bmatrix}\mathcal{R}(z_{\omega,k}) \}
\end{aligned}
\label{eq_Delta}
\end{equation}

Dans les équations ci-dessus, les matrices \Gv et \Hv dépendent
  linéairement de $z$. Il est remarquable de constater que la
  structure des matrices $\mathbf{\Delta}_{\omega,k}^p$ et
$\mathbf{\Delta}_{\omega,k}^s$ est indépendante du choix précis des
  variables $x_p$, $x_s$ et $z$.

\subsection{Critère}\label{part_CritereGeneral}
\subsubsection{Forme générale}
L'objectif de l'inversion est de déterminer les valeurs des différentes variables qui sont solution des équations de couplage et d'observation. Pour cela, nous allons construire un critère à partir d'une pénalisation quadratique de l'erreur sur chacune de ces deux équations. C'est en cherchant les valeurs des variables qui minimisent ce critère, et donc la somme des erreurs sur chacune des équations, que nous trouverons la solution au problème d'inversion.

Minimiser l'erreur sur l'équation d'observation permet de s'assurer
que la solution sera compatible avec les données mesurées par les
capteurs, tandis que l'erreur sur l'équation de couplage doit être vue
comme une contrainte que les variables doivent satisfaire pour que le
modèle physique décrivant le problème soit respecté en tout point du
domaine. Cette contrainte est relâchée, \cad qu'on tolère que
l'équation de couplage ne soit pas strictement vérifiée. Pour
contrôler l'importance du respect de la contrainte et régir le
compromis entre l'erreur qui touche l'équation de couplage et celle
qui affecte l'équation d'observation, nous modifions le critère en
introduisant une pondération de l'erreur sur l'équation de couplage
par un hyperparamètre $\gamma_c$.


Pour faire face au caractère mal posé du problème, il est possible d'introduire des connaissances \aprio sur la nature et le comportement spatial des différentes variables à inverser. Ceci peut être fait par le biais d'une fonction de régularisation qui est composée de plusieurs termes, chacun traduisant mathématiquement une information connue sur une des variables. Chaque terme est pondéré par un hyperparamètre qui rend compte de la confiance en l'information qu'il apporte. Le critère que nous utilisons dans le cadre des méthodes fondées sur une approche bilinéaire est donc constitué de 3 termes :
\begin{align}
\mathcal{C}(\mathit{x}_p,\mathit{x}_s,\mathit{z}_{\omega,k}) = \mathcal{C}_d(\mathit{x}_p,\mathit{x}_s,\mathit{z}_{\omega,k}) + \gamma_c(\mathit{x}_p,\mathit{x}_s,\mathit{z}_{\omega,k}) \mathcal{C}_c(\mathit{x}_p,\mathit{x}_s,\mathit{z}_{\omega,k}) + \mathcal{\phi}(\mathit{x}_p,\mathit{x}_s,\mathit{z}_{\omega,k})
\label{eqCritereXZ}
\end{align}
où :
\begin{itemize}
\item $\mathcal{C}_d$ correspond à l'erreur de mesure ; son expression se déduit de l'équation d'observation
\begin{align*}
 \mathcal{C}_d(\mathit{x}_p,\mathit{x}_s,\mathit{z}_{\omega,k}) = \| \mathit{y}_{\omega,k} - \mathbf{E}_1 \mathcal{R}(\mathit{z}_{\omega,k}) \|^2
\end{align*}
\item $\mathcal{C}_c$ correspond à l'erreur sur l'équation de couplage ; ce terme peut s'écrire de trois façons en faisant apparaitre explicitement chaque ensemble de variables grâce aux jeux d'équations~\eqref{eq_Def_R}, \eqref{eq_Def_H} et \eqref{eq_EgaliteBilineaire} :
\begin{align*}
 \mathcal{C}_c(\mathit{x}_p,\mathit{x}_s,\mathit{z}_{\omega,k}) &= \| \mathcal{H}(\mathit{z}_{\omega,k}) - \mathbf{K}(\mathit{x}_p, \mathit{x}_s)\mathcal{R}(\mathit{z}_{\omega,k}) \|^2\\
&= \| \mathcal{H}(\mathit{z}_{\omega,k}) - \mathbf{\Delta}_{\omega,k}^p \mathit{x}_p - \mathbf{\Delta}_{\omega,k}^s \mathit{x}_s - \mathbf{K}^\omega\mathcal{R}(\mathit{z}_{\omega,k}) \|^2\\
&= \| (\mathbf{M}_c-\mathbf{K}(\mathit{x}_p, \mathit{x}_s)\mathbf{M}_d)\mathit{z}_{\omega,k} + \mathit{u}_c - \mathbf{K}(\mathit{x}_p, \mathit{x}_s)\mathit{u}_d \|^2
\end{align*}
\item $\gamma_c$ est l'hyperparamètre qui quantifie le relâchement de la contrainte. Il peut dépendre des variables ou être constant.
\item $\mathcal{\phi}$ est la fonction de régularisation qui permet d'introduire de l'information \aprio sur les variables $\mathit{x}_p$, $\mathit{x}_s$ et $\mathit{z}_{\omega,k}$.
\end{itemize}

\subsubsection{Régularisation employée}\label{part_CritereGeneral_Reg}

La fonction de régularisation que nous avons employée pour toutes les formulations bilinéaires présentées est composée de trois termes : un rappel aux valeurs caractéristiques de la terre ($\mathit{x}_p^T$ et $\mathit{x}_s^T$) sur les variables $\mathit{x}_p$ et $\mathit{x}_s$, un autre de rappel à zéro sur les variables $\mathit{z}_{\omega,k}$ et un terme de pénalisation quadratique des différences premières de $\mathit{x}_p$ et $\mathit{x}_s$.

Le terme de rappel à la valeur de la terre  de la variable d'intérêt (qui devient un rappel à zéro dans le cas où l'on utilise des contrastes) correspond à la traduction d'une connaissance \aprio sur la distribution spatiale de cette variable.
Nous faisons l'hypothèse que la taille de l'objet diffractant enfoui est petite par rapport à celle de la zone d'étude, et donc que la plus grande partie des pixels correspondent à de la terre. Dans le cas de l'utilisation d'un contraste sur la variable d'intérêt, cela revient à dire que l'on considère que la plus grande partie de la zone d'étude est identique au milieu de référence et donc de contraste nul.

Le terme de rappel à zéro sur les variables auxiliaires correspond à
une connaissance \aprio peu précise sur le champ de
vitesse. Il pénalise les valeurs élevées du champ de vitesse, et donc permet d'éviter l'apparition de vitesses arbitrairement grandes ; il
permet en outre d'améliorer le conditionnement de la méthode d'inversion.

Le terme de pénalisation quadratique des différences premières a pour effet de pénaliser les grandes variations spatiales des variables d'intérêt et favorise ainsi l'apparition de zones homogènes. Toutefois, ce terme présente certaines limitations dans notre cas, puisque les valeurs des caractéristiques physiques recherchées peuvent varier fortement d'un objet à l'autre, et que pénaliser quadratiquement la variation spatiale de ces variables ne favorise pas la reconstruction d'une image aux transitions franches entre objets, introduisant ainsi une incertitude sur le positionnement des frontières. Cependant, les changements de variables proposés dans la section~\ref{Part_IntroChgtVariable} permettent de réduire l'écart entre les valeurs des vitesses des ondes P et S des différents objets, et donc minimisent l'importance de cette affirmation.
Pour mieux prendre en compte cet aspect du problème, nous avons envisagé de remplacer ce terme par une pénalisation de type $L_2L_1$, mais le temps nous a manqué pour mettre en \oe{uvre} et tester cette modification.

\subsection{Gradients}
Toutes les méthodes d'optimisation envisagées pour résoudre le problème inverse sont décrites dans la section~\ref{Part_IntroAlgoMinimisation} et utilisent le gradient du critère. Dans le cas des méthodes fondées sur l'approche bilinéaire, la minimisation du critère~\eqref{eqCritereXZ} se fait suivant une direction de descente qui dépend du gradient par rapport aux variables d'intérêt et aux variables auxiliaires. Cela nécessite donc d'établir l'expression du gradient du critère pour chaque jeu de variables.

Tous calculs faits, le gradient relativement à $\mathit{z}_{\omega,k}$ est donné par l'expression suivante :
\begin{align}
\mathcal{G}_{\mathit{z}_{\omega,k}}(\omega,k) = \nabla_{\mathit{z}_{\omega,k}}\mathcal{C}(\mathit{z}_{\omega,k}) &= 2((\mathbf{M}_d^\dagger\mathbf{E}_1^\dagger\mathbf{E}_1 \mathbf{M}_d+\gamma_c(\mathbf{M}_c - \mathbf{K}(\mathit{x}_p, \mathit{x}_s)\mathbf{M}_d)^\dagger(\mathbf{M}_c - \mathbf{K}(\mathit{x}_p, \mathit{x}_s)\mathbf{M}_d))\mathit{z}_{\omega,k} \label{eq_GradientGeneralz}\\
&-(\mathbf{M}_d^\dagger\mathbf{E}_1^\dagger(\mathit{y}_{\omega,k} + \mathbf{E}_1 \mathit{u}_d) - \gamma_c(\mathbf{M}_c - \mathbf{K}(\mathit{x}_p, \mathit{x}_s)\mathbf{M}_d)^\dagger(\mathit{u}_c-\mathbf{K}(\mathit{x}_p, \mathit{x}_s)\mathit{u}_d))) + \nabla_{\mathit{z}_{\omega,k}}\phi(\mathit{z}_{\omega,k})\notag
\end{align}

Les gradients par rapport à $\mathit{x}_p$ et $\mathit{x}_s$ sont calculés en utilisant l'expression $\mathcal{C}_c$ faisant explicitement intervenir $\mathit{x}_p$ et $\mathit{x}_s$. Ces deux gradients sont donnés par les équations suivantes :
\begin{align}
\mathcal{G}_{\mathit{x}_p} = \nabla_{\mathit{x}_p}(\gamma_c\mathcal{C}_c(\mathit{x}_p)+\phi(\mathit{x}_p)) &= 2\gamma_c(\sum_k \sum_\omega\mathbf{\Delta}_{\omega,k}^{p\dagger}\mathbf{\Delta}_{\omega,k}^p\mathit{x}_p -  \mathbf{\Delta}_{\omega,k}^{p\dagger}(\mathcal{H}(\mathit{z}_{\omega,k})- \mathbf{\Delta}_{\omega,k}^s \mathit{x}_s)) + \nabla_{\mathit{x}_p}\phi(\mathit{x}_p)
\label{eq_GradientGeneralxp} \\
\mathcal{G}_{\mathit{x}_s} = \nabla_{\mathit{x}_s}(\gamma_c\mathcal{C}_c(\mathit{x}_s)+\phi(\mathit{x}_s)) &= 2\gamma_c(\sum_k \sum_\omega\mathbf{\Delta}_{\omega,k}^{s\dagger}\mathbf{\Delta}_{\omega,k}^{s}\mathit{x}_s -  \mathbf{\Delta}_{\omega,k}^{s\dagger}(\mathcal{H}(\mathit{z}_{\omega,k})-\mathbf{\Delta}_{\omega,k}^p \mathit{x}_p))  + \nabla_{\mathit{x}_s}\phi(\mathit{x}_s)
\label{eq_GradientGeneralxs}
\end{align}

L'expression des gradients de $\phi$ par rapport à chacune des variables est détaillée dans la section~\ref{part_RegularisationDetail}.

\subsection{Choix des variables à inverser}
\label{sec:choix-variables}

Les formulations bilinéaires du problème ne diffèrent les unes des
autres que par les variables sur lesquelles on travaille. On a le
choix de travailler directement sur les variables physiques, à savoir
le carré des vitesses des ondes P et S ou les composantes du champ de
vitesse, ou de travailler sur les contrastes et les sources de
contrastes définis dans la section~\ref{partDef_contrastes}. Le choix
des variables à inverser implique des différences techniques et
numériques qu'il est important de souligner pour comprendre pleinement
les avantages et inconvénients de chaque méthode.

\bigskip
\textbf{Implications techniques}

Dans le cas où les $\mathit{x}_p$ et $\mathit{x}_s$ correspondent aux contrastes des variables d'intérêt par rapport à un milieu de référence, il est alors possible de restreindre l'estimation de cette variable à une zone d'étude comme nous l'avons signalé dans la section~\ref{partReductionZE}. En utilisant une connaissance \aprio sur la position et la taille de l'objet enfoui, nous pouvons considérer que les caractéristiques physiques du sous-sol ne diffèrent de l'arrière-plan de référence que dans une petite partie du milieu $D$. Les contrastes ne prennent donc des valeurs non nulles que dans une zone restreinte de l'image, et ils ne doivent plus être calculés que dans la zone d'étude ce qui diminue d'autant le nombre d'inconnues à déterminer. La réduction de la zone de travail permet aussi de s'affranchir de la dépendance fréquentielle des matrices d'impédance puisqu'on fait l'hypothèse que les contrastes sont nuls dans la zone des PML. Il ne faut donc plus stocker en mémoire et mettre à jour qu'une seule matrice de contraste $\mathbf{X}_p$ et $\mathbf{X}_s$ pour chaque caractéristique physique du sol, ce qui représente un gain intéressant en place mémoire et en temps de calcul.

Par contre, si l'on choisit d'inverser le carré des vitesses des ondes P et S, il est alors nécessaire de faire l'estimation de ces variables en tout point du milieu $D$, incluant les PML. Cela nous oblige à mettre à jour et conserver en mémoire une matrice d'impédance par fréquence.

À titre d'exemple, dans le cas du milieu de petite taille présenté à la section~\ref{Part_DonneesPetites}, on passe de $2604$ inconnues à $150$ si on n'estime les contrastes que dans la zone d'étude suggérée. La matrice d'impédance est creuse et ne possède que $18$ lignes non nulles, comme nous pouvons le voir à la figure~(\ref{FigMatrice}). Nous travaillons en double précision donc une matrice d'impédance occupe environ $6$ Mb. L'utilisation des contrastes permet de ne stocker que deux matrices de $6$ Mb contre une matrice $6$ Mb par fréquence dans le cas où l'on utilise le carré des vitesses des ondes P et S (les résultats présentés dans ce document ont été obtenus en utilisant $15$ fréquences).

Le choix des variables auxiliaires à inverser est aussi un point
important de la méthode d'inversion. Les variables auxiliaires
$\mathit{z}_{\omega,k}$ introduites dans la formulation bilinéaire
dépendent de la fréquence et de la position de la source. Le nombre
total de composantes de $\mathit{z}_{\omega,k}$ à déterminer est donc
élevé, ce qui va demander un temps de calcul important. L'utilisation
des sources de contrastes, définies dans la
section~\ref{partDef_contrastes}, permet également d'en restreindre
l'estimation à une zone d'étude, d'où un gain important en nombre
total d'inconnues et en volume de calcul. Par contre, si l'on choisit
de travailler directement avec les composantes du champ des vitesses, on est
obligé de les estimer en tout point du domaine $D$. Si on reprend le
cas du domaine de petite taille, on passe de $300 \times N_f \times
N_k$ inconnues à déterminer si on utilise les sources de contraste sur
une zone restreinte, à $5208 \times N_f \times N_k$ dans le cas où on
ne réduit pas la zone d'étude.

Un autre avantage de travailler sur une petite zone, outre les gains en temps de calcul et en place mémoire, est que cela a tendance à réduire la sous-détermination du problème. Effectivement, le nombre d'inconnues diminue tandis que le nombre de mesures ne change pas. Cependant, pour que l'utilisation de la zone d'étude soit justifiée, il faut disposer d'un milieu de référence qui soit assez proche de la réalité, ce qui reste jusqu'à présent un problème ouvert.

\bigskip
\textbf{Implications numériques}

L'utilisation de tel ou tel jeu de variable modifie les équations de couplage et d'observation, et définit donc une nouvelle méthode d'inversion. Le critère diffère suivant la méthode, et l'expression du gradient par rapport à l'une ou l'autre des variables fait intervenir des formes algébriques différentes plus ou moins simples et rapides à calculer. Le choix des variables à estimer a donc un impact direct sur la complexité calculatoire de l'inversion et le comportement numérique des méthodes d'optimisation. En particulier, les estimateurs associés à chacune des méthodes d'inversion ont des conditionnements différents, ce qui influe fortement sur la vitesse de convergence et la qualité du résultat.

\subsection{Détail de la mise en \oe{uvre} des algorithmes d'inversion}\label{partDetails_Algo}
Nous avons vu que le choix des variables à inverser a des conséquences
importantes sur les caractéristiques numériques d'un algorithme
d'inversion basé sur une formulation bilinéaire du problème. Les
différents choix de couples de variables à inverser permettent de créer
autant de variantes de la méthode d'inversion bilinéaire, dont la mise
en \oe{uvre} nécessite un certain nombre de choix techniques tels que
la forme de la fonction de régularisation utilisée, les stratégies
d'optimisation possibles, les changements de variables envisagés, la
prise en compte progressive des fréquences dans l'inversion, la
détermination des hyperparamètres, les données utilisées pour les
tests, les critères de convergence employés et les problèmes de
conditionnement de la matrice d'impédance du problème direct. Nous allons discuter dans cette section des divers aspects numériques et techniques soulevés par ces choix de mise en \oe{uvre}.

\bigskip
\textbf{Détail de la régularisation employée}\label{part_RegularisationDetail}

Le premier choix à faire concerne la forme de la fonction de
régularisation. De celle-ci va dépendre l'expression des estimateurs
de chaque variable, ce qui va influer sur le type de méthode
d'optimisation à employer. Nous avons vu dans la
section~\ref{part_CritereGeneral_Reg} que la fonction de
régularisation était composée de trois termes. Cette fonction, qui
sera utilisée dans toutes les méthodes d'inversion basées sur une
approche bilinéaire abordées dans ce document, est donnée par
l'expression suivante :

\begin{align}
 \mathcal{\phi}(\mathit{x}_p,\mathit{x}_s,\mathit{z}) = \gamma_{r0}^\mathit{x} (\|\mathit{x}_p-(\mathit{x}_p)_T\|^2 + \|\mathit{x}_s-(\mathit{x}_s)_T\|^2) + \gamma_{r0}^\mathit{z}\sum_k \sum_\omega \|\mathit{z}_{\omega,k}\|^2 + \gamma_{r1}^\mathit{x} (\|(\mathbf{D}_1 + \mathbf{D}_2)\mathit{x}_p\|^2 + \|(\mathbf{D}_1 + \mathbf{D}_2)\mathit{x}_s\|^2)
\label{eq_Regularisation}
\end{align}
où la notation $(\cdotp)_T$ fait référence aux caractéristiques de la terre, $\mathbf{D}_1$ et $\mathbf{D}_2$ sont les matrices des différences premières dans les directions verticale et horizontale, et les $\gamma_{r0}^\mathit{x}$, $\gamma_{r0}^\mathit{z}$ et $\gamma_{r1}^\mathit{z}$ sont les hyperparamètres qui pondèrent l'importance des différents \aprio introduits par la régularisation ($\gamma_{r0}^\mathit{x}$ pour le rappel à la terre des variables d'intérêt, $\gamma_{r0}^\mathit{z}$ pour le rappel à zéro sur les variables auxiliaires et $\gamma_{r1}^\mathit{x}$ pour la pénalisation quadratique des différences premières des variables d'intérêt).

La fonction de régularisation~\eqref{eq_Regularisation} est quadratique par rapport à chacune des variables. Le gradient d'une telle fonction par rapport à chacune des variables se calcule simplement. On obtient :
\begin{align}
 \nabla_{\mathit{x}_p}\phi(\mathit{x}_p) &= 2\gamma_{r0}^\mathit{x}(\mathit{x}_p-(\mathit{x}_p)_T) + 2\gamma_{r1}^\mathit{x}(\mathbf{D}_1 + \mathbf{D}_2)^{t}(\mathbf{D}_1 + \mathbf{D}_2)\mathit{x}_p\\
 \nabla_{\mathit{x}_s}\phi(\mathit{x}_s) &= 2\gamma_{r0}^\mathit{x}(\mathit{x}_s-(\mathit{x}_s)_T) + 2\gamma_{r1}^\mathit{x}(\mathbf{D}_1 + \mathbf{D}_2)^{t}(\mathbf{D}_1 + \mathbf{D}_2)\mathit{x}_s\\
 \nabla_{\mathit{z}_{\omega,k}}\phi(\mathit{z}_{\omega,k}) &= 2\gamma_{r0}^\mathit{z}\sum_k \sum_\omega \mathit{z}_{\omega,k}
\label{eq_RegularisationGradient}
\end{align}

\bigskip
\textbf{Stratégie d'optimisation employée}

Pour réaliser une inversion basée sur une formulation bilinéaire, nous pouvons choisir d'optimiser les deux ensembles de variables soit de façon alternée, soit simultanément. De plus, il est possible d'estimer conjointement, ou de façon alternée, les différentes variables de chaque jeu de variables. Plus précisément, l'estimation des deux variables d'intérêt $\mathit{x}_p$ et $\mathit{x}_s$ peut être fait alternativement ou conjointement, et il en va de même pour les variables auxiliaires $\mathit{z}_{\omega,k}$ correspondant à chaque fréquence et chaque position de tir.

Nous avons choisi d'inverser les deux variables d'intérêt conjointement. Les valeurs de $\mathit{x}_p$ et $\mathit{x}_s$ sont du même ordre de grandeur, et leur estimation respective partage un certain nombre de calculs, comme on peut le voir sur les équations~\eqref{eq_GradientGeneralxp} et \eqref{eq_GradientGeneralxs}. Il est donc avantageux de réaliser l'estimation de ces deux variables simultanément. Il suffit pour cela de concaténer les systèmes linéaires de chacun des estimateurs.
Cette stratégie a été appliquée à toutes les méthodes présentées dans ce chapitre.

De même, il existe deux façons de voir l'estimation des variables
auxiliaires $\mathit{z}_{\omega,k}$.  Soit on effectue successivement
la résolution d'autant de systèmes linéaires qu'il y a de positions de
la source et de fréquences, soit on estime toutes les fréquences
conjointement en un seul bloc.  Des tests nous ont montré que les deux
approches donnent des résultats similaires. Nous avons donc choisi
d'utiliser l'algorithme d'inversion alternée qui est détaillé dans le
pseudo-code de l'algorithme~(\ref{Algo_General}).

\bigskip

L'estimation alternée des deux jeux de variables repose sur la construction itérative et alternée de séquences $(\mathit{\hat{x}}_s,\mathit{\hat{x}}_p)^{(n)}$ et $\{(\mathit{\hat{z}}_{\omega,k})^{(n)};\omega\in [1\dots N_f] \textit{ et } k\in [1\dots N_k]\}$. Les variables sont estimées alternativement en utilisant une méthode d'optimisation tronquée.
\begin{algorithm}
\begin{algorithmic}
\REQUIRE Initialisation des valeurs $\mathit{x}^{(0)}_p$, $\mathit{x}^{(0)}_s$ et $\mathit{z}^{(0)}_{\omega,k}$
\STATE $n=1$
\REPEAT
  \FOR[\textit{Boucle sur le nombre de fréquences $N_f$}]{$\omega = \omega_1, \dots, \omega_{N_f}$}
     \FOR[\textit{Boucle sur le nombre de positions de la source $N_k$}]{$k = 1, \dots, N_k$}
        \STATE
        \STATE Déterminer $(\mathit{\hat{z}}_{\omega,k})^{(n)}\in\argmin{\mathit{z}_{\omega,k}}{\mathcal{C}(\mathit{z}_{\omega,k})}$ à l'aide de $(\mathit{\hat{z}}_{\omega,k})^{(n-1)}$
     \ENDFOR
  \ENDFOR
  \STATE
   \STATE Déterminer $\mathit{\hat{x}}^{(n)}_p\in\argmin{\mathit{x}_p}{\mathcal{C}(\mathit{x}_p)}$ à l'aide de $\mathit{\hat{x}}^{(n-1)}_p$
   \STATE Déterminer $\mathit{\hat{x}}^{(n)}_s\in\argmin{\mathit{x}_s}{\mathcal{C}(\mathit{x}_s)}$ à l'aide de $\mathit{\hat{x}}^{(n-1)}_s$
   \STATE
   \STATE Mise à jour des matrices d'impédances
   \STATE $n=n+1$
\UNTIL{Convergence}
\end{algorithmic}
\caption{Algorithme d'inversion alternée}
\label{Algo_General}
\end{algorithm}
chaque inconnue est estimée avec seulement quelques itérations de l'algorithme d'optimisation. Une résolution complète serait coûteuse en temps de calcul et n'est pas forcément utile dans le cadre d'une optimisation par blocs.

Comme nous l'avons vu dans la
section~\ref{Part_IntroAlgoMinimisation}, divers algorithmes
d'optimisation sont disponibles pour résoudre un système
linéaire. Nous avons constaté que les résultats obtenus à l'aide de
LBFGSB et du gradient conjugué sont similaires, mais que le gradient
conjugué est généralement plus rapide. Nous avons donc choisi
d'utiliser ce dernier pour résoudre les systèmes linéaires intervenant
dans l'estimation de chaque variable.

\bigskip

La méthode d'optimisation alternée peut s'avérer inefficace si le
minimum de la fonction de coût se trouve dans une vallée très
étroite. Dans ce cas, la recherche du minimum global en suivant
alternativement la direction donnée par le gradient associé à chaque
jeu de variables se fait selon une trajectoire en zigzag, et un nombre
très important de petits pas est requis pour l'atteindre
\cite{Press92}. Pour remédier à ce problème, il est possible de
réaliser l'optimisation simultanée de toutes les variables, \cad de construire itérativement une suite
d'estimées $(\mathit{\hat{x}}_p, \mathit{\hat{x}}_s,
\{\mathit{\hat{z}}_{\omega,k};\omega\in [1\dots N_f] \textit{ et }
k\in [1\dots N_k]\})^{(n)}$. De cette façon, on n'effectue la descente du gradient que suivant une seule
direction ; on limite ainsi les problèmes de zigzag liés à la
méthode alternée. L'optimisation simultanée des deux ensembles de variables est réalisée en concaténant les différents
systèmes algébriques.

Le pseudo code de la méthode est donné dans l'algorithme~(\ref{Algo_General_conjoint}).

\begin{algorithm}
\begin{algorithmic}
\REQUIRE Initialisation des valeurs $\mathit{x}^0_p$, $\mathit{x}^0_s$ et $\mathit{z}^0$
\STATE $n=1$
\REPEAT
   \STATE Déterminer $(\mathit{\hat{x}}_p, \mathit{\hat{x}}_s, \mathit{\hat{z}}_{\omega,k})^{(n)}\in\argmin{\mathit{x}_p, \mathit{x}_s,\mathit{z}_{\omega,k}}{\mathcal{C}(\mathit{x}_p, \mathit{x}_s, \mathit{z}_{\omega,k})}$ à l'aide de $(\mathit{\hat{x}}_p, \mathit{\hat{x}}_s, \mathit{\hat{z}}_{\omega,k})^{(n-1)}$
   \STATE
   \STATE Mise à jour des matrices d'impédances
   \STATE $n=n+1$
\UNTIL{Convergence}
\end{algorithmic}
\caption{Algorithme d'inversion conjointe}
\label{Algo_General_conjoint}
\end{algorithm}

Lorsque l'on réalise l'estimation simultanée des $\mathit{x}_p$, $\mathit{x}_s$ et $\mathit{z}_{\omega,k}$, le pas de mise à jour de la méthode d'optimisation ne s'obtient plus en résolvant un système linéaire \cite{Barriere08}. Comme indiqué dans la section~\ref{part_PasProgression}, nous utilisons l'algorithme de Moré Thuente pour calculer le pas de descente et la méthode d'optimisation choisie est le LBFGSB qui semble converger plus rapidement que la méthode du gradient conjugué non linéaire. L'inconvénient de cette méthode d'inversion simultanée est qu'il faut s'assurer que toutes les variables inversées simultanément ont la même échelle pour que la convergence ne soit pas trop lente.

\bigskip
\textbf{Influence des changements de variables}

Les changements de variables proposés dans la section~\ref{Part_IntroChgtVariable} ont pour particularité de modifier significativement l'intervalle des valeurs que peuvent prendre $\mathit{x}_p$ et $\mathit{x}_s$, tout en n'entraînant qu'une petite modification du calcul du gradient. On tente ainsi de modifier le conditionnement du problème
sans augmenter significativement le volume de calcul.

Chaque changement de variable de $\mathit{x}$ en $\mathit{\sigma}$
étant de nature scalaire, la modification apportée au calcul du gradient consiste simplement à multiplier ce dernier par une matrice diagonale :

\begin{eqnarray*}
\nabla_\mathit{\sigma}\mathcal{C}(\mathit{\sigma}) =  \Diag(\nabla_\mathit{\sigma}\mathit{x})\nabla_\mathit{x} \mathcal{C}(\mathit{x})
\end{eqnarray*}

\bigskip
\textbf{Introduction progressive des fréquences}\label{part_Multifrequences}

Les méthodes présentées précédemment réalisent l'inversion en tenant compte de toutes les fréquences simultanément, comme dans \cite{Hu07}. Cependant, les travaux de Pratt \etal \cite{Gelis05,Sirgue04,Pratt88} ont montré qu'il pouvait être intéressant d'utiliser une stratégie d'incorporation progressive des fréquences pour effectuer l'inversion. Les basses fréquences ont un comportement plus linéaire \vav des perturbations du modèle que les hautes fréquences \cite{Sirgue04}. Il est donc suggéré par Pratt \cite{Pratt88} d'effectuer l'inversion en partant d'un certain nombre de basses fréquences et d'introduire, au fur et à mesure de l'inversion, de nouvelles fréquences plus élevées pour ajouter l'information qu'elles contiennent. Cette technique a aussi l'avantage d'accélérer les premières itérations en réduisant considérablement le nombre de variables à estimer.

L'algorithme~(\ref{Algo_multi_frequence}) décrit la méthode utilisée pour introduire les fréquences progressivement. Les fréquences sont introduites par paquets de deux, des plus basses vers les plus hautes.

\begin{algorithm}
\begin{algorithmic}
\REQUIRE Initialisation de la méthode d'inversion
\REQUIRE Regroupement des $N_f$ fréquences en $N_G$ groupes $G_i$ de $2$ fréquences, $i \in [1,N_G]$
\STATE $G = \varnothing$
  \FOR[\textit{Boucle sur les groupes de fréquences, des plus basses fréquences vers les plus hautes}]{$i = 1, \dots, N_G$}
		 \STATE $n = 1$
		 \STATE Construction du groupe de fréquences utilisé pour l'inversion $G = G\cup G_{i}$
     \REPEAT
        \STATE Estimation des contrastes $\chi$ avec une itération de la méthode d'inversion et les fréquences du groupe $G$
        \STATE Calcule du déplacement relatif de la solution $var = \frac{norm(\chi_n-\chi_{n-1})}{norm(\chi_{n-1})}$
   			\STATE $n = n + 1$
     \UNTIL{$n>n_{max}$ ou $var<seuil$}
  \ENDFOR
\end{algorithmic}
\caption{Algorithme d'inversion avec introduction progressive des fréquences}
\label{Algo_multi_frequence}
\end{algorithm}

\bigskip
\textbf{Détermination des hyperparamètres}

Les hyperparamètres sont tous fixés empiriquement. Ils pondèrent l'importance de la régularisation et de la contrainte imposée par l'équation de couplage \vav de l'adéquation aux données.

\bigskip
\textbf{Données utilisées}\label{part_DonneesUtilisees}

Tous les résultats présentés dans ce chapitre ont été obtenus en utilisant des données synthétiques. Les mesures sont issues du modèle direct développé par EdF auxquelles on a ajouté un bruit blanc gaussien tel que le rapport signal sur bruit soit de 30 dB. La géométrie et les caractéristiques du milieu utilisées pour les générer sont décrites dans la section~\ref{Part_DonneesPetites}.

Nous nous sommes servi de deux jeux de données pour initialiser les méthodes lors des tests. Le premier jeu est obtenu en utilisant les caractéristiques physiques de la terre, et le deuxième correspond à la solution exacte du problème. L'initialisation à la solution permet d'apprécier la dégradation introduite par la méthode d'inversion et sert de référence.

\bigskip
\textbf{Critère de convergence employé}\label{part_CritereConvergence}

Le critère d'arrêt que nous avons choisi d'employer est un seuil maximal sur la valeur de la norme du gradient qui est fixé empiriquement. Toutefois, nous utilisons également le déplacement relatif de l'estimée pour stopper l'algorithme. Cette métrique mesure la variation de l'estimateur considéré entre les itérations $n$ et $n-1$, et permet de voir si l'estimée que l'on calcule à chaque itération stagne à cause de problèmes numériques ou continue à évoluer. Par exemple, dans le cas du déplacement relatif de la variable $\mathit{x}_p$, il est défini comme :
\begin{align*}
\frac{\|\mathit{x}_p^n-\mathit{x}_p^{n-1}\|}{\|\mathit{x}_p^{n-1}\|}
\end{align*}

\bigskip
\textbf{Problèmes de conditionnement de la matrice d'impédance du problème direct} \label{partPbConditionnementA}

La matrice d'impédance, définie dans la section~\ref{Part_SystLinPbDirect}, est construite en utilisant les équations physiques~\eqref{Eq_MilieuAprDiscr} décrivant le problème de propagation des ondes dans un milieu élastique. Ces équations lient les inconnues du problème entre elles, les valeurs des vitesses $V_p$ et $V_s$ aux composantes du champ de vitesse $V_x$ et $V_y$. Ce lien est algébriquement réalisé par l'introduction d'une matrice d'impédance. À partir des équations de la physique, nous avons formulé une méthode pour générer cette matrice en utilisant une décomposition de l'opérateur de discrétisation en plusieurs matrices assimilables à des filtres (voir partie~\ref{PartStructMatrImpedance}).

La construction de la matrice d'impédance du modèle direct, telle que mise en \oe{uvre} initialement par EdF et décrite dans \cite{Kerzale09}, est basée sur un domaine (composé de la zone d'étude et des PML) ceinturé par une bande de 1 pixel de large de valeur $1$ autour du domaine. Ceci se traduit, pour une image de taille $M\times N$, par l'ajout de $2(M+N-2)$ valeurs égales à $-1$ sur la diagonale de la matrice d'impédance. Cet ajout n'affecte pas le problème direct et donne des résultats similaires à ceux obtenus en utilisant la matrice d'impédance construite tel que nous le présentons dans la section~\ref{PartStructMatrImpedance}. L'erreur quadratique moyenne relative entre les résultats du problème direct générés en utilisant la matrice d'impédance que nous proposons et celle fournie par EdF est de $10^{-14}$ en moyenne sur toutes les fréquences.

Toutefois, nous avons observé des différences significatives au niveau algébrique entre les deux matrices d'impédance. La présence de ces valeurs supplémentaires égales à $-1$ sur la diagonale de la matrice fournie par EdF est responsable d'une importante dégradation du conditionnement de la matrice d'impédance $\mathbf{A}_\omega$.
Une estimation du nombre de conditionnement des matrices calculées par le code fourni par EdF nous indique qu'il est de l'ordre de $10^{12}$ (en fixant les variables d'intérêt à la solution). Cependant, en utilisant les matrices d'impédances construites en utilisant la méthode montrée dans la section~\ref{PartStructMatrImpedance}, le nombre de conditionnement de $\mathbf{A}_{\omega,p,s}$ tombe à $10^{6}$.

Une autre remarque, de moindre importance, peut être faite sur la structure des matrices d'impédance. Des collaborateurs au projet, travaillant chez EdF, nous ont suggéré d'effectuer la permutation des lignes et colonnes de ces matrices pour faire des gains de temps lors de certaines opérations. Notamment, si l'on permute les lignes et colonnes de ces matrices de telle sorte que l'on alterne les composantes $\mathit{V}_x$ et $\mathit{V}_y$ de la vitesse dans le vecteur $\mathit{V}_{\omega,k}$, on obtient des matrices d'impédance dont les valeurs non nulles sont concentrées autour de la diagonale, ce qui permet une accélération significative du temps de calcul d'un préconditionneur de type décomposition de Cholesky tronquée (10 fois plus rapide). Nous avons donc effectué le réarrangement des lignes et colonnes de ces matrices à chaque fois qu'un tel préconditionnement était employé.

Les valeurs du nombre de conditionnement données dans ce chapitre sont toutes estimées à l'aide de la fonction condest de \matlab.

\section{Méthode CSI}
\subsection{Description}
L'obtention des équations utilisées dans la méthode "Contraste Source Inversion" (CSI), \cite{VanDenBerg97}, est détaillée dans la partie~\ref{part_FormulationBilineaire1}. La CSI est la méthode bilinéaire correspondant au choix de variables suivant :

\begin{description}
 \item[Les variables d'intérêt] $\mathit{x}$ sont les contrastes $\mathit{\chi}_p$ et $\mathit{\chi}_s$ du carré des vitesses $\mathit{V}_p$ et $\mathit{V}_s$ qui sont estimés sur une partie réduite du domaine $D$, et ont été définis dans la section~\ref{partDef_contrastes}
 \item[Les variables auxiliaires] $\mathit{z}_{\omega,k}$ sont les sources de contraste $\mathit{W}_{\omega,k}$, définies dans la section~\ref{partDef_contrastes}, qui sont estimées sur la même partie réduite du domaine $D$
\end{description}

Nous rappelons ici les équations d'observation et de couplage, présentées dans la section~\ref{part_FormulationBilineaire1}, qui sont à la base de la méthode CSI :
\begin{description}
\item[Équation d'observation] $\mathit{y}_{\omega,k} = \mathbf{E}_1\mathit{V}_{\omega,k}^0 - \mathbf{B}^d_\omega \mathit{W}_{\omega,k}$
\item[Équation de couplage] $\mathit{W}_{\omega,k} = (\mathbf{X}_p + \mathbf{X}_s) (\mathbf{E}_2\mathit{V}_{\omega,k}^0 - \mathbf{B}^c_\omega \mathit{W}_{\omega,k})$
\end{description}
où les matrices $\mathbf{B}^c_\omega$ et $\mathbf{B}^d_\omega$ sont définies dans la section~\ref{Part_SimplifEcr} et $\mathbf{E}_2$ est la matrice d'échantillonnage sur la zone d'étude définie dans la section~\ref{partReductionZE}.

En reprenant les notations de la section~\ref{partIdeeGenerale}, on établit que :
\begin{align}
 \mathcal{H}(\mathit{W}_{\omega,k}) &= \mathit{W}_{\omega,k} & \text{  avec } \textbf{M}_c=\indentit \text{ et } \textit{u}_c = 0 \notag\\
 \mathcal{R}(\mathit{W}_{\omega,k}) &= \mathit{V}_{\omega,k}^0 - (\mathbf{A}_{\omega,p,s})_0^{-1} \mathbf{E}_2^t \mathit{W}_{\omega,k} & \text{  avec } \textbf{M}_d=-(\mathbf{A}_{\omega,p,s})_0^{-1} \mathbf{E}_2^t \text{ et } \textit{u}_d = \mathit{V}_{\omega,k}^0
\label{eq_DefNotationsCSI}\\
 \mathbf{K}(\mathit{\chi}_p, \mathit{\chi}_s) &= (\mathbf{X}_p + \mathbf{X}_s)\mathbf{E}_2 \notag
\end{align}
où $(\mathbf{A}_{\omega,p,s})_0^{-1}$ est l'inverse de la matrice d'impédance du modèle direct lié aux caractéristiques du domaine de référence qui a été définie dans la section~\ref{part_FormulationBilineaire1}.

À partir de ce système d'équations, et en nous basant sur ce qui a été dit dans la section~\ref{part_CritereGeneral}, nous définissons la fonction de coût de la méthode CSI comme suit :
\begin{align}
\mathcal{C}    = & \sum_k \sum_\omega \| y_{\omega,k} - \mathbf{E}_1\mathit{V}_{\omega,k}^0 + \mathbf{B}^d_\omega \mathit{W}_{\omega,k} \|^2  \notag\\
							&  + \gamma_c\sum_k \sum_\omega \| \mathit{W}_{\omega,k} - (\mathbf{X}_p + \mathbf{X}_s) (\mathbf{E}_2\mathit{V}_{\omega,k}^0 - \mathbf{B}^c_\omega \mathit{W}_{\omega,k}) \|^2 + \phi
\label{eq_CritereCSI}
\end{align}

À partir de l'expression du critère CSI~\eqref{eq_CritereCSI}, et en substituant les variables à inverser spécifiées par la méthode CSI et les notations données par~\eqref{eq_DefNotationsCSI}, nous pouvons réécrire les formules du gradient par rapport aux variables d'intérêt et aux variables auxiliaires données par les équations~\eqref{eq_GradientGeneralz}, \eqref{eq_GradientGeneralxp} et \eqref{eq_GradientGeneralxs}.
L'expression du gradient du critère par rapport à chaque variable est donné ci-dessous :
\begin{itemize}
\item Gradient par rapport à $\mathit{W}_{\omega,k}$ :
\begin{align*}
\mathcal{G}_{\mathit{W}_{\omega,k}} = \nabla_{\mathit{W}_{\omega,k}} \mathcal{C}(\mathit{W}_{\omega,k})= & 2((\mathbf{B}_\omega^{d\dagger} \mathbf{B}^d_\omega + \gamma_c (I + (\mathbf{X}_p + \mathbf{X}_s) \mathbf{B}^c_\omega)^\dagger (I + (\mathbf{X}_p + \mathbf{X}_s) \mathbf{B}^c_\omega) + \gamma_{r0}^W)\mathit{W}_{\omega,k})\\
& +\mathbf{B}_\omega^{d\dagger} (y_{\omega,k} - \mathbf{E}_1\mathit{V}_{\omega,k}^0) - \gamma_c(I + (\mathbf{X}_p + \mathbf{X}_s) \mathbf{B}^c_\omega)^\dagger (\mathbf{X}_p + \mathbf{X}_s) \mathbf{E}_2\mathit{V}_{\omega,k}^0)
\end{align*}

\item Gradient par rapport à $\mathit{\chi}_p$ :
\begin{align*}
\mathcal{G}_{\mathit{\chi}_p} = \nabla_{\mathit{\chi}_p} \mathcal{C}(\mathit{\chi}_p)= &
2(\sum_k \sum_\omega ((\gamma_c\mathbf{\Delta}_{\omega,k}^{p\dagger}\mathbf{\Delta}_{\omega,k}^p + \gamma_{r0}^\chi + \gamma_{r1}^\chi(\mathbf{D}_1+\mathbf{D}_2)^t(\mathbf{D}_1+\mathbf{D}_2))\mathit{\chi}_p\\
& - \gamma_c\mathbf{\Delta}_{\omega,k}^{p\dagger}(\mathit{W}_{\omega,k} - \mathbf{\Delta}_{\omega,k}^s\mathit{\chi}_s)))
\end{align*}
où $\mathbf{\Delta}_{\omega,k}^p$ et $\mathbf{\Delta}_{\omega,k}^s$ sont telles que :
\begin{align*}
& \mathbf{\Delta}_{\omega,k}^p \mathit{\chi}_p = \mathbf{X}_p [\mathit{V}_{\omega,k}^0 - (\mathbf{A}_{\omega,p,s})_0^{-1} \mathbf{E}_2^t \mathit{W}_{\omega,k}]\\
\text{et} \quad & \mathbf{\Delta}_{\omega,k}^s \mathit{\chi}_s = \mathbf{X}_s [\mathit{V}_{\omega,k}^0 - (\mathbf{A}_{\omega,p,s})_0^{-1} \mathbf{E}_2^t \mathit{W}_{\omega,k}]
\end{align*}
Pour obtenir l'expression du gradient par rapport à $\chi_s$, il suffit d'intervertir les indices $p$ et $s$.
\end{itemize}

\subsection{Spécificités de la méthodes CSI}

\textbf{Aspects numériques}

On remarque que la matrice normale associée au calcul des $\mathit{W}_{\omega,k}$ fait intervenir les matrices $\mathbf{B}^d$ et $\mathbf{B}^c$ qui sont des versions échantillonnées de $(\mathbf{A}_{\omega,p,s})_0^{-1}$. $(\mathbf{A}_{\omega,p,s})_0$ est une matrice creuse et son inversion est coûteuse en temps de calcul et en place mémoire, surtout dans le cas d'un problème de dimension réelle. Nous avons décidé de ne pas inverser ces matrices, mais plutôt de résoudre les systèmes linéaires qui interviennent dans le calcul du gradient à l'aide de décompositions LU.

La méthode CSI permet une estimation rapide des contrastes, d'autant plus qu'il est possible de calculer la matrice normale et d'utiliser un préconditionneur de Jacobi pour accélérer le calcul.

\bigskip
\textbf{Aspects techniques}

La méthode CSI nous permet de restreindre l'estimation des variables $\mathit{\chi}_p, \mathit{\chi}_s$ et $\mathit{W}_{\omega,k}$ à une petite zone d'étude tant que le milieu de référence est identique à l'arrière-plan de l'objet à détecter. L'estimation des sources de contraste s'avère être très lourde en temps de calcul. Pour chaque itération de la méthode d'inversion, il faut résoudre $N_f\times N_s \times(2 + N_{GC})$ systèmes linéaires, où $N_f$ est le nombre de fréquences, $N_s$ le nombre de positions de tir et $N_{GC}$ le nombre d'itérations du gradient conjugué.

À chaque itération, il faut mettre à jour les matrices $\mathbf{\Delta}_{\omega,k}^p $, $\mathbf{\Delta}_{\omega,k}^s $, $\mathbf{X}_p$ et $\mathbf{X}_s$.

\bigskip
\textbf{Le facteur de pondération CSI}

Les articles de Abubakar et Van Den Berg \cite{VanDenBerg97,VanDenBerg01,Abubakar08} suggèrent d'utiliser une valeur de l'hyperparamètre telle que les deux composantes de la fonction de coût soient égales lorsque les sources de contraste sont nulles. On obtient ainsi l'expression suivante pour la valeur de l'hyperparamètre $\gamma_c$ :
\begin{eqnarray}
\gamma_c = \gamma_{CSI} = \frac{\sum_k \sum_\omega \| y_{\omega,k} - \mathbf{E}_1\mathit{V}_{\omega,k}^0 \|^2}{\sum_k \sum_\omega \| (\mathbf{X}_p + \mathbf{X}_s) V^0_{\omega,k} \|^2}
\end{eqnarray}
Il s'est avéré que l'utilisation de ce facteur de pondération n'est pas adaptée à notre problème. Nous avons donc opté pour une estimation empirique des valeurs des hyperparamètres, ce qui nécessite des réglages supplémentaires fastidieux.

\subsection{Résultats}

\textbf{Méthode d'optimisation alternée}

Les résultats que nous allons présenter ont été obtenus en utilisant les données décrites dans la section~\ref{part_DonneesUtilisees}, et, pour des raisons de temps de calcul, nous nous sommes limité au milieu de petite taille.

\begin{figure}[!htb]
\centering
\includegraphics[width=4.3in]{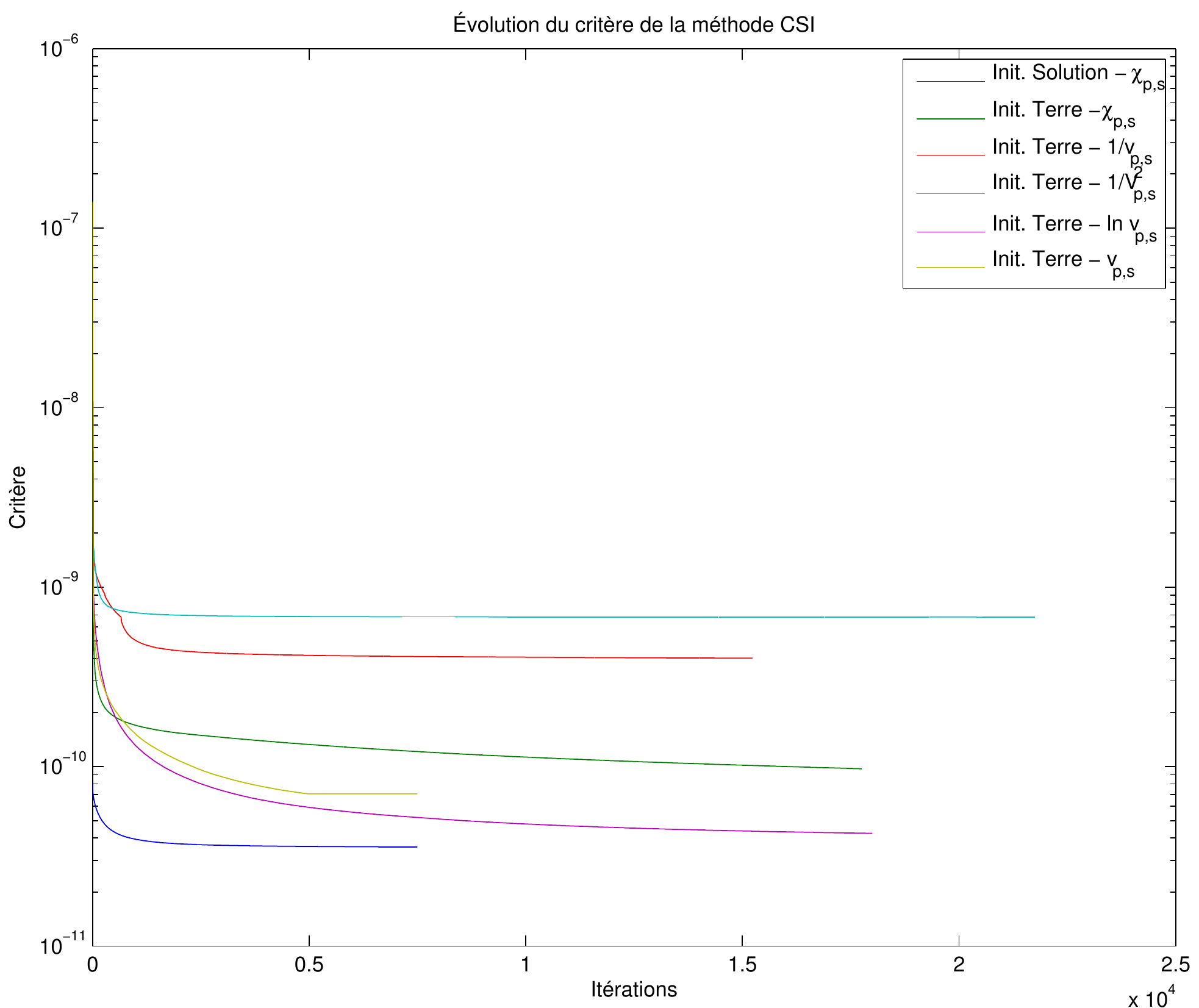}
\caption{Évolution du critère de la méthode CSI pour les différents changements de variables et deux initialisations}
\label{figCritereCSI}
\end{figure}

La figure~(\ref{figCritereCSI}) montre l'évolution du critère pour les différentes changement de variable proposés et pour deux initialisations différentes. On constate que, quel que soit le changement de variable, le critère décroît très lentement. Même après un grand nombre d'itérations, aucune des méthodes ne semble pouvoir rejoindre la courbe du critère obtenue en initialisant l'algorithme avec la solution.

\begin{figure}[!htb]
\centering
\includegraphics[width=4.3in]{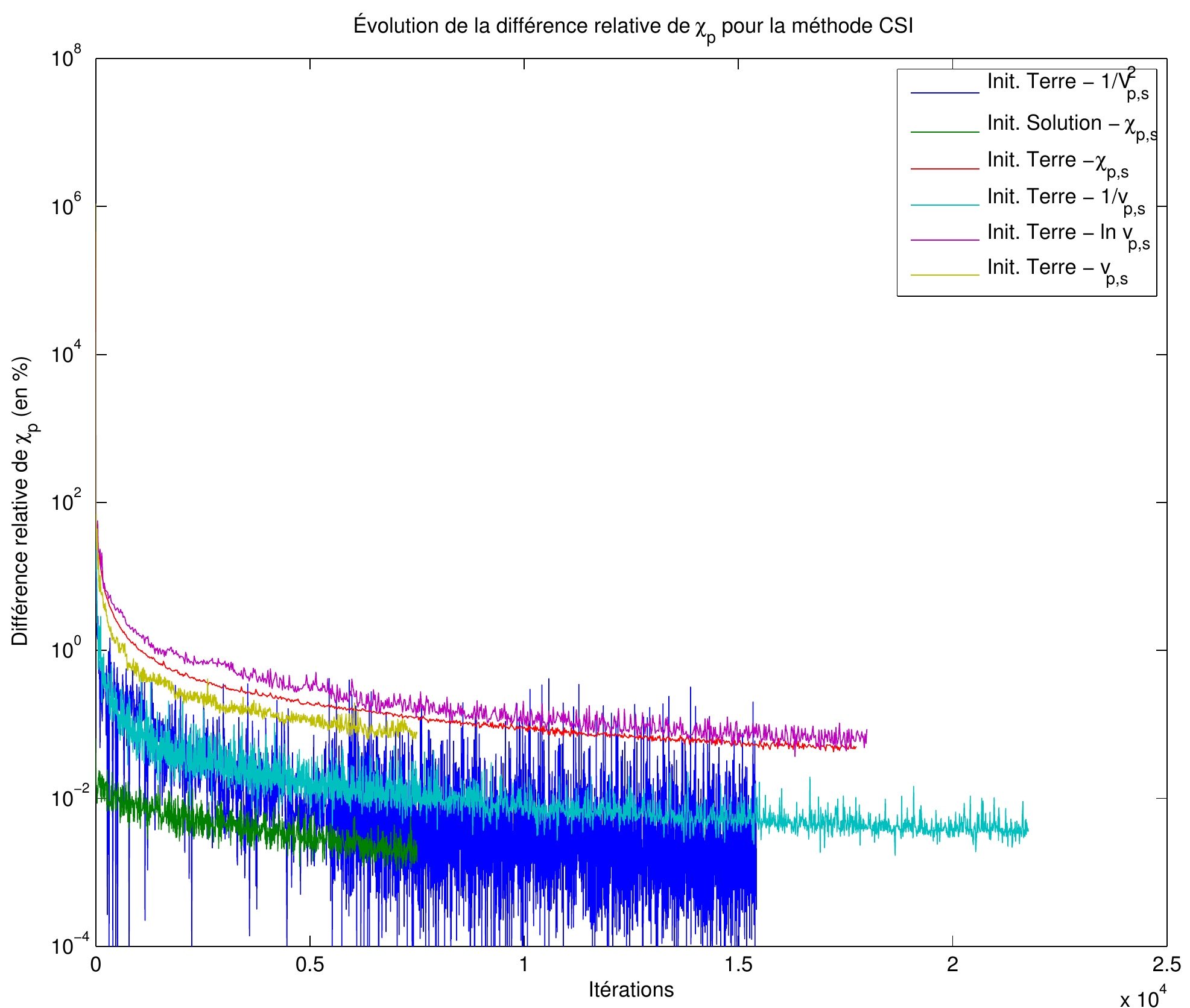}
\caption{Évolution du déplacement relatif de l'estimée des contrastes pour les différents changements de variables et deux initialisations}
\label{figDiffCSI}
\end{figure}

Les courbes de la figure~(\ref{figDiffCSI}) représentent l'évolution du déplacement relatif de l'estimée du contraste $\chi_p$ au cours des itérations. Une courbe semblable est obtenue dans le cas des $\chi_s$. On remarque que, assez rapidement, le déplacement du contraste d'une itération à l'autre est très petit, et qu'il ne cesse de décroître. L'évolution des variables se fait donc très lentement et de plus en plus lentement, ce qui explique pourquoi le critère décroit très lentement et semble stagner après un certain nombre d'itérations. Cependant, le déplacement de la solution reste toujours suffisamment significatif pour ne pas fanchir le seuil fixé comme critère d'arrêt de l'agorithme tel que décrit dans la section~\ref{part_CritereConvergence}. Il ne nous a donc pas été possible de faire converger cette méthode en un temps raisonnable.

\begin{figure}[!htb]
\centering
\subfloat[Initialisation à la solution, variable $\mathit{\chi}_{p,s}$]{\includegraphics[width=4.3in]{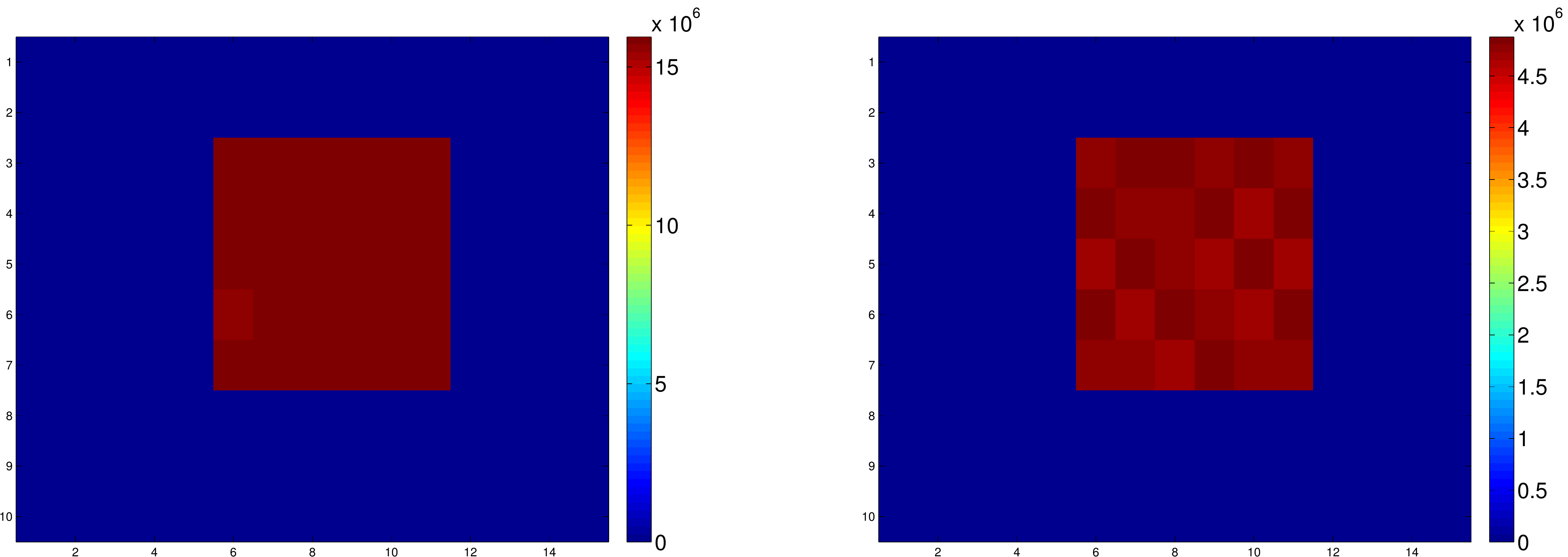}}\\
\subfloat[Initialisation aux caractéristiques de la terre, variable $\mathit{\chi}_{p,s}$]{\includegraphics[width=4.3in]{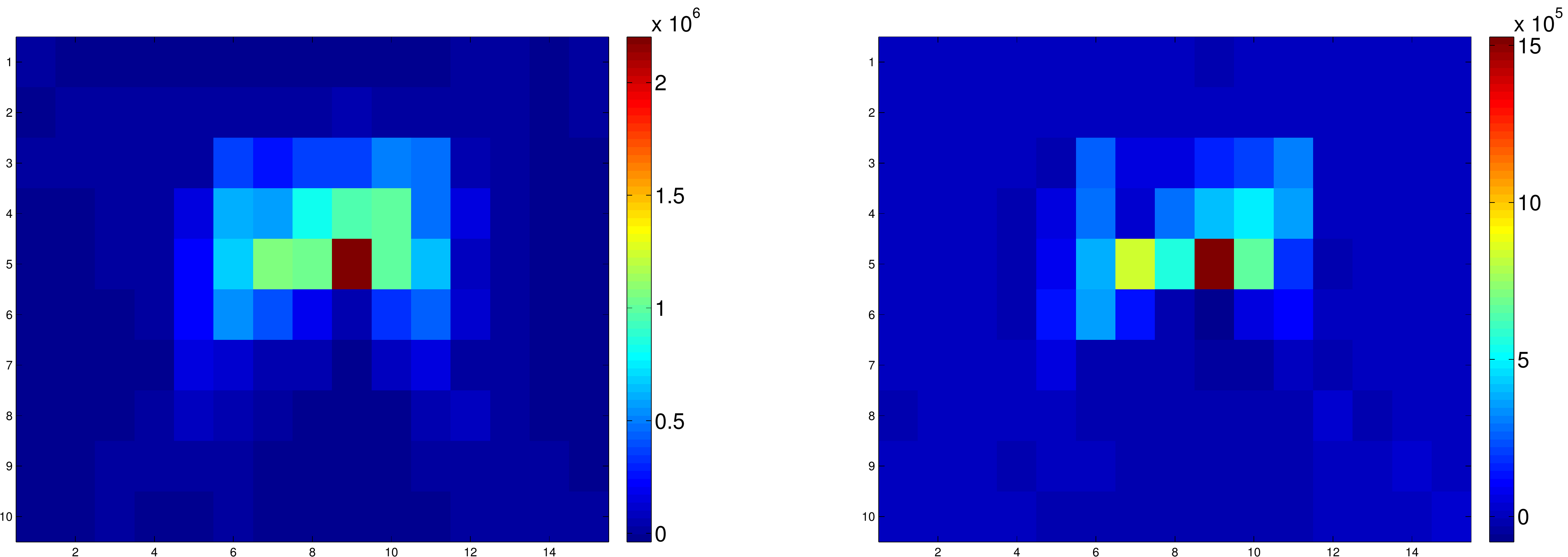}}\\
\subfloat[Changement de variable en $1/\mathit{v}_{p,s}$, initialisation aux caractéristiques de la terre]{\includegraphics[width=4.3in]{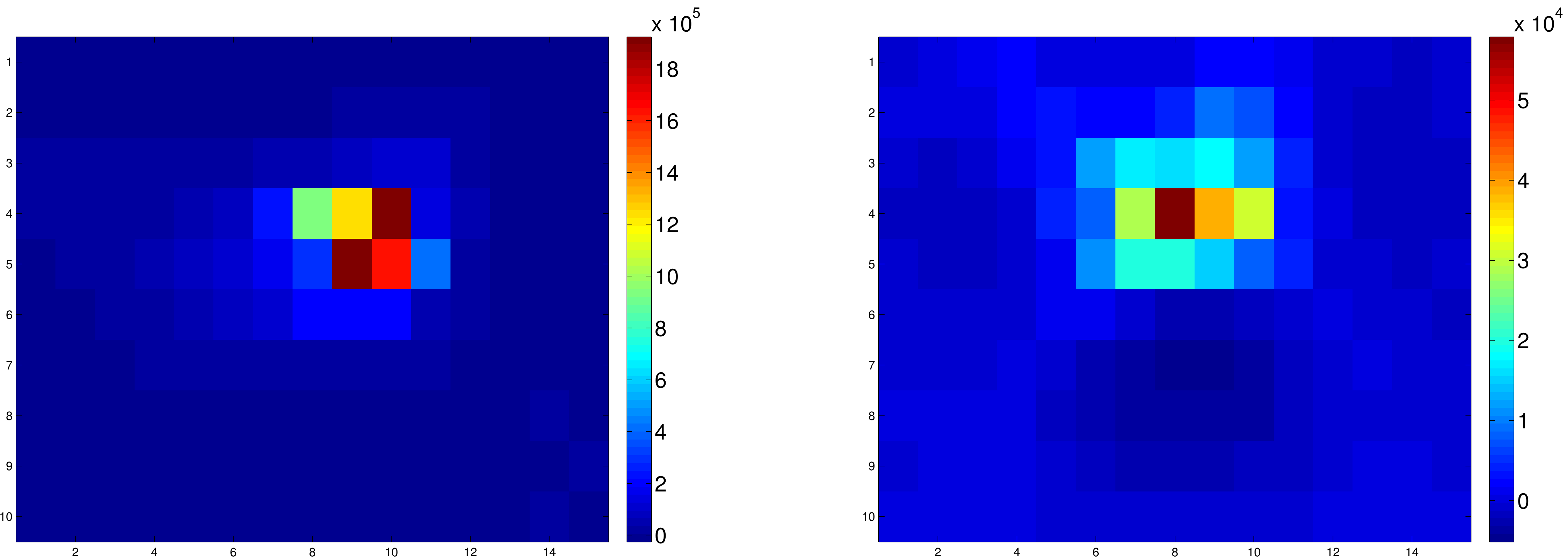}}\\
\subfloat[Changement de variable en $1/\mathit{v}_{p,s}^2$, initialisation aux caractéristiques de la terre]{\includegraphics[width=4.3in]{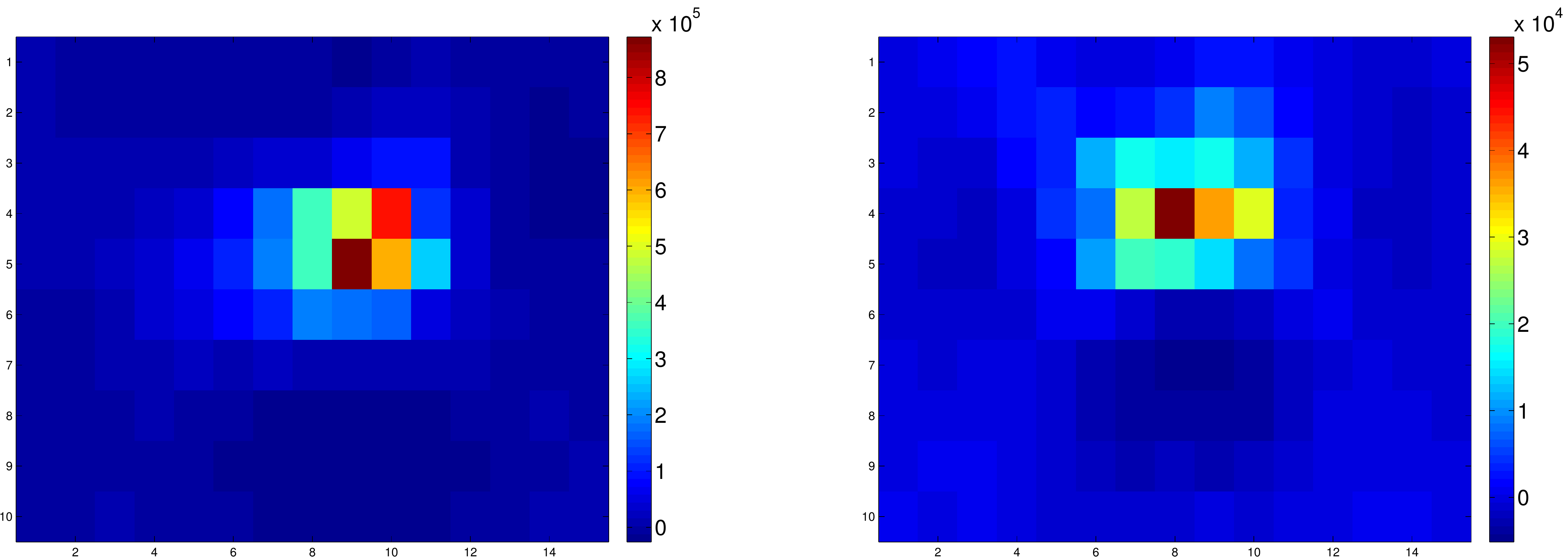}}\\
\subfloat[Changement de variable en $\ln(\mathit{v}_{p,s})$, initialisation aux caractéristiques de la terre]{\includegraphics[width=4.3in]{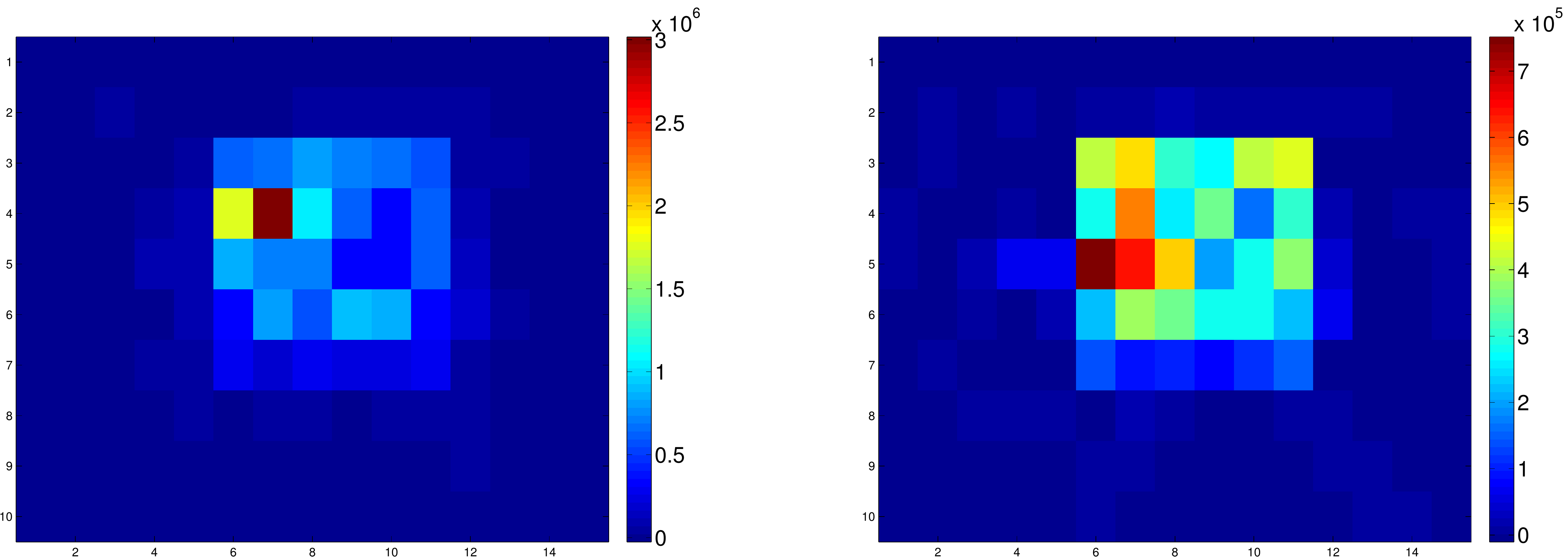}}
\caption{Résultats des contrastes $\chi_s$ (à droite) et $\chi_p$ (à gauche) obtenus avec la méthode CSI pour différents changements de variables et deux initialisations}
\label{figResultatsCSI}
\end{figure}

Les résultats obtenus pour la méthode CSI avec une optimisation alternée des deux jeux de variables, et pour les changements de variables indiqués à la section~\ref{Part_IntroChgtVariable}, sont présentés dans la figure~(\ref{figResultatsCSI}). Les cartes montrées dans cette figure ont été obtenues après $17600$ itérations (soit environ $3$ jours de calcul). On constate que, même après un grand nombre d'itérations, aucune des méthodes testées ne permet de reconstruire l'amplitude et la forme de l'objet recherché. On peut noter toutefois, que tous les changements de variable testés ont un effet différent sur la convergence de la méthode. Notamment, le changement de variable en $\ln(\mathit{v}_{p,s})$ donne les meilleurs résultats en termes de décroissance du critère et de différence relative, et les cartes obtenues semblent visuellement supérieures.

\begin{figure}[!htb]
\centering
\subfloat[Changement de variable en $\mathit{v}_{p,s}$, initialisation aux caractéristiques de la terre]{\includegraphics[width=4.3in]{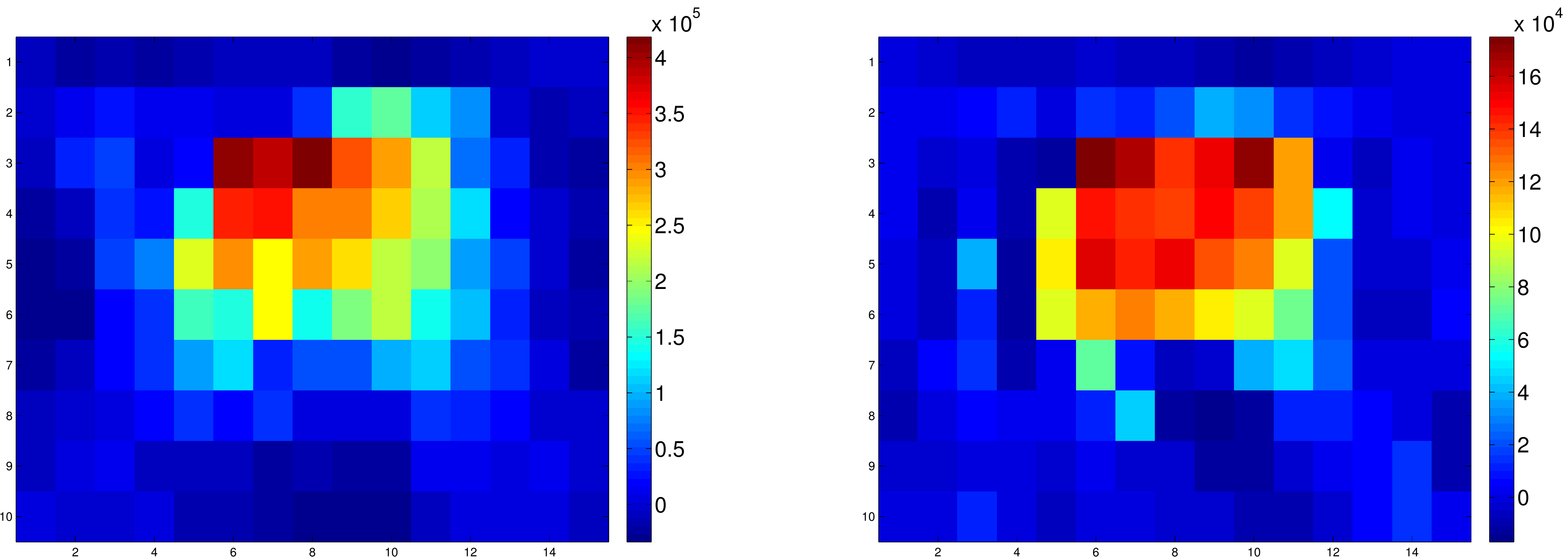}}
\caption{Résultats des contrastes $\chi_s$ (à droite) et $\chi_p$ (à gauche) obtenus avec la méthode CSI pour différents changements de variables et deux initialisations (suite)}
\end{figure}

\begin{figure}[!htb]
\centering
\subfloat[Évolution du conditionnement de l'estimateur des variables auxiliaires pour plusieurs fréquences]{\includegraphics[scale=0.35]{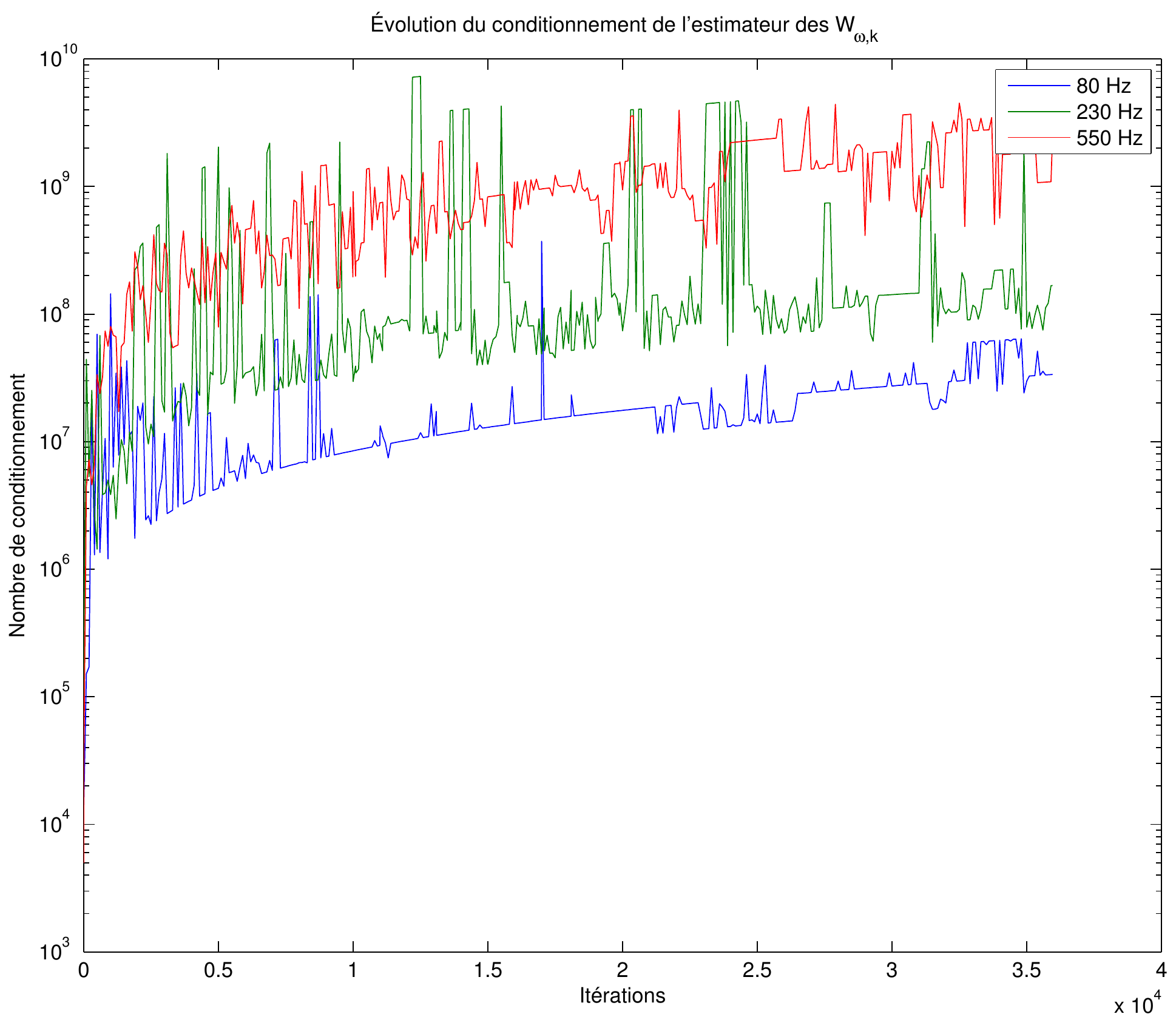}}
\subfloat[Évolution du conditionnement de l'estimateur des contrastes]{\includegraphics[scale=0.35]{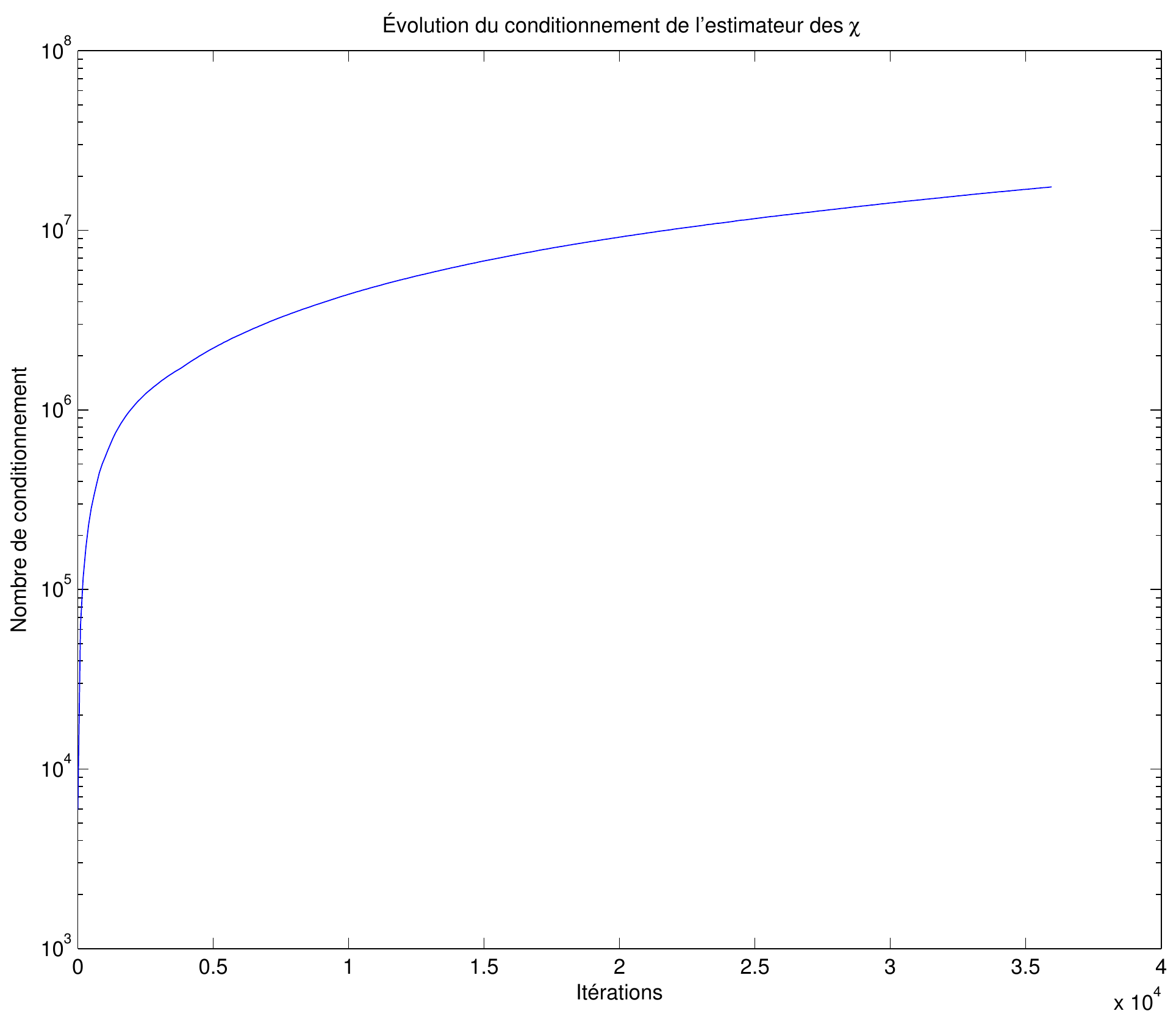}}
\caption{Évolution du conditionnement des estimateurs de la méthode CSI}
\label{figConditionnement_CSI}
\end{figure}

En raison de la lenteur de l'évolution de la solution, il nous est impossible de dire si la méthode converge véritablement. La convergence des méthodes de type descente de gradient dépend fortement du nombre de conditionnement du hessien \cite{Shewchuk94}. Or, on voit sur la figure~(\ref{figConditionnement_CSI}) que le nombre de conditionnement du hessien des estimateurs CSI augmente au fil des itérations. Cette détérioration du conditionnement pourrait tout à fait expliquer les problèmes de stagnation du déplacement relatif des estimées et du critère mis en évidence dans les figures~(\ref{figCritereCSI}) et (\ref{figDiffCSI}). Il semble donc légitime de penser que la lenteur de la convergence de la méthode CSI est due à un mauvais conditionnement des matrices normales associées aux estimateurs.

\bigskip
\textbf{Méthode d'optimisation conjointe}

Pour tenter de remédier aux problèmes de convergence de la méthode CSI alternée, nous avons mis en oeuvre une version d'estimation conjointe de la CSI, voir \cite{Barriere08}.

La mise en \oe{uvre} de cette méthode fait apparaître un problème de
différence d'échelle entre les deux jeux d'inconnues, \cad
entre les contrastes et les sources de contrastes ; l'amplitude des
premiers est de l'ordre de $10^6$, tandis que pour les deuxièmes elle
ne dépasse pas $10^2$. Cela nuit au calcul du pas de descente utilisé
par la méthode d'optimisation, qui sera un compromis entre les pas
optimaux de chacune des deux variables. On résout  traditionnellement
ce type de problème en utilisant un préconditionneur de Jacobi ou un
préconditionneur diagonal empirique basé sur une connaissance \aprio de l'ordre de grandeur de chaque jeu de
variables. Toutefois, les matrices normales des estimateurs de la
méthode CSI ne sont pas à diagonale dominante, et il est reconnu que
les préconditionneurs diagonaux ne sont pas efficaces dans ce cas
\cite{Saad00}. Nous avons d'ailleurs effectué des tests qui nous ont
convaincu que les préconditionneurs de Jacobi et empiriques ne sont
pas adaptés à notre problème, car ils n'améliorent pas les résultats.

\begin{figure}[!htb]
\centering
\subfloat[Initialisation à la solution, variable $\mathit{\chi}_{p,s}$]{\includegraphics[width=4.3in]{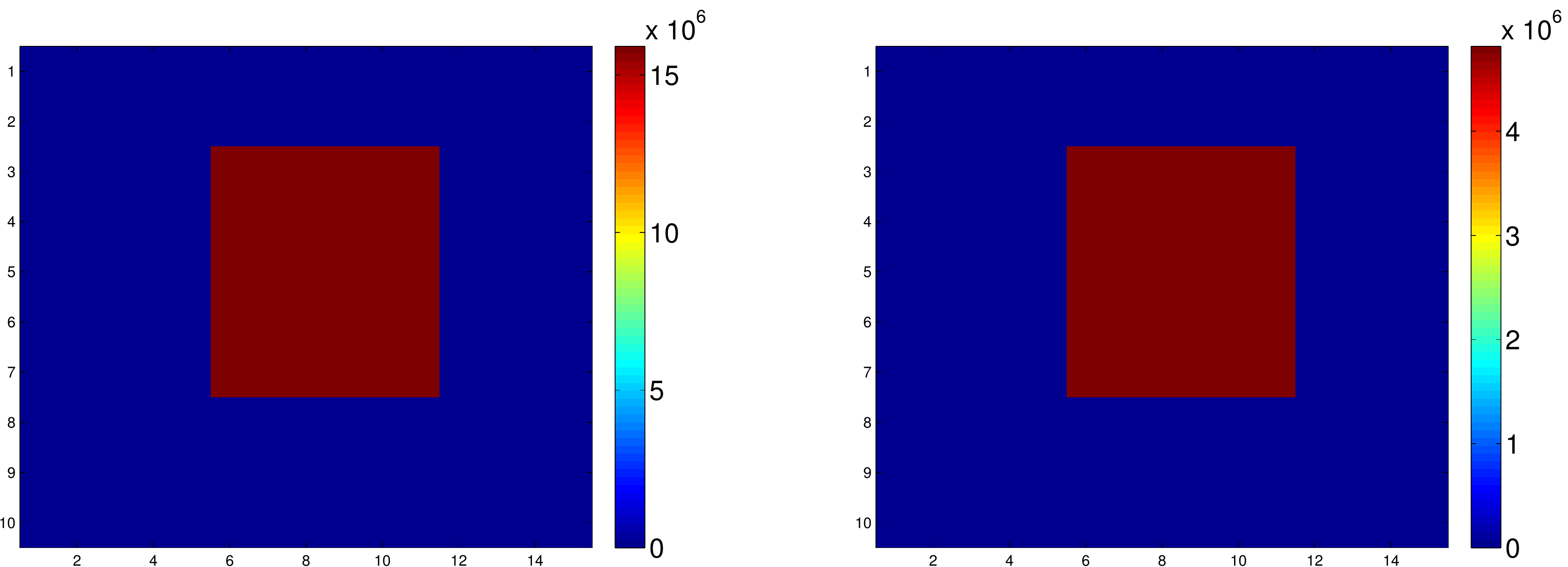}}\\
\subfloat[Initialisation aux caractéristiques de la terre, variable $\mathit{\chi}_{p,s}$]{\includegraphics[width=4.3in]{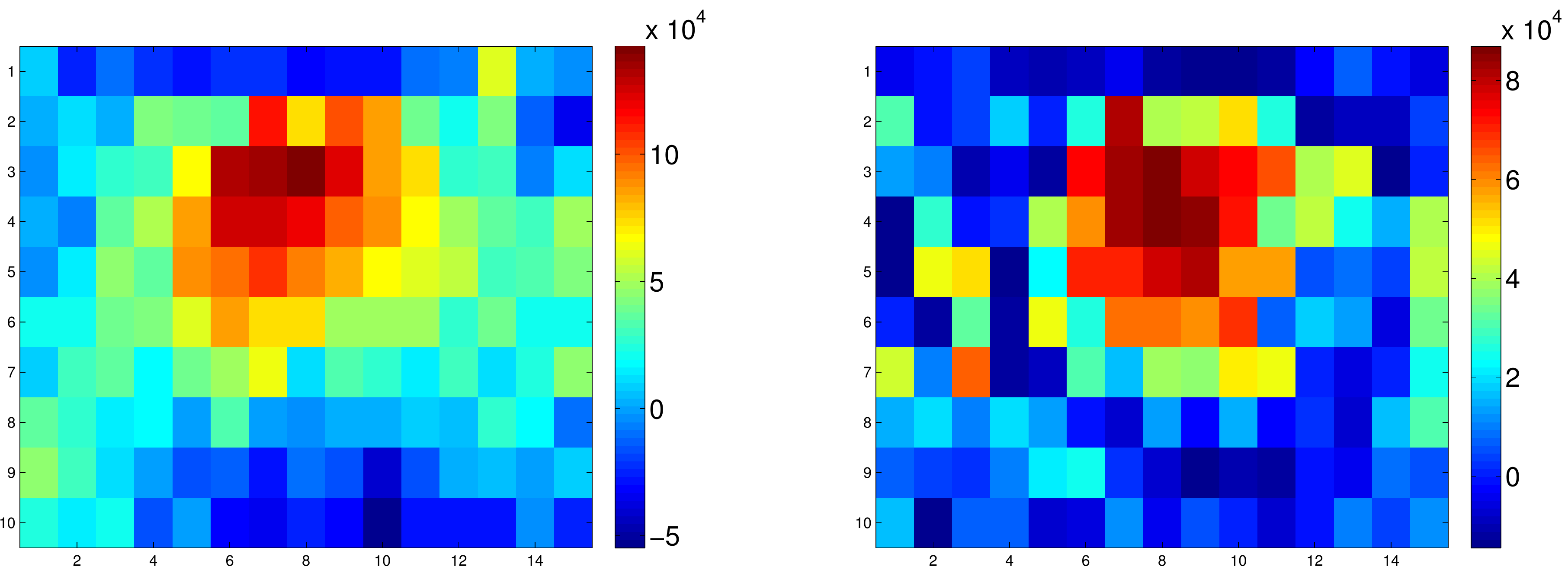}}\\
\subfloat[Changement de variable en $1/\mathit{v}_{p,s}$, initialisation aux caractéristiques de la terre]{\includegraphics[width=4.3in]{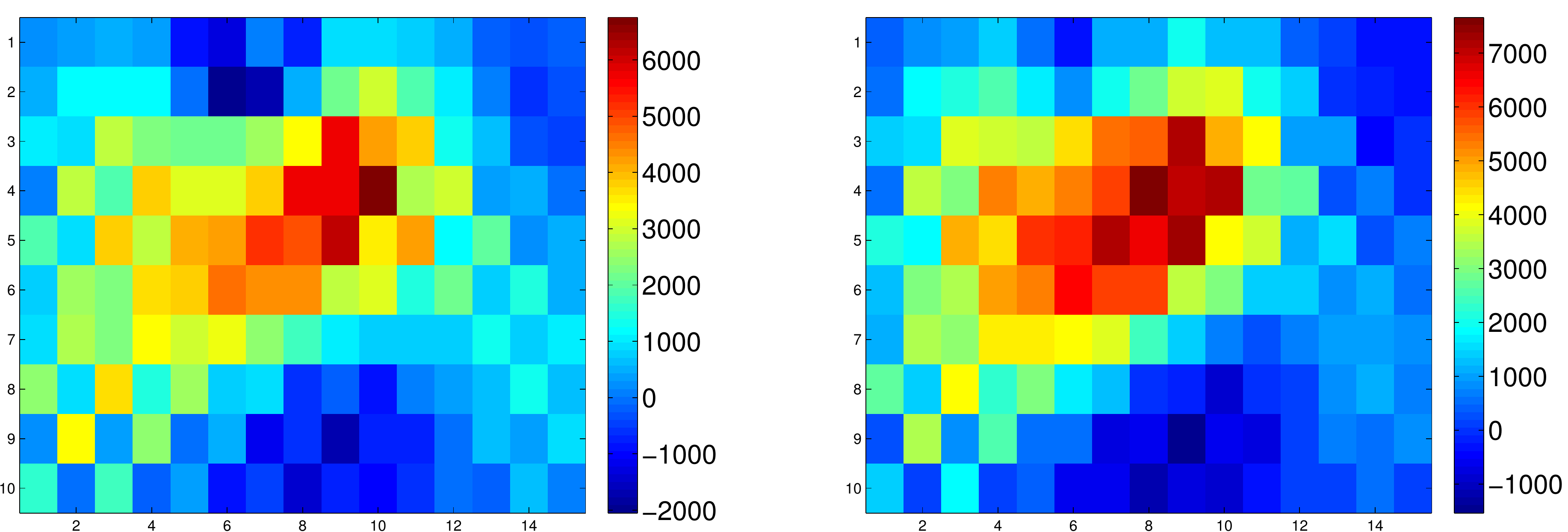}}\\
\subfloat[Changement de variable en $\log(v_{p,s})$, initialisation aux caractéristiques de la terre]{\includegraphics[width=4.3in]{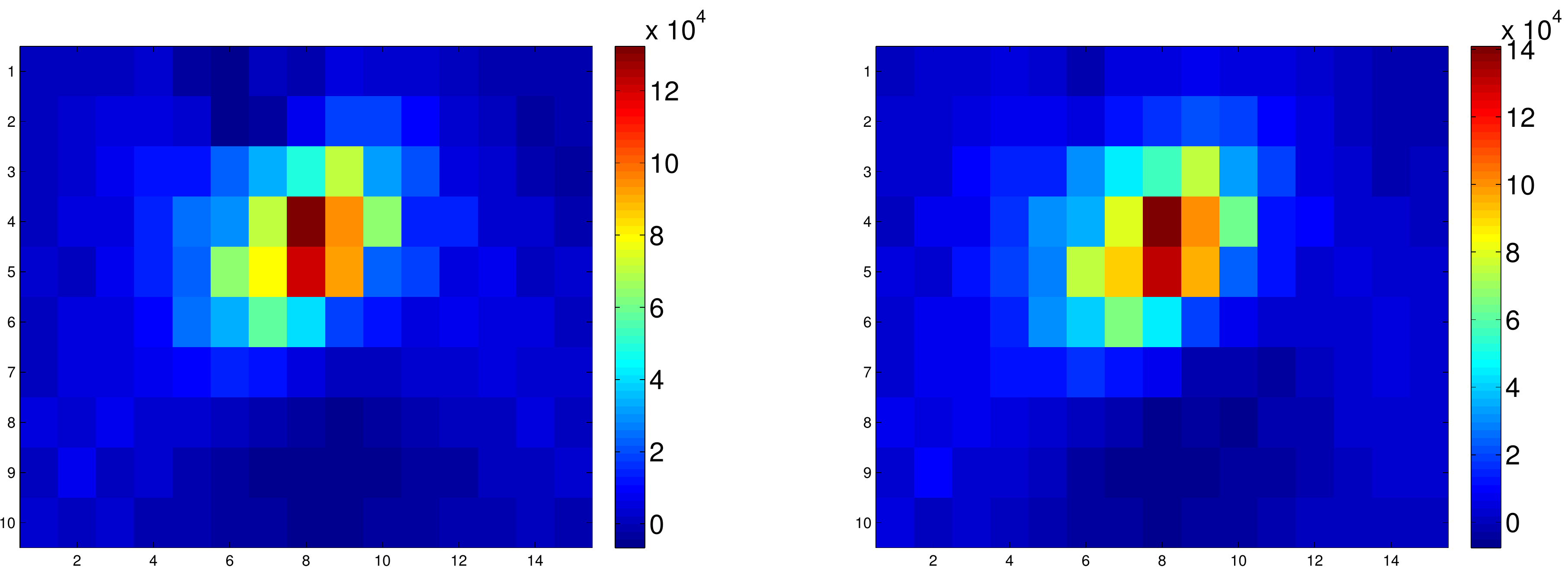}}\\
\caption{Résultats de la méthode CSI conjointe}
\label{figResultatsConjointe}
\end{figure}

\begin{figure}[!htb]
\centering
\includegraphics[width=4.3in]{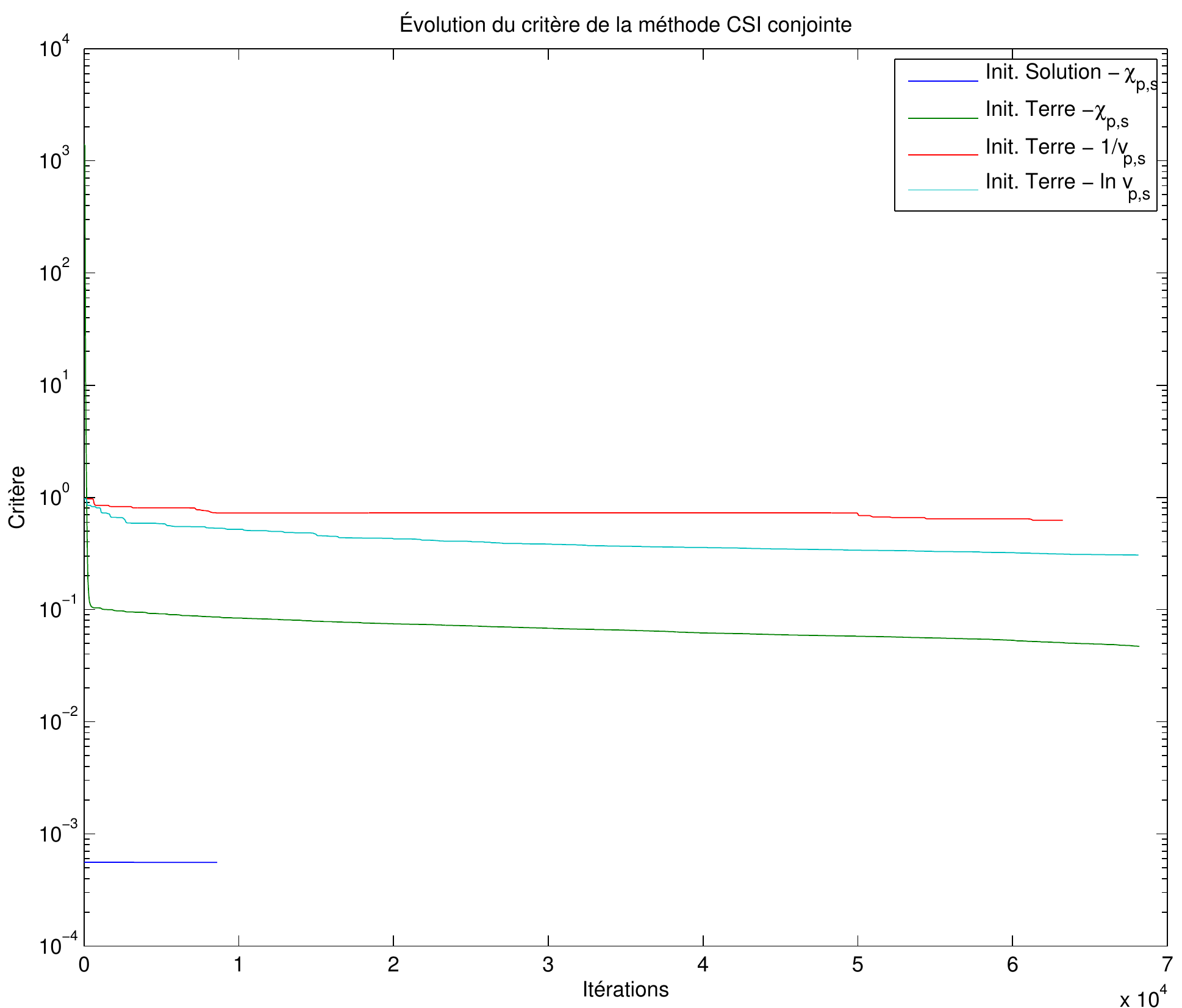}
\caption{Évolution du critère de la méthode CSI conjointe}
\label{figCritereConjointe}
\end{figure}

Les cartes de contraste calculées avec la méthode CSI conjointe sont présentées dans les figures~(\ref{figResultatsConjointe}) et (\ref{figCritereConjointe}). Elles ont été obtenues en pondérant le terme d'adéquation aux données du critère par la norme des mesures. Cette pondération permet de réduire l'importance du problème d'échelle entre les deux jeux de variables. Sans elle, le pas de progression de l'algorithme d'optimisation est trop petit, et l'évolution de la solution est excessivement lente.


Les résultats montrent que cette méthode ne permet pas de régler les problèmes de lenteur de la convergence, elle semble même converger beaucoup plus lentement que la CSI classique. Les contrastes montrés dans le figure~(\ref{figResultatsConjointe}) sont qualitativement et quantitativement moins bons que ceux de la CSI après un temps de calcul équivalent. Cette méthode atteint plus rapidement le palier de stagnation de la solution, et ce, pour tous les changements de variables testés. On en déduit donc que, dans ce cas-ci, les meilleurs résultats sont obtenus sans effectuer de changement de variable.

\section{Méthode du gradient modifié}\label{part_GM}

Pour pallier les problèmes de conditionnement qui surviennent dans les méthodes de type CSI, nous avons exploré d'autres approches basées sur des formulations bilinéaires. Nous avons commencé par étudier la méthode du gradient modifié (GM). Elle est antérieure à la CSI et a été proposée la première fois par Kleinman et van den Berg \cite{Kleinman92}.

\subsection{Description}

Dans cette méthode, la variable d'intérêt est le contraste des caractéristiques physiques, et la variable auxiliaire représente directement les composantes du champ de vitesse. Plus précisément, on a :

\begin{description}
 \item[Les variables d'intérêt] $\mathit{x}$ sont les contrastes $\mathit{\chi}_p$ et $\mathit{\chi}_s$ des carrés des vitesses $\mathit{V}_p$ et $\mathit{V}_s$ qui sont estimés sur une partie réduite du domaine $D$, et ont été définis dans la section~\ref{partDef_contrastes}
 \item[Les variables auxiliaires] $\mathit{z}_{\omega,k}$ sont les composantes du champ de vitesse $\mathit{V}_{\omega,k}$ qui doivent être estimées sur tout le domaine $D$
\end{description}

L'adéquation du modèle aux mesures et la relation bilinéaire entre les deux ensembles de variables, les contrastes $\chi_p$ et $\chi_s$ et les variables auxiliaires $\mathit{V}_{\omega,k}$, sont données dans le système d'équations suivant :

\begin{description}
\item [Équation d'observation] $\mathit{y}_{\omega,k} = \mathbf{E}_1\mathit{V}_{\omega,k}$
\item [Équation de couplage] $(\mathbf{A}_{\omega,p,s})_0(\mathit{V}_{\omega,k}^0 - \mathit{V}_{\omega,k}) = (\mathbf{X}_p + \mathbf{X}_s)\mathit{V}_{\omega,k}$
\end{description}

En reprenant les notations de la section~\ref{partIdeeGenerale}, on établit que :
\begin{align}
 \mathcal{H}(\mathit{V}_{\omega,k}) &= (\mathbf{A}_{\omega,p,s})_0(\mathit{V}_{\omega,k}^0 - \mathit{V}_{\omega,k}) & \text{ avec } \mathbf{M}_c = -(\mathbf{A}_{\omega,p,s})_0 \text{ et } \mathit{u}_c = (\mathbf{A}_{\omega,p,s})_0 \mathit{V}_{\omega,k}^0 \notag\\
 \mathcal{R}(\mathit{V}_{\omega,k}) &= \mathit{V}_{\omega,k} & \text{ avec } \mathbf{M}_d = \indentit \text{ et } \mathit{u}_d = 0
\label{eq_DefNotationsGM}\\
 \mathbf{K}(\mathit{\chi}_p, \mathit{\chi}_s) &= \mathbf{X}_p + \mathbf{X}_s\notag
\end{align}

En se référant à la section~\ref{part_CritereGeneral} et en utilisant les équations de couplage et d'observation propres à la méthode GM, nous établissons que le critère prend la forme :

\begin{align}
\mathcal{C} &= \sum_k \sum_\omega \| \mathit{y}_{\omega,k} - \mathbf{E}_1\mathit{V}_{\omega,k} \|^2  \\
& + \gamma_c\sum_k \sum_\omega \| (\mathbf{A}_{\omega,p,s})_0(\mathit{V}_{\omega,k}^0 - \mathit{V}_{\omega,k}) - (\mathbf{X}_p + \mathbf{X}_s)\mathit{V}_{\omega,k} \|^2 + \phi
\label{eq_CritereGM}
\end{align}

À partir de l'expression du critère, nous pouvons réécrire les formules des gradients donnés dans les équations~\eqref{eq_GradientGeneralz}, \eqref{eq_GradientGeneralxp} et \eqref{eq_GradientGeneralxs} en y substituant les équations~\eqref{eq_DefNotationsGM}, et en tenant compte du choix des variables fait par cette méthode.

\begin{itemize}
\item Le gradient du critère par rapport à $\mathit{V}_{\omega,k}$ est :
\begin{align*}
\mathcal{G}_{\mathit{V}_{\omega,k}} = \nabla_{\mathit{V}_{\omega,k}} \mathcal{C}(\mathit{V}_{\omega,k})= &2((\gamma_c (\mathbf{X}_p+\mathbf{X}_s+(\mathbf{A}_{\omega,p,s})_0)^{\dagger}(\mathbf{X}_p+\mathbf{X}_s+(\mathbf{A}_{\omega,p,s})_0) + \mathbf{E}_1^\dagger\mathbf{E}_1 + \gamma^V_{r0})\mathit{V}_{\omega,k}\\
 & - ((\mathbf{E}_1^\dagger\mathit{y}_{\omega,k} +\gamma_c( (\mathbf{X}_p+\mathbf{X}_s+(\mathbf{A}_{\omega,p,s})_0)^{\dagger}(\mathbf{A}_{\omega,p,s})_0\mathit{V}_{\omega,k}^0))))\\
\end{align*}

\item Le gradient du critère par rapport à $\mathit{\chi}_p$ est :
\begin{align*}
\mathcal{G}_{\mathit{\chi}_p} = \nabla_{\mathit{\chi}_p} \mathcal{C}(\mathit{\chi}_p)= &
2(\gamma_c\sum_k \sum_\omega ((\mathbf{\Delta}_{\omega,k}^{p\dagger}\mathbf{\Delta}_{\omega,k}^p + \gamma_{r0}^\chi + \gamma_{r1}^\chi(\mathbf{D}_1+\mathbf{D}_2)^T(\mathbf{D}_1+\mathbf{D}_2))\mathit{\chi}_p\\
& -\mathbf{\Delta}_{\omega,k}^{p\dagger}((\mathbf{A}_{\omega,p,s})_0(\mathit{V}_{\omega,k}^0-\mathit{V}_{\omega,k})+\mathbf{\Delta}_{\omega,k}^s\mathit{\chi}_s)))
\end{align*}
avec
\begin{align*}
\mathbf{\Delta}_{\omega,k}^p\mathit{\chi}_p&=\mathbf{X}_p\mathit{V}_{\omega,k}\\
\mathbf{\Delta}_{\omega,k}^s\mathit{\chi}_s&=\mathbf{X}_s\mathit{V}_{\omega,k}\\
\end{align*}
L'expression du gradient du critère de la méthode GM par rapport à $\chi_s$ est obtenu en interchangeant les indices $p$ et $s$. L'estimation des $\chi_p$ et $\chi_s$ est faite conjointement.
\end{itemize}

\subsection{Spécificités de la méthode GM}\label{part_SpecGM}

Le calcul des $\mathbf{\chi}_p$ et $\mathbf{\chi}_s$ est similaire au cas de la CSI ; seule l'expression de $\mathcal{H}$ change.

Le calcul du gradient par rapport au champ de vitesse ne fait intervenir l'inverse d'aucune matrice et nécessite seulement l'évaluation de sommes de matrices et de produits matrice-vecteur. Le choix du champ de vitesse comme variable auxiliaire nous impose l'inversion d'un grand nombre de variables puisqu'on ne peut se restreindre à une zone d'étude pour la détermination de cette variable. Cet inconvénient est toutefois compensé par la rapidité et la simplicité du calcul du gradient par rapport aux variables auxiliaires.

Lorsque l'on teste isolément l'estimateur des composantes du champ de vitesse de la méthode GM, en initialisant la méthode avec les contrastes solution, on constate que la convergence est extrêmement lente. Cet estimateur fait intervenir la matrice normale suivante : $\mathbf{D}_\omega = \mathbf{A}^\dagger_{\omega,p,s}\mathbf{A}_{\omega,p,s} + \mathbf{E}^{\dagger}_1\mathbf{E}_1$ où $\mathbf{A}_{\omega,p,s} =(\mathbf{X}_p+\mathbf{X}_s+(\mathbf{A}_{\omega,p,s})_0)$ est la matrice d'impédance du problème direct. Nous avons vu dans la section~\ref{partPbConditionnementA} que cette matrice est très mal conditionnée. Le nombre de conditionnement associé à la matrice normale de l'estimateur du champ de vitesse de la méthode GM, en initialisant les contrastes à la solution, est de $10^{25}$ lorsqu'on la génère avec le code fourni par EdF. Si on utilise la méthode suggérée dans la section~\ref{PartStructMatrImpedance} pour la construire, le conditionnement tombe à $10^{12}$. Même si cela reste élevé, il est maintenant possible de résoudre le système intervenant dans l'estimateur du champ de vitesse à l'aide d'une décomposition LU, ou d'utiliser des méthodes de préconditionnement basées sur une factorisation.

Le choix d'un préconditionneur pour l'estimation du champ de vitesse est limité par le fait que les matrices normales associées ne sont pas à diagonale dominante. Les préconditionneurs diagonaux, comme celui de Jacobi, ou, quand ils existent, de type factorisation simple sont reconnus pour ne pas être efficaces pour ce type de problème \cite{Saad00}. Nous avons testé les préconditionneurs de Jacobi, LU incomplète et Cholesky incomplète, et nous avons remarqué qu'ils n'apportaient aucune amélioration de la convergence et pouvaient même parfois nuire.

Il est cependant possible d'utiliser un préconditionneur issu de la décomposition de Cholesky incomplète décalée \cite{Manteuffel80} qui est basé sur la factorisation de la matrice normale augmentée $\mathbf{D}_\omega + \alpha\:\mathbf{\indentit}$, où $\alpha$ est un paramètre assurant la stabilité de la factorisation qui doit être fixé manuellement et que l'on doit garder le plus petit possible. L'emploi de ce préconditionneur permet d'accélérer significativement la convergence des premières itérations de la méthode GM, mais l'évolution du critère finit par stagner.


\bigskip

Il existe une autre formulation du gradient modifié qui ne diffère de celle qu'on vient de présenter que dans son équation de couplage qui prend la forme suivante : $\mathit{V}_{\omega,k}^0 - \mathit{V}_{\omega,k} = (\mathbf{A}_{\omega,p,s})_0^{-1}(\mathbf{X}_p + \mathbf{X}_s)\mathit{V}_{\omega,k}$. Cette variante de la méthode GM s'obtient facilement à partir de la méthode présentée en multipliant les deux membres de l'équation de couplage par $(\mathbf{A}_{\omega,p,s})_0^{-1}$. La présence de la matrice $(\mathbf{A}_{\omega,p,s})_0^{-1}$ dans l'expression des matrices normales des deux estimateurs ne permet pas de faire le calcul explicite du Hessien, et la complexité calculatoire de cette méthode est la même que celle de la CSI. Le calcul du gradient par rapport au champ de vitesses nécessite la résolution d'autant de systèmes linéaires que dans le cas de la méthode CSI, mais il doit être fait pour tous les points du domaine contrairement à la CSI. Pour ces raisons, nous avons choisi de ne pas tester cette méthode.

\subsection{Résultats}

Pour des raisons de temps de calcul, nous ne présentons les résulats de la méthode GM que pour le milieu de petit taille décrit dans la section~\ref{part_DonneesUtilisees}. Les résultats ont été obtenus en utilisant une inversion directe par décomposition LU pour l'estimation du champ de vitesse et un gradient conjugué non linéaire pour l'estimation des contrastes. L'utilisation de la décomposition LU entraîne un gain de temps important par rapport à l'emploi d'un gradient conjugué préconditionné avec un préconditionneur tel que suggéré dans la section~\ref{part_SpecGM}.

\begin{figure}[!htb]
\centering
\subfloat[Initialisation à la solution, variable $\mathit{\chi}_{p,s}$]{
\includegraphics[width=4.3in]{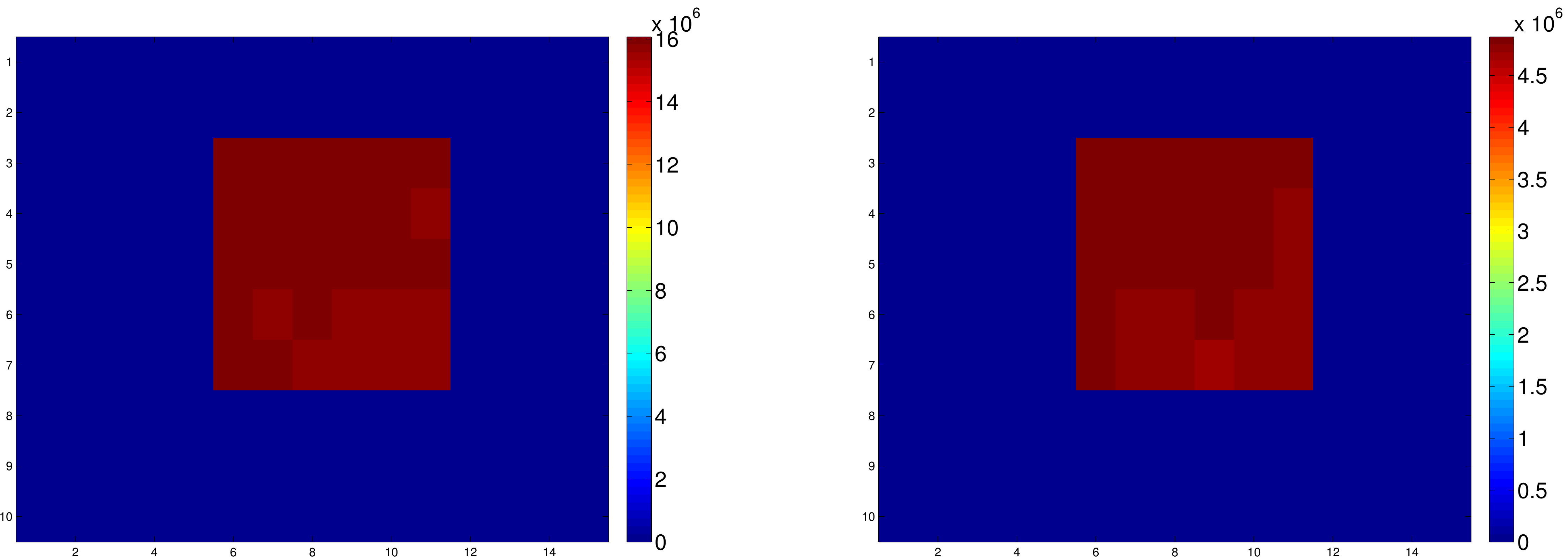}}\\
\subfloat[Initialisation à la terre, variable $\mathit{\chi}_{p,s}$]{
\includegraphics[width=4.3in]{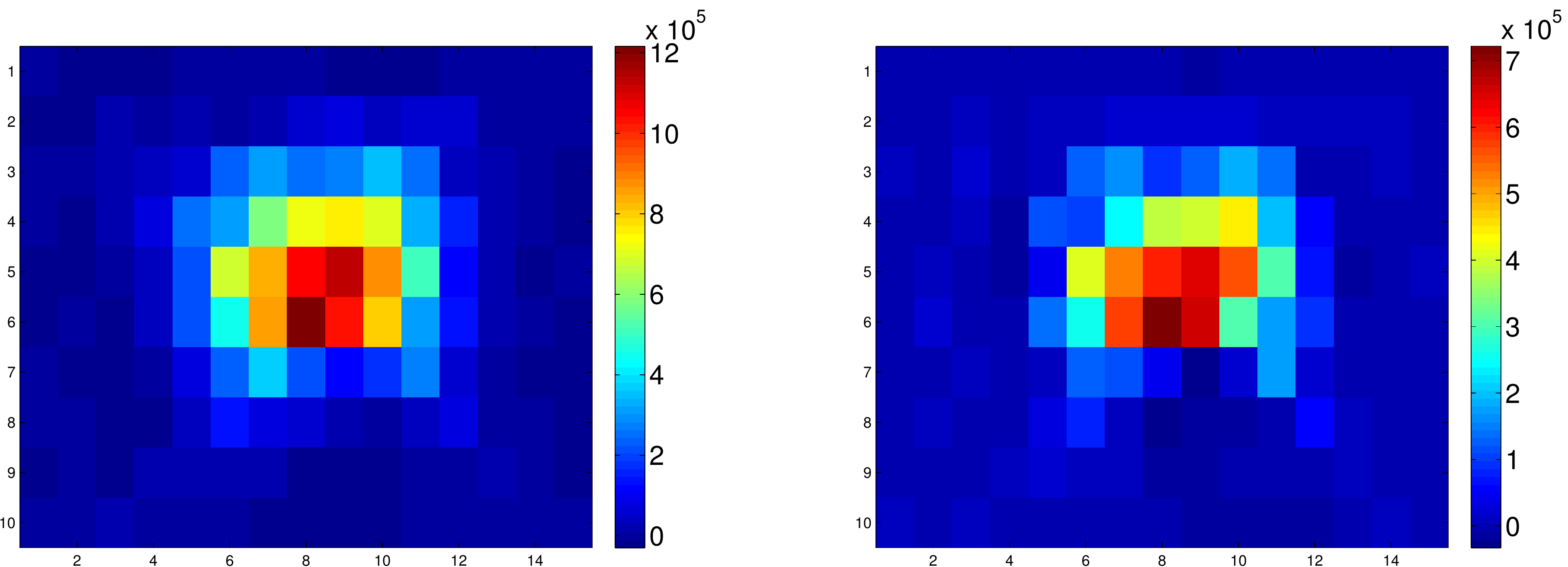}}\\
\subfloat[Changement de variable en $1/v_{p,s}$, initialisation aux caractéristiques de la terre]{
\includegraphics[width=4.3in]{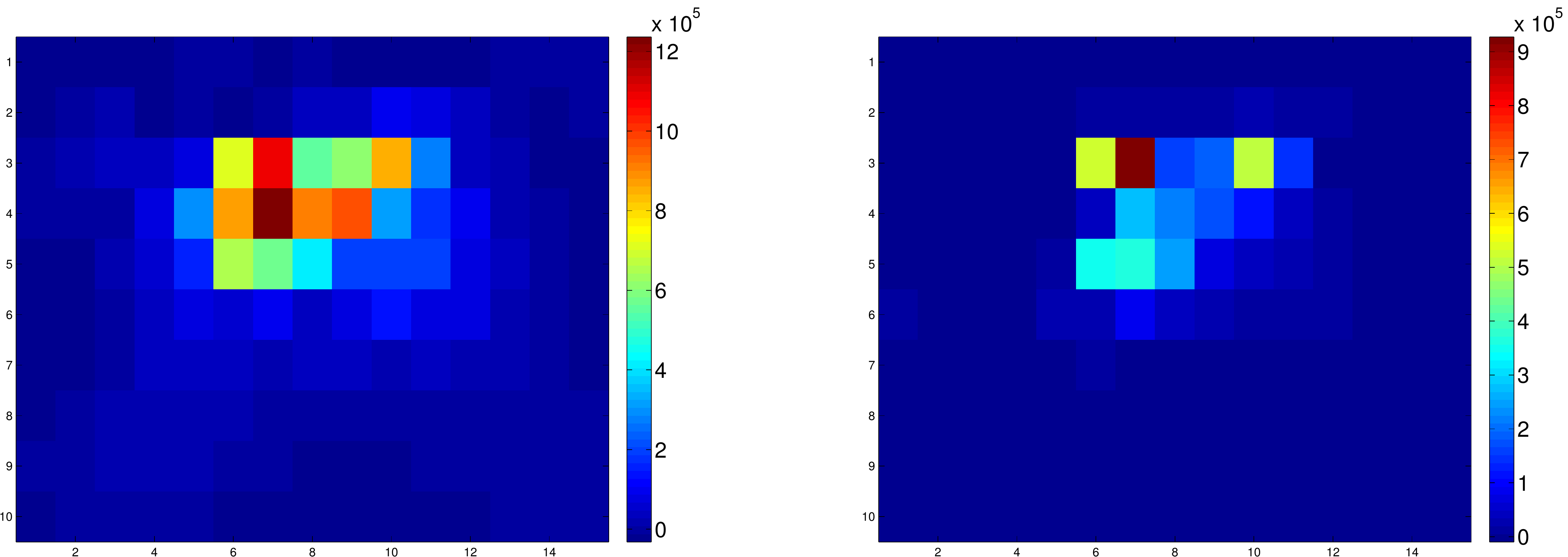}}\\
\subfloat[Changement de variable en $1/v_{p,s}^2$, initialisation aux caractéristiques de la terre]{
\includegraphics[width=4.3in]{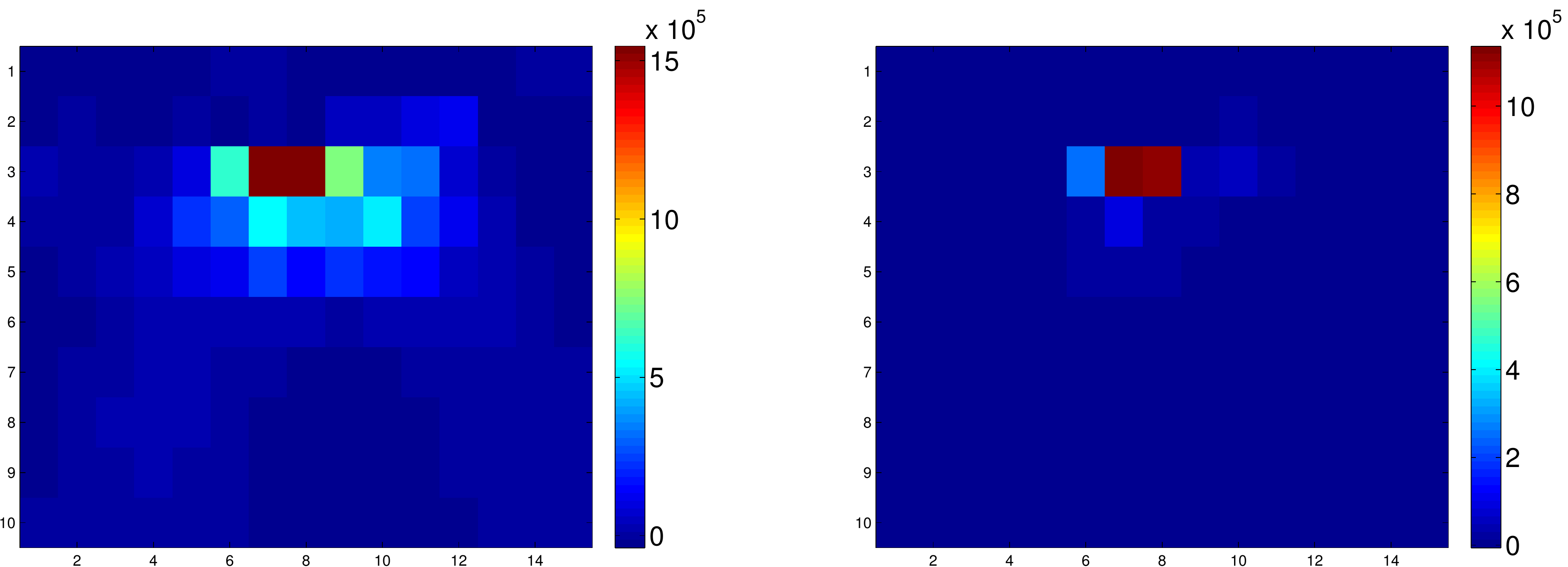}}\\
\subfloat[Changement de variable en $\ln(v_{p,s})$, initialisation aux caractéristiques de la terre]{
\includegraphics[width=4.3in]{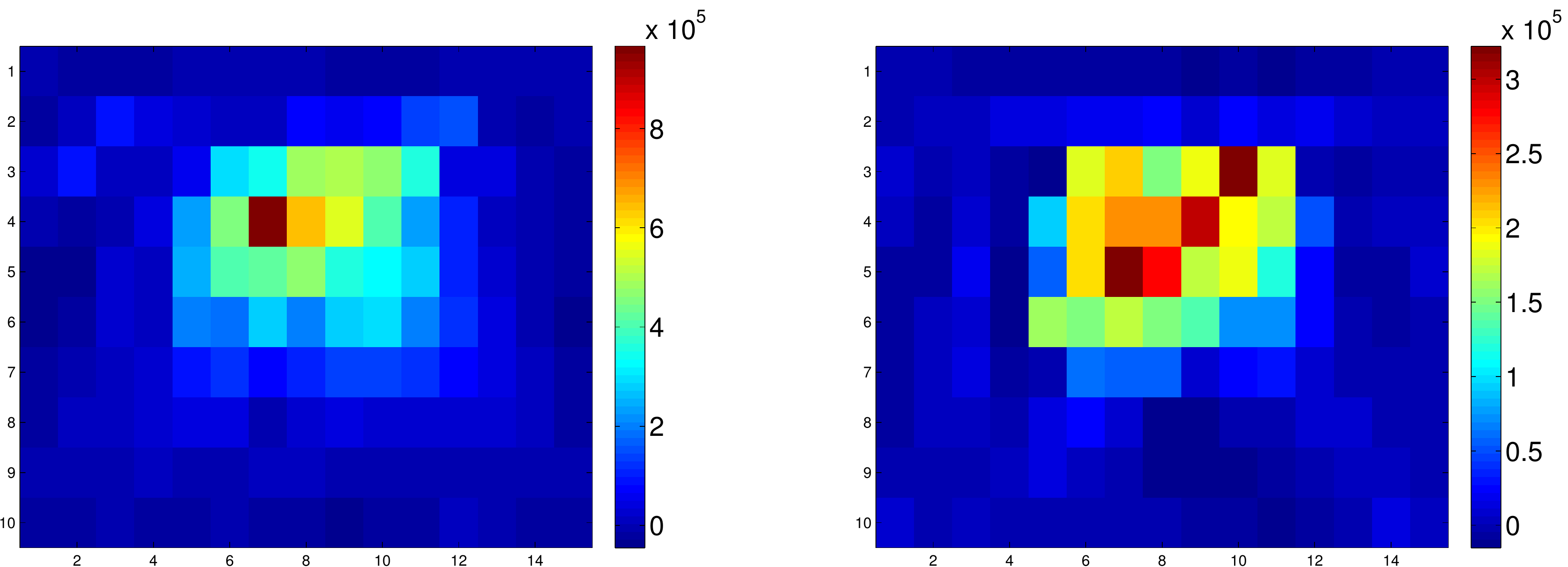}}
\caption{Résultats obtenus avec la méthode GM pour différents changements de variables et deux initialisations}
\label{figResultats_GM}
\end{figure}

\begin{figure}[!htb]
\centering
\includegraphics[width=4.3in]{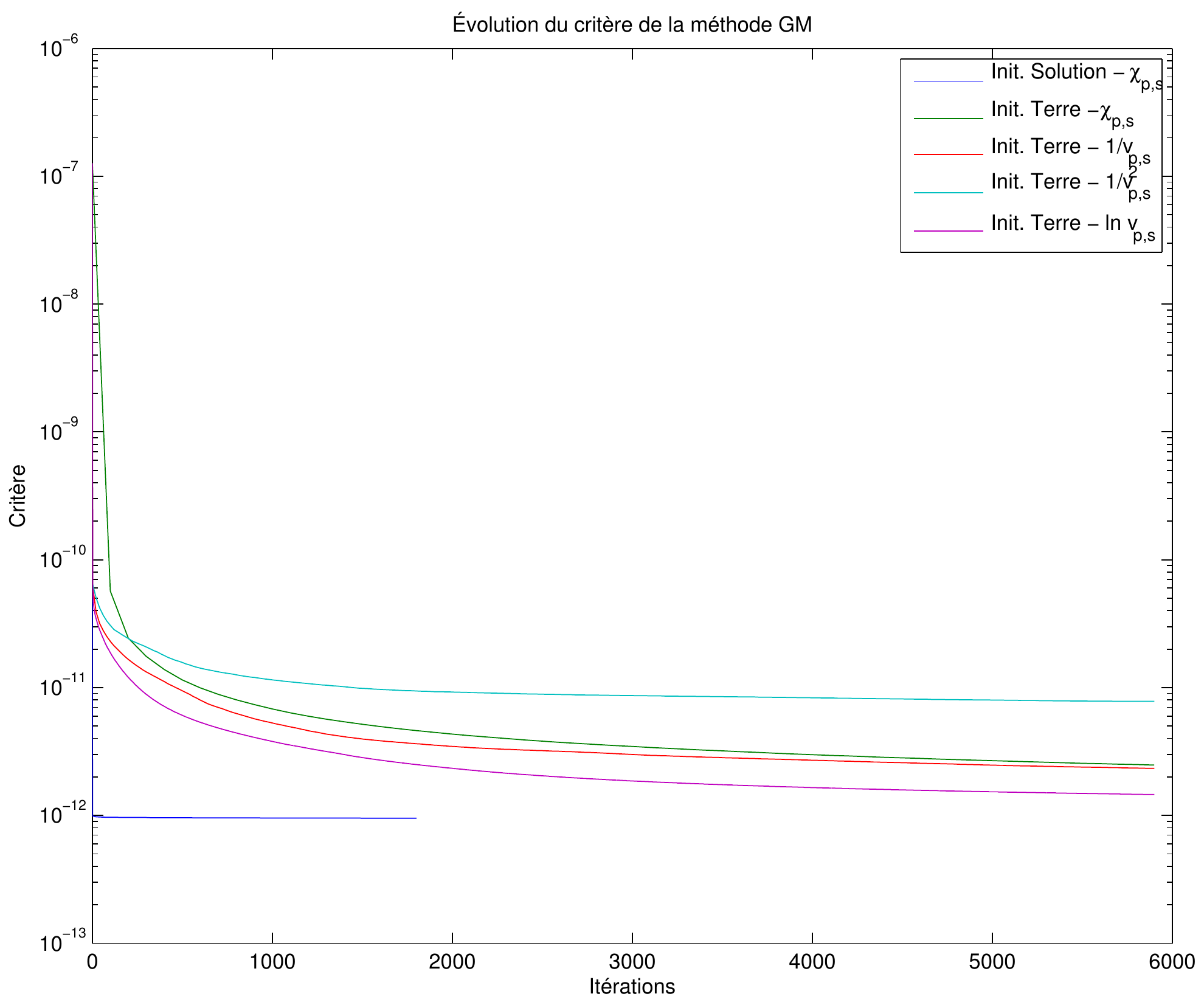}
\caption{Évolution du critère de la méthode GM pour différents changements de variables et deux initialisations}
\label{figCritere_GM}
\end{figure}

Les résultats de la méthode GM après $6000$ itérations sont montrés dans les figures~(\ref{figResultats_GM}) et (\ref{figCritere_GM}).
On observe que le critère de la méthode GM décroit très lentement, et que l'utilisation des changements de variables décrits dans la section~\ref{Part_IntroChgtVariable} influe peu sur cette situation.

Cependant, le changement de variable en $\ln(v_{p,s})$ donne la meilleure décroissance du critère, mais les cartes de contraste obtenues sans changement de variable semblent plus intéressantes, car c'est dans ces conditions que l'erreur quadratique moyenne par rapport aux contrastes de la solution est la plus faible.

\begin{figure}[ht]
\centering
\subfloat[Évolution du conditionnement de l'estimateur de la variable auxiliaire pour plusieurs fréquences]{\includegraphics[scale=0.35]{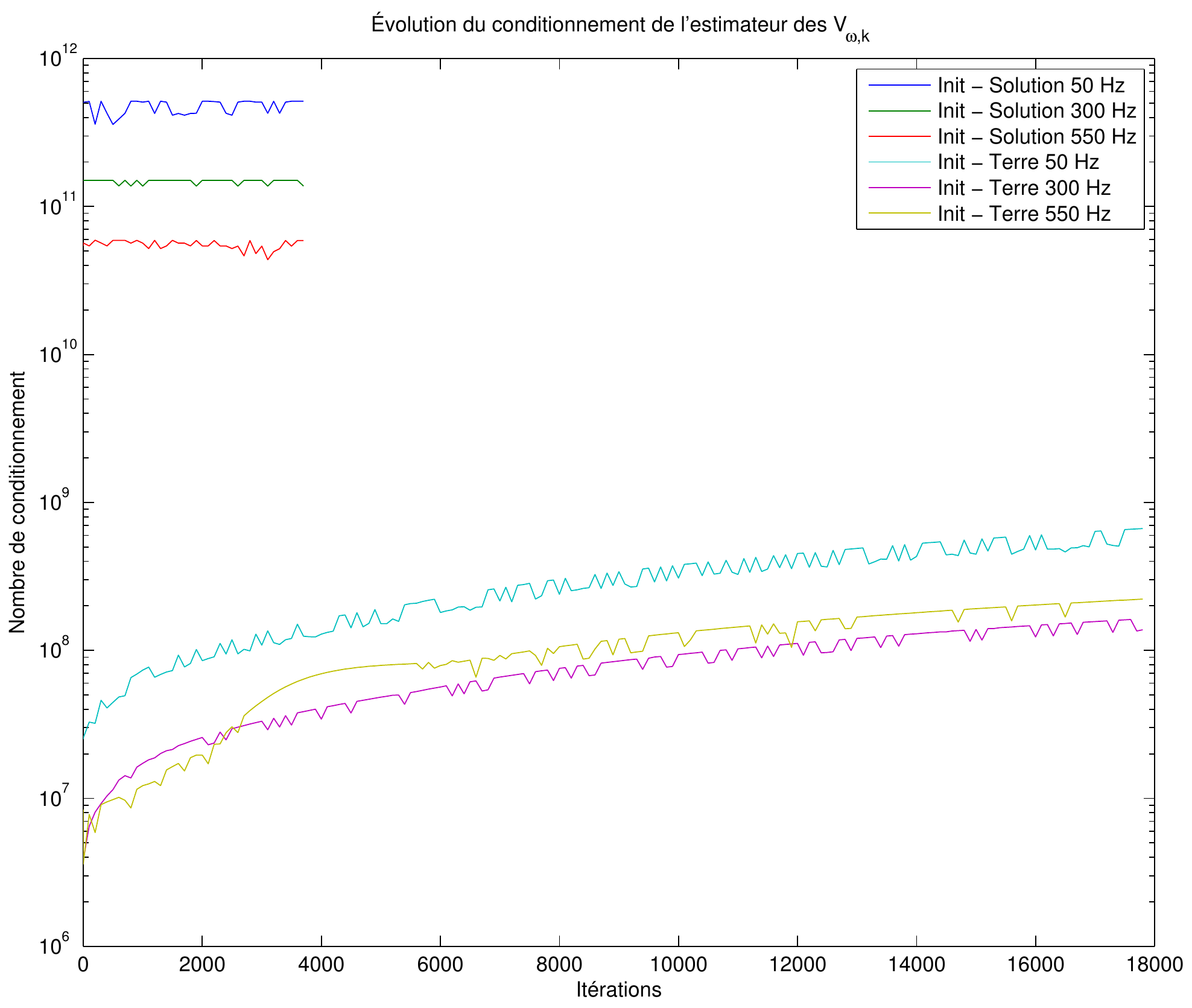}}
\subfloat[Évolution du conditionnement de l'estimateur des contrastes]{\includegraphics[scale=0.35]{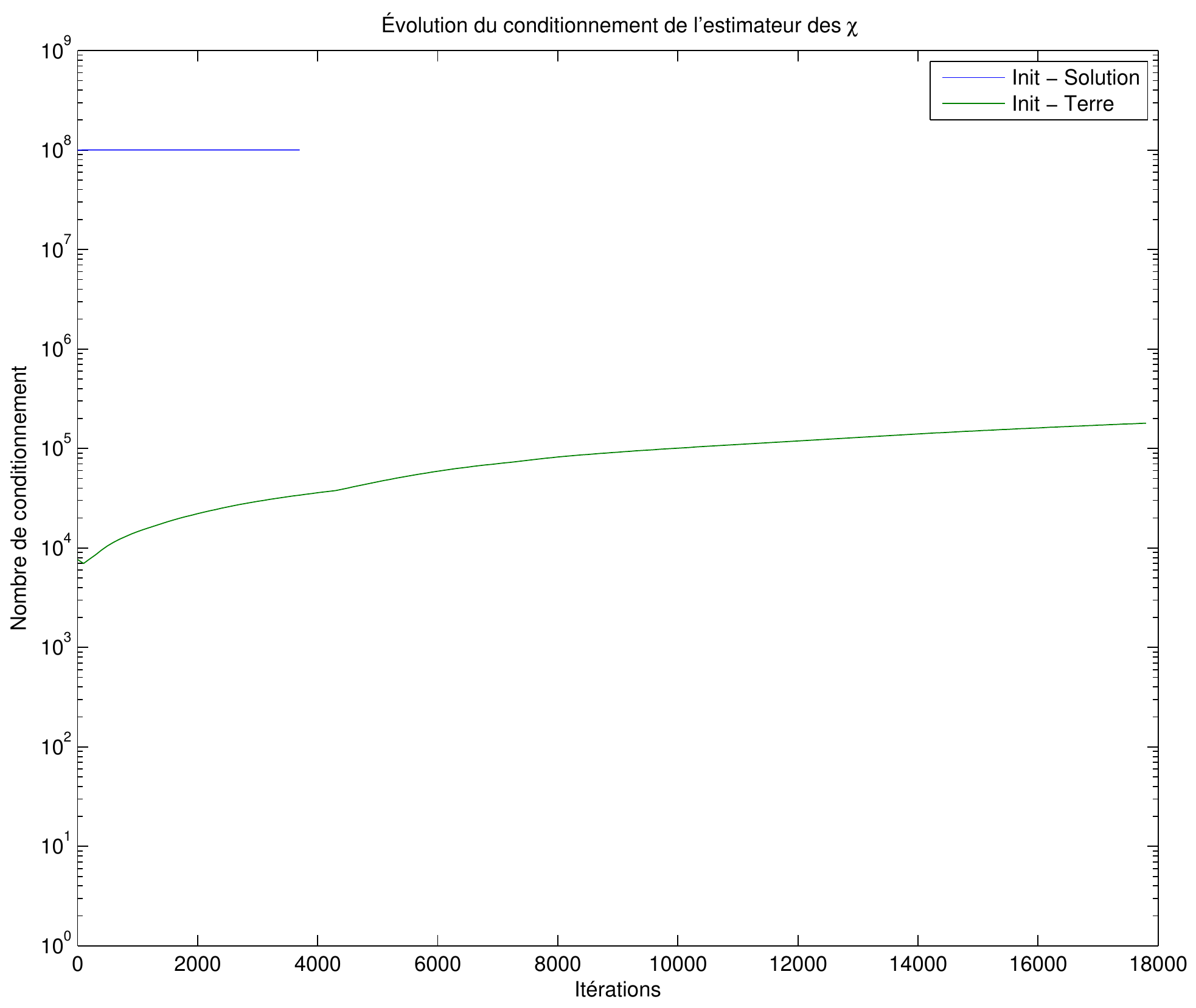}}
\caption{Évolution du conditionnement des estimateurs de la méthode GM en initialisant à la solution et à la terre}
\label{figConditionnement_GM}
\end{figure}

La figure~(\ref{figConditionnement_GM}) montre le conditionnement de
la matrice normale de l'estimateur du champ de vitesse. On remarque
que celui-ci se dégrade fortement après seulement quelques itérations. Le système est donc mal conditionné, ce qui pourrait expliquer la lenteur de convergence de la méthode GM.

La méthode GM a un comportement analogue à celui observé pour la méthode CSI en ce qui concerne la décroissance du critère, l'évolution du déplacement de la solution et le conditionnement des estimateurs. La méthode GM a donc le même problème de convergence que la méthode CSI ; elle s'avère néanmoins plus intéressante que cette dernière puisque le coût de calcul d'une itération est deux fois plus faible et que l'erreur quadratique moyenne atteinte après $17600$ itérations est plus petite (avec $90.32\%$ pour $\chi_p$ et $82.27\%$ pour $\chi_s$ contre respectivement $93.56\%$ et $90.43\%$ dans le cas de la méthode CSI).

\section{Méthode sans contraste}
La méthode bilinéaire que nous présentons dans cette section ne fait pas appel à un milieu de référence et repose sur l'estimation directe des caractéristiques du milieu. Pour cette raison nous l'avons appelée méthode sans contraste. Nous avons étudié cette méthode car elle permet de faire intervenir des formes algébriques simples dans le calcul des gradients.

\subsection{Description}
Cette méthode travaille directement sur les grandeurs physiques. Plus précisément :

\begin{description}
 \item[Les variables d'intérêt] $\mathit{x}_p$ et $\mathit{x}_s$ sont les carrés des vitesses des ondes P et S, $\mathit{V}_p^2$ et $\mathit{V}_s^2$, qui sont estimés sur tout le domaine $D$
 \item[Les variables auxiliaires] $\mathit{z}_{\omega,k}$ sont les composantes du champ de vitesse $\mathit{V}_{\omega,k}$ qui doivent être estimées sur tout le domaine $D$
\end{description}

L'équation de couplage utilisée dans ce modèle est exactement celle utilisée par le modèle direct, et l'équation d'observation correspond à un simple rééchantillonnage du champ des vitesses.

\begin{description}
\item [Équation d'observation] $\mathit{y}_{\omega,k} = \mathbf{E}_1 \mathit{V}_{\omega,k}$
\item [Équation de couplage] $\mathit{F}_{\omega,k} = \mathbf{A}_{\omega,p,s} \mathit{V}_{\omega,k}$
\end{description}
où $\mathit{F}_{\omega,k}$ est un vecteur formé par les composantes de la fonction source, tel que défini dans la section~\ref{Part_SystLinPbDirect}.

En reprenant les notations de la section~\ref{partIdeeGenerale}, on établit que :

\begin{align}
 \mathcal{H}(\mathit{V}_{\omega,k}) &= \mathit{F}_{\omega,k}  & \text{ avec } \mathbf{M}_c = 0 \text{ et } \mathit{u}_c = \mathit{F}_{\omega,k} \notag\\
 \mathcal{R}(\mathit{V}_{\omega,k}) &= \mathit{V}_{\omega,k}  & \text{ avec } \mathbf{M}_d = \indentit \text{ et } \mathit{u}_d = 0
\label{eq_DefNotationsSC} \\
 \mathbf{K}(\mathit{\chi}_p, \mathit{\chi}_s) &= \mathbf{A}_{\omega,p,s} & \notag
\end{align}

 En se référant à la section~\ref{part_CritereGeneral} et à partir des équations de couplage et d'observation propres à la méthode sans contraste, nous établissons le critère comme :

\begin{align*}
\mathcal{C} &= \sum_k \sum_\omega \| \mathit{y}_{\omega,k} - \mathbf{E}_1 \mathit{V}_{\omega,k} \|^2  + \gamma_c\sum_k \sum_\omega \| \mathit{F}_{\omega,k} - \mathbf{A}_\omega \mathit{V}_{\omega,k} \|^2 + \phi
\end{align*}

À partir de l'expression du critère, nous pouvons réécrire les formules des gradients donnés dans les équations~\eqref{eq_GradientGeneralz}, \eqref{eq_GradientGeneralxp} et \eqref{eq_GradientGeneralxs} en y substituant les équations~\eqref{eq_DefNotationsSC} et en tenant compte du choix des variables fait par cette méthode.

\begin{itemize}
\item Le gradient du critère par rapport à $\mathit{V}_{\omega,k}$ est :
\begin{align*}
\mathcal{G}_{\mathit{V}_{\omega,k}} = \nabla_{\mathit{V}_{\omega,k}} \mathcal{C}(\mathit{V}_{\omega,k})= &2((\gamma_c \mathbf{A}_{\omega,p,s}^{\dagger}\mathbf{A}_{\omega,p,s} + \mathbf{E}_1^\dagger\mathbf{E}_1 + \gamma^V_{r0})\mathit{V}_{\omega,k}\\
 & - (\mathbf{E}_1^\dagger\mathit{y}_{\omega,k} +\gamma_c( \mathbf{A}_{\omega,p,s}^{\dagger}\mathit{F}_{\omega,k})))\\
\end{align*}

\item Le gradient du critère par rapport à $\mathit{\chi}_p$ est :
\begin{align*}
\mathcal{G}_{\mathit{\chi}_p} = \nabla_{\mathit{\chi}_p} \mathcal{C}(\mathit{\chi}_p)= &
2(\gamma_c\sum_k \sum_\omega ((\mathbf{\Delta}_{\omega,k}^{p\dagger}\mathbf{\Delta}_{\omega,k}^p + \gamma_{r0}^\chi + \gamma_{r1}^\chi(\mathbf{D}_1+\mathbf{D}_2)^T(\mathbf{D}_1+\mathbf{D}_2))\mathit{V}_p\\
& -\mathbf{\Delta}_{\omega,k}^{p\dagger}(\mathit{F}_{\omega,k}-\mathbf{A}_{\omega}\mathit{V}_{\omega,k}-\mathbf{\Delta}_{\omega,k}^s\mathit{V}_s)))
\end{align*}
avec
\begin{align*}
\mathbf{\Delta}_{\omega,k}^p\mathit{V}_p&=\mathbf{A}_p\mathit{V}_{\omega,k}\\
\mathbf{\Delta}_{\omega,k}^s\mathit{V}_s&=\mathbf{A}_s\mathit{V}_{\omega,k}\\
\end{align*}
On obtient une expression du gradient semblable pour $\mathit{V}_s$, il suffit d'échanger les indices $p$ et $s$.
L'estimation des $\mathit{V}_p^2$ et $\mathit{V}_s^2$ est faite conjointement.
\end{itemize}

\section{Spécificités de la méthode sans contraste}

\textbf{Aspects numériques}

La méthode sans contraste est basée sur deux estimateurs très simples, et le calcul des gradients ne fait intervenir aucune résolution de système linéaire. Dans les deux cas, le Hessien peut être calculé, ce qui permet d'utiliser des méthodes de préconditionnement simples. Cependant, toutes les inconnues doivent être estimées sur tout le domaine, y compris les PML.
L'estimateur du champ de vitesse est exactement le même que pour la méthode GM. Cette méthode souffre donc des mêmes problèmes de conditionnement.

\bigskip
\textbf{Aspects techniques}

La méthode d'inversion bilinéaire la plus simple ne fait pas intervenir de milieu de référence. Elle consiste à travailler directement sur les variables d'intérêt et à utiliser les composantes du champs de vitesse $\mathit{V}_{\omega,k}$ comme ensemble de variables auxiliaires. L'avantage de cette méthode est que l'on s'affranchit de la difficulté de trouver un domaine de référence valable. Par contre, il faut estimer les deux jeux de variables sur tout le domaine, car elles sont \aprio connues en aucun point du domaine, y compris dans les PML. Les matrices $\mathbf{A}_p$ et $\mathbf{A}_s$ dépendent donc de la fréquence, et il faut stocker et mettre à jour à chaque itération deux matrices $\mathbf{A}_{\omega,p}$ et $\mathbf{A}_{\omega,s}$ pour chaque fréquence.

\subsection{Résultats}
L'estimateur des $\mathit{V}_s$ et $\mathit{V}_p$, lorsque testés séparemment en fixant les $\mathit{V}_{\omega,k}$ à la solution, se comporte bien et converge vers la solution en une centaine d'itérations du gradient conjugué, si l'on utilise un préconditionneur de Jacobi. Sans le préconditionneur de Jacobi, il faut compter autour de 100 fois plus d'itérations pour que cet estimateur converge vers la solution. Le critère de convergence atteint est la décroissance de la norme de la solution qui est inférieure à un seuil, et l'erreur quadratique moyenne obtenue après convergence est de l'ordre de $10^{-5}$.

L'estimateur $\hat{\mathit{V}}_{\omega,k}$ converge très lentement. La matrice $\mathbf{A}^\dagger_{\omega,p,s}\mathbf{A}_{\omega,p,s} + \mathbf{E}^{\dagger}_1\mathbf{E}_1$, qui intervient dans le calcul de cet estimateur, est mal conditionnée. Son nombre de conditionnement est de l'ordre de $10^{12}$. Un préconditionnement simple basé sur la diagonale du Hessien ne permet pas de résoudre le problème de mauvais conditionnement. Cependant, le problème de conditionnement peut être traité en utilisant la même approche que celle exposée dans la partie~\ref{part_GM}. Dans ce cas, l'estimateur des champs de vitesses converge rapidement vers la solution.

\begin{figure}[!htb]
\centering
\subfloat[Initialisation à la solution, variable $\mathit{\chi}_{p,s}$]{
\includegraphics[width=4.3in]{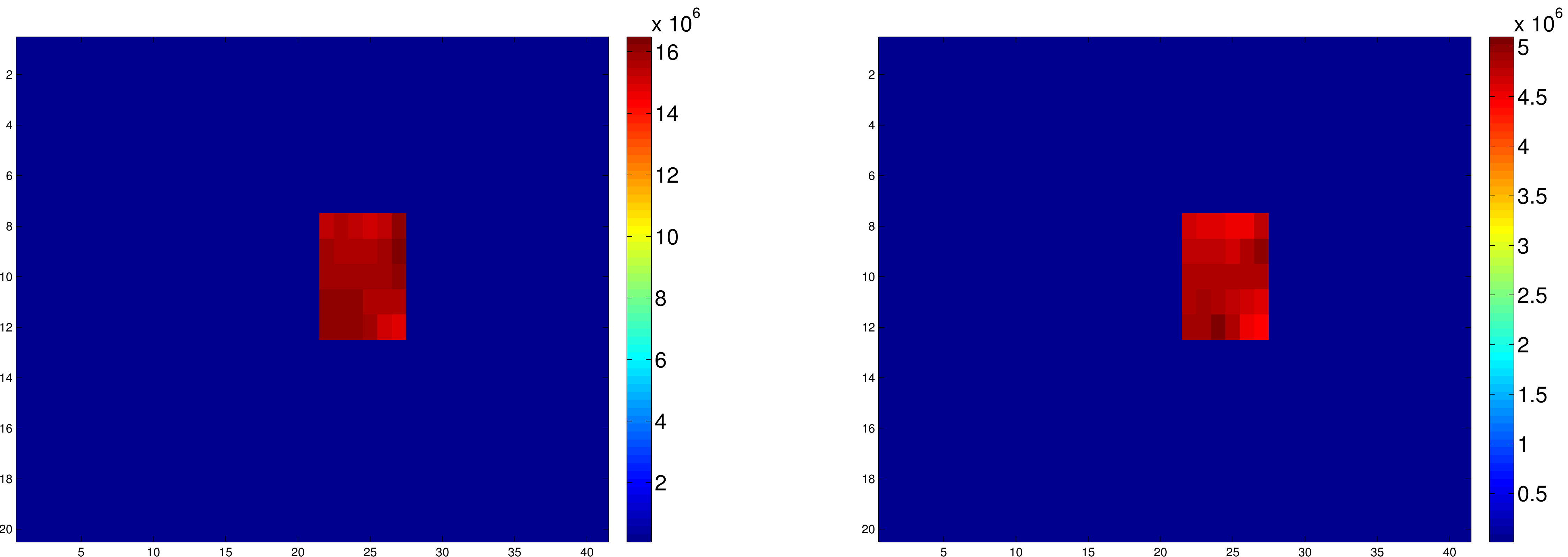}}\\
\subfloat[Initialisation à la terre, variable $\mathit{\chi}_{p,s}$]{
\includegraphics[width=4.3in]{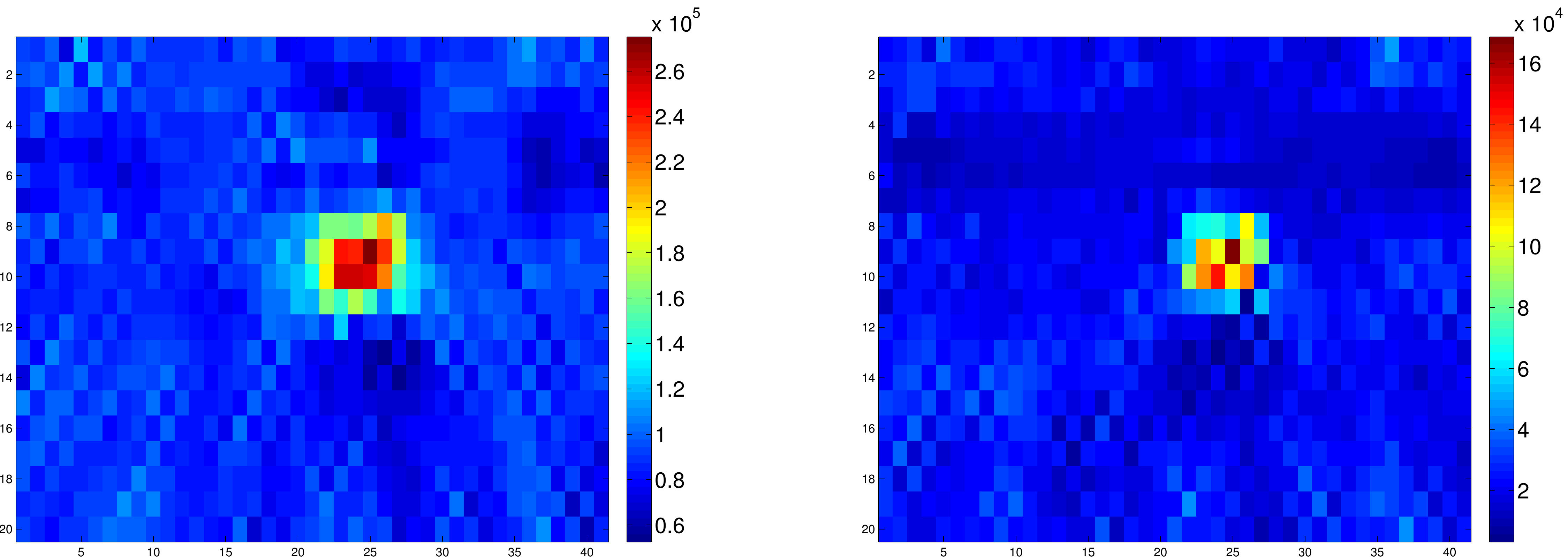}}
\caption{Résultats obtenus avec la méthode sans contraste}
\label{figResultats_basique}
\end{figure}

Les résultats présentés dans la figure~(\ref{figResultats_basique}) correspondent aux itérations $21000$ pour l'initialisation à la terre et $8200$ pour l'initialisation à la solution. Le calcul des variables d'intérêt se fait sur tout le domaine, donc les cartes des caractéristiques de la terre sont données pour tout le domaine sauf les PML.
\begin{figure}[!htb]
\centering
\includegraphics[width=4.3in]{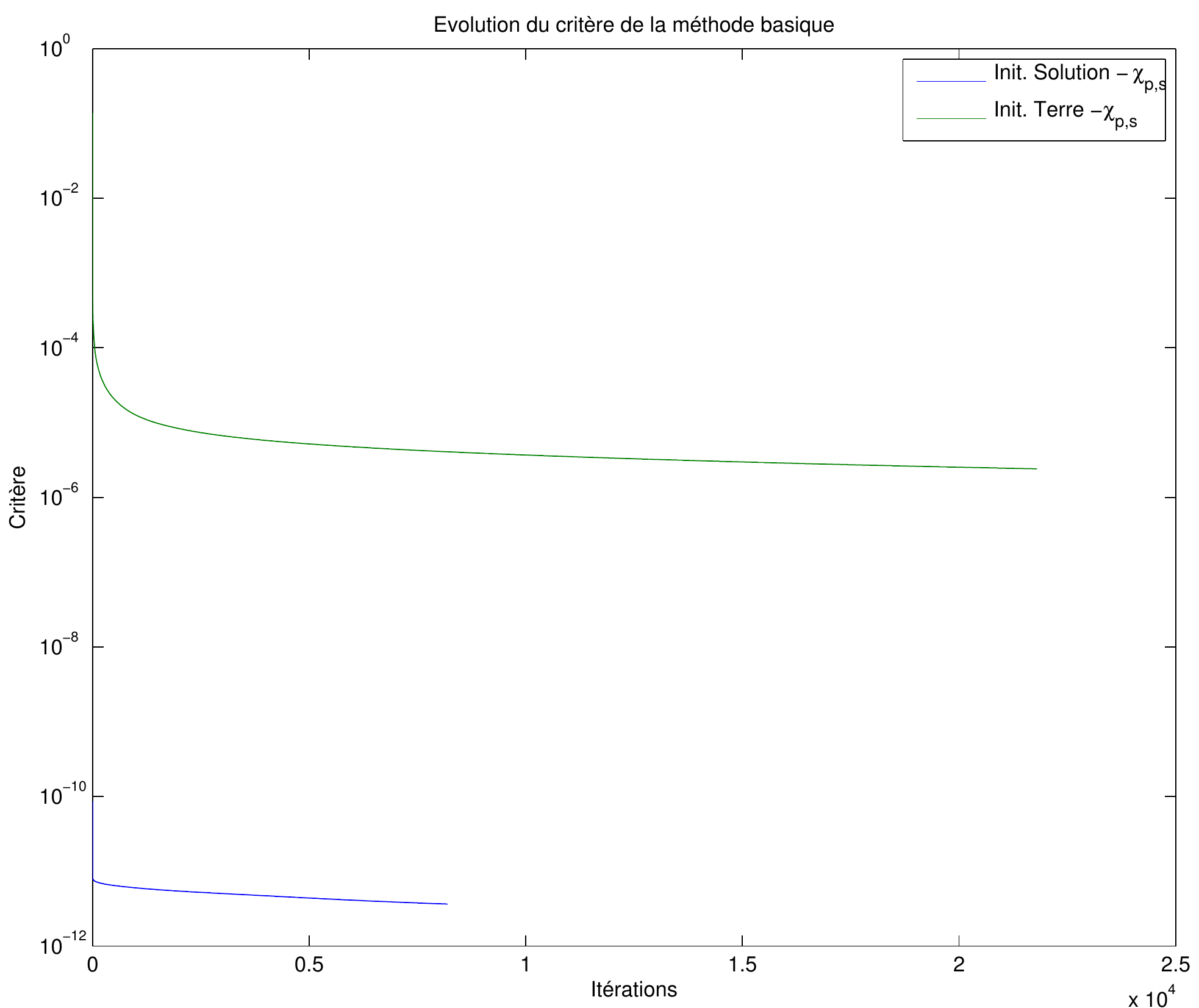}
\caption{Évolution du critère de la méthode sans contraste}
\label{figCritere_basique}
\end{figure}

Même si les deux estimateurs se comportent bien lorsqu'ils sont testés séparément, les résultats obtenus par cette méthode sont moins intéressants que ceux obtenus avec la méthode CSI. Après $21000$ itérations, l'amplitude maximale du signal reconstruit n'est que de $2,5 \times 10^{5}$ dans le cas de $\chi_p$ et $1.6 \times 10^{5}$ pour $\chi_s$, alors qu'on atteignait $2 \times 10^{6}$ et $1,5 \times 10^{6}$ dans le cas de la CSI. Il faut cependant signaler que le calcul d'une itération est très rapide, deux fois plus rapide que pour la CSI. On constate également, voir figure~(\ref{figCritere_basique}), que cette méthode souffre du même problème de lenteur de convergence que toutes les autres.

 \section{La méthode CFSI}
 La méthode CFSI \cite{Barriere08} (contraste field source inversion) est une version augmentée de la CSI dans laquelle on relâche la contrainte sur la définition des sources de contrastes,  $\mathit{W}_{\omega,k} = (\mathbf{X}_p + \mathbf{X}_s) \mathit{V}_{\omega,k}$, en incluant le terme d'erreur correspondant dans l'expression du critère. On introduit ainsi un jeu de variables auxiliaires supplémentaire : les composantes du champ de vitesses. L'estimation de ces composantes ne peut être retreinte à une zone d'étude, car, à l'exception du tour du domaine sur lequel elles doivent satisfaire les contraintes de bord de l'équation~\eqref{eqContraintesDomaine}, elles ont une valeur non nulle en tout point du domaine. Du fait de l'introduction d'une nouvelle contrainte et d'un nouveau jeu de variables, cette méthode ne rentre pas tout à fait dans le cadre général défini dans la section~\ref{partIdeeGenerale}, mais l'approche employée pour résoudre le problème et les choix techniques de la mise en \oe{uvre} sont analogues.

 \subsection{Description}
Trois jeux de variables doivent être estimés :

\begin{description}
\item [Les variables d'intérêt] $\mathit{x}_p$ et $\mathit{x}_s$ sont les contrastes $\mathit{\chi}_p$ et $\mathit{\chi}_s$ des vitesses $\mathit{V}_p^2$ et $\mathit{V}_s^2$ qui sont estimés sur une partie réduite du domaine $D$, et ont été définis dans la section~\ref{partDef_contrastes}.
\item[Les variables auxiliaires] $\mathit{z}_{\omega,k}$ sont les sources de contraste qui sont estimées sur une zone restreinte du domaine $D$, et ont été définies dans la section~\ref{partDef_contrastes}
\item[Les variables auxiliaires supplémentaires] sont les composantes du champ de vitesses $\mathit{V}_{\omega,k}$ qui sont à estimer sur tout le domaine $D$
\end{description}

La CFSI repose sur trois jeux d'équations qui se déduisent directement de la formulation CSI, l'adéquations aux données et deux contraintes :
\begin{description}
\item [Équation d'observation] $\mathit{y}_{\omega,k} = \mathbf{E}_1\mathit{V}_{\omega,k}^0 - \mathbf{B}_\omega^d \mathit{W}_{\omega,k}$
\item [Équation de couplage] $\mathit{V}_{\omega,k} = \mathit{V}_{\omega,k}^0 - \mathbf{B}_\omega^c \mathit{W}_{\omega,k}$
\item [Contrainte sur les sources de contraste] $\mathit{W}_{\omega,k} = (\mathbf{X}_p + \mathbf{X}_s) \mathit{V}_{\omega,k}$
\end{description}


Cette méthode diffère des autres méthodes bilinéaires par l'introduction d'un quatrième terme dans l'expression du critère qui correspond à l'erreur quadratique sur la définition des sources de contraste. À partir de ces équations on forme donc le critère suivant :
\begin{align*}
\mathcal{C} &= \sum_k \sum_\omega \| \mathit{y}_{\omega,k} - [\mathbf{E}_1\mathit{V}_{\omega,k}^0 - \mathbf{B}_\omega^d (\mathit{W}_{\omega,k})] \|^2  \\
& + \gamma_c\sum_k \sum_\omega \| \mathit{V}_{\omega,k} - \mathit{V}_{\omega,k}^0 + \mathbf{B}_\omega^c \mathit{W}_{\omega,k} \|^2 \\
& + \gamma_w\sum_k \sum_\omega \| \mathit{W}_{\omega,k} - (\mathbf{X}_p + \mathbf{X}_s) \mathit{V}_{\omega,k}\|^2 + \gamma_r \phi
\end{align*}
Cette méthode entraîne l'introduction d'un nouvel hyperparamètre $\gamma_w$ servant à pondérer l'importance de l'erreur sur l'équation des sources de contraste. Comme pour les hyperparamètres des méthodes précédentes, celui-ci est estimé empiriquement.

 La minimisation du critère de la CFSI par rapport à chaque variable nécessite de formuler l'expression des gradients par rapport à chacun des trois jeux d'inconnues :
\begin{itemize}
\item Gradient du critère par rapport à $\mathit{V}_{\omega,k}$ :
\begin{align*}
\mathcal{G}_{\mathit{V}_{\omega,k}} = \nabla_{\mathit{V}_{\omega,k}} \mathcal{C}(\mathit{V}_{\omega,k})&=2((\gamma_c \indentit + \gamma_w(\mathbf{X}_p+\mathbf{X}_s)^{\dagger}(\mathbf{X}_p+\mathbf{X}_s))\mathit{V}_{\omega,k}\\
 & - (\gamma_w(\mathbf{X}_p+\mathbf{X}_s)^{\dagger}\mathit{W}_{\omega,k}+\gamma_c(\mathit{V}_{\omega,k}^0 - \mathbf{B}_\omega^c \mathit{W}_{\omega,k})))
\end{align*}
\item Gradient du critère par rapport à $\mathit{W}_{\omega,k}$ :
\begin{align*}
\mathcal{G}_{\mathit{W}_{\omega,k}} = \nabla_{\mathit{W}_{\omega,k}} \mathcal{C}(\mathit{W}_{\omega,k})&= 2((\gamma_c \mathbf{B}_\omega^{c\dagger}\mathbf{B}_\omega^c + \gamma_w \indentit + \mathbf{B}_\omega^{d\dagger}\mathbf{B}_\omega^d)\mathit{W}_{\omega,k}\\
 & - (\gamma_w(\mathbf{X}_p+\mathbf{X}_s)\mathit{V}_{\omega,k}-\gamma_c\mathbf{B}_\omega^{c\dagger}(\mathit{V}_{\omega,k} - \mathit{V}_{\omega,k}^0) - \mathbf{B}_\omega^{d\dagger}(\mathit{y}_{\omega,k} - \mathbf{E}_1\mathit{V}_{\omega,k}^0)))\\
\end{align*}
\item Gradient du critère par rapport à $\mathit{\chi}_p$ :
\begin{align*}
\mathcal{G}_{\mathit{\chi}_p} = \nabla_{\mathit{\chi}_p} \mathcal{C}(\mathit{\chi}_p)&= 2(\gamma_w \sum_k \sum_\omega \mathbf{\Delta}_{\omega,k}^{p\dagger}\mathbf{\Delta}_{\omega,k}^p\mathit{\chi}_p - \mathbf{\Delta}_{\omega,k}^{p\dagger}(\mathit{W}_{\omega,k} - \mathbf{\Delta}_{\omega,k}^s\mathit{\chi}_s))
\end{align*}
avec
\begin{align*}
\mathbf{\Delta}_{\omega,k}^p\mathit{\chi}_p&=\mathbf{X}_p\mathit{V}_{\omega,k}\\
\mathbf{\Delta}_{\omega,k}^s\mathit{\chi}_s&=\mathbf{X}_s\mathit{V}_{\omega,k}\\
\end{align*}
L'expression du gradient du critère de la méthode CFSI par rapport à $\mathit{\chi}_s$ est obtenu en interchangeant les indices $p$ et $s$.
L'estimation des $\mathit{\chi}_p$ et $\mathit{\chi}_s$ est faite conjointement.
\end{itemize}

\subsection{Spécificités de la méthode CFSI}

\textbf{Aspects numériques}

L'intérêt de la CFSI par rapport à la CSI est qu'elle permet d'avoir une structure algébrique du calcul des $\hat{\mathit{W}}_{\omega,k}$ diffèrente. Cependant, cela nécessite l'introduction de nouvelles variables, ce qui ajoute un nombre important d'inconnues au problème. On se trouve ainsi devoir estimer les contrastes et sources de contraste sur la zone d'intérêt, et les composantes du champ de vitesse sur tout le domaine.

Il est néanmoins important de souligner que l'estimateur $\hat{\mathit{V}}_{\omega,k}$ a pour avantage de ne faire intervenir l'inverse d'aucune matrice. Le hessien peut donc être calculé facilement, ce qui permet de résoudre le système à l'aide d'une décomposition LU, ou d'utiliser un algorithme de type gradient conjugué préconditionné. Le nouveau jeu de variables introduit par la méthode CFSI n'ajoute donc pas une charge de calcul trop importante puisque celui-ci peut s'estimer de façon relativement simple et rapide.

La complexité du calcul de $\hat{\mathit{W}}_{\omega,k}$ est sensiblement la même que pour la CSI, mais fait intervenir des structures algébriques différentes. L'estimateur des contrastes de la CFSI est très semblable à celui de la méthode CSI, sauf pour ce qui est du calcul des matrices $\mathbf{\Delta}_{\omega,k}^p$ et $\mathbf{\Delta}_{\omega,k}^s$.

\begin{figure}[!htb]
\centering
\subfloat[Conditionnement de $\hat{V}_{\omega,k}$]{
\includegraphics[width=3.8in]{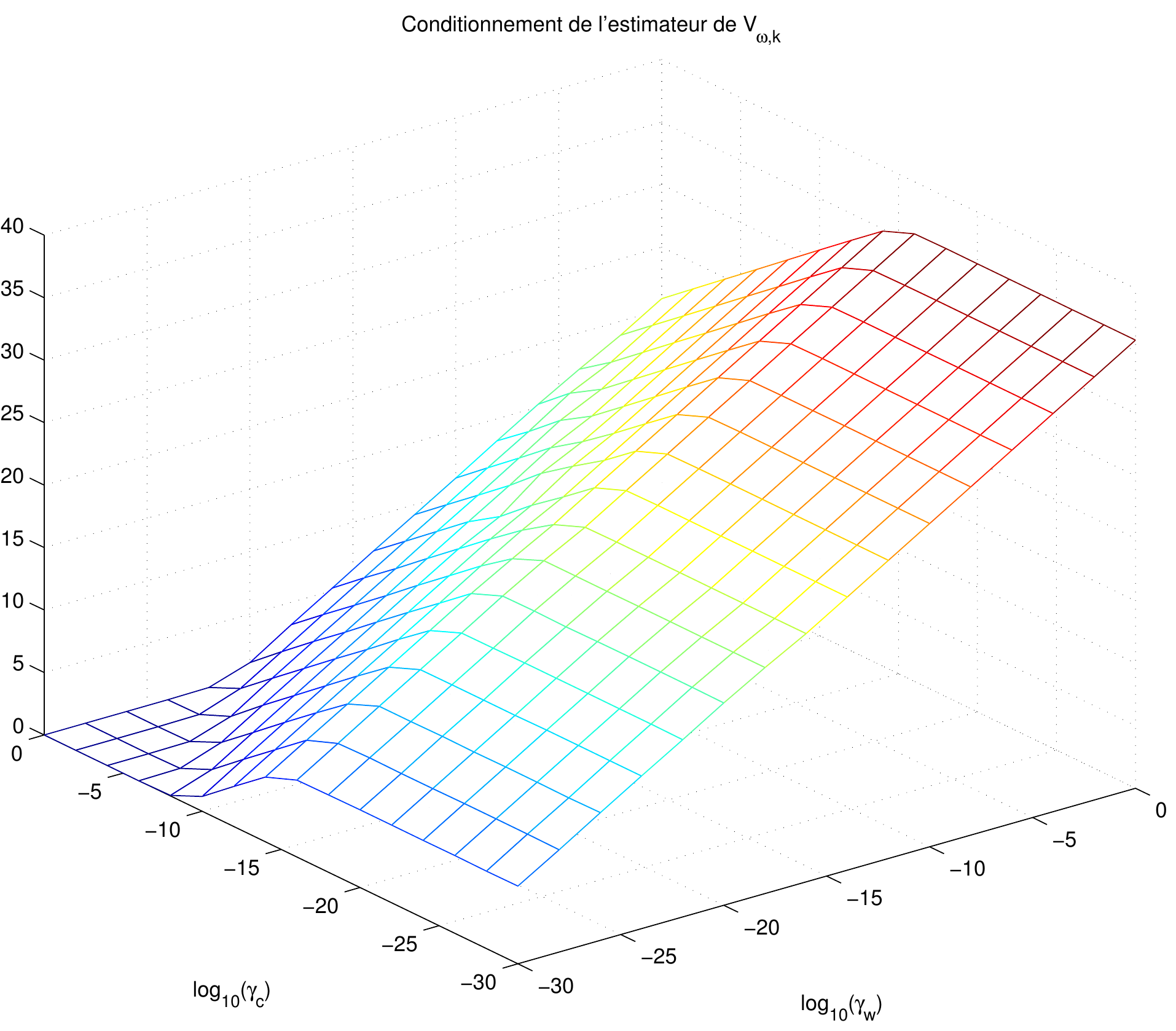}}\\
\subfloat[Conditionnement de $\hat{W}_{\omega,k}$]{
\includegraphics[width=3.8in]{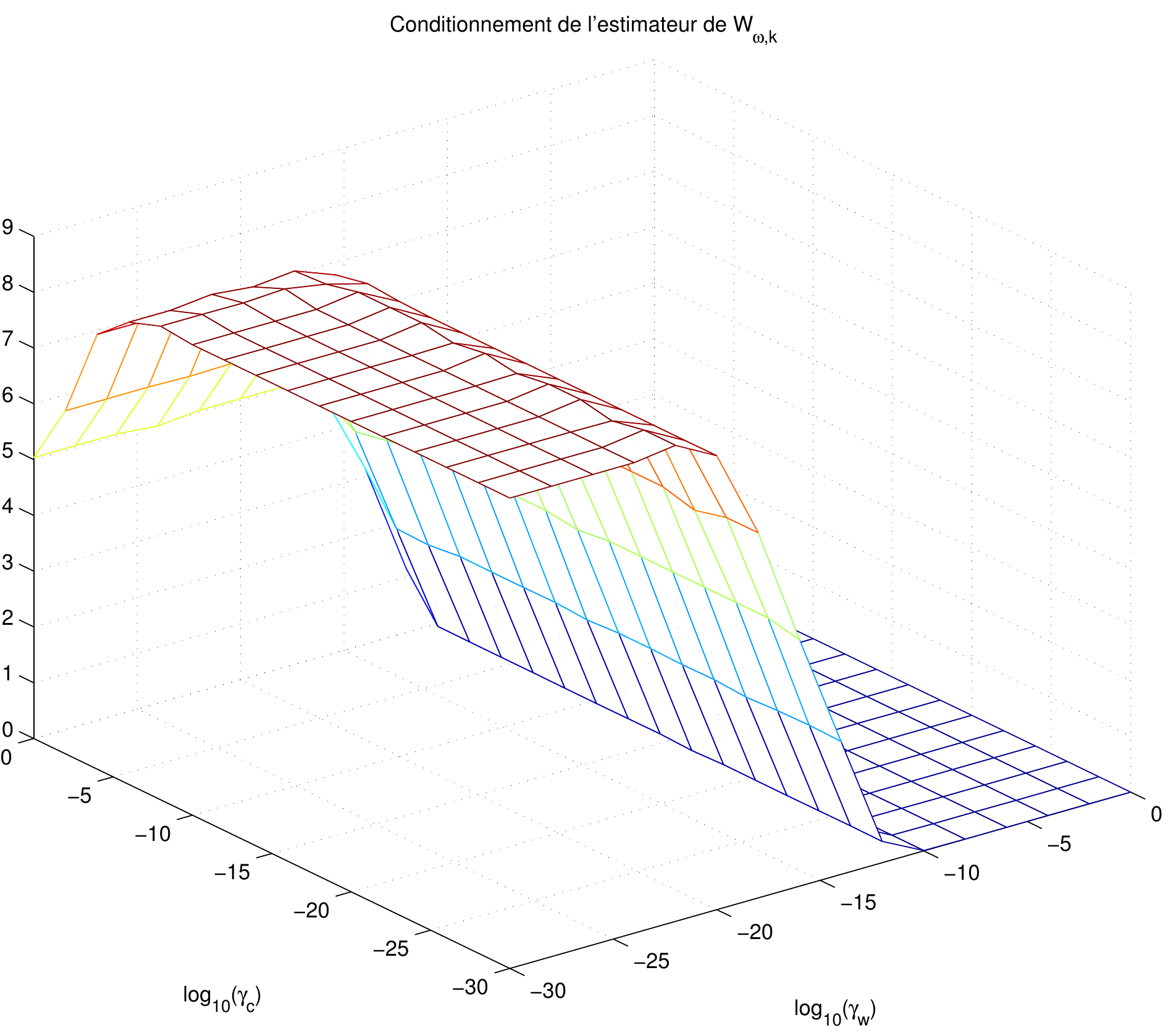}}
\caption{Logarithme du conditionnement des matrices normales en fonction des hyperparamètres}
\label{figConditionnement_CFSI}
\end{figure}

Le principal inconvénient de la CFSI est qu'elle introduit un hyperparamètre supplémentaire qu'il est difficile de fixer empiriquement. Si on teste séparemment chaque estimateur, \cad en initialisant avec la solution des deux autres variables, alors $\mathit{\hat{W}}_{\omega,k}$, $\mathit{\hat{\chi}}_p$ et $\mathit{\hat{\chi}}_s$ convergent rapidement vers la solution. Cependant, le conditionnement des matrices normales de ces estimateurs dépend directement du choix des hyperparamètres. Par exemple, le calcul de $\mathit{V}_{\omega,k}$ fait intervenir la matrice normale $\gamma_c \indentit + \gamma_w(\mathbf{X}_p + \mathbf{X}_s)^{\dagger}(\mathbf{X}_p + \mathbf{X}_s)$ dont le conditionnement est de l'ordre de $10^{20}$ si on utilise les contrastes solution pour construire les matrices $\mathbf{X}_s$ et $\mathbf{X}_p$ et que les hyperparamètres sont égaux à $1$. Dans le même cas de figure, la matrice normale de $\mathit{\hat{W}}_{\omega,k}$ est, quant-à elle, bien conditionnée. Les courbes de la figure~(\ref{figConditionnement_CFSI}) montrent le logarithme du conditionnement des matrices normales des estimateurs $\mathit{V}_{\omega,k}$ et $\mathit{W}_{\omega,k}$ en fonction des valeurs des hyperparamètres $\gamma_c$ et $\gamma_w$. On peut voir que le conditionnement de ces deux matrices varie fortement suivant la valeur des hyperparamètres.

D'une manière générale, il est préférable que $\gamma_w$ soit inférieur à $\gamma_c$. Le choix des ces paramètres résulte d'un compromis entre le conditionnement des deux matrices normales. Les deux ne peuvent être simultanément bien conditionnées, il faut donc choisir les hyperparamètres pour que le conditionnement de chacune des matrices ne soit pas trop mauvais et que la méthode ne diverge pas au bout d'un certain nombre d'itérations. Les tests empiriques ont montré que l'intervalle de valeurs qu'ils peuvent prendre est relativement étroit et qu'en dehors de cet intervalle la méthode diverge ou converge excessivement lentement. Si $\gamma_c$ est choisi dans l'intervalle $[10^{-5}, 1]$ et $\gamma_w$ dans $[10^{-18}, 10^{-10}]$, alors la méthode semble converger.

L'utilisation d'un rappel à zéro comme fonction de régularisation permet, dans une certaine mesure, de réduire le problème de conditionnement. Cependant, le poids que l'on peut donner à ce terme de rappel à zéro est limité par la dégradation qu'il introduit sur la solution. L'amélioration du conditionnement ainsi obtenue est donc limitée.

\subsection{Résultats}
\begin{figure}[!htb]
 \centering
 \subfloat[Initialisation à la solution, variable $\mathit{\chi}_{p,s}$]{\includegraphics[width=4.3in]{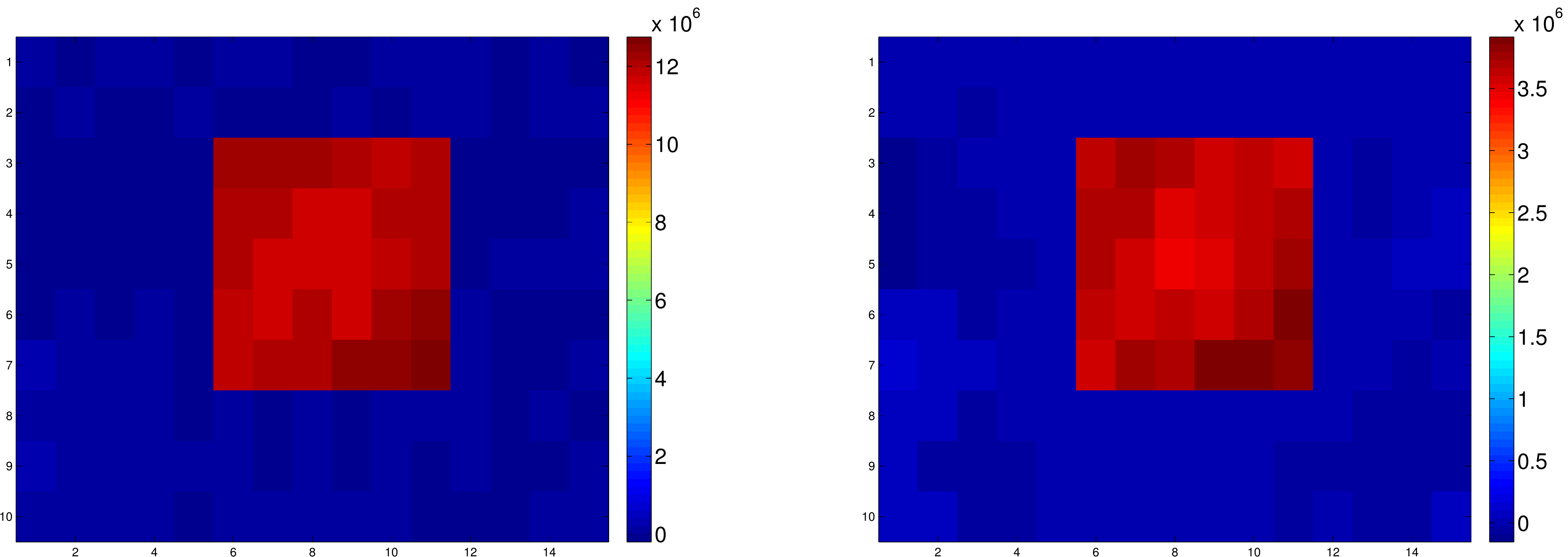}}\\
 \subfloat[Initialisation aux caractéristiques de la terre, variable $\mathit{\chi}_{p,s}$]{\includegraphics[width=4.3in]{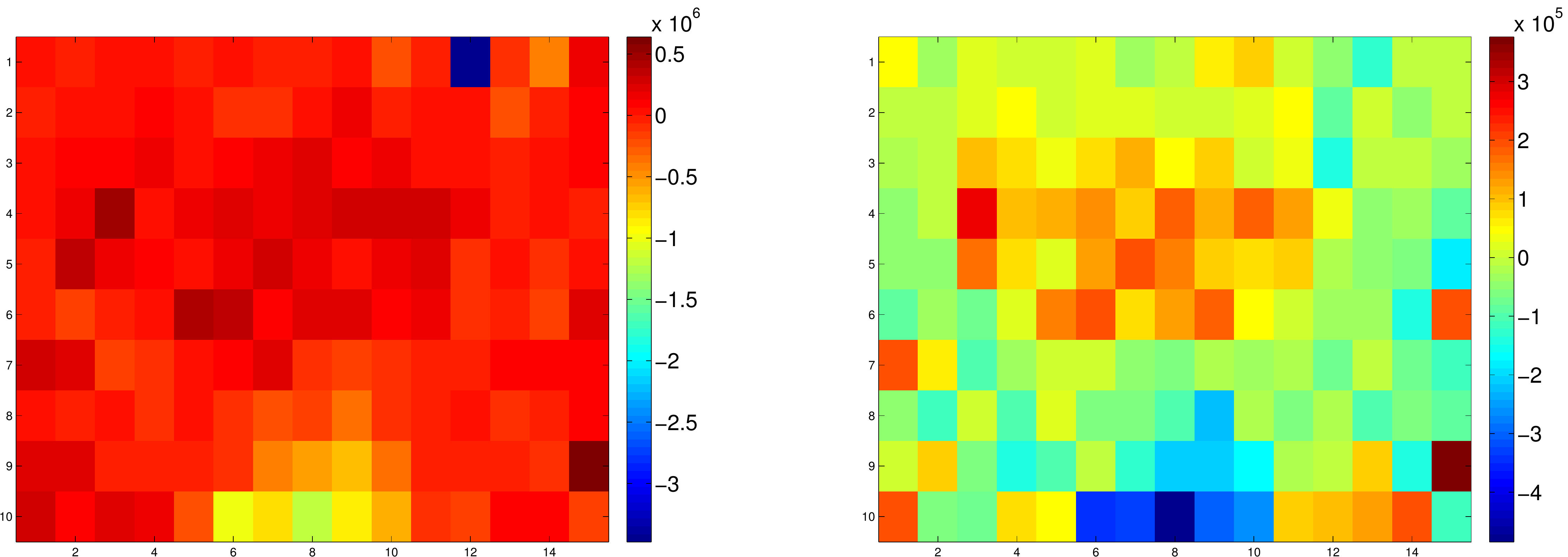}}
 \caption{Résultats de la méthode CFSI}
 \label{figResultat_CFSI}
 \end{figure}

\begin{figure}[!htb]
 \centering
 \includegraphics[width=4.3in]{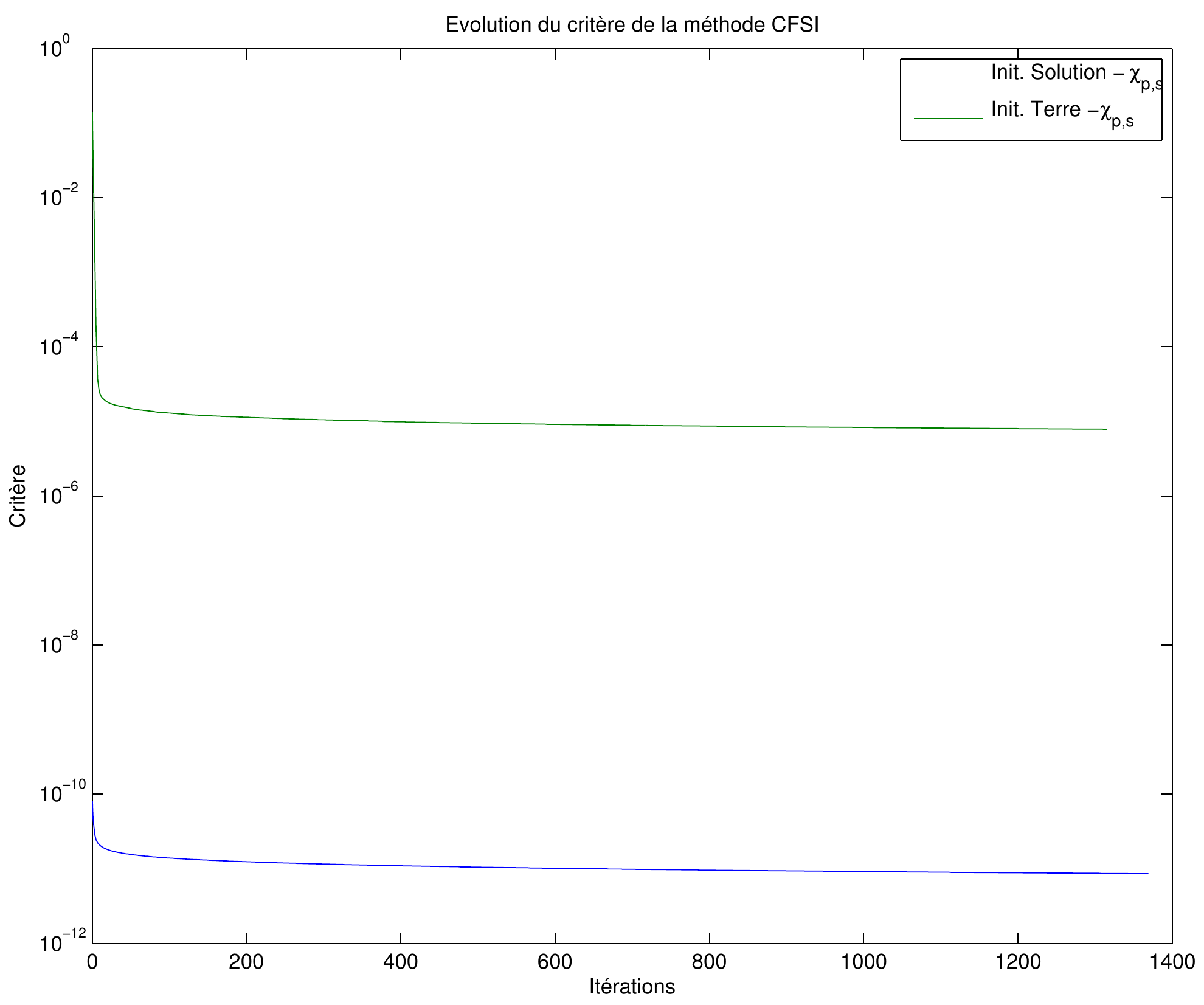}
 \caption{Évolution du critère de la méthode CFSI}
 \label{figCritere_CFSI}
 \end{figure}

Les résultats obtenus pour la méthode CFSI après $1380$ itérations sont présentés aux figures~(\ref{figResultat_CFSI}) et (\ref{figCritere_CFSI}).

 Les cartes de la figure (\ref{figResultat_CFSI}) montrent qui existe un certain nombre de pixels possédant une forte amplitude négative qui rendent difficile l'appréciation des résultats.

 Une fois de plus, nous constatons que l'évolution de la solution est lente. De plus, le temps de calcul d'une itération est deux fois plus important que pour la CSI.

 \subsection{Inversion en multifréquence}

\begin{figure}[!b]
\centering
\subfloat[Initialisation avec la solution]{
\includegraphics[width=4.3in]{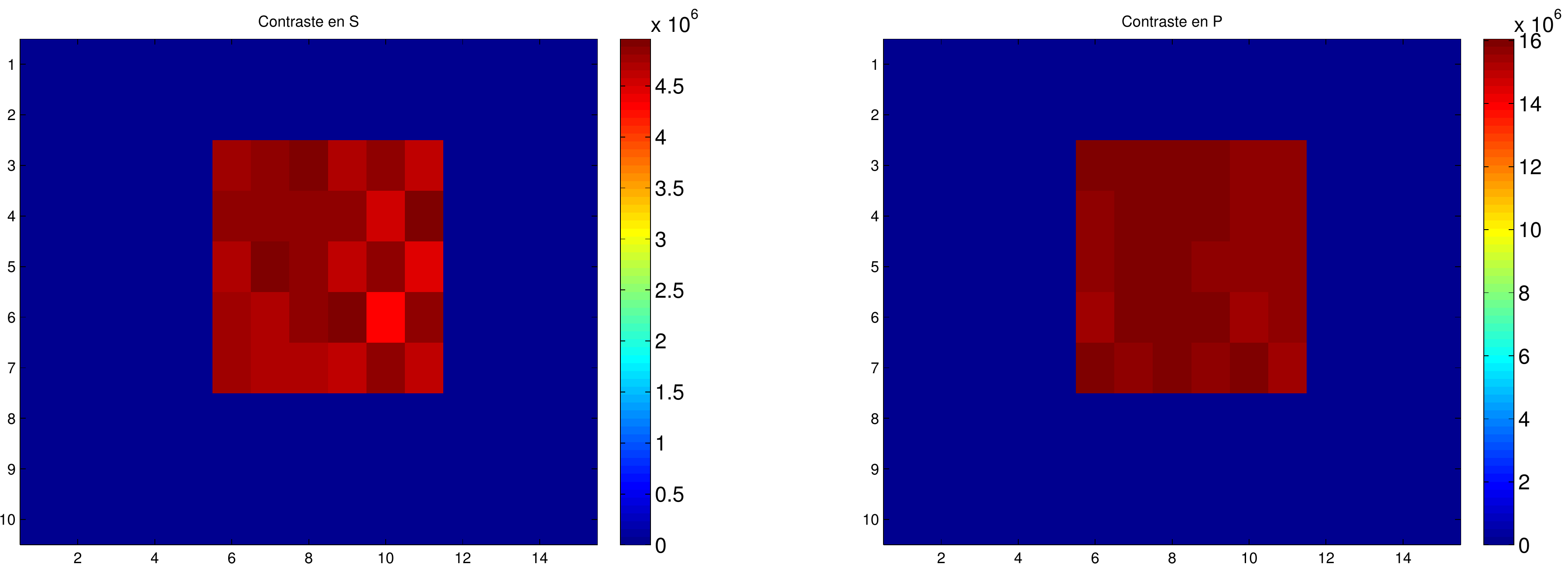}}\\
\subfloat[Initialisation à la terre]{
\includegraphics[width=4.3in]{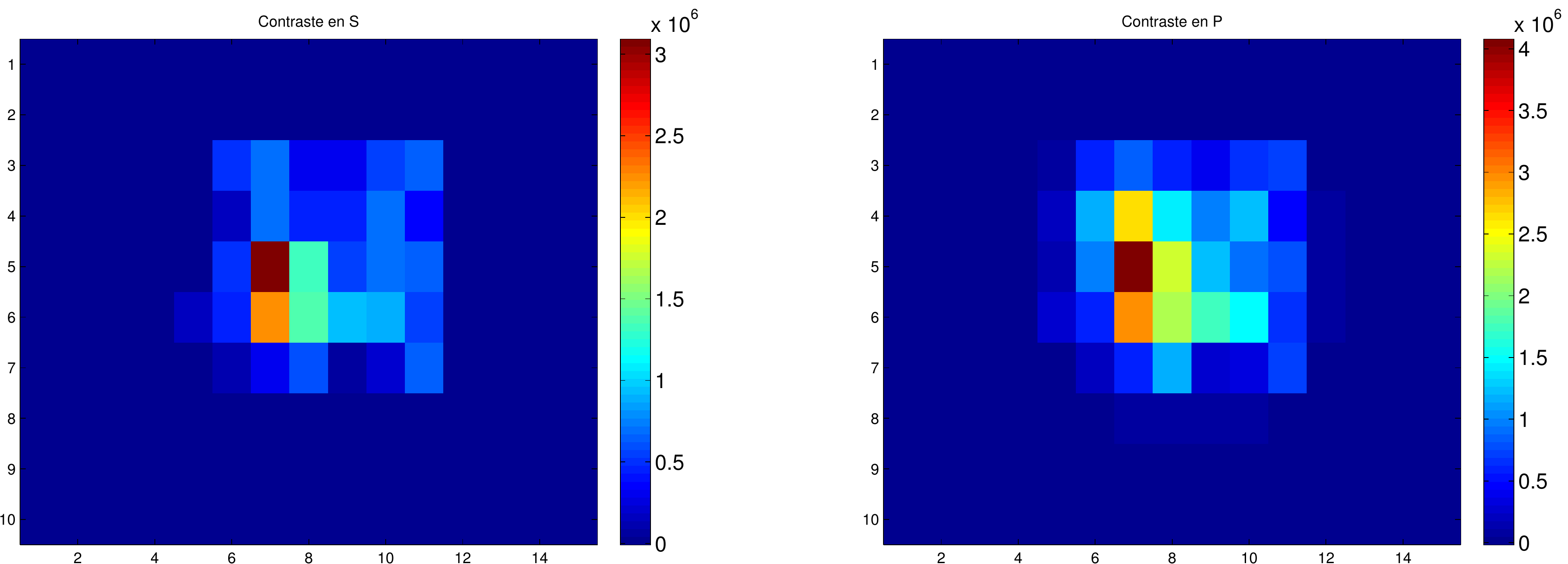}}\\
\subfloat[Évolution du critère]{
\includegraphics[width=4.3in]{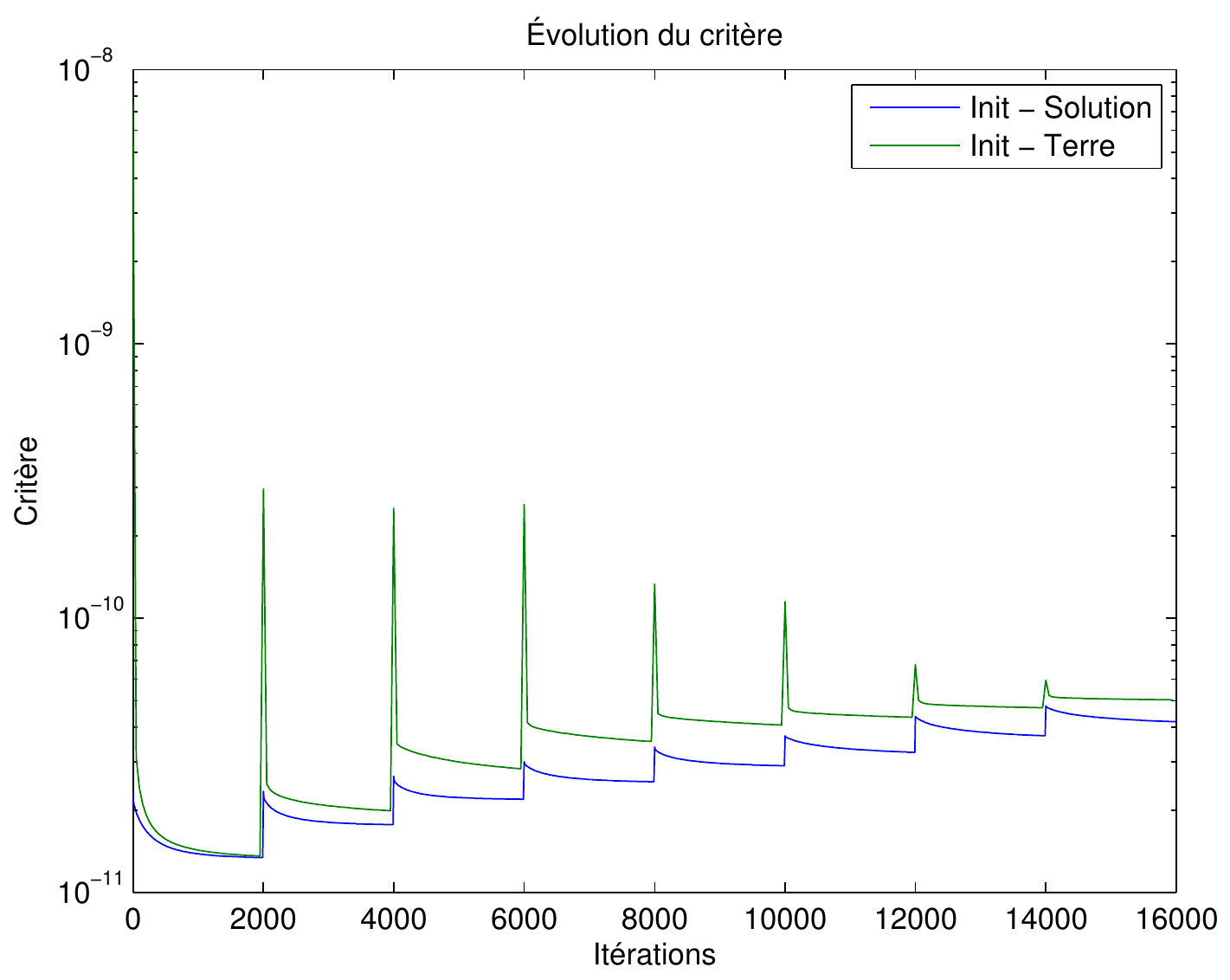}}
\caption{Résultats obtenus par la méthode CSI en introduisant progressivement les fréquences}
\label{figResultats_paquets_CSI}
\end{figure}

\begin{figure}[!b]
\centering
\subfloat[Initialisation avec la solution]{
\includegraphics[width=4.3in]{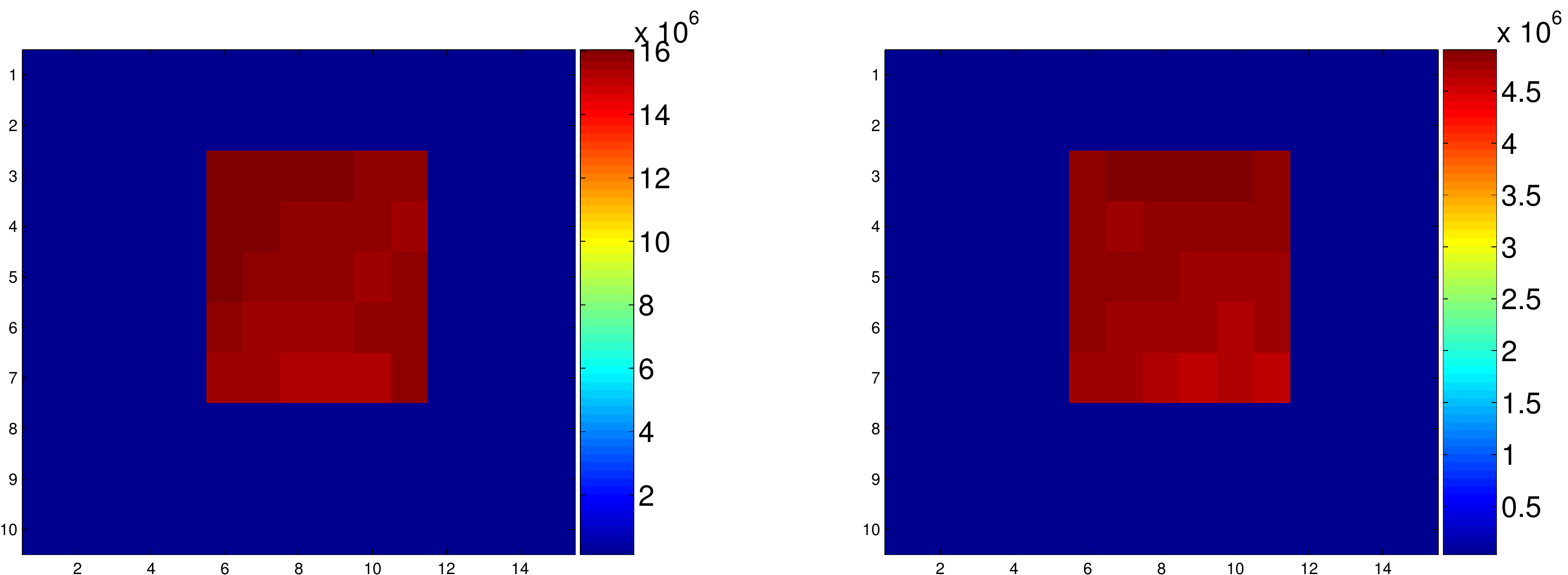}}\\
\subfloat[Initialisation à la terre]{
\includegraphics[width=4.3in]{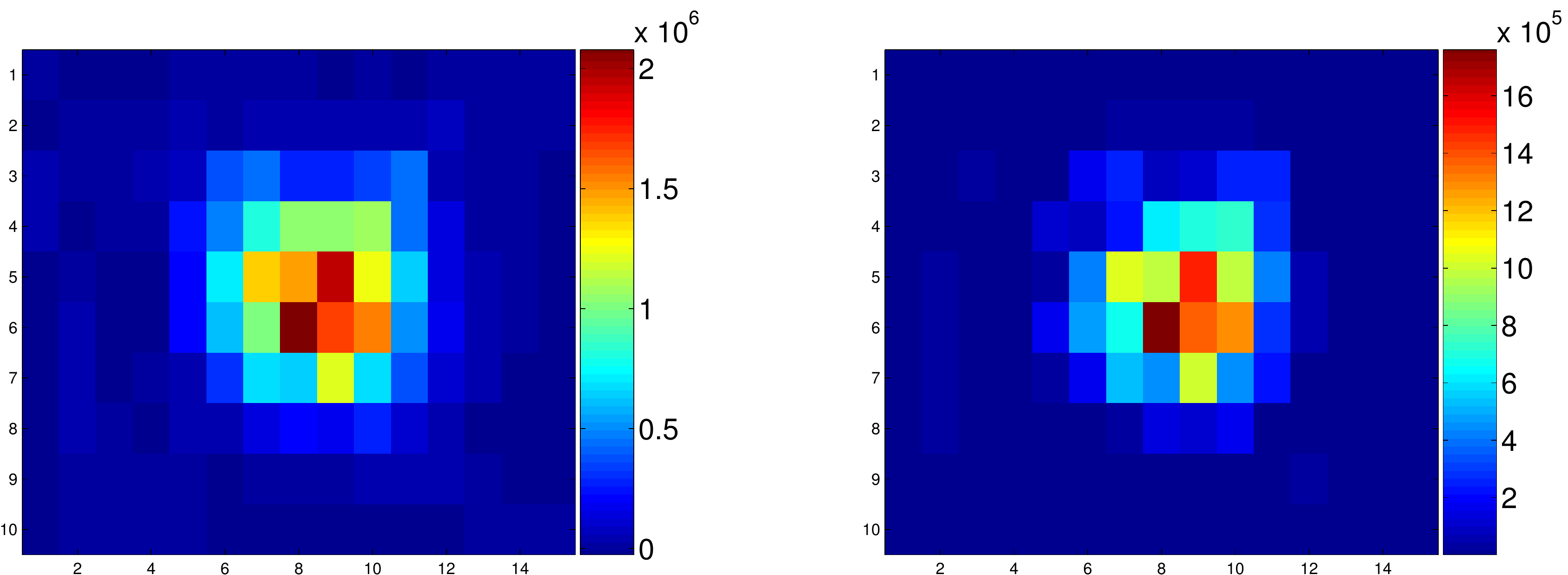}}\\
\subfloat[Évolution du critère]{
\includegraphics[width=4.3in]{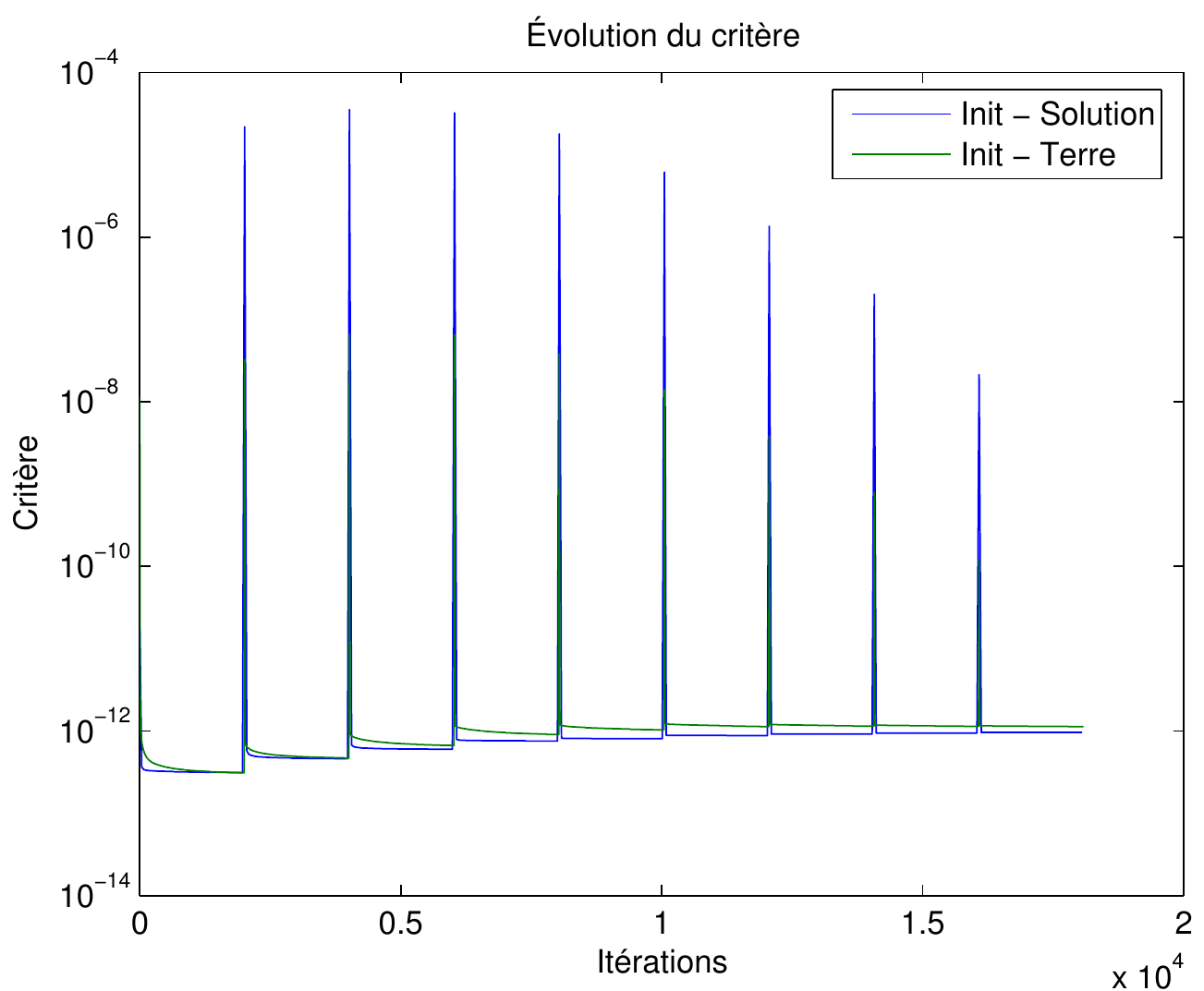}}
\caption{Résultats obtenus par la méthode GM en introduisant progressivement les fréquences}
\label{figResultats_paquets_GM}
\end{figure}

 Dans la section~\ref{part_Multifrequences}, nous avons décrit une stratégie d'incorporation progressive des fréquences pour réaliser l'inversion. Nous avons testé cette stratégie sur les méthodes CSI et GM sans utiliser de changement de variable. Les résultats sont présentés aux figures~(\ref{figResultats_paquets_CSI}) et (\ref{figResultats_paquets_GM}). Comme les méthodes testées convergent très lentement, il était exclu d'attendre la convergence de la méthode pour un nombre de fréquences donné avant de procéder à l'introduction de nouvelles fréquences dans l'algorithme d'inversion. Nous avons choisi d'incorporer une fréquence à chaque fois que la différence relative des contrastes serait inférieure à un certain seuil, tout en assurant un minimum de 2000 itérations.

\begin{figure}[!htb]
\centering
\subfloat[Évolution du déplacement relatif du contraste $\chi_p$]{
\includegraphics[scale=0.6]{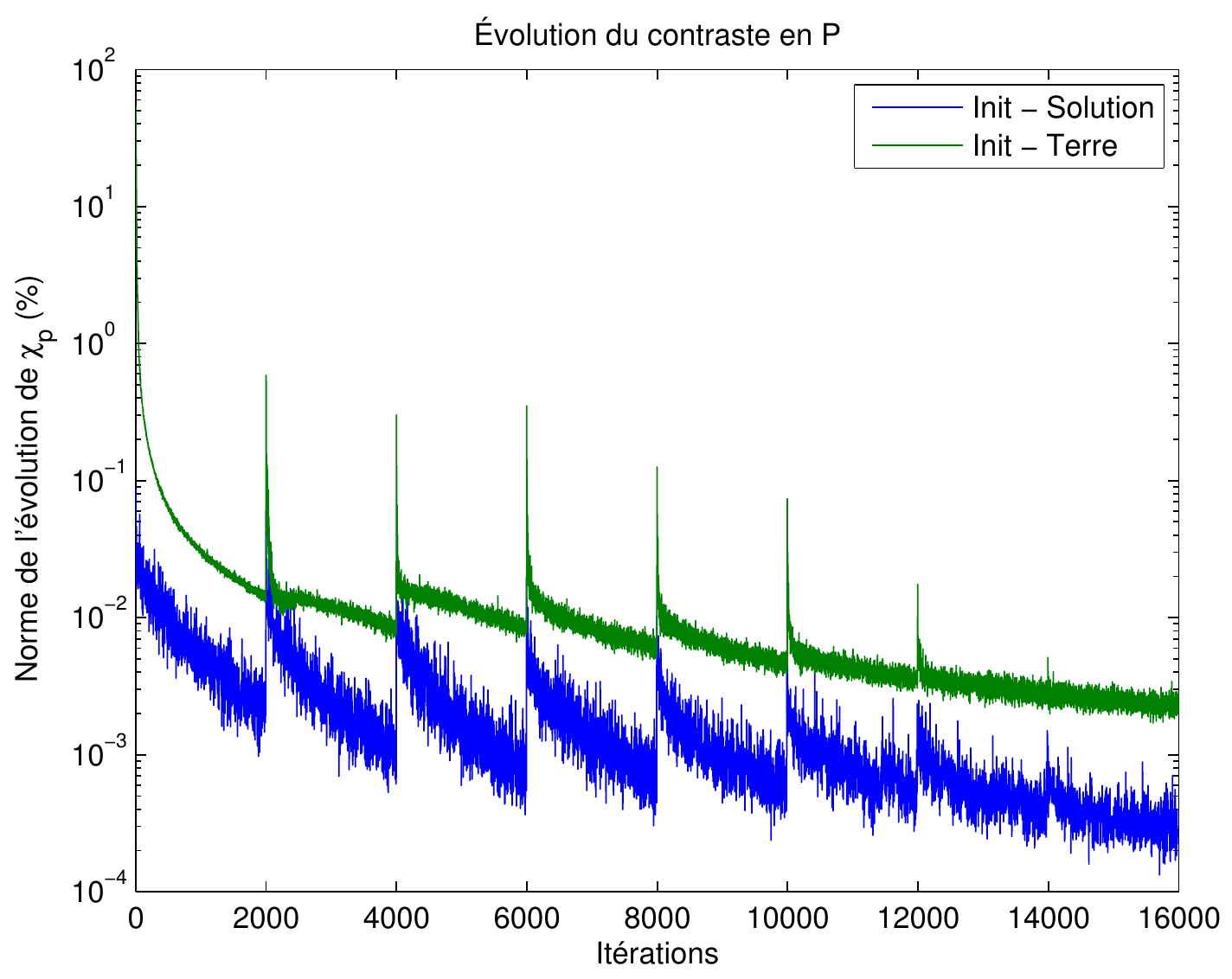}}
\subfloat[Évolution du déplacement relatif du contraste $\chi_s$]{
\includegraphics[scale=0.6]{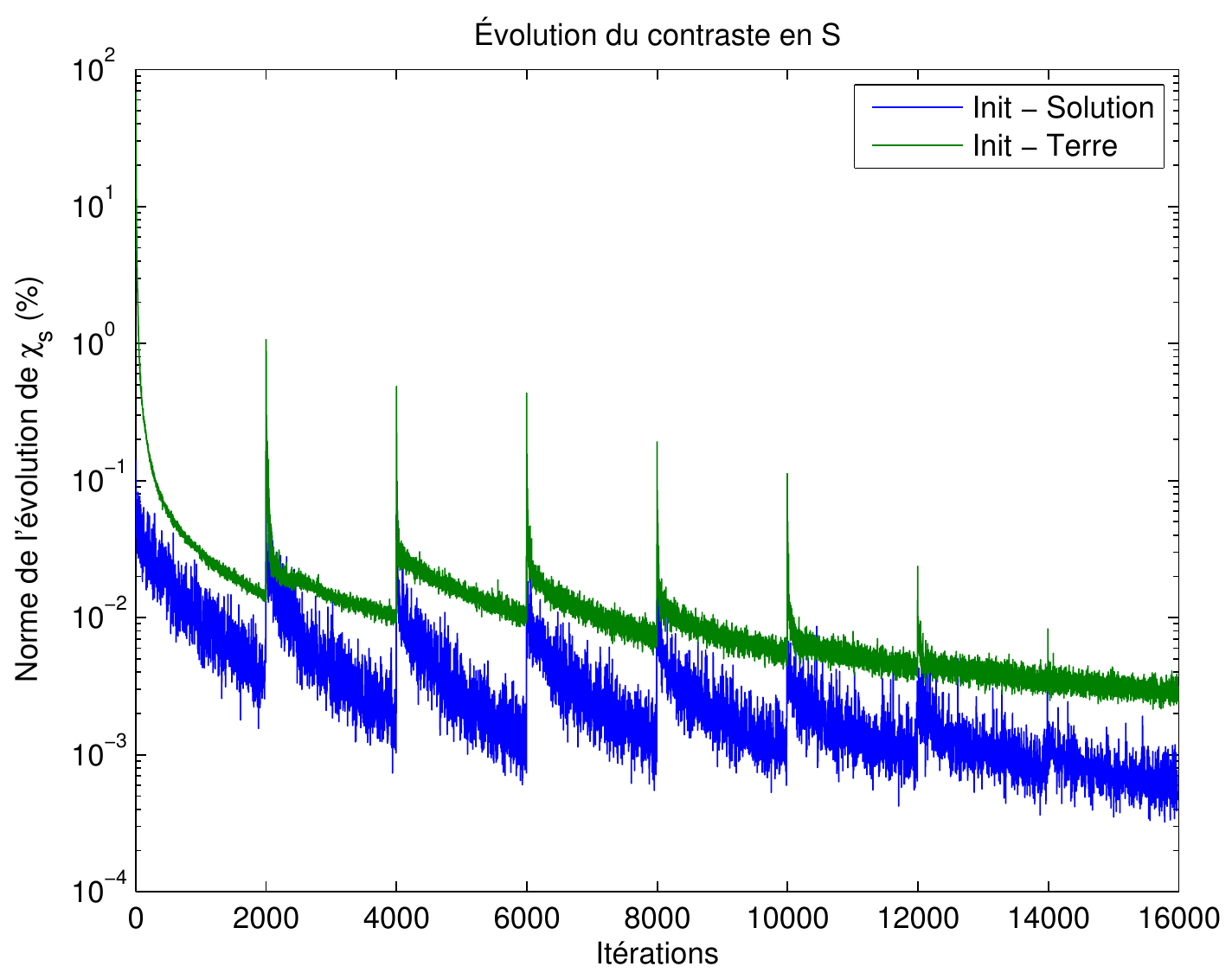}}
\caption{Évolution du déplacement relatif des contrastes obtenus par la méthode GM en introduisant progressivement les fréquences}
\label{figResultats_paquets_CSI_diff}
\end{figure}
Les courbes de la figure~(\ref{figResultats_paquets_CSI_diff}) montrent le déplacement relatif des contrastes. On remarque que l'introduction d'un nouveau groupe de fréquences entraîne un déplacement de la solution plus important pour les basses fréquences que pour les hautes fréquences. Ceci se traduit par une accélération de la convergence initiale ; cependant, la méthode semble atteindre un palier à partir duquel son comportement est similaire à celui observé lors de l'inversion simultanée de toutes les fréquences.


\chapter{Méthodes inverses fondées sur la formulation primale du probleme direct}

Dans le cas où l'on ne considère que le terme d'adéquation aux données, le critère associé à la formulation primale s'écrit de la façon suivante :
\begin{equation}
\mathcal{C}(\chi_p,\chi_s) = \frac{1}{2} \sum_\omega \sum_k \Vert y_{\omega,k} - g_{\omega,k}(\chi_p,\chi_s) \Vert ^2
\label{Eq_CriterePrimal}
\end{equation}
où :
\begin{itemize}
\item $\omega$ et $k$ désignent respectivement la fréquence et la position de la source considérées.
\item $y_{\omega,k}$ désigne le jeu de données mesurées ;
\item $g_{\omega,k}(\chi_p,\chi_s)$ correspond à la formulation primale du problème direct et désigne la fonction de construction des données synthétiques pour une distribution de contrastes donnée ;
\begin{equation} g_{\omega,k}(\chi_p,\chi_s) = (V_{\omega,k}^0)_{\capt} - \mathbf{B_\omega^d} \left(\mathbf{I} + (\mathbf{X_p} + \mathbf{X_s}) \mathbf{B_\omega^c} \right)^{-1} (\mathbf{X_p} + \mathbf{X_s}) V_{\omega,k}^0 \end{equation}
(la construction de cette formulation est détaillée dans le Chapitre \ref{Chap_IntroPbInv}, page \pageref{Chap_IntroPbInv})
\end{itemize}

\section{Les procédures itératives envisagées}

Pour minimiser le critère, nous envisageons d'utiliser une procédure itérative consistant à chaque itération à déterminer une direction de recherche dans l'espace de représentation puis à chercher un pas de progression efficace le long de cette direction de recherche (minimisation en une dimension).

En ce qui concerne la définition d'une direction de recherche, nous avons choisi de tester les algorithmes du gradient conjugué non linéaire et L-BFGS tel que décrit dans la Partie \ref{Part_IntroAlgoMinimisation} page \pageref{Part_IntroAlgoMinimisation} (le critère minimisé n'est pas quadratique).

Nous avons également envisagé de tester une méthode d'optimisation pixel par pixel. Cette méthode consiste à optimiser une des grandeurs d'intérêt ($\chi_p$ ou $\chi_s$) en un pixel, les autres caractéristiques restant constantes. Cela revient à considérer les variations du critère parallèlement à l'un des axes de l'espace d'état ; on ne prend donc pas en compte les variations locales du critère pour choisir la direction de descente. Malgré le fait qu'un nombre important de balayages de l'image soit nécessaire, cette méthode s'est avérée efficace pour certains problèmes d'imagerie par ondes diffractées car l'optimisation des caractéristiques en un pixel est très rapide \cite{Carfantan96}.

Concernant le choix d'un pas de progression, les valeurs des coefficients intervenant dans les conditions de Wolfe sont généralement telles que $c_1 \simeq 10^{-4}$ et $c_2 \simeq 0,1$ avec l'algorithme du gradient conjugué non linéaire \cite{Nocedal99}. Nous avons retenu ces valeurs pour les tests, les résultats seront présentés par la suite.

\section{Proposition d'une formulation primale optimisée}

Avec la formulation primale actuelle (Eq. \ref{Eq_FormulationPrimale}, page \pageref{Eq_FormulationPrimale}), le calcul du critère (terme d'adéquation aux données) et de son gradient en un point quelconque de l'espace de représentation nécessite la résolution de deux systèmes linéaires pour chaque fréquence et chaque position de la source. Les matrices normales de ces sytèmes sont égales à $\mathbf{I} + (\mathbf{X_p} + \mathbf{X_s}) \mathbf{B_\omega^c}$. Elles dépendent de la fréquence et des contrastes mais ne dépendent pas de la position de la source. Nous proposons donc de passer par la décomposition LU de ces matrices afin de diminuer le coût de calcul.

Cependant, la matrice $\mathbf{B_\omega^c}$ est une matrice pleine. Les matrices normales des systèmes linéaires sont donc pleines, ce qui nécessite une quantité d'espace mémoire importante. On s'attend donc à ce que la résolution des systèmes linéaires (décomposition LU des matrices et résolution des systèmes linéaires triangulaires qui s'ensuivent) soit longue. De plus, les matrices $\mathbf{B_\omega^d}$ qui interviennent dans l'expression du critère et du gradient sont également pleines.

Nous proposons d'utiliser une formulation primale optimisée ne faisant intervenir que des matrices creuses. Nous allègerons ainsi le temps et l'espace mémoire requis pour effectuer les calculs.

\subsection{Modifications apportées à la formulation primale}

\bigskip {\bf Nouvelle expression faisant intervenir une matrice normale creuse} \bigskip

Nous commençons par réécrire l'expression de la matrice normale des systèmes linéaires en faisant intervenir l'inverse de $\mathbf{B_\omega^c}$ :
\begin{equation} ( \mathbf{I} + (\mathbf{X_p} + \mathbf{X_s}) \mathbf{B_\omega^c} )^{-1} = (\mathbf{B_\omega^c})^{-1} ( (\mathbf{B_\omega^c})^{-1} + \mathbf{X_p} + \mathbf{X_s} )^{-1} \end{equation}

La matrice $(\mathbf{B_\omega^c})^{-1}$ est une matrice creuse. En effet, on sait que $\mathbf{B_\omega^c}$ est une sous-matrice carrée de $\mathbf{A}_{\omega,0}^{-1}$ (voir Partie \ref{Part_SimplifEcr}, page \pageref{Part_SimplifEcr}) et que $\mathbf{A_{\omega,0}}$ est creuse. A une permutation des lignes et des colonnes près, décomposons $\mathbf{A}_{\omega,0}^{-1}$ et $\mathbf{A_{\omega,0}}$ en quatre sous-matrices :
\begin{equation} (\mathbf{A_{\omega,0}})^{-1} = \left[ \begin{array}{c|c} \mathbf{B_\omega^c} & \dots \\ \hline \dots & \dots \end{array} \right]
\quad \text{et} \quad \mathbf{A_{\omega,0}} = \left[ \begin{array}{c|c} \mathcal{M}_{1,\omega} & \mathcal{M}_{2,\omega} \\ \hline \mathcal{M}_{3,\omega} & \mathcal{M}_{4,\omega} \end{array} \right] \end{equation}
où $\mathcal{M}_{1,\omega}$, $\mathcal{M}_{2,\omega}$, $\mathcal{M}_{3,\omega}$ et $\mathcal{M}_{4,\omega}$ sont creuses et $\mathcal{M}_{1,\omega}$ est de la taille de $\mathbf{B_\omega^c}$.

Appliquons ensuite le lemme d'inversion par bloc \footnote{Lemme d'inversion par bloc : $\begin{bmatrix} A & B \\ C & D \end{bmatrix}^{-1} = \begin{bmatrix} A^{-1} + A^{-1} B (D - C A^{-1} B)^{-1} C A^{-1} & - A^{-1} B (D - C A^{-1} B)^{-1} \\ - (D - C A^{-1} B)^{-1} C A^{-1} & (D - C A^{-1} B)^{-1} \end{bmatrix}$}. On obtient :
\begin{equation} (\mathbf{B_\omega^c})^{-1} = \mathcal{M}_{1,\omega} - \mathcal{M}_{2,\omega} \mathcal{M}_{4,\omega}^{-1} \mathcal{M}_{3,\omega} \end{equation}

Si l'on désigne par $n_{\ZE,x}$ et $n_{\ZE,y}$ les dimensions horizontales et verticales de la zone d'étude en nombre de pixels :
\begin{itemize}
\item $\mathcal{M}_{1,\omega}$ est une matrice creuse de taille $2 (n_{\ZE,x} + 1) (n_{\ZE,y} + 1) \times 2 (n_{\ZE,x} + 1) (n_{\ZE,y} + 1)$ (chaque ligne contient au plus 18 éléments non nuls) ;
\item $\mathcal{M}_{4,\omega}^{-1}$ est une matrice pleine ;
\item $\mathcal{M}_{2,\omega}$ est une matrice creuse. Elle compte $2 (n_{\ZE,x} + 1) (n_{\ZE,y} + 1)$ lignes. Seules $4 (n_{\ZE,x} + n_{\ZE,y})$ lignes de cette matrice contiennent des éléments non nuls. Il s'agit du nombre de composantes de $\mathit{F}_{\omega,k}$ qui s'expriment en fonction de composantes de $\mathit{V}_{\omega,k}^0$ n'appartenant pas à la zone d'étude (voir équation \ref{EqMilieuRef} page \pageref{EqMilieuRef}) ;
\item $\mathcal{M}_{3,\omega}$ est une matrice creuse. Elle compte $2 (n_{\ZE,x} + 1) (n_{\ZE,y} + 1)$ colonnes. Seules $4 (n_{\ZE,x} + n_{\ZE,y})$ colonnes de cette matrice contiennent des éléments non nuls. Il s'agit du nombre de composantes de $\mathit{V}_{\omega,k}^0$ appartenant à la zone d'étude et intervenant dans les expressions des composantes de $\mathit{F}_{\omega,k}$ qui n'appartiennent pas à la zone d'étude.
\end{itemize}

Si $n_{\ZE,x}$ et $n_{\ZE,y}$ sont suffisamment élevés, on en déduit que la matrice $\mathcal{M}_{2,\omega} \mathcal{M}_{4,\omega}^{-1} \mathcal{M}_{3,\omega}$ contient une majorité de coefficients nuls. La matrice $(\mathbf{B_\omega^c})^{-1}$ est donc creuse.

En tenant compte de cette nouvelle expression, le calcul du critère et du gradient passe maintenant par la résolution de systèmes linéaires dont les matrices normales sont creuses. On procèdera donc de la manière suivante :
\begin{itemize}
\item lors d'une phase d'initialisation, on calculera les matrices $(\mathbf{B_\omega^c})^{-1}$ associées aux différentes fréquences en utilisant la décomposition de la matrice $\mathbf{A_{\omega,0}}$ en quatre sous-matrices ;
\item pour chaque point de l'espace de représentation considéré, on calculera la matrice $(\mathbf{B_\omega^c})^{-1} + \mathbf{X_p} + \mathbf{X_s}$ puis on effectuera sa décomposition LU ;
\item pour résoudre les systèmes linéaires intervenant dans le calcul du critère et du gradient, on utilisera les facteurs $\mathbf{L_{(\mathbf{B_\omega^c})^{-1} + \mathbf{X_p} + \mathbf{X_s}}}$ et $\mathbf{U_{(\mathbf{B_\omega^c})^{-1} + \mathbf{X_p} + \mathbf{X_s}}}$ afin de se ramener à la résolution de systèmes linéaires triangulaires.
\end{itemize}
Comme la matrice $(\mathbf{B_\omega^c})^{-1} + \mathbf{X_p} + \mathbf{X_s}$ ne dépend pas de la position de la source, cette procédure est plus économe que l'inversion de chaque système pris indépendamment.

\bigskip {\bf Modification supplémentaire} \bigskip

Les matrices $\mathbf{B_\omega^d}$ qui interviennent dans les expressions du critère et du gradient sont des matrices pleines. Ce sont également des sous-matrices de $\mathbf{A}_{\omega,0}^{-1}$ (voir Partie \ref{Part_SimplifEcr}, page \pageref{Part_SimplifEcr}). On propose de les remplacer par leur expression faisant intervenir $\mathbf{A}_{\omega,0}^{-1}$ :
\begin{equation} \mathbf{B}^d_\omega = \mathbf{E}_1 (\mathbf{A}_{\omega,p,s})_0^{-1} \mathbf{E}_2^t \end{equation}

Ainsi, au lieu de multiplications par $\mathbf{B_\omega^d}$, nous effectuons des inversions de systèmes linéaires dont la matrice normale est creuse. Etant donné que la matrice $\mathbf{A_{\omega,0}}$ ne dépend pas du contraste, les mêmes matrices normales interviendront à plusieurs reprises. On procède donc de la manière suivante :
\begin{itemize}
\item lors d'une phase d'initialisation, on effectue la décomposition LU des matrices $\mathbf{A_{\omega,0}}$ ;
\item pour chaque système linéaire associé, on utilise les facteurs $\mathbf{L_{\mathbf{A_{\omega,0}}}}$ et $\mathbf{U_{\mathbf{A_{\omega,0}}}}$ afin de se ramener à la résolution de systèmes linéaires triangulaires.
\end{itemize}

Cela permet de gagner notamment en place mémoire et en temps de calcul lors de la phase d'initialisation (matrices $\mathbf{B_\omega^d}$ longues à calculer).

\subsection{Nouvelle écriture de la formulation primale}

Désormais, nous écrirons la relation $g_{\omega,k}(\chi_p,\chi_s)$ qui lie les contrastes aux données synthétiques de la façon suivante :
\begin{equation} \label{Eq_FormPrimaleOpt}
g_{\omega,k}(\chi_p,\chi_s) = (V_{\omega,k}^0)_{\capt} - \mathbf{E}_1 (\mathbf{A}_{\omega,p,s})_0^{-1} \mathbf{E}_2^t (\mathbf{B_\omega^c})^{-1} \mathbf{M}_\omega^{-1} (\mathbf{X_p} + \mathbf{X_s}) V_{\omega,k}^0
\end{equation}
avec : $\mathbf{M}_\omega = (\mathbf{B_\omega^c})^{-1} + \mathbf{X_p} + \mathbf{X_s}$

\subsection{Mise en évidence du gain en temps de calcul et en espace mémoire}

L'utilisation de la formulation optimisée permet un gain à la fois en temps de calcul et en espace mémoire occupé. Nous le montrons en utilisant les expressions avant et après modifications sur le cas suivant :
\begin{itemize}
\item le domaine considéré est constitué de 3700 pixels et la zone d'étude comprend 800 pixels ;
\item le nombre de fréquences retenues est égal à 15 et la source prend trois positions différentes.
\end{itemize}

\bigskip {\bf Mise en évidence du gain en temps de calcul} \bigskip

Nous présentons dans le tableau suivant le temps de calcul associé aux étapes les plus coûteuses pour le calcul du critère et du gradient en un point. Les temps de calculs affichés correspondent à la prise en compte de l'ensemble des fréquences et des positions de la source.

\begin{center} \begin{tabular}{|l|c||l|c|} \hline
\multicolumn{2}{|c||}{{\bf Avec la formulation initiale}}   &   \multicolumn{2}{c|}{{\bf Avec la formulation optimisée}}   \\ \hline
Initialisation   &   1 heure   &   Initialisation   &   35 secondes   \\
\quad {\scriptsize Calcul des $\mathbf{B_\omega^c}$ et des $\mathbf{B_\omega^d}$}
   &
   &   \quad {\scriptsize Calcul des $(\mathbf{B_\omega^c})^{-1}$}
   &   {\scriptsize 30 secondes}   \\
   &
   &   \quad {\scriptsize  Décomposition LU des $\mathbf{A_{\omega,0}}$}
   &   {\scriptsize 5 secondes} \\ \hline
Calcul du critère et du gradient
   &   > 1 minute
   &   Calcul du critère et du gradient
   &   6 secondes   \\
\quad {\scriptsize Calcul des $\mathbf{I} + (\mathbf{X_p} + \mathbf{X_s}) \mathbf{B_\omega^c}$}
   & {\scriptsize 10 secondes}
   & \quad {\scriptsize Décomposition LU des $(\mathbf{B_\omega^c})^{-1} + \mathbf{X_p} + \mathbf{X_s}$}
   & {\scriptsize 2 secondes}							\\
\quad {\scriptsize Décomposition LU des $\mathbf{I} + (\mathbf{X_p} + \mathbf{X_s}) \mathbf{B_\omega^c}$}
   & {\scriptsize 50 secondes}
   & \quad {\scriptsize Calcul du critère}
   & {\scriptsize 2 secondes}							\\
\quad {\scriptsize Calcul du critère}
   & {\scriptsize 2 secondes}
   & \quad \quad {\scriptsize (2 systèmes linéaires par fréq. et par pos.)}
   & 							\\
\quad \quad {\scriptsize (1 système linéaire par fréq. et par pos.)}
   & 
   & \quad {\scriptsize Calcul du gradient}
   & {\scriptsize 2 secondes}							\\
\quad {\scriptsize Calcul du gradient}
   & {\scriptsize 2 secondes}
   & \quad \quad {\scriptsize (2 systèmes linéaires par fréq. et par pos.)}
   & 							\\
\quad \quad {\scriptsize (1 système linéaire par fréq. et par pos.)}
   & 
   &
   &							\\ \hline
\end{tabular} \end{center}

\bigskip {\bf Mise en évidence du gain en mémoire} \bigskip

Nous mettons en évidence le gain en espace mémoire obtenu en utilisant les expressions avant et après modification sur le même domaine (nous n'affichons que les éléments occupant le plus d'espace mémoire) :

\begin{center} \begin{tabular}{|l|c||l|c|} \hline
{\bf Avec la formulation initiale}   &   > 120 Mo / freq   &   {\bf Avec la formulation optimisée}   &   $\sim$ 21 Mo / freq   \\ \hline
\quad {\scriptsize $\mathbf{B_\omega^c}$}
   & {\scriptsize 40 Mo / freq}
   & \quad {\scriptsize $\mathbf{L_{\mathbf{A_{\omega,0}}}}$}
   & {\scriptsize 7 Mo / freq}							\\
\quad {\scriptsize $\mathbf{B_\omega^d}$}
   & {\scriptsize 1 Mo / freq}
   & \quad {\scriptsize $\mathbf{U_{\mathbf{A_{\omega,0}}}}$}
   & {\scriptsize 7 Mo / freq}							\\
\quad {\scriptsize $\mathbf{I} + (\mathbf{X_p} + \mathbf{X_s}) \mathbf{B_\omega^c}$}
   & {\scriptsize 40 Mo / freq}
   & \quad {\scriptsize $(\mathbf{B_\omega^c})^{-1}$}
   & {\scriptsize 1.5 Mo / freq}							\\
\quad {\scriptsize $\mathbf{L_{\mathbf{I} + (\mathbf{X_p} + \mathbf{X_s}) \mathbf{B_\omega^c}}}$}
   & {\scriptsize 20 Mo / freq}
   & \quad {\scriptsize $(\mathbf{B_\omega^c})^{-1} + \mathbf{X_p} + \mathbf{X_s}$}
   & {\scriptsize 1.5 Mo / freq}							\\
\quad {\scriptsize $\mathbf{U_{\mathbf{I} + (\mathbf{X_p} + \mathbf{X_s}) \mathbf{B_\omega^c}}}$}
   & {\scriptsize 20 Mo / freq}
   & \quad {\scriptsize $\mathbf{L_{(\mathbf{B_\omega^c})^{-1} + \mathbf{X_p} + \mathbf{X_s}}}$}
   & {\scriptsize 2 Mo / freq}							\\
   &
   & \quad {\scriptsize $\mathbf{U_{(\mathbf{B_\omega^c})^{-1} + \mathbf{X_p} + \mathbf{X_s}}}$}
   & {\scriptsize 2 Mo / freq}							\\ \hline
\end{tabular} \end{center}

\section{Calcul du gradient du critère}

Pour les différentes méthodes d'optimisation proposées, il est nécessaire de savoir calculer le gradient en un point quelconque de l'espace d'état. Nous résumons ici une démarche permettant d'aboutir à l'expression du gradient du terme d'adéquation aux données en un point quelconque. Pour le terme de régularisation, il n'y a pas de difficulté.

\bigskip {\bf Expression du gradient du critère pour une seule fréquence et une seule source} \bigskip

Nous nous plaçons dans le cas où l'on ne considère qu'une seule fréquence et qu'une seule position de la source afin de simplifier le raisonnement et les écritures. On a donc :
\begin{equation} \mathcal{C}(\chi_p,\chi_s) = \dfrac{1}{2} \Vert y - g(\chi_p,\chi_s) \Vert ^2 \quad \text{avec} \quad g(\chi_p,\chi_s) = V^0_{\capt} - \mathbf{E}_1 \mathbf{A}_0^{-1} \mathbf{E}_2^t (\mathbf{B^c})^{-1} \mathbf{M}^{-1} (\mathbf{X_p} + \mathbf{X_s}) V^0 \end{equation}
où $\mathbf{M} = (\mathbf{B^c})^{-1} + \mathbf{X_p} + \mathbf{X_s}$.

Les gradients selon $\chi_p$ et $\chi_s$ sont égaux à :
\begin{equation} \frac{\partial \mathcal{C}}{\partial \chi_p} = - \Re \{ \mathbf{J_p}(\chi_p,\chi_s)^\dagger (y - g(\chi_p,\chi_s)) \}
 \quad \text{ et } \quad
 \frac{\partial \mathcal{C}}{\partial \chi_s} = - \Re \{ \mathbf{J_s}(\chi_p,\chi_s)^\dagger (y - g(\chi_p,\chi_s)) \} \end{equation}
où $\mathbf{J_p}$ et $\mathbf{J_s}$ sont les matrices jacobiennes de la fonction $g$ par rapport aux variables $\chi_p$ et $\chi_s$ :
\begin{equation} ( \mathbf{J_p}(\chi_p,\chi_s) )_{i,j} = \frac{\partial ( g(\chi_p,\chi_s) )_i}{\partial ( \chi_p )_j} \quad \text{ et } \quad ( \mathbf{J_s}(\chi_p,\chi_s) )_{i,j} = \frac{\partial ( g(\chi_p,\chi_s) )_i}{\partial ( \chi_s )_j} \end{equation}

\bigskip {\bf Calcul de la matrice jacobienne $\mathbf{J_p}$} \bigskip

Pour calculer la matrice jacobienne $\mathbf{J_p}$, on effectue un développement de Taylor à l'ordre 1 de la fonction $g$ par rapport à $\chi_p$. La matrice recherchée est telle que $ g(\chi_p + \delta \chi_p,\chi_s) = g(\chi_p,\chi_s) + \mathbf{J_p} \delta \chi_p + \mathcal{O}(\Vert \delta \chi_p \Vert^2) $. Or, on a :
\begin{equation} g(\chi_p + \delta \chi_p,\chi_s) = V^0_{\capt} - \mathbf{E}_1 \mathbf{A}_0^{-1} \mathbf{E}_2^t (\mathbf{B^c})^{-1} ( (\mathbf{B^c})^{-1} + \mathbf{X_p} + \delta \mathbf{X_p} + \mathbf{X_s} )^{-1} (\mathbf{X_p} + \delta \mathbf{X_p} + \mathbf{X_s}) V^0 \end{equation}

\begin{itemize}
\item[$\bullet$] On commence par effectuer un développement de $( (\mathbf{B^c})^{-1} + \mathbf{X_p} + \delta \mathbf{X_p} + \mathbf{X_s} )^{-1}$ à l'ordre 1 en utilisant le lemme d'inversion matricielle \footnote{Lemme d'inversion matricielle : $ (\mathbf{P} + \mathbf{Q} \mathbf{R} \mathbf{S})^{-1} = \mathbf{P}^{-1} - \mathbf{P}^{-1} \mathbf{Q} (\mathbf{R}^{-1} + \mathbf{S} \mathbf{P}^{-1} \mathbf{Q})^{-1} \mathbf{S} \mathbf{P}^{-1} $}. On obtient :
\begin{equation} ( (\mathbf{B^c})^{-1} + \mathbf{X_p} + \delta \mathbf{X_p} + \mathbf{X_s} )^{-1} = \mathbf{M}^{-1} \left[ \mathbf{I} - \delta \mathbf{X_p} \mathbf{M}^{-1} \right] + \mathcal{O}(\Vert \delta \mathbf{X_p} \Vert^2) \end{equation}
\item[$\bullet$] On revient alors à l'expression de $g(\chi_p + \delta \chi_p,\chi_s)$ pour obtenir le développement de Taylor à l'ordre 1 de la fonction $g$ par rapport à $\chi_p$. En utilisant l'expression de $\mathbf{X_p}$ en fonction de $\chi_p$ (voir Partie \ref{Part_SimplifEcr}, page \pageref{Part_SimplifEcr}), on obtient :
\begin{equation} g(\chi_p + \delta \chi_p,\chi_s) = g(\chi_p,\chi_s) - \mathbf{E}_1 \mathbf{A}_0^{-1} \mathbf{E}_2^t (\mathbf{B^c})^{-1} \mathbf{M}^{-1} \mathbf{H^p} \Diag \{ \delta \chi_p \} \mathbf{G^p} ( \mathbf{I} - \mathbf{M}^{-1} (\mathbf{X_p} + \mathbf{X_s}) ) V^0 + \mathcal{O}(\Vert \delta \chi_p \Vert^2) \end{equation}
\item[$\bullet$] On en déduit l'expression de la matrice jacobienne $\mathbf{J_p}$ :
\begin{equation} \mathbf{J_p} = - \mathbf{E}_1 \mathbf{A}_0^{-1} \mathbf{E}_2^t (\mathbf{B^c})^{-1} \mathbf{M}^{-1} \mathbf{H^p} \Diag \left\lbrace \mathbf{G^p} ( \mathbf{I} - \mathbf{M}^{-1} (\mathbf{X_p} + \mathbf{X_s}) ) V^0 \right\rbrace \end{equation}
(en effet, pour deux vecteurs $w_1$ et $w_2$ de même taille, on a : $\Diag \{ w_1 \} w_2 = \Diag \{ w_2 \} w_1$)
\end{itemize}

\bigskip {\bf Calcul de la matrice jacobienne $\mathbf{J_s}$} \bigskip

En reprenant la même démarche que pour le calcul de $\mathbf{J_p}$, on obtient l'expression de la matrice jacobienne $\mathbf{J_s}$ :
\begin{equation} \mathbf{J_s} = - \mathbf{E}_1 \mathbf{A}_0^{-1} \mathbf{E}_2^t (\mathbf{B^c})^{-1} \mathbf{M}^{-1} \sum_{i=1}^3 \mathbf{H_i^s} \Diag \left\lbrace \mathbf{G_i^s} ( \mathbf{I} - \mathbf{M}^{-1} (\mathbf{X_p} + \mathbf{X_s}) ) V^0 \right\rbrace \end{equation}

\bigskip {\bf Retour à l'expression du gradient} \bigskip

Les gradients selon $\chi_p$ et $\chi_s$ sont égaux à :

\begin{equation} \frac{\partial \mathcal{C}}{\partial \chi_p}
= \Re \left\lbrace \Diag \left\lbrace \overline{\mathbf{G^p} ( \mathbf{I} - \mathbf{M}^{-1} (\mathbf{X_p} + \mathbf{X_s}) ) V^0} \right\rbrace (\mathbf{H^p})^t (\mathbf{M}^{-1})^\dagger ((\mathbf{B^c})^{-1})^\dagger \mathbf{E}_2 (\mathbf{A}_0^{-1})^\dagger \mathbf{E}_1^t (y - g(\chi_p,\chi_s)) \right\rbrace \end{equation}

\begin{equation} \frac{\partial \mathcal{C}}{\partial \chi_s}
= \Re \left\lbrace \sum_{i=1}^3 \Diag \left\lbrace \overline{\mathbf{G_i^s} ( \mathbf{I} - \mathbf{M}^{-1} (\mathbf{X_p} + \mathbf{X_s}) ) V^0} \right\rbrace (\mathbf{H_i^s})^t (\mathbf{M}^{-1})^\dagger ((\mathbf{B^c})^{-1})^\dagger \mathbf{E}_2 (\mathbf{A}_0^{-1})^\dagger \mathbf{E}_1^t (y - g(\chi_p,\chi_s)) \right\rbrace \end{equation}

Lorsque l'on considère plusieurs fréquences $\omega$ et plusieurs positions de la source $k$, le gradient correspond à la somme des expressions obtenues pour chaque couple $\{ \omega,k\}$. On a donc :

\begin{multline}
\frac{\partial \mathcal{C}}{\partial \chi}
   = \begin{bmatrix} \frac{\partial \mathcal{C}}{\partial \chi_p} \\ \frac{\partial \mathcal{C}}{\partial \chi_s} \end{bmatrix}
   = \sum_\omega \sum_k \Re \Bigg\lbrace
   \begin{bmatrix} \Diag \left\lbrace \overline{\mathbf{G^p} ( \mathbf{I} - \mathbf{M_\omega}^{-1} (\mathbf{X_p} + \mathbf{X_s}) ) V_{\omega,k}^0} \right\rbrace (\mathbf{H^p})^t \\
\sum_{i=1}^3 \Diag \left\lbrace \overline{\mathbf{G_i^s} ( \mathbf{I} - \mathbf{M_\omega}^{-1} (\mathbf{X_p} + \mathbf{X_s}) ) V_{\omega,k}^0} \right\rbrace (\mathbf{H_i^s})^t \end{bmatrix} \\
(\mathbf{M}_\omega^{-1})^\dagger ((\mathbf{B_\omega^c})^{-1})^\dagger \mathbf{E}_2 ((\mathbf{A}_{\omega,p,s})_0^{-1})^\dagger \mathbf{E}_1^t [y_{\omega,k} - g_{\omega,k}(\chi_p,\chi_s)] \Bigg\rbrace
\label{Eq_GradientPrimal}
\end{multline}

\section{Analyse de la méthode d'optimisation pixel par pixel}

La méthode d'optimisation pixel par pixel consiste à faire décroître la valeur du critère en ne faisant évoluer qu'une seule composante de l'espace d'état à chaque étape de minimisation : pour un point initial donné, on considère un pixel de la zone d'étude et l'on fait évoluer une de ses caractéristiques ($\chi_p$ ou $\chi_s$). A priori, nous cherchons à minimiser le critère selon l'axe considéré. Après optimisation de la caractéristique, le contraste obtenu fait office de nouveau point initial et l'on passe à une autre composante.

Pour certains problèmes d'imagerie à ondes diffractées, il est possible de définir de telles variations de façon analytique \cite{Carfantan96} ce qui permet de déterminer un minimiseur exact de façon rapide. Cependant, dans le cadre de notre étude, le fait que nous travaillons en multifréquentiel et que nous recherchons deux caractéristiques en chaque pixel rend une telle démarche inenvisageable. Pour chaque composante considérée, nous rechercherons donc un pas de progression vérifiant les conditions de Wolfe.

Nous détaillons ci-dessous les étapes de calcul associées à cette méthode d'optimisation (calcul de la valeur du critère et du gradient au point initial et après variation en un pixel) ainsi que les coûts de calculs correspondants. Nous comparons ensuite cette méthode avec les méthodes de type gradient.

\subsection{Calculs à un point initial}

\bigskip {\bf Calcul du critère} \bigskip

\begin{align}
\mathcal{C}(\chi_p,\chi_s) & = \frac{1}{2} \sum_\omega \sum_k \Vert y_{\omega,k} - g_{\omega,k}(\chi_p,\chi_s) \Vert ^2 \nonumber \\
& = \frac{1}{2} \sum_\omega \sum_k \Vert y_{\omega,k} - (V_{\omega,k}^0)_{\capt} + \mathbf{E}_1 (\mathbf{A}_{\omega,p,s})_0^{-1} \mathbf{E}_2^t (\mathbf{B_\omega^c})^{-1} \mathbf{M}_\omega^{-1} (\mathbf{X_p} + \mathbf{X_s}) V_{\omega,k}^0 \Vert ^2
\end{align}
avec $\mathbf{M}_\omega = (\mathbf{B_\omega^c})^{-1} + \mathbf{X_p} + \mathbf{X_s}$

On remarque que pour calculer le critère à un point initial, deux systèmes linéaires doivent être résolus pour chaque fréquence et chaque position de la source. Or, les matrices normales de ces systèmes linéaires sont $(\mathbf{A}_{\omega,p,s})_0$ et $\mathbf{M_\omega}$. Elles ne dépendent donc pas de la position de la source et la première est indépendante du contraste. Afin de gagner en coût de calcul, nous choisissons d'exploiter le fait qu'une même matrice normale est commune à plusieurs systèmes linéaires en procédant de la façon suivante :
\begin{description}
\item[Lors d'une phase d'initialisation de la méthode :]
\end{description}
\begin{itemize}
\item pour chaque fréquence, on effectue la factorisation LU de la matrice $(\mathbf{A}_{\omega,p,s})_0$ : $(\mathbf{A}_{\omega,p,s})_0 = \mathbf{L_{(\mathbf{A}_{\omega,p,s})_0}} \mathbf{U_{(\mathbf{A}_{\omega,p,s})_0}}$ (coût de calcul : $\mathcal{O}(n_A^3)$) ;
\end{itemize}
\begin{description}
\item[Pour chaque point initial considéré :]
\end{description}
\begin{itemize}
\item pour chaque fréquence, on effectue la factorisation LU de la matrice $\mathbf{M_\omega}$ : $\mathbf{M_\omega} = \mathbf{L_{\mathbf{M_\omega}}} \mathbf{U_{\mathbf{M_\omega}}}$ (coût de calcul : $\mathcal{O}(n_M^3)$) ;
\item pour chaque fréquence et chaque position de la source, on résout quatre systèmes linéaires triangulaires (coût de calcul : $2 \times \mathcal{O}(n_A^2) + 2 \times \mathcal{O}(n_M^2)$). On pose : $w_0 = \mathbf{U}_\mathbf{M_\omega}^{-1} \mathbf{L}_\mathbf{M_\omega}^{-1} (\mathbf{X_p} + \mathbf{X_s}) V_{\omega,k}^0$.
\end{itemize}
($n_A$ (resp. $n_M$) désigne la taille de la matrice $(\mathbf{A}_{\omega,p,s})_0$ (resp. $\mathbf{M_\omega}$).)

\bigskip {\bf Calcul du gradient} \bigskip

\begin{multline}
\frac{\partial \mathcal{C}}{\partial \chi}
   = \begin{bmatrix} \frac{\partial \mathcal{C}}{\partial \chi_p} \\ \frac{\partial \mathcal{C}}{\partial \chi_s} \end{bmatrix}
   = \sum_\omega \sum_k \Re \Bigg\lbrace
   \begin{bmatrix} \Diag \left\lbrace \overline{\mathbf{G^p} (V_{\omega,k}^0 - w_0} \right\rbrace (\mathbf{H^p})^t \\
\sum_{i=1}^3 \Diag \left\lbrace \overline{\mathbf{G_i^s} (V_{\omega,k}^0 - w_0} \right\rbrace (\mathbf{H_i^s})^t \end{bmatrix} \\
(\mathbf{M}_\omega^{-1})^\dagger ((\mathbf{B_\omega^c})^{-1})^\dagger \mathbf{E}_2 ((\mathbf{A}_{\omega,p,s})_0^{-1})^\dagger \mathbf{E}_1^t [y_{\omega,k} - g_{\omega,k}(\chi_p,\chi_s)] \Bigg\rbrace
\end{multline}

Le calcul du gradient du critère à un point initial nécessite la résolution de deux systèmes linéaires supplémentaires pour chaque fréquence et chaque position de la source. Or, les matrices normales sont $(\mathbf{A}_{\omega,p,s})_0^\dagger$ et $\mathbf{M}_\omega^\dagger$. Nous exploitons donc à nouveau la décomposition LU des matrices $(\mathbf{A}_{\omega,p,s})_0$ et $\mathbf{M_\omega}$ pour se ramener à la résolution de quatre systèmes linéaires triangulaires pour chaque fréquence et chaque position de la source.

\subsection{Variation d'une caractéristique en un pixel}

\bigskip {\bf Calcul du critère} \bigskip

Si l'on considère un pixel et que l'on fait varier la caractéristique $\chi_p$ de ce pixel d'un pas $\alpha$, la matrice de contraste $\mathbf{X_p}$ devient $\mathbf{X_p} + \alpha \mathbf{h^p_1} \mathbf{g^p_1}$ où $\mathbf{g^p_1}$ est un vecteur ligne et $\mathbf{h^p_1}$ est un vecteur colonne. En utilisant le lemme d'inversion matricielle, on obtient :

\begin{equation} g_{\omega,k}(\chi_p + \delta \chi_p,\chi_s) = g_{\omega,k}(\chi_p,\chi_s) - \mathbf{E}_1 (\mathbf{A}_{\omega,p,s})_0^{-1} \mathbf{E}_2^t (\mathbf{B_\omega^c})^{-1} \mathbf{M}_\omega^{-1} \mathbf{h^p_1} \underbrace{ \left( \frac{1}{\alpha} + \mathbf{g^p_1} \mathbf{M}_\omega^{-1} \mathbf{h^p_1} \right)^{-1} }_{= \Delta_1(\alpha) \text{, un scalaire}} \mathbf{g^p_1} (V_{\omega,k}^0 - w_0) \end{equation}

Par conséquent, la résolution de deux systèmes linéaires liée au choix du pixel (et donc au choix des vecteurs $\mathbf{g^p_1}$ et $\mathbf{h^p_1}$) mais indépendante de la valeur du pas $\alpha$ est nécessaire pour chaque fréquence. Les matrices normales étant $(\mathbf{A}_{\omega,p,s})_0$ et $\mathbf{M_\omega}$, on utilise à nouveau la factorisation LU de ces deux matrices et on résout quatre systèmes linéaires triangulaires. Un changement de la valeur de $\alpha$ n'implique pas la résolution de systèmes linéaires supplémentaires puisque ce coefficient n'influe que sur la valeur de $\Delta_1(\alpha)$. On pose : $w_1 = \mathbf{U}_\mathbf{M_\omega}^{-1} \mathbf{L}_\mathbf{M_\omega}^{-1} \mathbf{h^p_1}$.

Le raisonnement est similaire pour $\chi_s$ mais le nombre de systèmes linéaires triangulaires à résoudre est multiplié par trois.

\bigskip {\bf Calcul du gradient} \bigskip

En utilisant le lemme d'inversion matricielle, on obtient :

\begin{multline} \nabla \mathcal{C}(\chi_p + \delta \chi_p,\chi_s)
	= \sum_\omega \sum_k \Re \Bigg\lbrace \begin{bmatrix}
				\Diag \left\lbrace \mathbf{G^p} (\overline{V_{\omega,k}^0 - w_0 - w_1 \Delta_1(\alpha) \mathbf{g^p_1} (V_{\omega,k}^0 - w_0)}) \right\rbrace (\mathbf{H^p})^t   \\
				\sum_{i=1}^3 \Diag \left\lbrace \mathbf{G_i^s} (\overline{V_{\omega,k}^0 - w_0 - w_1 \Delta_1(\alpha) \mathbf{g^p_1} (V_{\omega,k}^0 - w_0)}) \right\rbrace (\mathbf{H_i^s})^t   \end{bmatrix}   \\
	(\mathbf{M_\omega^\dagger})^{-1} (I - w_1 \Delta_1(\alpha) \mathbf{g^p_1})^\dagger ((\mathbf{B_\omega^c})^{-1})^\dagger \mathbf{E}_2 ((\mathbf{A}_{\omega,p,s})_0^{-1})^\dagger \mathbf{E}_1^t [y_{\omega,k} - g_{\omega,k}(\chi_p + \delta \chi_p,\chi_s)] \Bigg\rbrace \end{multline}

Par conséquent, la résolution de deux systèmes linéaires qui dépendent cette fois-ci de la valeur du pas $\alpha$ est nécessaire pour chaque fréquence et chaque position de la source. Les matrices normales étant $(\mathbf{A}_{\omega,p,s})_0^\dagger$ et $\mathbf{M}_\omega^\dagger$, on utilise à nouveau les facteurs $\mathbf{L_{(\mathbf{A}_{\omega,p,s})_0}}$, $\mathbf{U_{(\mathbf{A}_{\omega,p,s})_0}}$, $\mathbf{L_{\mathbf{M_\omega}}}$ et $\mathbf{U_{\mathbf{M_\omega}}}$ correspondants.

Le raisonnement est similaire pour $\chi_s$.

\subsection{Comparaison avec les méthodes de type gradient}

On résume tout d'abord dans le tableau ci-dessous les étapes de calcul les plus coûteuses associées à la méthode d'optimisation pixel par pixel. On ne tient pas compte des calculs effectués lors de la phase d'initialisation de la méthode (décomposition LU de la matrice $(\mathbf{A}_{\omega,p,s})_0$) car ils n'entrent pas dans le processus itératif.

\begin{center} \begin{tabular}{|l|l|c|} \hline
\multicolumn{2}{|l|}{\bf Evolution de $\chi_p$ en un pixel}
		& {\bf Coût}	\\ \hline
{\sl Pour le point initial}
	& Décomposition LU des matrices $\mathbf{M_\omega}$
		& \\
	& \quad $N_f$ décompositions LU
		& $N_f \times \mathcal{O}(n_M^3)$	\\
	&	Calcul du critère au point initial
		& \\
	& \quad $4 N_f N_k$ résolutions de syst. lin. triangulaires
		& $N_f \times N_k \left( 2 \times \mathcal{O}(n_A^2) + 2 \times \mathcal{O}(n_M^2) \right)$   \\
	& Calcul du gradient au point initial
		& \\
	& \quad $4 N_f N_k$ résolutions de syst. lin. triangulaires
		& $N_f \times N_k \left( 2 \times \mathcal{O}(n_A^2) + 2 \times \mathcal{O}(n_M^2) \right)$   \\
	& Choix du pixel
		& \\
	& \quad $4 N_f$ résolutions de syst. lin. triangulaires
		& $N_f \left( 2 \times \mathcal{O}(n_A^2) + 2 \times \mathcal{O}(n_M^2) \right)$   \\ \hline
{\sl Pour chaque pas testé}
	& Calcul du gradient
		& \\
	& \quad $4 N_f N_k$ résolutions de syst. lin. triangulaires
		& $N_f \times N_k \left( 2 \times \mathcal{O}(n_A^2) + 2 \times \mathcal{O}(n_M^2) \right)$   \\ \hline
\multicolumn{3}{c}{} \\ \hline
\multicolumn{2}{|l|}{\bf Evolution de $\chi_s$ en un pixel}
		& {\bf Coût}	\\ \hline
{\sl Pour le point initial}
	& Décomposition LU des matrices $\mathbf{M_\omega}$
		& \\
	& \quad $N_f$ décompositions LU
		& $N_f \times \mathcal{O}(n_M^3)$	\\
	&	Calcul du critère au point initial
		& \\
	& \quad $4 N_f N_k$ résolutions de syst. lin. triangulaires
		& $N_f \times N_k \left( 2 \times \mathcal{O}(n_A^2) + 2 \times \mathcal{O}(n_M^2) \right)$   \\
	& Calcul du gradient au point initial
		& \\
	& \quad $4 N_f N_k$ résolutions de syst. lin. triangulaires
		& $N_f \times N_k \left( 2 \times \mathcal{O}(n_A^2) + 2 \times \mathcal{O}(n_M^2) \right)$   \\
	& Choix du pixel
		& \\
	& \quad $12 N_f$ résolutions de syst. lin. triangulaires
		& $3 \times N_f \left( 2 \times \mathcal{O}(n_A^2) + 2 \times \mathcal{O}(n_M^2) \right)$\\ \hline
{\sl Pour chaque pas testé}
	& Calcul du gradient
		& \\
	& \quad $4 N_f N_k$ résolutions de syst. lin. triangulaires
		& $N_f \times N_k \left( 2 \times \mathcal{O}(n_A^2) + 2 \times \mathcal{O}(n_M^2) \right)$   \\ \hline
\end{tabular} \end{center}

Comme pour la méthode d'optimisation pixel par pixel, une méthode de type gradient nécessite le calcul de la valeur du critère et du gradient en différents points de l'espace de représentation. Pour chaque point considéré, on est alors amené à résoudre plusieurs systèmes linéaires dont les matrices normales sont $(\mathbf{A}_{\omega,p,s})_0$ (ou $(\mathbf{A}_{\omega,p,s})_0^\dagger$) et $\mathbf{M_\omega}$ (ou $\mathbf{M}_\omega^\dagger$). On procède donc de la façon suivante :
\begin{description}
\item[Lors d'une phase d'initialisation de la méthode :]
\end{description}
\begin{itemize}
\item pour chaque fréquence, on effectue la factorisation LU de la matrice $(\mathbf{A}_{\omega,p,s})_0$ (coût de calcul : $N_f \times \mathcal{O}(n_A^3)$) ;
\end{itemize}
\begin{description}
\item[Pour chaque point de l'espace de représentation considéré :]
\end{description}
\begin{itemize}
\item pour chaque fréquence, on effectue la factorisation LU de la matrice $\mathbf{M_\omega}$ (coût de calcul : $N_f \times \mathcal{O}(n_M^3)$) ;
\item pour calculer la valeur du critère, on utilise les facteurs LU pour se ramener à la résolution de $4 N_f N_k$ systèmes linéaires triangulaires (coût de calcul : $N_f \times N_k ( 2 \times \mathcal{O}(n_A^2) + 2 \times \mathcal{O}(n_M^2) )$) ;
\item pour calculer le gradient du critère, on utilise les facteurs LU pour se ramener à la résolution de $4 N_f N_k$  systèmes linéaires triangulaires supplémentaires (coût de calcul : $N_f \times N_k ( 2 \times \mathcal{O}(n_A^2) + 2 \times \mathcal{O}(n_M^2) )$).
\end{itemize}

Etant donné que l'algorithme de Moré et Thuente ne nécessite que peu d'essais pour obtenir un pas de progression vérifiant les conditions de Wolfe, le temps gagné à chaque itération en utilisant la méthode d'optimisation pixel par pixel n'est pas significatif par rapport à une méthode de type gradient.

De plus, pour les méthodes de type gradient, la direction de recherche est choisie en fonction des variations locales du critère dans l'espace de représentation, ce qui n'est pas le cas pour la méthode d'optimisation pixel par pixel (à chaque itération, on se restreint aux variations du critère le long d'un des axes principaux). On s'attend donc à ce qu'une méthode de type gradient converge en un nombre d'itérations beaucoup plus réduit.

L'utilisation d'une méthode de type gradient semble donc plus adaptée à notre problème.

\section{Choix d'une méthode pour définir la direction de recherche}

Pour définir une direction de recherche, nous avons proposé :
\begin{itemize}
\item le gradient conjugué non linéaire ;
\item l'algorithme L-BFGS ;
\item une méthode d'optimisation pixel par pixel.
\end{itemize}

\begin{figure}[!ht]
\begin{center}
\includegraphics[scale=0.6]{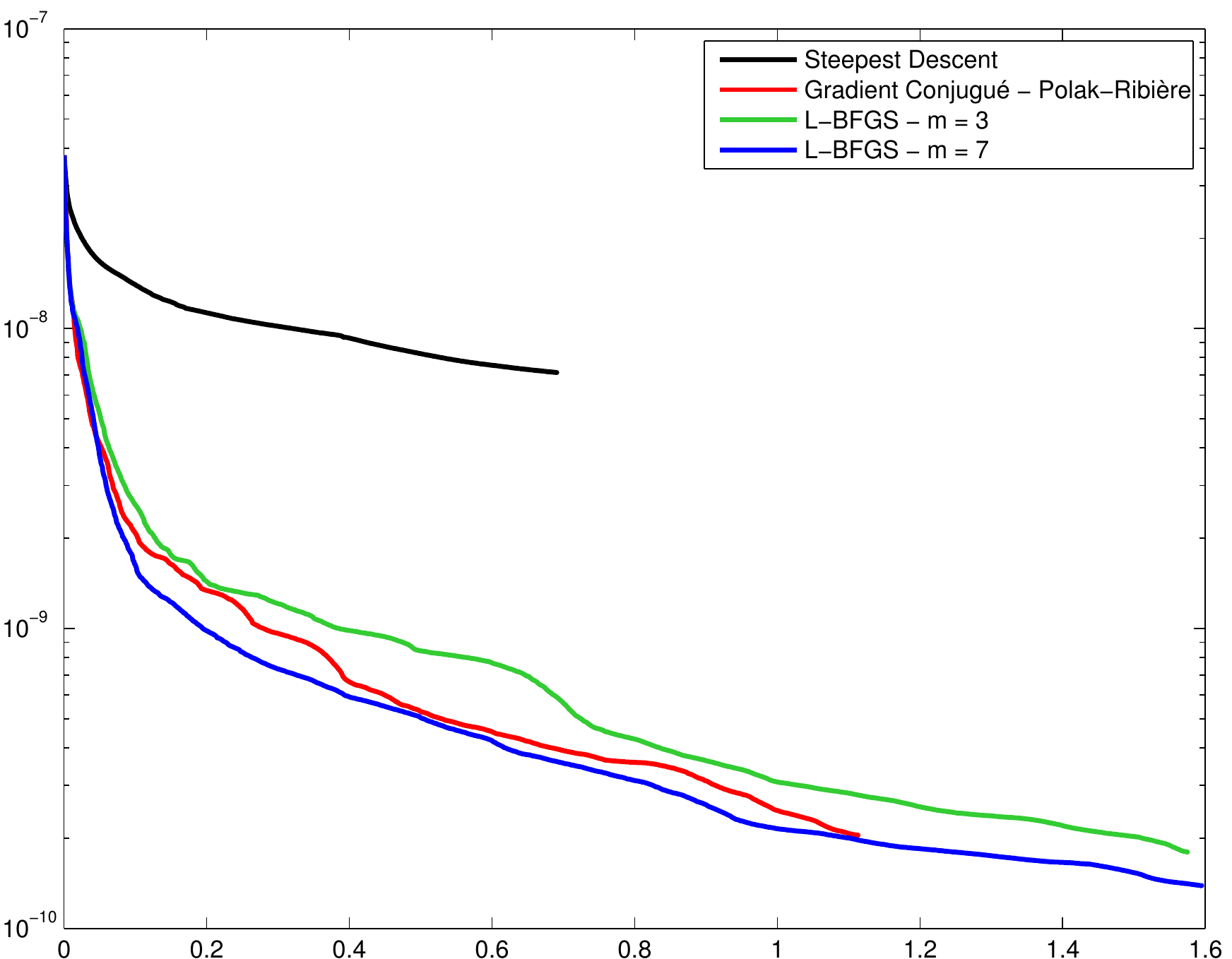}
\caption{Choix d'une méthode pour définir la direction de recherche. Les courbes montrent l'évolution temporelle du critère pour quatre méthodes de type gradient : plus forte pente, gradient conjugué non linéaire (Polak Ribière), L-BFGS à l'ordre 3 et à l'ordre 7.}
\label{Fig_ChoixMethodeDirectionDescente}
\end{center}
\end{figure}

Une analyse de la méthode d'optimisation pixel par pixel nous a montré qu'elle serait moins efficace que les autres pour le problème traité ici. Pour faire un choix parmi les méthodes restantes, nous avons appliqué les algorithmes d'inversion pour chacune de ces méthodes (nous avons utilisé le milieu test de petite taille présenté Partie \ref{Part_MilieuxTests} page \pageref{Part_MilieuxTests} ; les données mesurées ne sont pas bruitées et il n'y a pas de terme de régularisation). Nous présentons sur la Figure \ref{Fig_ChoixMethodeDirectionDescente} l'évolution du critère pour chaque cas au cours des 2000 premières itérations ; nous y incluons également l'évolution du critère obtenue avec la méthode de plus forte pente.

\bigskip

{\bf Remarque :} Il existe plusieurs variantes de l'algorithme du gradient conjugué non linéaire. Nous avons retenu l'algorithme de Polak-Ribière qui s'avère généralement plus efficace \cite{Nocedal99}.

\bigskip

Hormis l'algorithme de plus forte pente, les évolutions temporelles de la valeur du critère obtenues avec les différentes méthodes sont similaires. Etant donné que l'algorithme L-BFGS nécessite de stocker en mémoire un nombre de variables plus important que l'algorithme du gradient conjugué (l'ordre $m$ désigne le nombre de directions de recherche utilisées pour définir la direction de recherche à l'itération courante), nous choisissons finalement de retenir l'algorithme du gradient conjugué non linéaire pour définir la direction de descente à chaque itération.

\section{Premiers résultats}

Nous avons appliqué l'algorithme d'inversion au milieu de petite taille présenté dans la Partie \ref{Part_MilieuxTests}, page \pageref{Part_MilieuxTests}. Les données mesurées correspondent aux données obtenues par résolution du problème direct auxquelles nous avons ajouté un bruit blanc gaussien tel que le rapport signal à bruit soit égal à 30dB. Le critère comprend désormais deux termes de régularisation :
\begin{itemize}
\item un premier terme de rappel aux caractéristiques de la terre (rappel à zéro), on utilise la norme L1. On pénalise ainsi les valeurs de contraste d'amplitude trop grande. Le coefficient associé à ce terme est fixé à $10^{-19}$.
\item un second terme de différence entre pixels voisins, on utilise une norme L1L2 ($\mathcal{C}_{L1L2} = \sqrt{\Vert \chi \Vert^2 + \delta^2}$ où le vecteur $\chi$ est la concaténation des vecteurs $\chi_p$ et $\chi_s$). Le paramètre $\delta$ est fixé à $10^5$ et le coefficient associé à ce terme est fixé à $10^{-18}$.
\end{itemize}

\begin{figure}[!p]
\begin{center}
\subfloat[Cartes obtenues en initialisant aux caractéristiques de la terre (à gauche : $\chi_p$, à droite : $\chi_s$)]
{\includegraphics[scale=0.5]{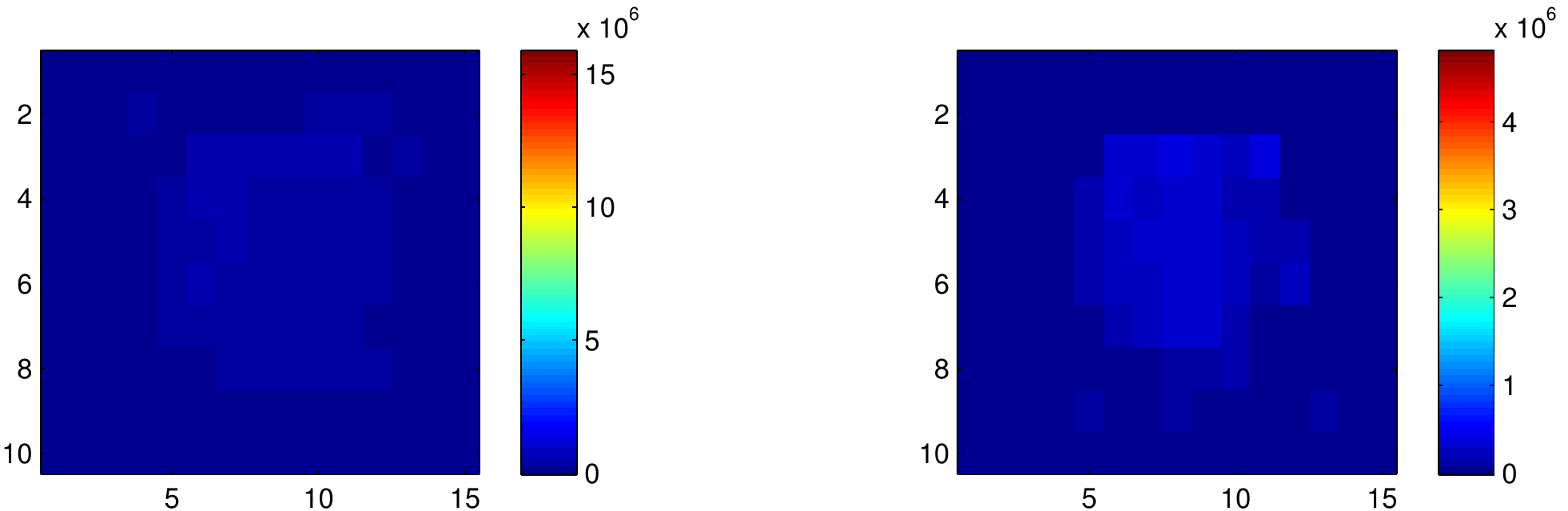}} \\
\subfloat[Cartes obtenues en initialisant à la solution (à gauche : $\chi_p$, à droite : $\chi_s$)]
{\includegraphics[scale=0.5]{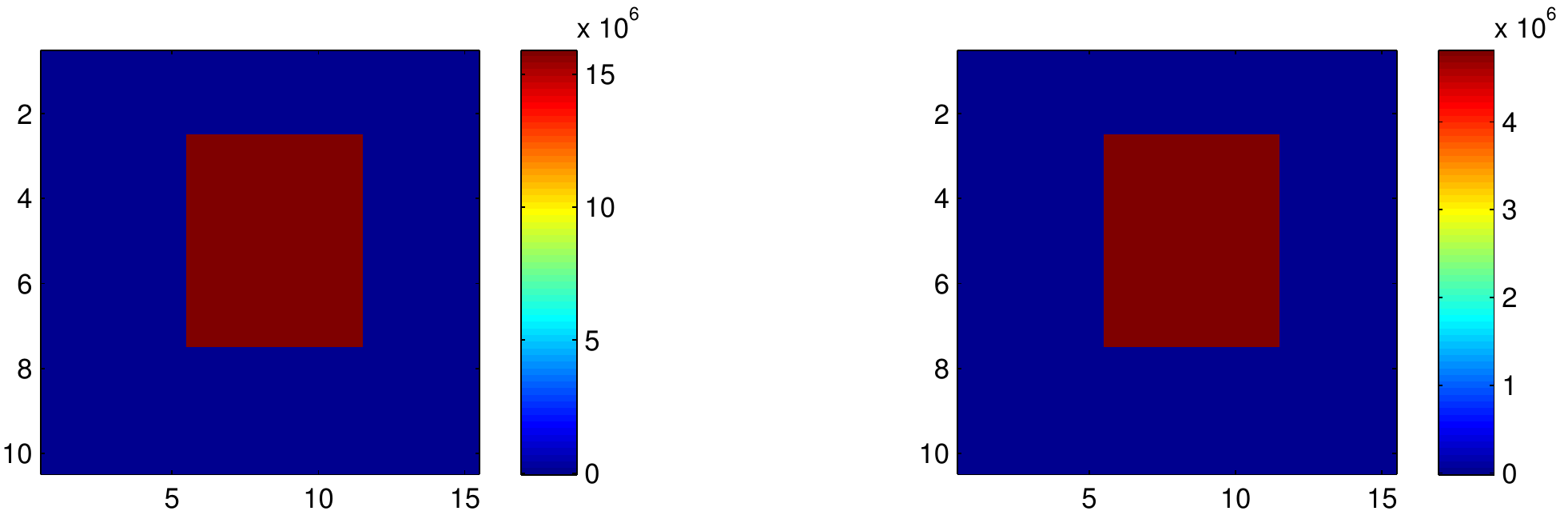}} \\
\subfloat[Evolution temporelle du critère (critère total et terme d'adéquation aux données)]
{\includegraphics[scale=0.5]{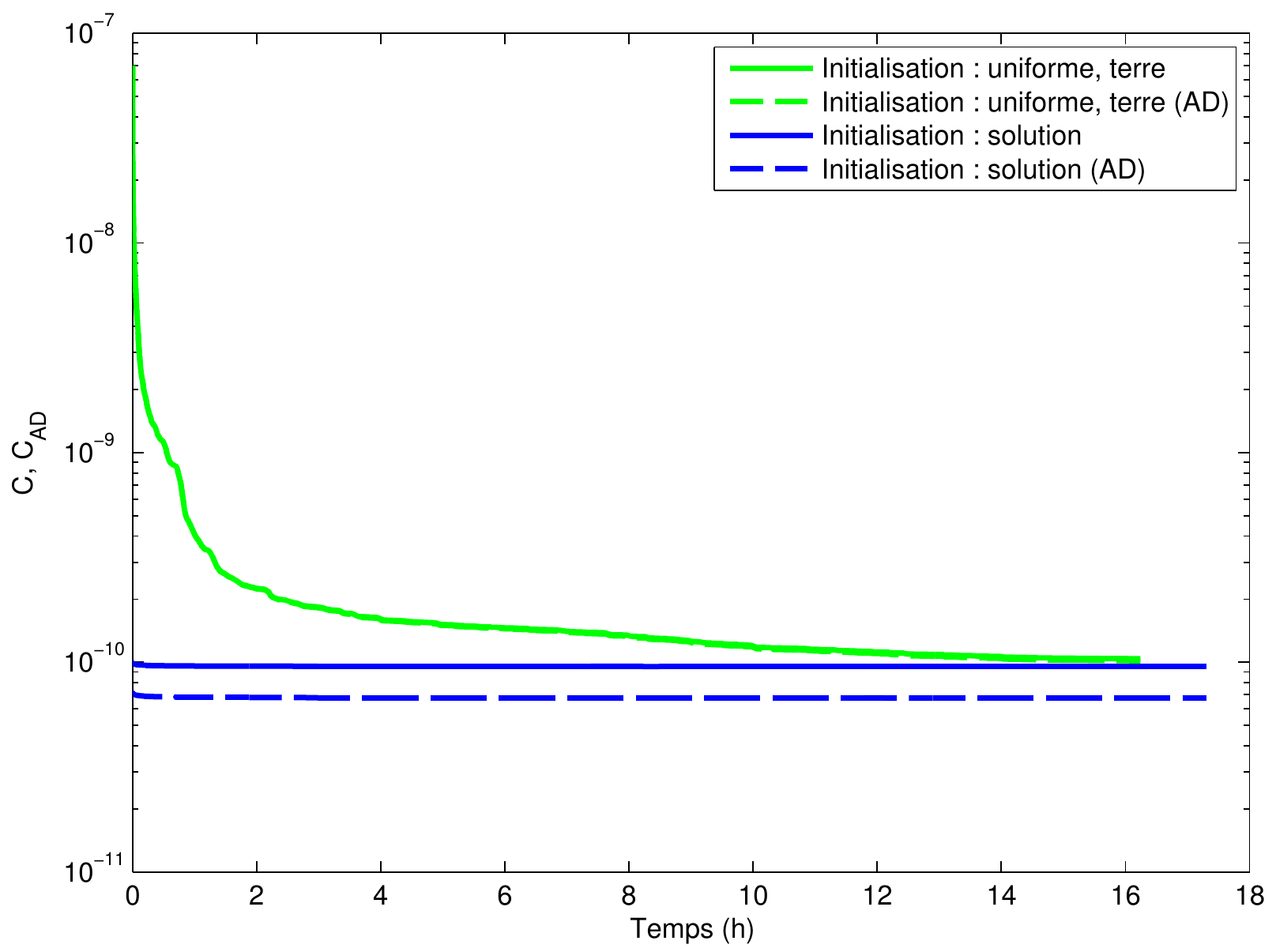}} \quad
\subfloat[Evolution temporelle de la norme du gradient]
{\includegraphics[scale=0.5]{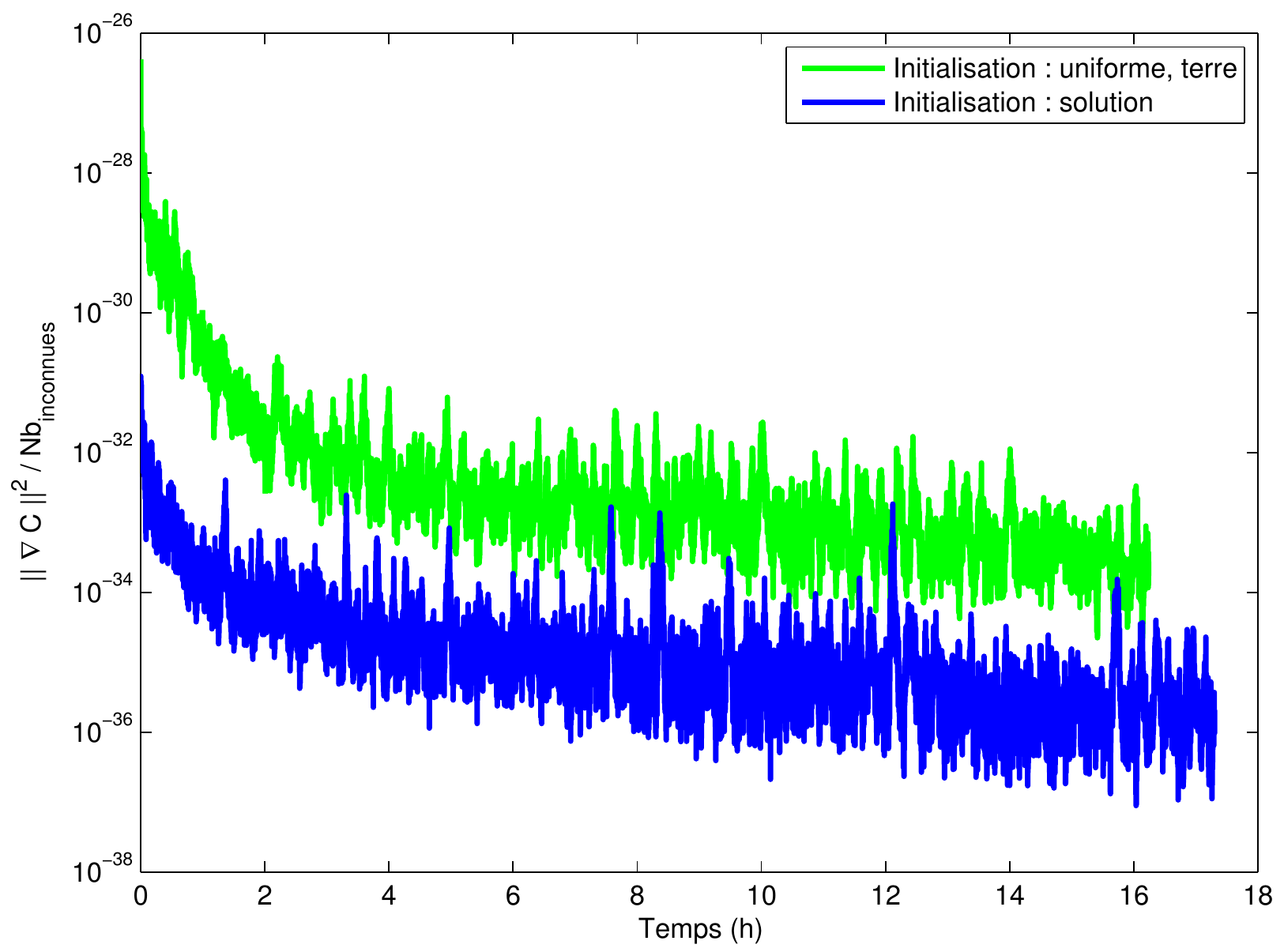}} \\
\end{center}
\caption{Premiers résultats obtenus sur le milieu de petite taille}
\label{Fig_ResultatsPetit_PrimalChi}
\end{figure}

Nous présentons sur la Figure \ref{Fig_ResultatsPetit_PrimalChi} les résultats obtenus pour deux initialisations différentes : pour la première initialisation, les caractéristiques de la zone d'étude sont égales à celles de la terre (les deux contrastes $\chi_p$ et $\chi_s$ initiaux sont nuls dans toute la zone d'étude) et pour la seconde, les contrastes sont égaux à la solution recherchée. Nous montrons les cartes obtenues ainsi que l'évolution temporelle de la valeur du critère (critère total et terme d'adéquations aux données) et de la norme du gradient pour ces deux initialisations.

On remarque tout d'abord un problème de convergence : les résultats présentés ici ont été obtenus après 30000 itérations (soit 16 heures de calcul environ) et la convergence n'a toujours pas été atteinte. Pour l'initialisation de la zone d'étude à des contrastes nuls (initialisation aux caractéristiques de la terre), le critère prend des valeurs de plus en plus proches de celles obtenues en initialisant à la solution mais il diminue de façon très lente.

On note également que pour l'initialisation à des contrastes nuls, les valeurs de contraste obtenues après 30000 itérations sont élevées mais encore très éloignées de celles du béton : le contraste $\chi_p$ atteint la valeur maximale de $9,9.10^5 m^2/s^2$ au lieu de $15,9.10^6 m^2/s^2$ et le contraste $\chi_s$ atteint la valeur maximale de $6,6.10^5 m^2/s^2$ au lieu de $4,8.10^6 m^2/s^2$. Les contrastes obtenus en partant de contrastes nuls sont donc encore loin de la solution recherchée.

Si l'on s'intéresse aux contrastes obtenus en initialisant les contrastes à zéro au bout de 10000 itérations (nombre d'itérations à partir duquel le critère évolue lentement), on remarque que les cartes obtenues sont déjà proches de celles obtenues au bout de 30000 itérations et que les valeurs de contrastes atteintes sont du même ordre de grandeur : $6,6.10^5 m^2/s^2$ pour $\chi_p$ et $4,2.10^5 m^2/s^2$ pour $\chi_s$. La lenteur de l'évolution du critère semble donc coïncider avec la forte amplitude des contrastes obtenus.

\section{Accélération de la convergence avec un changement de variable}

\subsection{Expression du critère en fonction d'un autre jeu de variables caractéristiques}

Les observations précédentes semblent mettre en évidence un problème de sensibilité du critère vis-à-vis des variations de contraste $\chi_p$ et $\chi_s$ lorsque ces derniers prennent des valeurs élevées. En effet, pour l'initialisation aux caractéristiques de la terre (contrastes nuls), on observe une rapide décroissance du critère lors des premières itérations puis le critère décroît plus lentement et les valeurs des contrastes correspondants sont plus élevées, bien qu'encore éloignées des valeurs recherchées. De même, pour l'initialisation à la solution, $\chi_p$ et $\chi_s$ prennent des valeurs élevées et le critère évolue lentement dès les premières itérations.

Afin d'améliorer la sensibilité du critère, nous proposons d'effectuer un changement de variable : au lieu d'exprimer les termes d'adéquation aux données et de régularisation du critère en fonction des variables $\chi_p$ et $\chi_s$, on les exprime en fonction des variables $\sigma_p$ et $\sigma_s$ qui sont choisies de sorte que de faibles variations de $\sigma_p$ et $\sigma_s$ induisent de fortes variations de $\chi_p$ et $\chi_s$ pour des valeurs de contraste élevées. Ainsi, le critère à minimiser reste identique mais les directions de recherches sélectionnées diffèrent d'une variable à l'autre (le changement de variable a une incidence sur le calcul du gradient).

Les différentes variables utilisées ont été présentées dans la Partie \ref{Part_IntroChgtVariable}, page \pageref{Part_IntroChgtVariable}. Nous avons effectué ces changements de variable en reprenant le même cas d'étude que précédemment :
\begin{itemize}
\item on considère le milieu de petite taille ;
\item les données mesurées sont obtenues par résolution du problème direct avec ajout de bruit blanc gaussien (le rapport signal à bruit est égal à 30dB) ;
\item le critère comprend deux termes de régularisation : un premier terme de rappel aux caractéristiques de la terre utilisant la norme L1 avec un coefficient de pondération de $10^{-19}$ et un second terme de différence entre pixels voisins utilisant une norme L1L2 dont le paramètre $\delta$ est fixé à $10^5$ avec un coefficient de pondération de $10^{-18}$.
\end{itemize}

\begin{figure}[!ht]
\begin{center}
\includegraphics[scale=0.6]{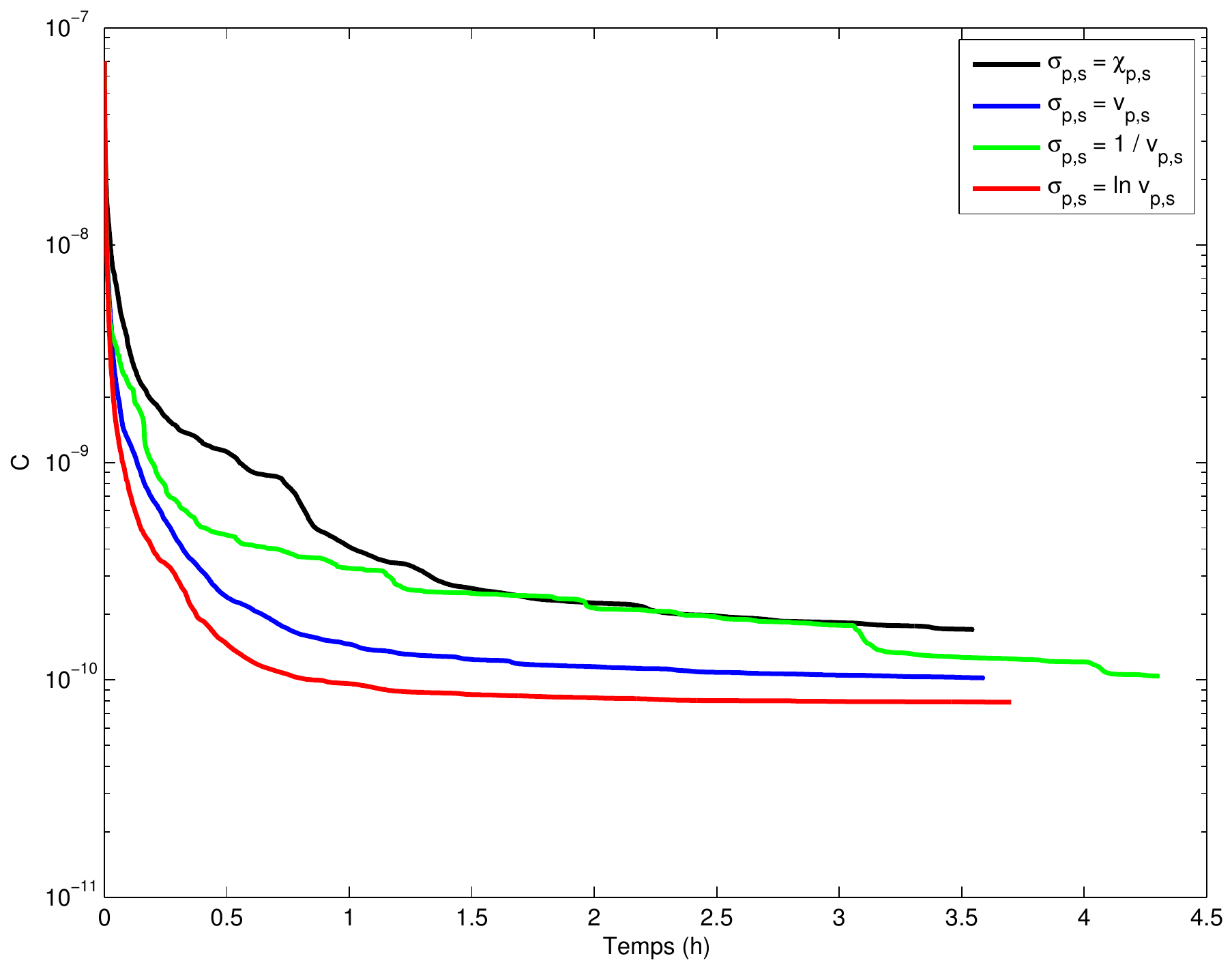}
\caption{Evolution temporelle du critère pour les différents changements de variable proposés en initialisant aux caractéristiques de la terre (en noir : $\sigma_{p,s} = \chi_{p,s}$, en bleu : $\sigma_{p,s} = \mathit{v}_{p,s}$, en vert : $\sigma_{p,s} = 1/\mathit{v}_{p,s}$, en rouge : $\sigma_{p,s} = \ln{\mathit{v}_{p,s}}$)}
\label{Fig_EvolCritChangementsVariable}
\end{center}
\end{figure}

On représente sur la Figure \ref{Fig_EvolCritChangementsVariable} les évolutions du critère obtenues pour les différents changements de variable lors des 5000 premières itérations. On remarque tout d'abord que pour les différents changements de variable proposés, le critère décroît plus rapidement qu'avec les variables $\chi_p$ et $\chi_s$. Cela confirme le fait qu'un problème de sensibilité du critère vis-à-vis de ces variables ralentissait la convergence de l'algorithme dans le cas précédent. La décroissance la plus rapide est observée avec l'utilisation des variables $\sigma_p = \ln{\mathit{v}_p}$ et $\sigma_s = \ln{\mathit{v}_s}$. Nous conserverons donc ce changement de variable par la suite.

\bigskip

{\bf Remarque sur la prise en compte des contraintes de positivité sur les vitesses :}

\bigskip

Les vitesses de propagation des ondes en pression ($\mathit{v}_p$) et en cisaillement ($\mathit{v}_s$) sont des grandeurs positives. Or, pour certains changements de variable, on risque de passer par des valeurs négatives de $\mathit{v}_p$ et $\mathit{v}_s$, ce qui peut engendrer un comportement pathologique de l'algorithme. C'est le cas par exemple lors de l'utilisation des variables $\sigma_p = \frac{1}{\mathit{v}_p}$ et $\sigma_s = \frac{1}{\mathit{v}_s}$ qui ne devraient pas prendre de valeurs négatives.

\begin{figure}[!htbp]
\begin{center}
\includegraphics[scale=0.6]{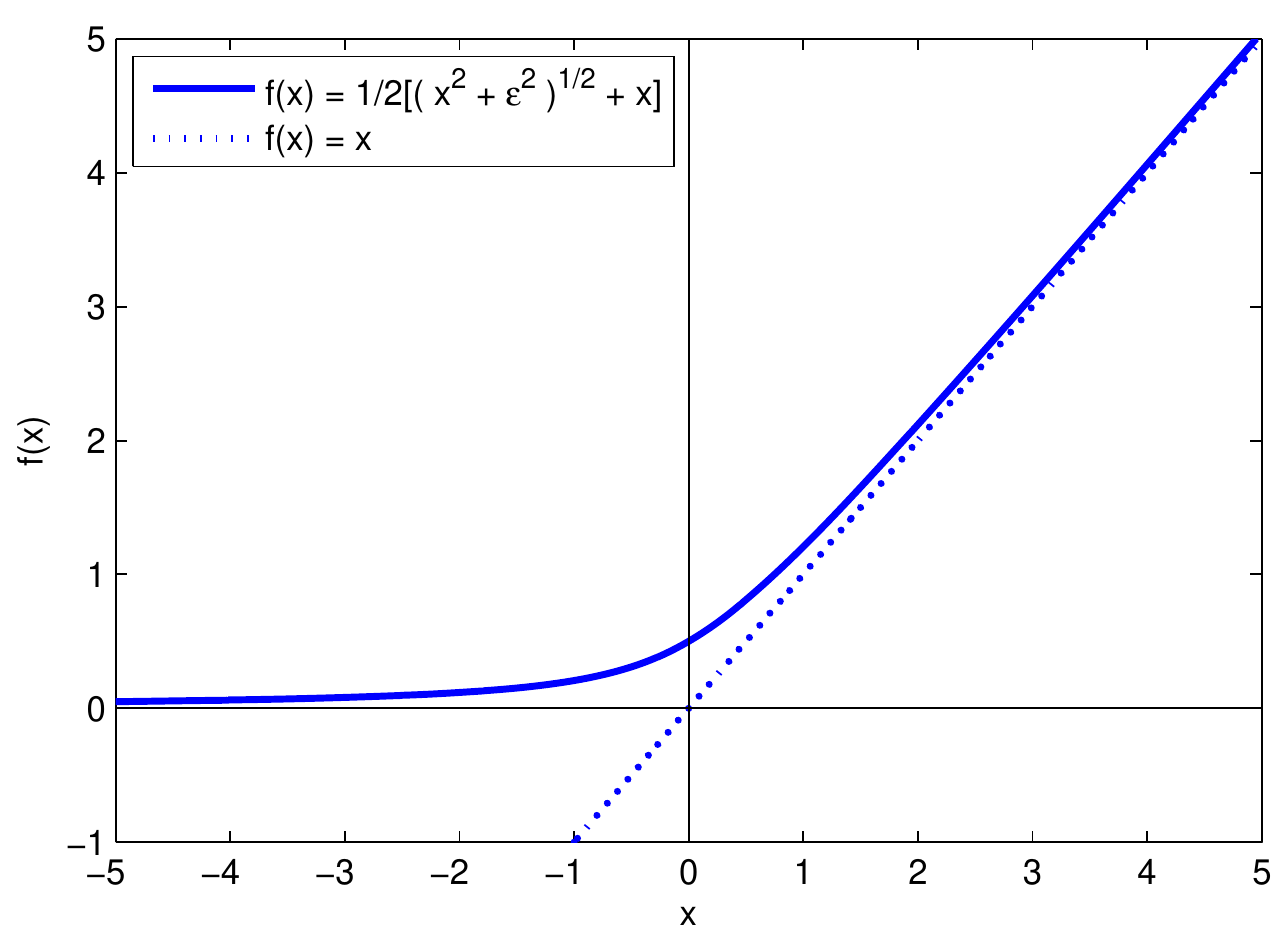}
\caption{Tracé de la fonction utilisée pour vérifier la contrainte de positivité sur les vitesses avec $\epsilon = 1$ (trait plein) et comparaison avec la fonction identité (pointillés)}
\label{Fig_ContraintePositivite}
\end{center}
\end{figure}

Dans ce cas, il est possible de respecter les contraintes de positivité tout en évitant une incidence notable sur le comportement de l'algorithme d'inversion en effectuant un changement de variable supplémentaire. On propose d'utiliser les variables $\psi_p$ ou $\psi_s$ telles que :
\begin{equation} \sigma_p = \frac{1}{2} \sqrt{\psi_p^2 + \epsilon^2} + \frac{\psi_p}{2} \quad \text{et} \quad \sigma_s = \frac{1}{2} \sqrt{\psi_s^2 + \epsilon^2} + \frac{\psi_s}{2} \end{equation}
Ainsi, on s'assure que la contrainte de positivité est respectée et si le coefficient $\epsilon$ est choisi suffisamment petit, le changement de variable n'a qu'une faible incidence au-delà d'un certain seuil strictement positif (la fonction est proche de la fonction identité). On représente sur la figure \ref{Fig_ContraintePositivite} le tracé de la fonction utilisée pour passer de $\psi_{p,s}$ à $\sigma_{p,s}$.

Pour le changement de variable retenu, les variables $\sigma_p$ et $\sigma_s$ sont forcément associées à des valeurs de $\mathit{v}_p$ et $\mathit{v}_s$ positives (on a $\mathit{v}_p = \exp{\sigma_p}$ et $\mathit{v}_s = \exp{\sigma_s}$). Un tel changement de variable est donc inutile dans ce cas.

\bigskip

{\bf Remarque sur l'affichage des résultats :}

\bigskip

Sur la Figure \ref{Fig_ResultatsPetit_PrimalChi}, nous avons présenté les premiers résultats obtenus sur le milieu de petite taille en affichant les contrastes $\chi_p$ et $\chi_s$. Nous afficherons maintenant les cartes correspondant aux nouvelles variables utilisées, c'est-à-dire $\sigma_p = \ln{\mathit{v}_p}$ et $\sigma_s = \ln{\mathit{v}_s}$.

\subsection{Pénalisation d'un autre jeu de variables}

Pour les premiers résultats obtenus, les deux termes de régularisation du critère portaient sur les contrastes $\chi_p$ et $\chi_s$. Cependant, cela peut induire des problèmes de conditionnement (on a $\chi_{p,s} = \exp{2 \sigma_{p,s}} - \exp{2 \sigma_{p,s,0}}$ pour $\sigma_{p,s} = \ln{v_{p,s}}$). Par exemple, des variations de $\sigma_p$ ou $\sigma_s$ en un pixel autour d'une grande valeur auront une forte incidence sur le terme de rappel aux caractéristiques de la terre contrairement à des variations autour d'une valeur plus faible. C'est pourquoi il peut être préférable d'appliquer la régularisation à un autre jeu de variables.

Nous avons lancé l'algorithme d'inversion sur le milieu de petite taille en portant la régularisation sur les variables $\sigma_p = \ln{\mathit{v}_p}$ et $\sigma_s = \ln{\mathit{v}_s}$ et en exprimant le critère en fonction de ces mêmes variables. Comme précédemment, nous utilisons les données obtenues par résolution du problème direct auxquelles nous avons ajouté un bruit blanc gaussien (rapport signal à bruit égal à 30dB) et le critère comprend deux termes de régularisation :
\begin{itemize}
\item un terme de rappel aux caractéristiques de la terre pour lequel on utilise la norme L1 avec un coefficient de pondération égal à $10^{-13}$ ;
\item un terme de différence entre pixels voisins pour lequel on utilise une norme L1L2 avec un paramètre $\delta$ égal à $0,1$ et un coefficient de pondération égal à $10^{-11}$.
\end{itemize}

\begin{figure}[!p]
\begin{center}
\subfloat[Cartes obtenues en initialisant aux caractéristiques de la terre (à gauche : $\ln{\mathit{v}_p}$, à droite : $\ln{\mathit{v}_s}$)]
{\includegraphics[scale=0.5]{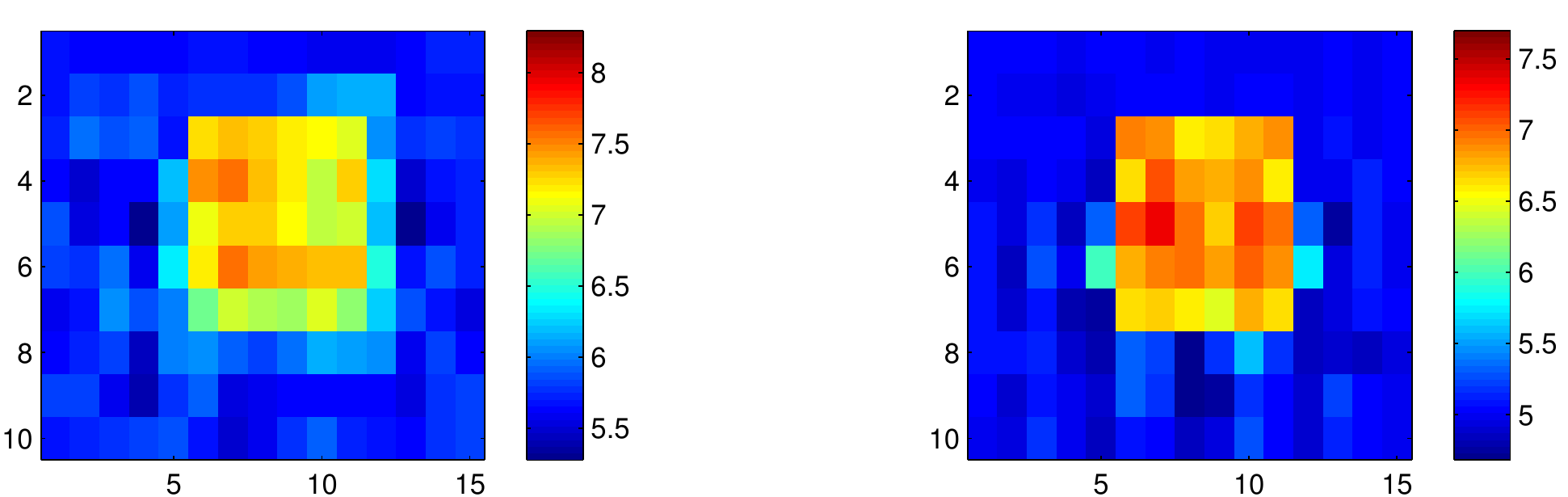}} \\
\subfloat[Cartes obtenues en initialisant à la solution (à gauche : $\ln{\mathit{v}_p}$, à droite : $\ln{\mathit{v}_s}$)]
{\includegraphics[scale=0.5]{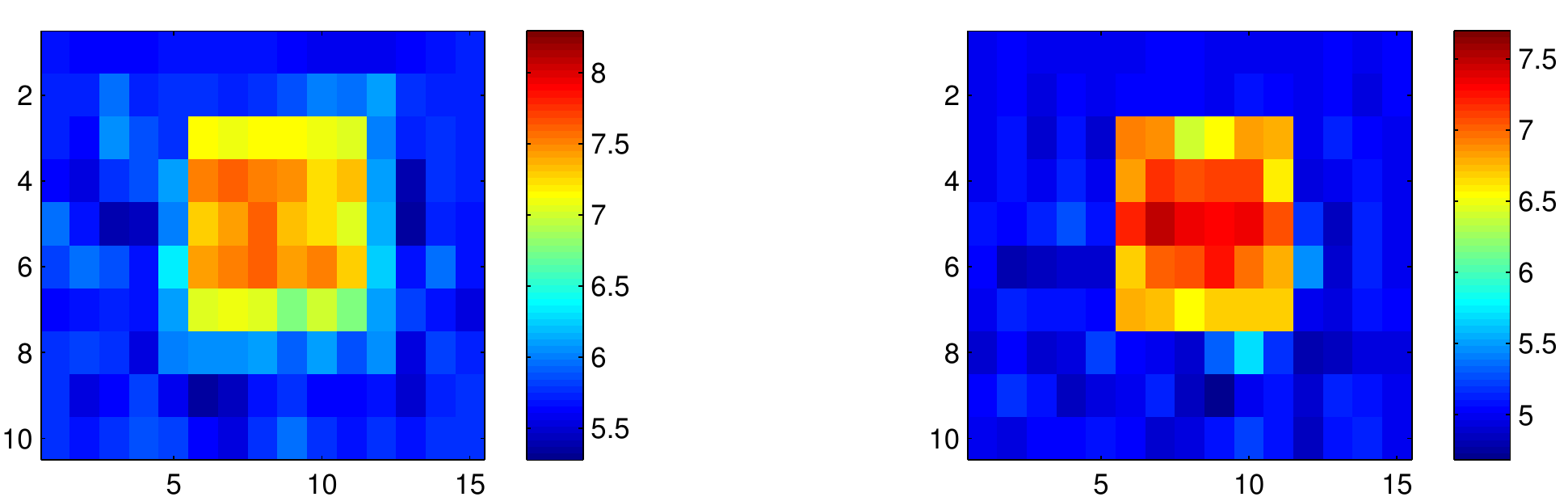}} \\
\subfloat[Evolution temporelle du critère (critère total et terme d'adéquation aux données)]
{\includegraphics[scale=0.5]{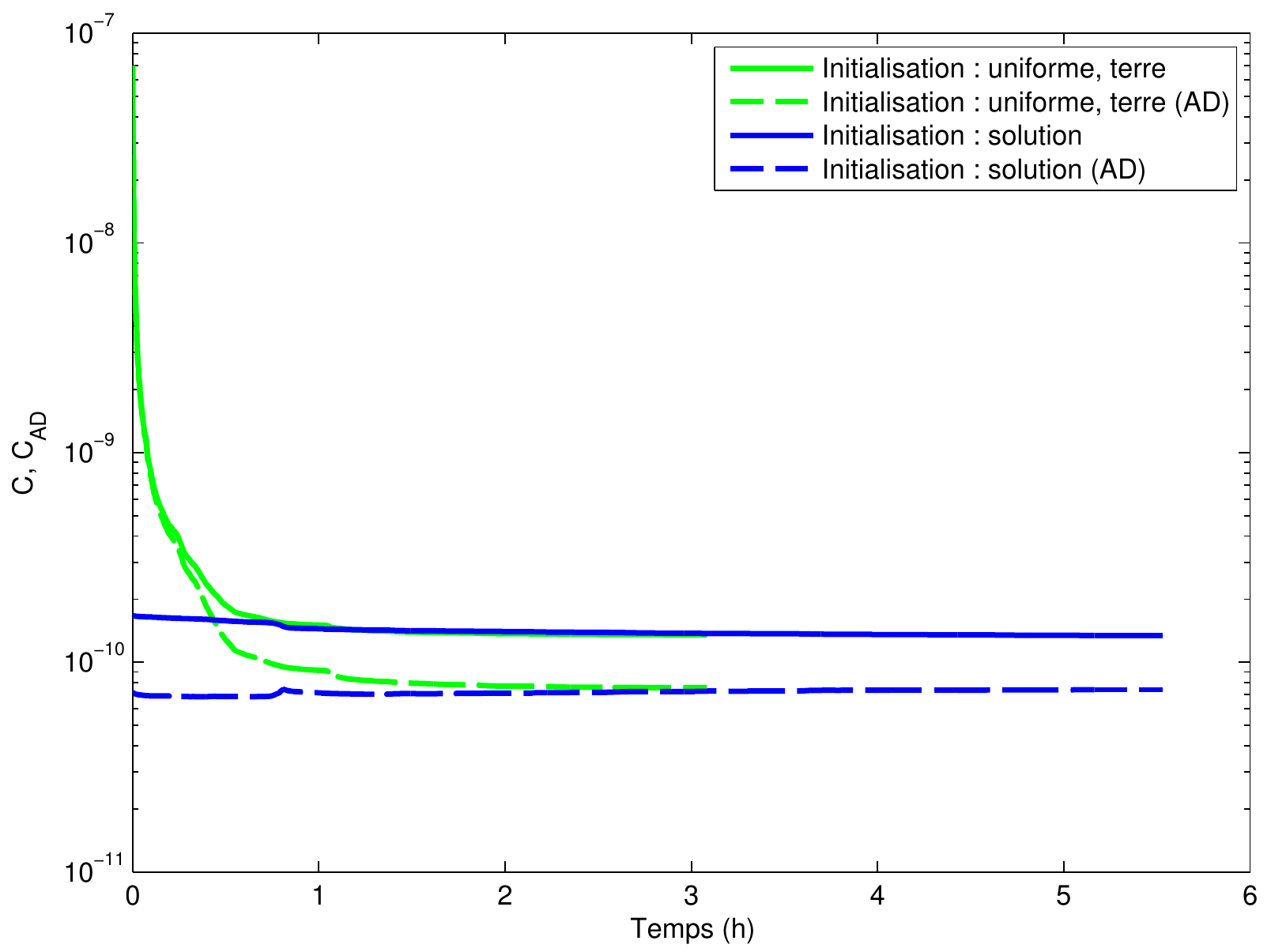}} \quad
\subfloat[Evolution temporelle de la norme du gradient]
{\includegraphics[scale=0.5]{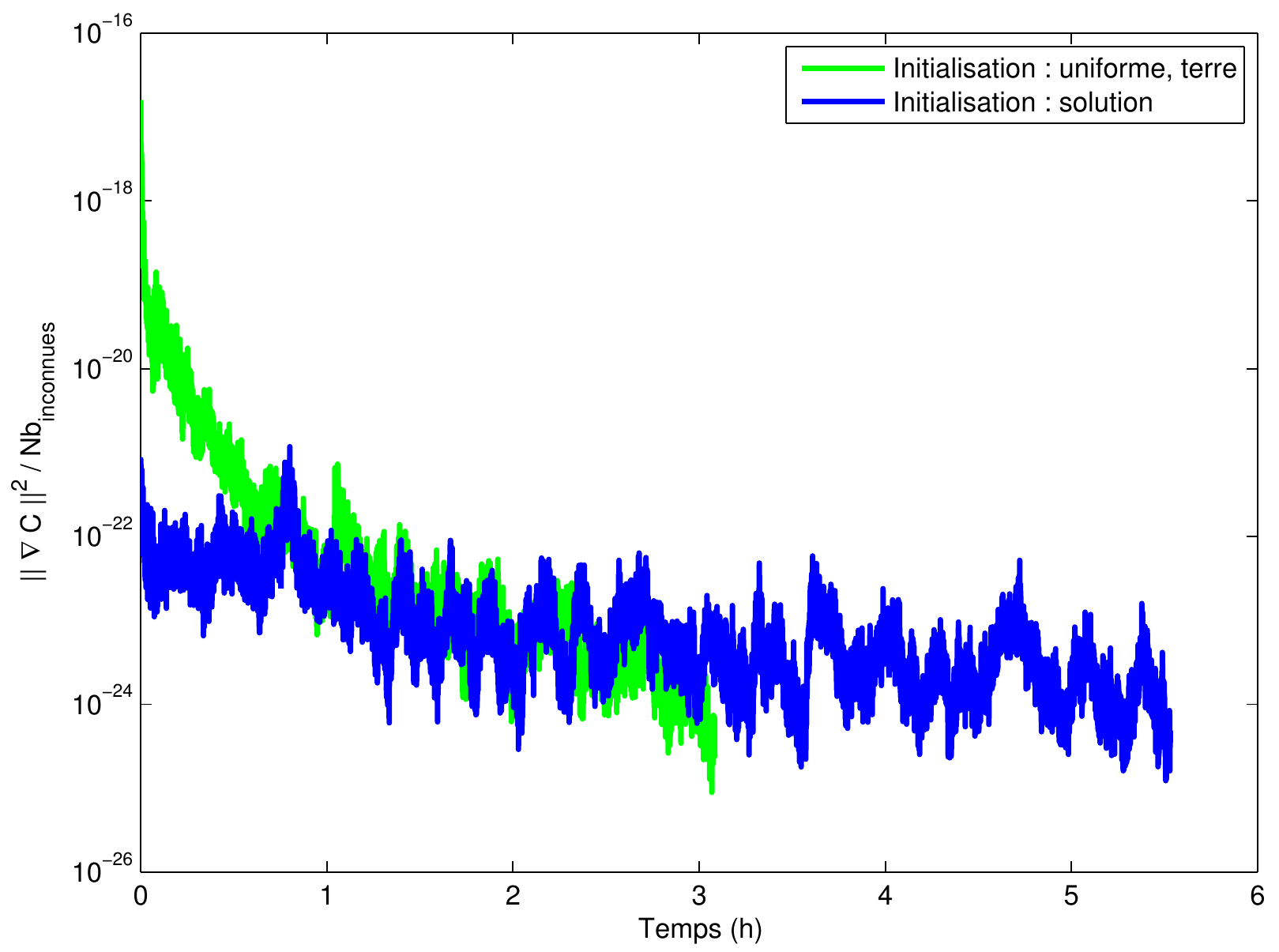}} \\
\end{center}
\caption{Résultats obtenus sur le milieu de petite taille en utilisant les variables $\sigma_p = \ln{\mathit{v}_p}$ et $\sigma_s = \ln{\mathit{v}_s}$}
\label{Fig_ResultatsPetit_PrimalLnVpVs}
\end{figure}

Nous considérons que l'algorithme est arrivé à convergence lorsque la norme du gradient divisée par le nombre d'inconnues (égal à $300$ ici) est inférieure à $10^{-24}$ pendant 50 itérations successives. Nous présentons sur la Figure \ref{Fig_ResultatsPetit_PrimalLnVpVs} les résultats obtenus en initialisant les caractéristiques de la zone d'étude à celles de la terre d'une part et à la solution recherchée d'autre part. Nous montrons les cartes obtenues ainsi que l'évolution temporelle de la valeur du critère et de la norme du gradient.

Le fait d'exprimer le critère en fonction de $\sigma_p = \ln{\mathit{v}_p}$ et $\sigma_s = \ln{\mathit{v}_s}$ et d'utiliser ces variables pour la régularisation a permis d'améliorer le comportement de l'algorithme d'inversion. La sensibilité du critère vis-à-vis des variables optimisées est améliorée puisqu'en moins de six heures, le critère a maintenant convergé vers la même valeur pour les deux initialisations.

On note également que pour les deux initialisations, les deux cartes obtenues sont similaires. Cependant, les valeurs maximales n'atteignent pas les valeurs recherchées ($7,6$ au lieu de $8,3$ pour $\ln{\mathit{v}_p}$ et $7,4$ au lieu de $7,7$ pour $\ln{\mathit{v}_s}$). Cela est dû aux termes de régularisation du critère qui tendent à diminuer les amplitudes (on remarque que cette diminution des valeurs maximales n'affecte quasiment pas la valeur du terme d'adéquation aux données du critère pour l'initialisation à la solution).

\section{Introduction des fréquences de façon progressive}

Jusqu'à maintenant, nous avons utilisé l'algorithme d'inversion en introduisant dès le départ toute l'information fréquentielle. Or, il peut être préférable d'introduire les informations contenues dans les basses fréquences dans un premier temps puis d'ajouter les informations contenues dans les plus hautes fréquences de manière progressive : les basses fréquences apportent des informations sur les variations spatiales lentes et permettent d'obtenir une image lisse du milieu ; l'introduction des plus hautes fréquences permet ensuite d'affiner cette image \cite{Gelis05}.

Nous avons testé cette démarche en procédant de la façon suivante : tout d'abord, nous n'avons introduit que les informations associées à la plus basse fréquence (46,7 Hz). L'algorithme étant arrivé à convergence, nous avons ensuite ajouté les informations associées à la fréquence suivante (93,3 Hz) puis nous avons procédé de la même manière pour les fréquences plus élevées jusqu'à prendre en compte toutes les fréquences (15 fréquences allant de 46,7 Hz à 700 Hz).

Pour chaque groupe de fréquences, nous considérons que l'algorithme est arrivé à convergence lorsque la norme du gradient divisée par le nombre d'inconnues est inférieure à $5.10^{-23}$ pendant 50 itérations successives. Lorsque ce critère d'arrêt est vérifié, on utilise les cartes obtenues pour initialiser l'algorithme d'inversion que l'on relance en introduisant une fréquence supplémentaire.

\begin{figure}[!p]
\begin{center}
\subfloat[Cartes obtenues en initialisant aux caractéristiques de la terre (à gauche : $\ln{\mathit{v}_p}$, à droite : $\ln{\mathit{v}_s}$)]
{\includegraphics[scale=0.5]{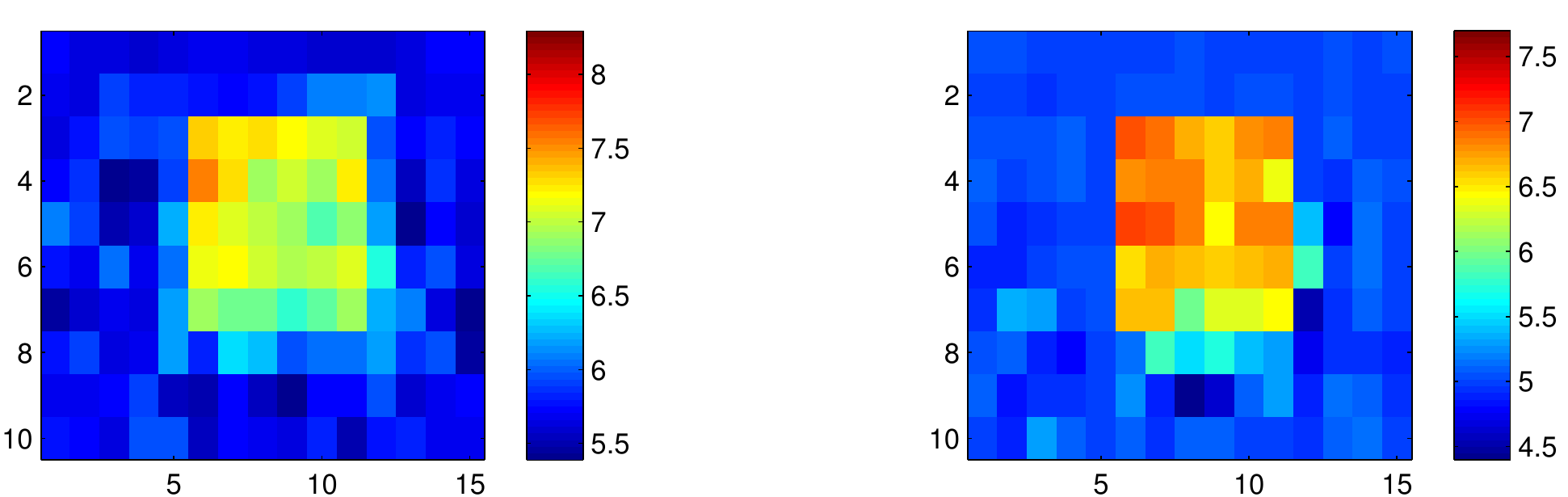}} \\
\subfloat[Cartes obtenues en initialisant à la solution (à gauche : $\ln{\mathit{v}_p}$, à droite : $\ln{\mathit{v}_s}$)]
{\includegraphics[scale=0.5]{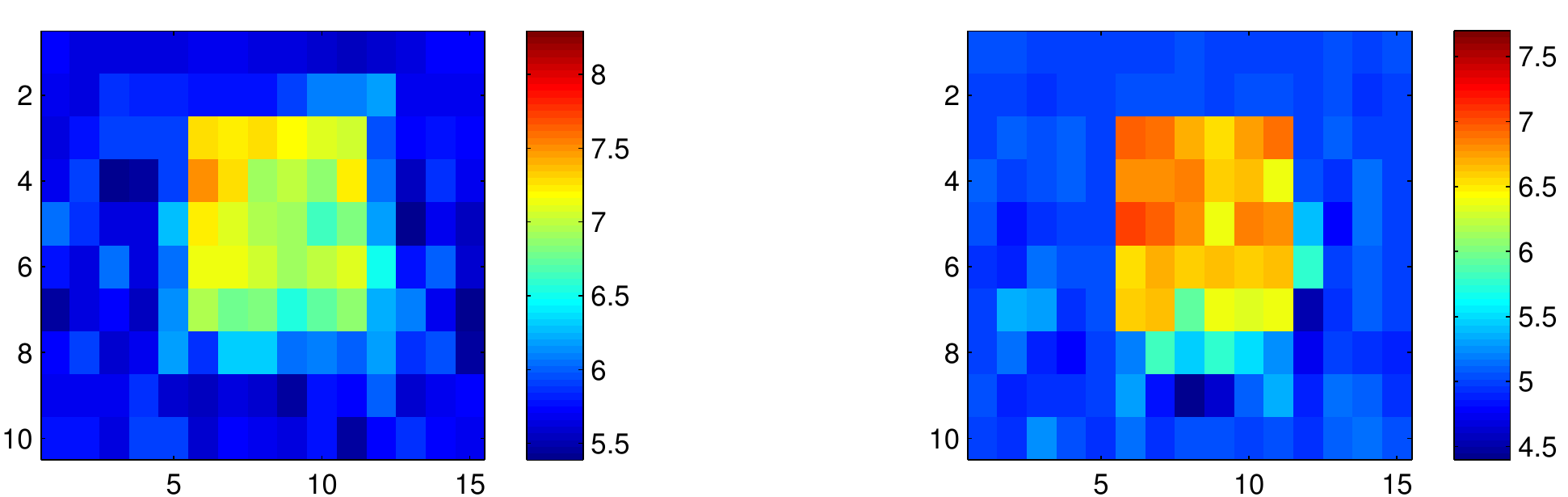}} \\
\subfloat[Evolution temporelle du critère (critère total et terme d'adéquation aux données)]
{\includegraphics[scale=0.5]{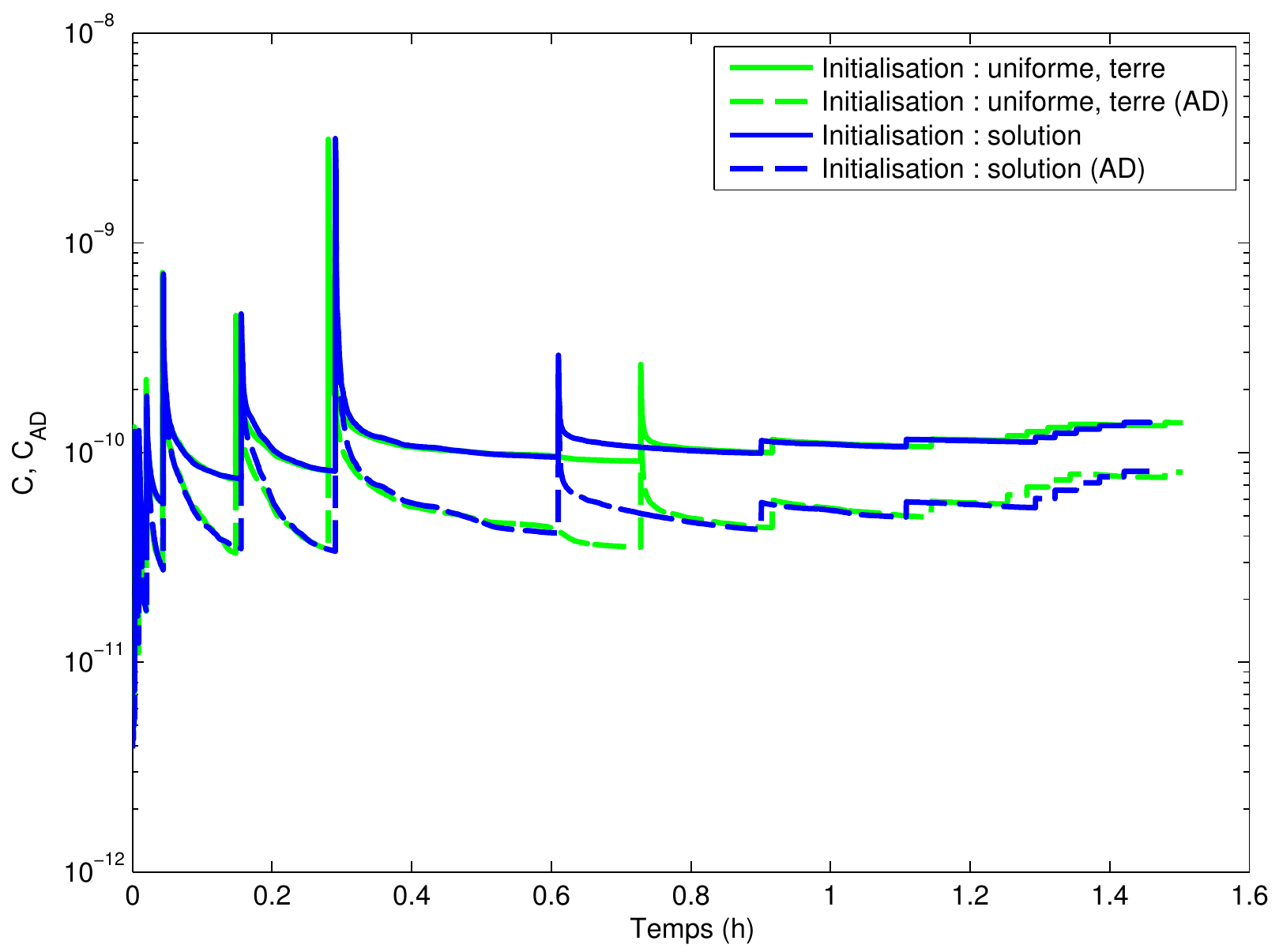}} \quad
\subfloat[Evolution temporelle de la norme du gradient]
{\includegraphics[scale=0.5]{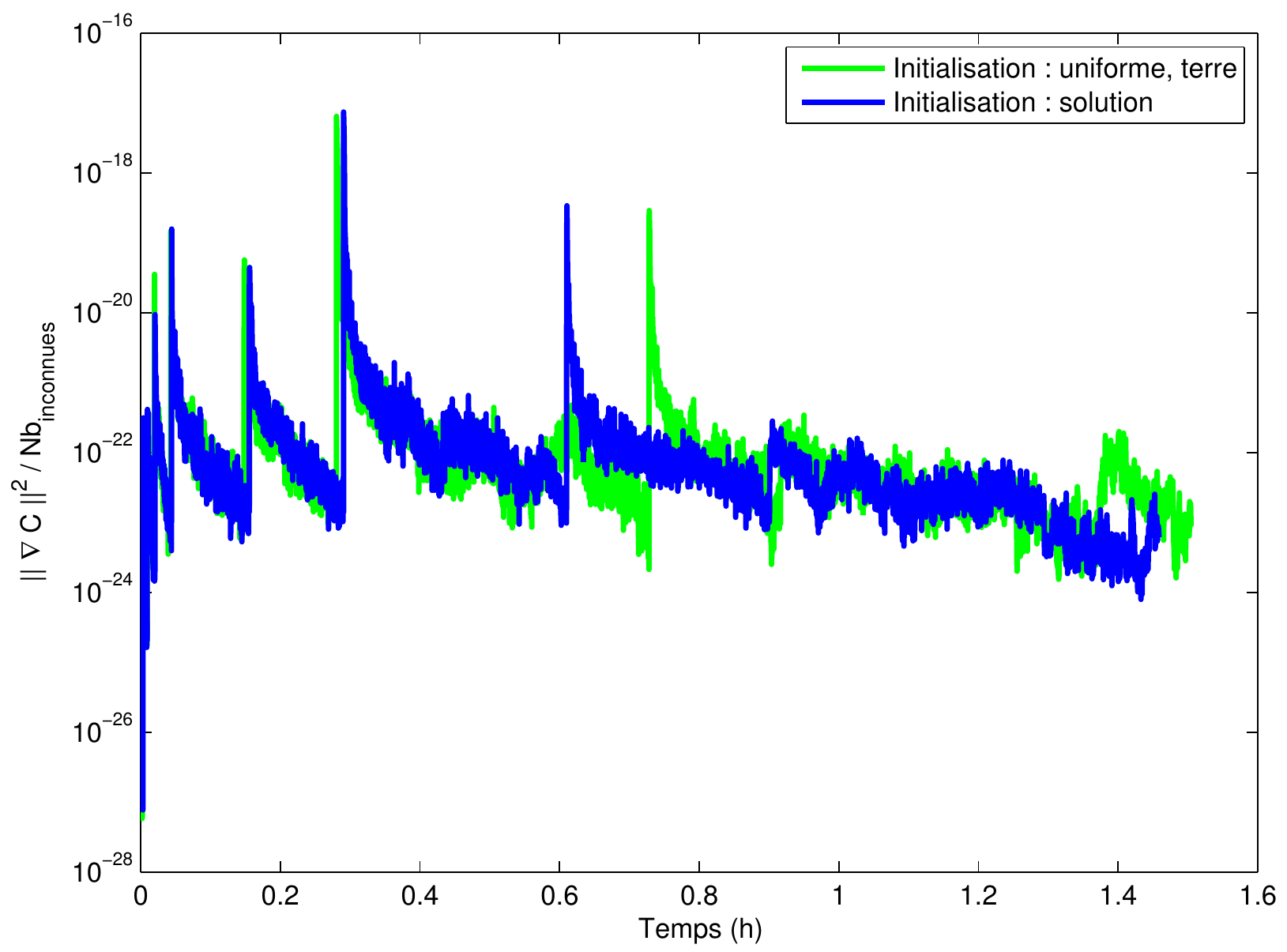}} \\
\end{center}
\caption{Résultats obtenus sur le milieu de petite taille en utilisant les variables $\sigma_p = \ln{\mathit{v}_p}$ et $\sigma_s = \ln{\mathit{v}_s}$ et en introduisant les fréquences de façon progressive}
\label{Fig_IntroductionProgressiveFreq}
\end{figure}

\begin{figure}[!p]
\begin{center}
\subfloat[Pour chaque fréquence considérée, amplitude des données bruitées utilisées pour l'inversion (rouge pointillé) et amplitude des mesures correspondant à la solution recherchée (rouge), aux cartes obtenues en initialisant à la terre (vert) et aux cartes obtenues en initialisant à la solution (bleu) en chaque capteur]
{\includegraphics[scale=0.6]{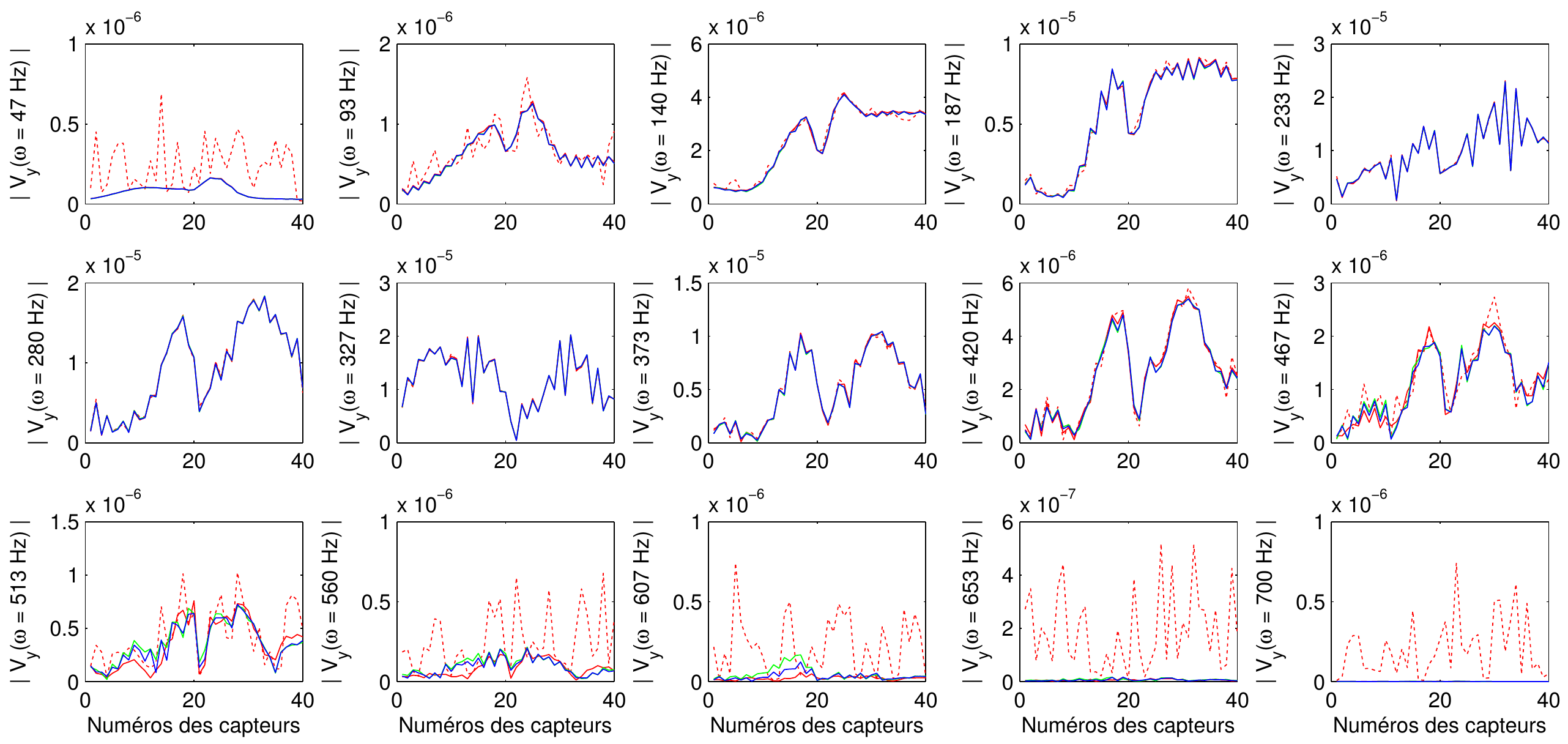}} \\
\subfloat[Sismogrammes reconstruits à partir des cartes obtenues en initialisant aux caractéristiques de la terre]
{\includegraphics[scale=0.55]{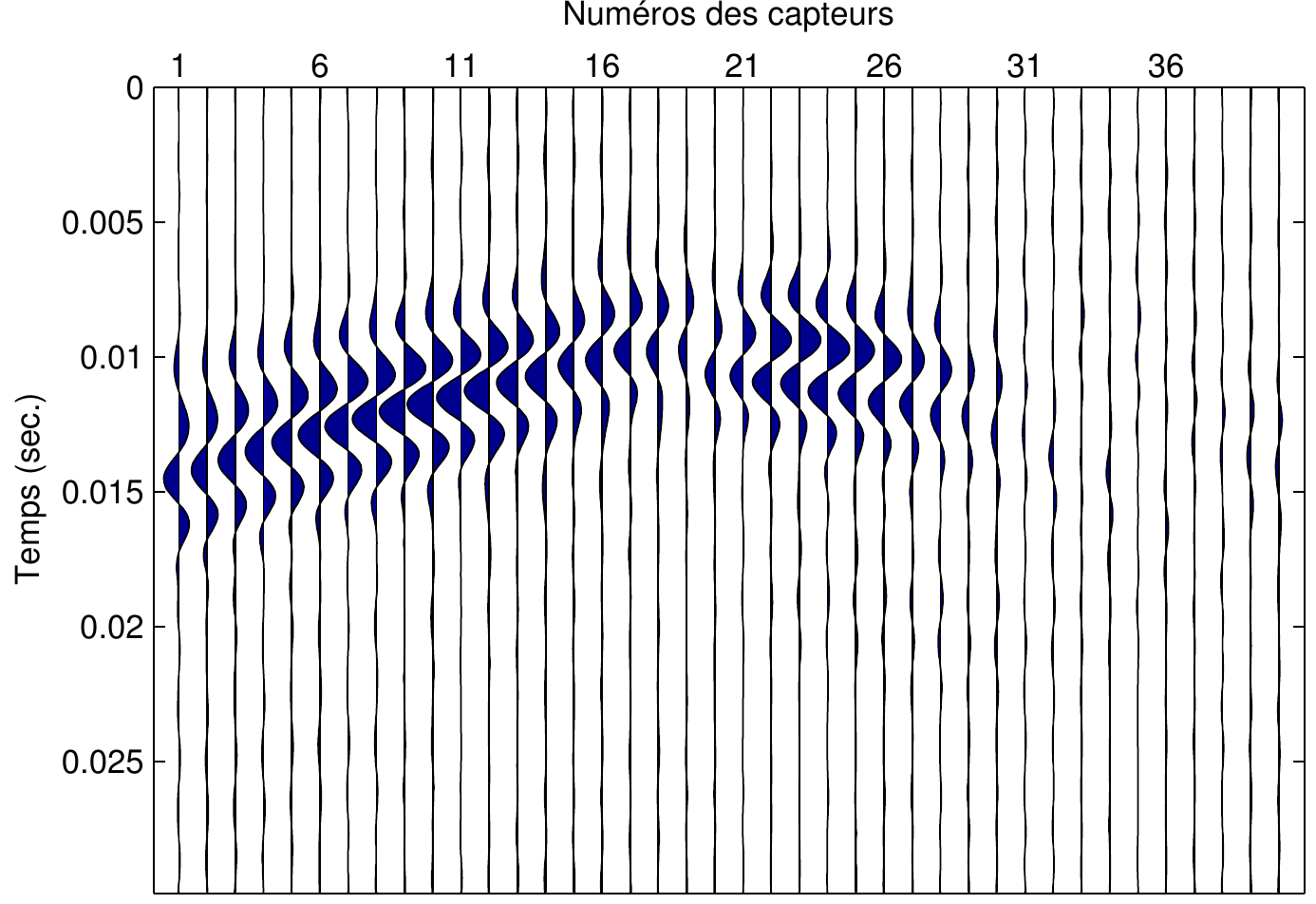}} \quad
\subfloat[Sismogrammes reconstruits à partir des cartes obtenues en initialisant à la solution]
{\includegraphics[scale=0.55]{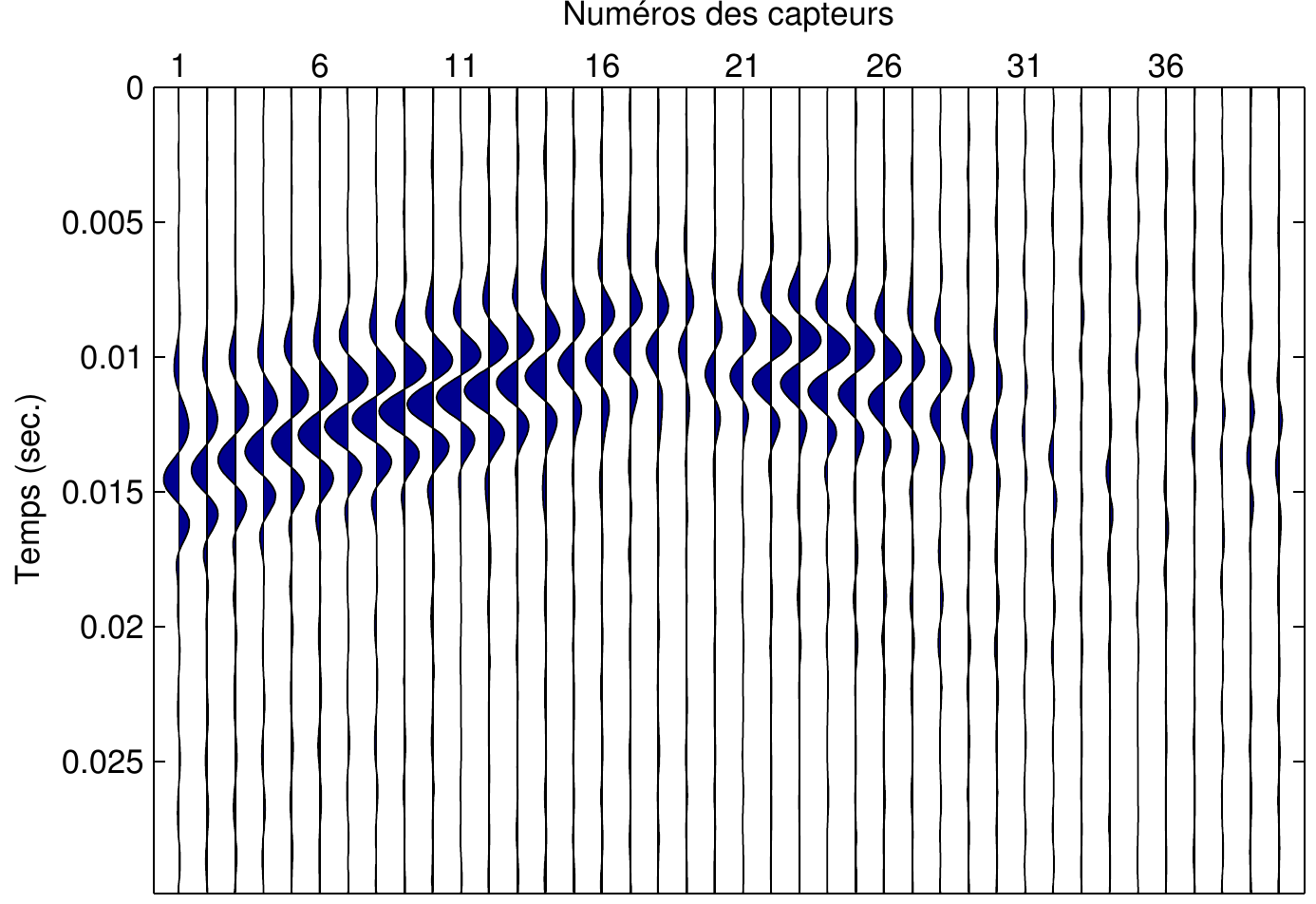}} \\
\subfloat[Sismogrammes correspondant à la solution recherchée]
{\includegraphics[scale=0.55]{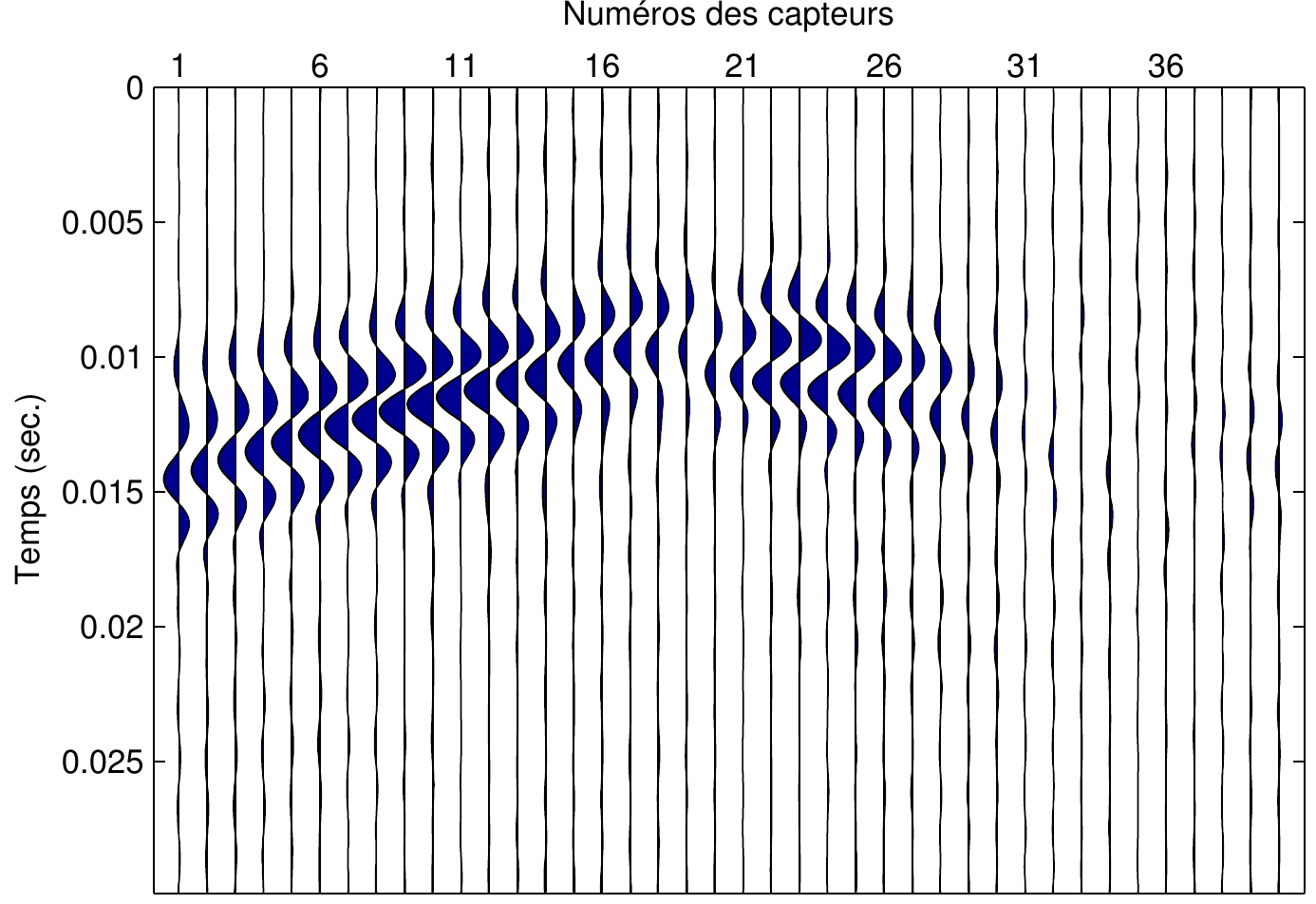}}
\end{center}
\caption{Comparaison des données capteurs correspondant aux cartes solutions (sans et avec ajout de bruit blanc) et aux cartes obtenues après inversion (initialisation à la terre et à la solution) dans le cas où la source est positionnée à gauche de l'objet diffractant (première position sur la Figure \ref{Fig_PetitMilieuTest}, page \pageref{Fig_PetitMilieuTest}). On affiche les données dans le domaine fréquentiel et dans le domaine temporel (sismogrammes).}
\label{Fig_SismoPetitMilieu}
\end{figure}

Les valeurs des coefficients intervenant dans les deux termes de régularisation sont les mêmes que précédemment (le coefficient de pondération associé au terme de rappel aux caractéristiques de la terre est égal à $10^{-13}$ et le coefficient de pondération associé au terme de différence entre pixels voisins est égal à $10^{-11}$ ; le paramètre $\delta$ est égal à $0,1$). Les résultats obtenus sont présentés sur la Figure \ref{Fig_IntroductionProgressiveFreq}. Nous affichons également sur la Figure \ref{Fig_SismoPetitMilieu} les données mesurées par les capteurs lorsque les cartes correspondent à la solution recherchée (sans et avec ajout de bruit blanc) et aux résultats obtenus pour les deux initialisations.

Les résultats obtenus avec l'introduction progressive des fréquences restent satisfaisants : l'algorithme arrive à convergence et les cartes obtenues sont similaires à celles obtenues précédemment (voir Figure \ref{Fig_ResultatsPetit_PrimalLnVpVs}). De plus, la comparaison des données reconstruites montrent que pour les deux initialisations, les mesures sont semblables. On remarque cependant que l'on ne parvient pas à retrouver le contenu hautes fréquences des données utilisées pour l'inversion. Cela s'explique par le fait que les composantes hautes fréquences sont de faible amplitude et n'ont donc qu'une faible influence sur la valeur du critère.

L'intérêt principal de cette démarche est la diminution du temps de calcul : sans l'introduction progressive des fréquences, il fallait environ trois heures de calcul pour l'initialisation aux caractéristiques de la terre et presque six heures de calcul pour l'initialisation à la solution. Il faut maintenant une heure et demie environ à l'algorithme pour arriver à convergence pour les deux initialisations. Cela s'explique surtout par le fait que le coût de calcul par itération pour les premiers groupes de fréquences est plus faible (le temps de calcul du critère et du gradient est quasiment proportionnel au nombre de fréquences considérées).

\section{Résultats obtenus sur le milieu de taille intermédiaire}

Nous présentons maintenant les résultats obtenus sur le milieu de taille intermédiaire présenté dans la Partie \ref{Part_MilieuxTests}, page \pageref{Part_MilieuxTests} en tenant compte des conclusions auxquelles nous sommes parvenus d'après les résultats obtenus sur le petit milieu :
\begin{itemize}
\item le critère à minimiser s'exprime en fonction des variables $\sigma_p = \ln{\mathit{v}_p}$ et $\sigma_s = \ln{\mathit{v}_s}$ ;
\item les termes de régularisation du critère portent sur cette même variable ;
\item nous introduisons les fréquences de manière progressives, des basses fréquences vers les hautes fréquences.
\end{itemize}

Comme pour le petit milieu, les données mesurées sont obtenues par résolution du problème direct puis ajout de bruit blanc gaussien (le rapport signal à bruit est égal à 30dB) et le critère comprend deux termes de régularisation : un terme de rappel aux caractéristiques de la terre pour lequel on utilise la norme L1 avec un coefficient de pondération égal à $10^{-12}$ et un terme de différence entre pixels voisins pour lequel on utilise une norme L1L2 avec un paramètre $\delta$ est fixé à $0,1$ et un coefficient de pondération égal à $10^{-10}$.

Nous avons considéré que l'algorithme était arrivé à convergence lorsque la norme du gradient divisée par le nombre d'inconnues (environ $1400$ ici) est inférieure à $5.10^{-23}$ pendant 50 itérations successives.

\begin{figure}[!p]
\begin{center}
\subfloat[Carte recherchée]
{\includegraphics[scale=0.5]{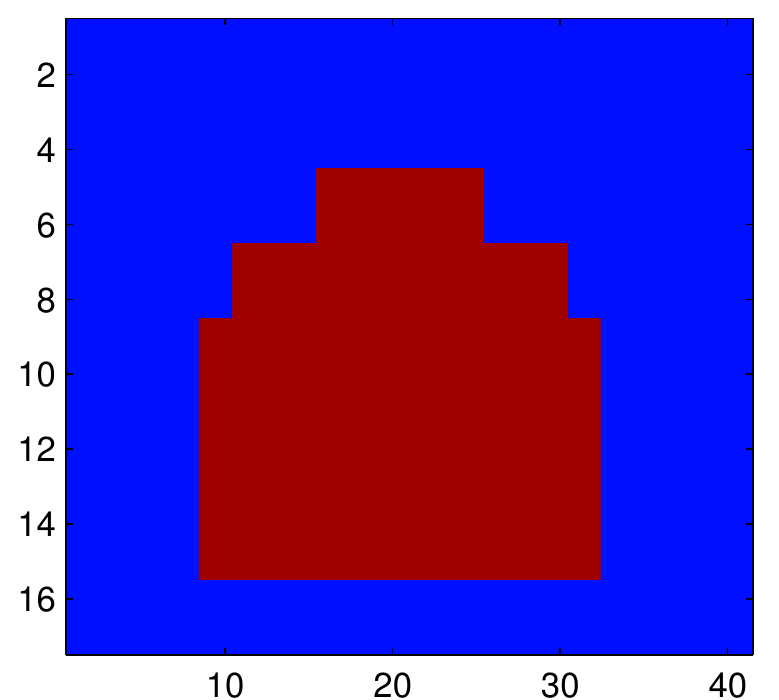}} \\
\subfloat[Cartes obtenues en initialisant aux caractéristiques de la terre (à gauche : $\ln{\mathit{v}_p}$, à droite : $\ln{\mathit{v}_s}$)]
{\includegraphics[scale=0.5]{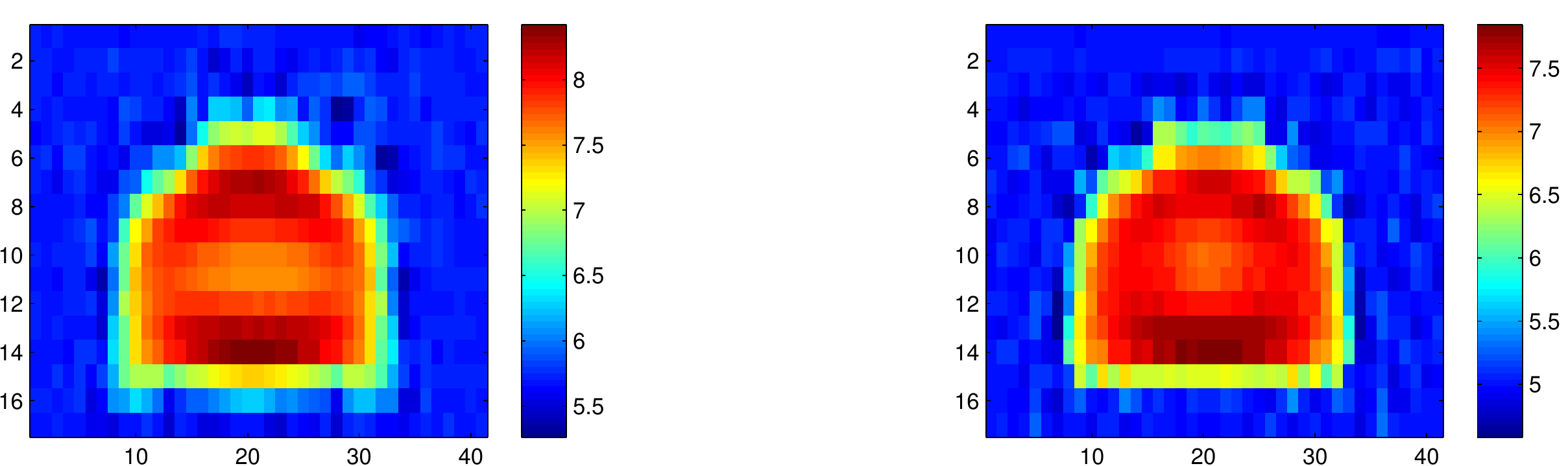}} \\
\subfloat[Cartes obtenues en initialisant à la solution (à gauche : $\ln{\mathit{v}_p}$, à droite : $\ln{\mathit{v}_s}$)]
{\includegraphics[scale=0.5]{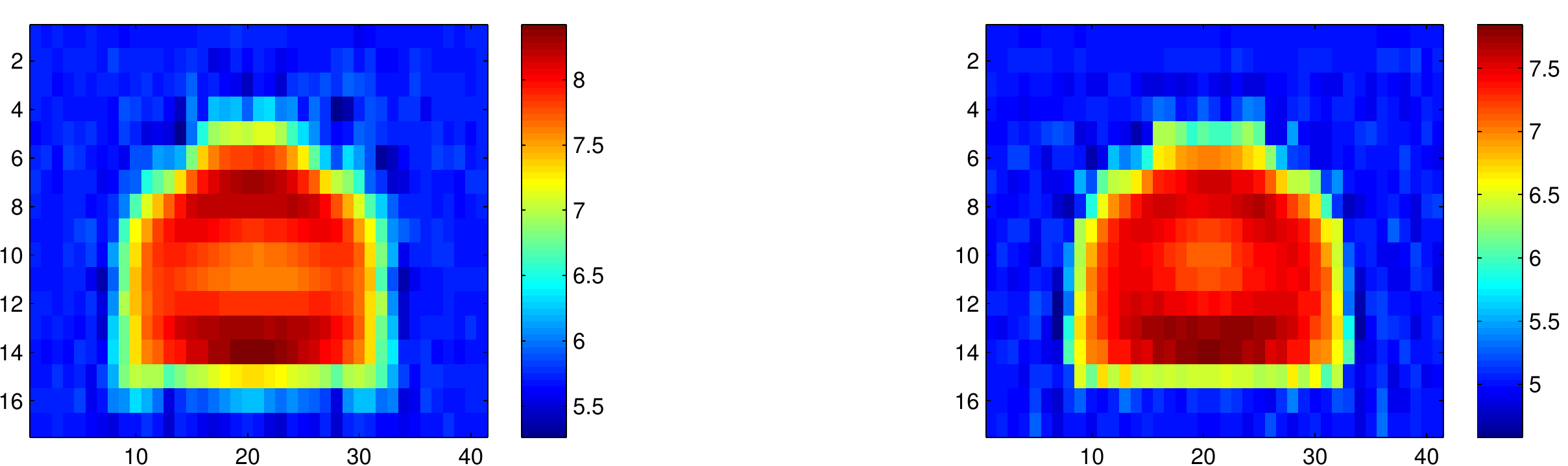}} \\
\subfloat[Evolution temporelle du critère (critère total et terme d'adéquation aux données)]
{\includegraphics[scale=0.5]{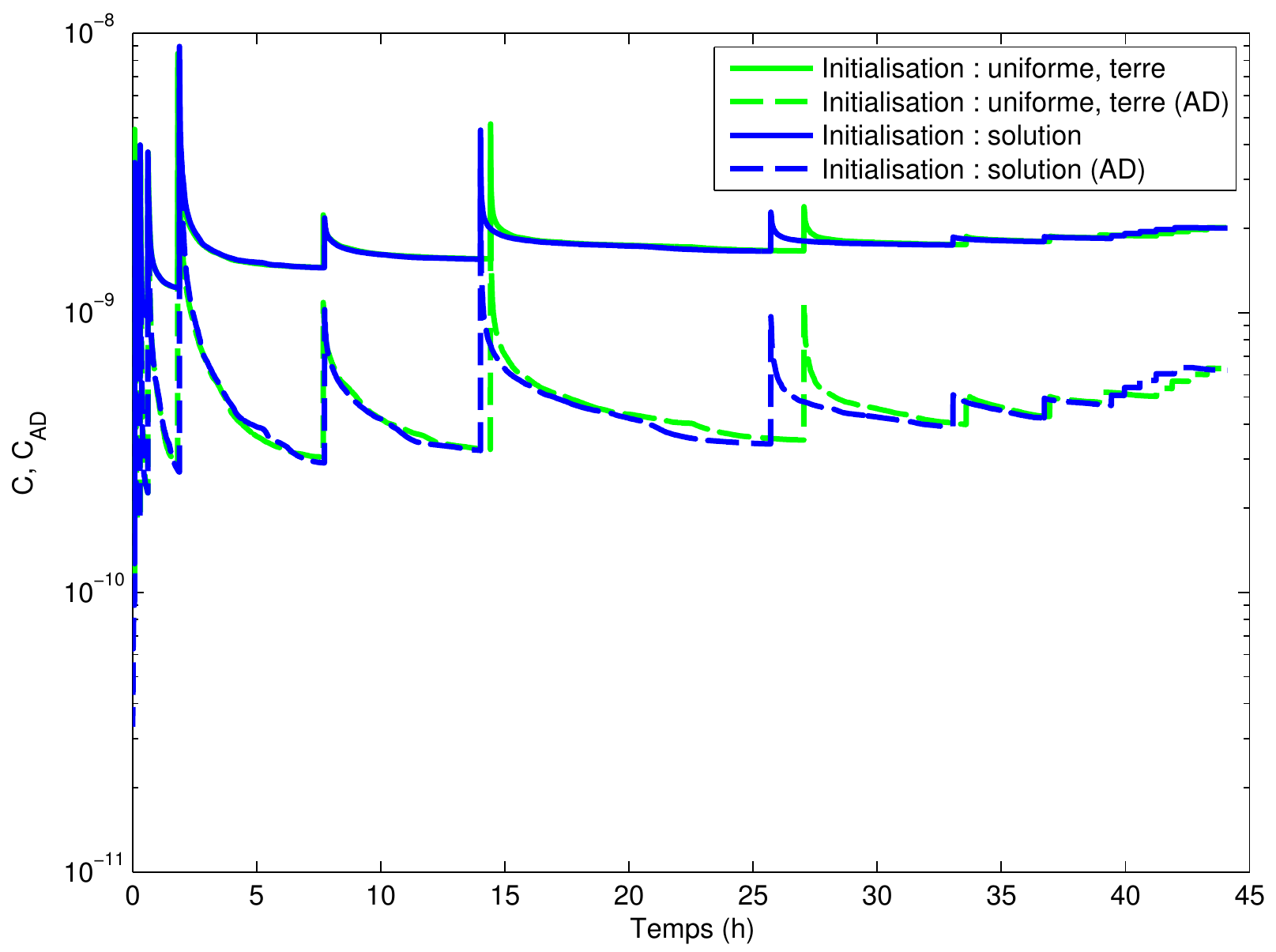}} \quad
\subfloat[Evolution temporelle de la norme du gradient]
{\includegraphics[scale=0.5]{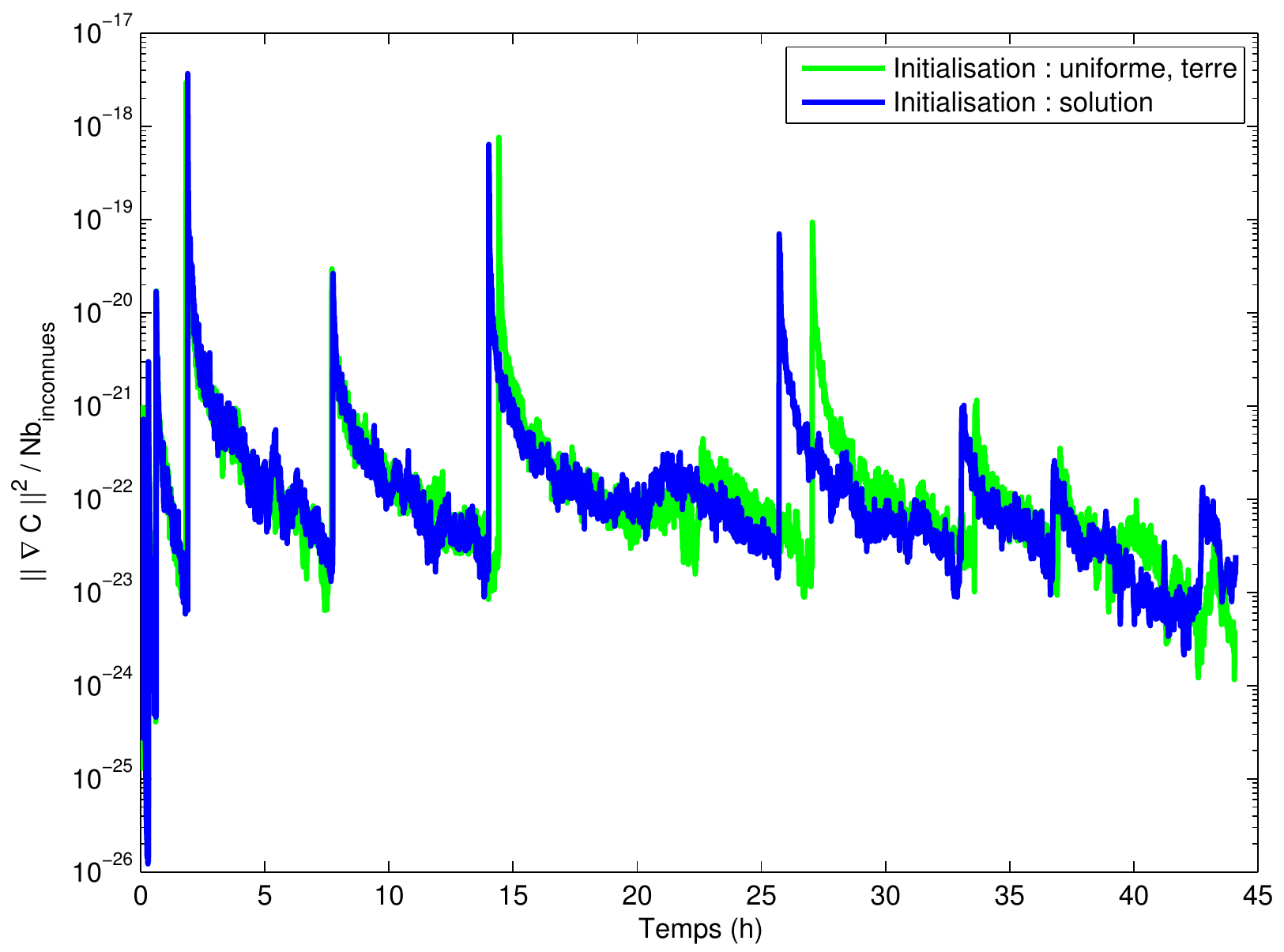}} \\
\end{center}
\caption{Résultats obtenus sur le milieu de taille intermédiaire}
\label{Fig_GrandMilieu_IntroductionProgressiveFreq}
\end{figure}

\begin{figure}[!p]
\begin{center}
\subfloat[Pour chaque fréquence considérée, amplitude des données bruitées utilisées pour l'inversion (rouge pointillé) et amplitude des mesures correspondant à la solution recherchée (rouge), aux cartes obtenues en initialisant à la terre (vert) et aux cartes obtenues en initialisant à la solution (bleu) en chaque capteur]
{\includegraphics[scale=0.6]{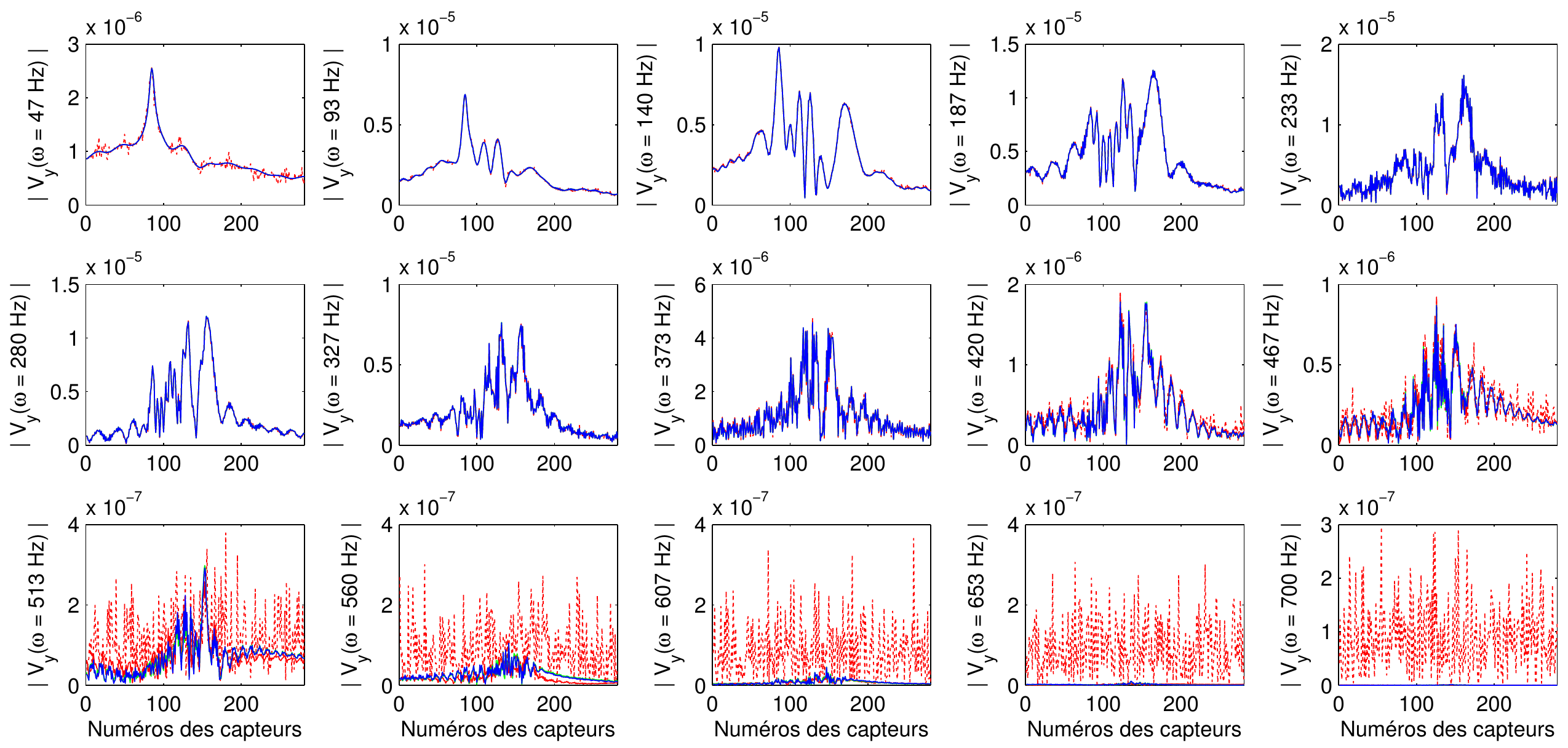}} \\
\subfloat[Sismogrammes reconstruits à partir des cartes obtenues en initialisant aux caractéristiques de la terre]
{\includegraphics[scale=0.55]{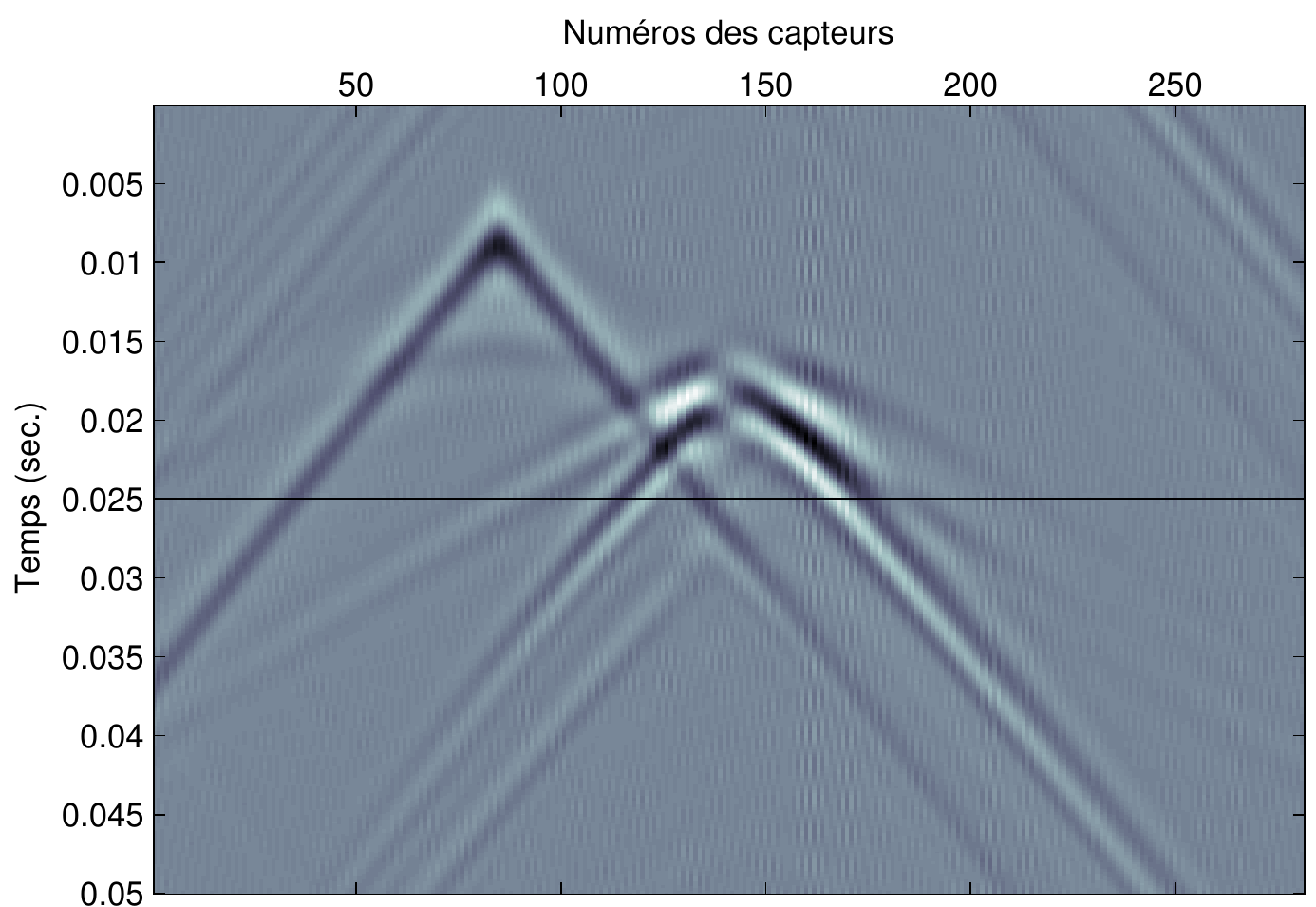}} \quad
\subfloat[Sismogrammes reconstruits à partir des cartes obtenues en initialisant à la solution]
{\includegraphics[scale=0.55]{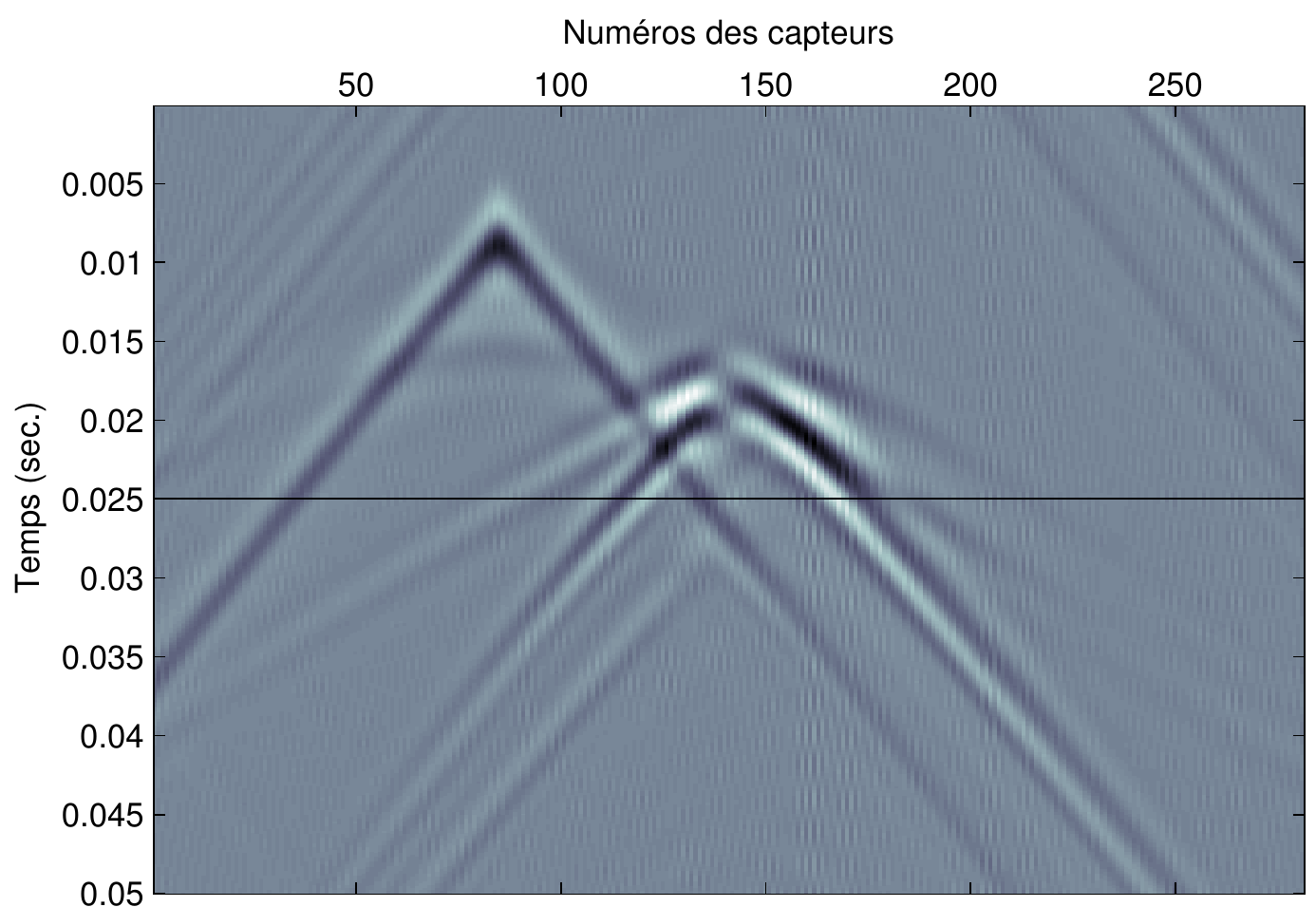}} \\
\subfloat[Sismogrammes correspondant à la solution recherchée]
{\includegraphics[scale=0.55]{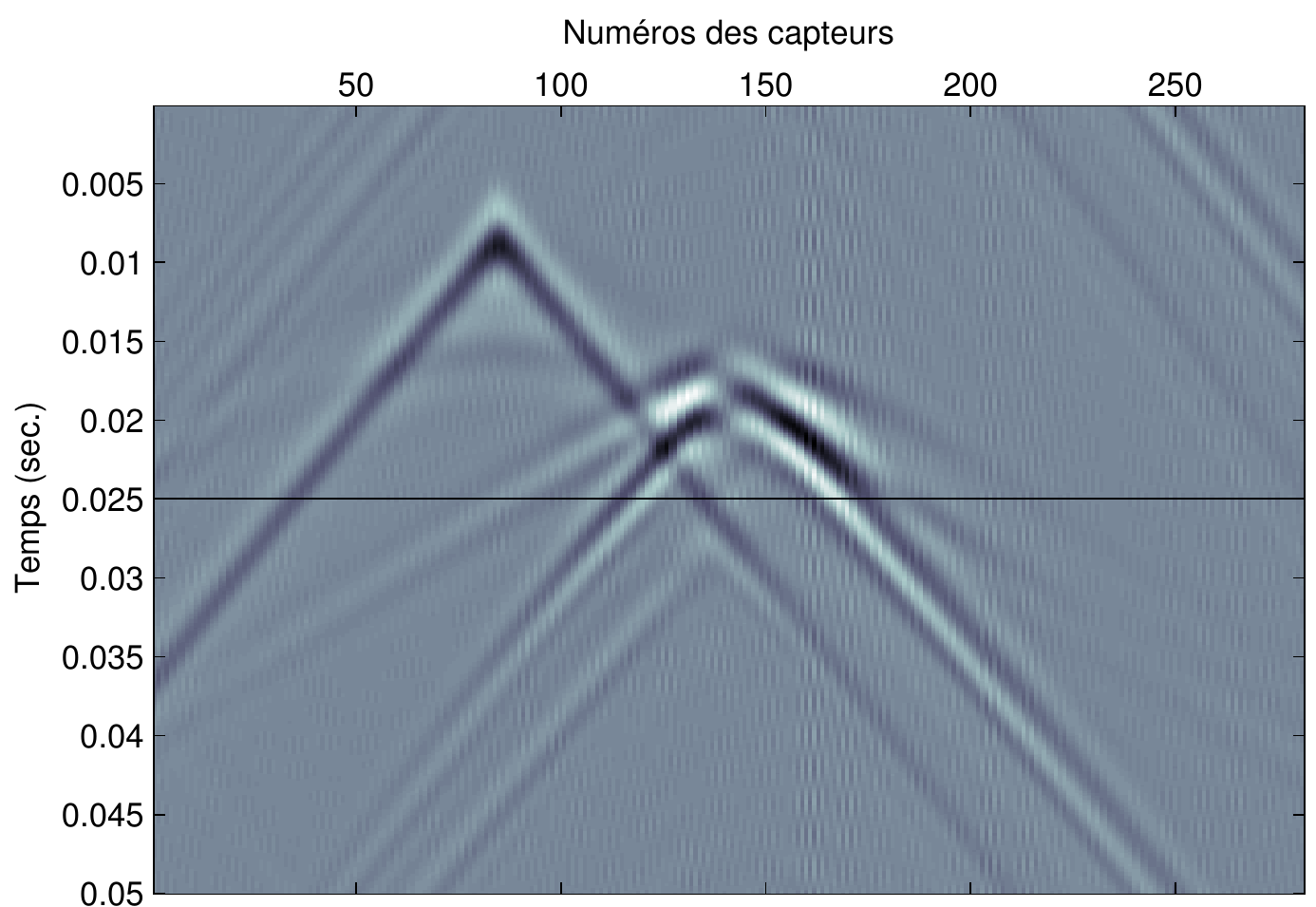}} \quad
\subfloat[Comparaison des mesures à l'instant t = 25 ms (correspond au trait noir affiché sur les sismogrammes)]
{\includegraphics[scale=0.55]{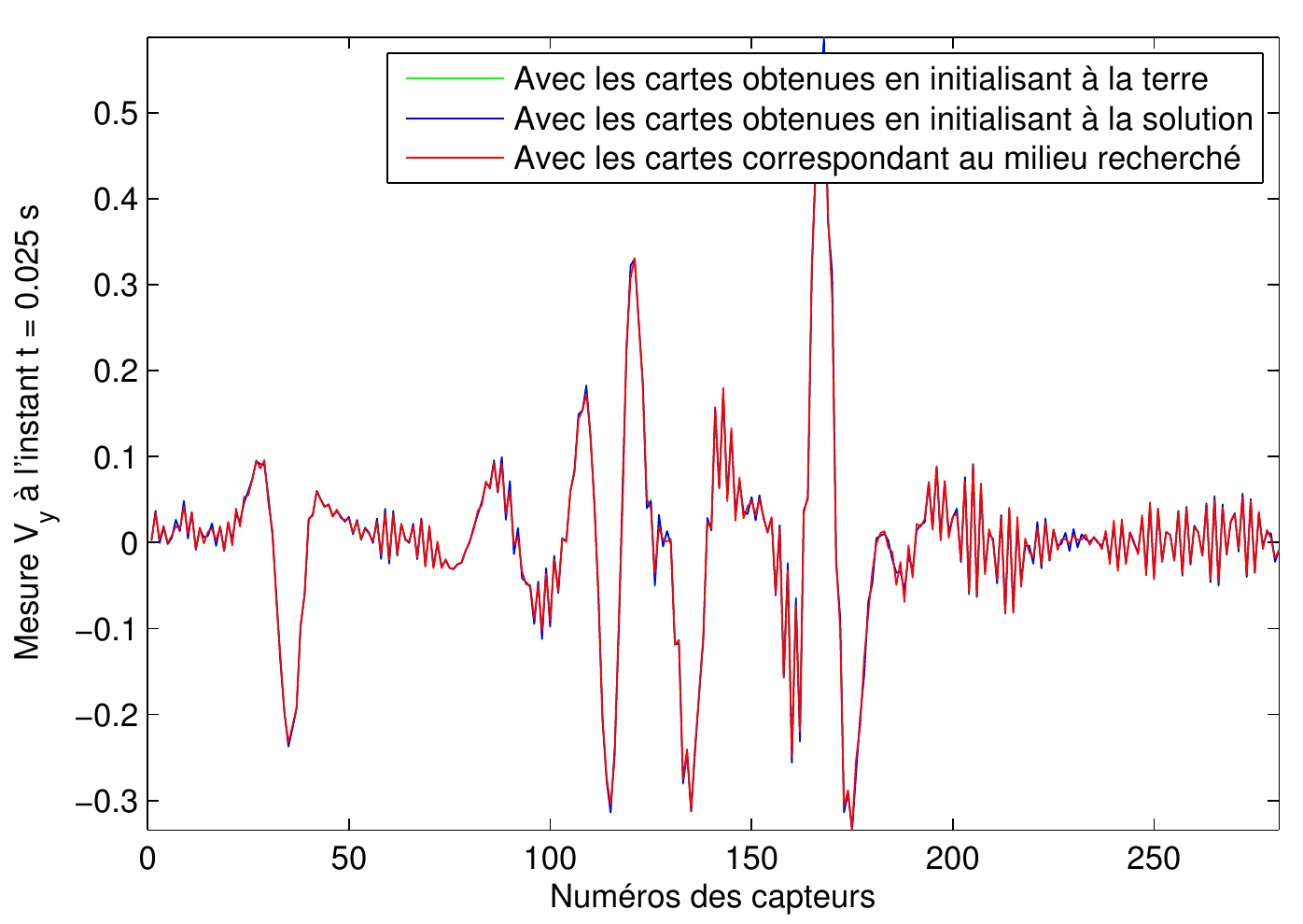}}
\end{center}
\caption{Comparaison des données capteurs correspondant aux cartes solutions (sans et avec ajout de bruit blanc) et aux cartes obtenues après inversion (initialisation à la terre et à la solution) dans le cas où la source est positionnée à gauche de l'objet diffractant (septième position sur la Figure \ref{Fig_GrandMilieuTest}, page \pageref{Fig_GrandMilieuTest}). On affiche les données dans le domaine fréquentiel et dans le domaine temporel (sismogrammes).}
\label{Fig_SismoGrandMilieu}
\end{figure}

Nous présentons sur la Figure \ref{Fig_GrandMilieu_IntroductionProgressiveFreq} les résultats obtenus pour deux initialisations différentes (pour la première initialisation, les caractéristiques de la zone d'étude sont égales à celles de la terre et pour la seconde, les contrastes sont égaux à la solution recherchée). Nous présentons également sur la Figure \ref{Fig_SismoGrandMilieu} une comparaison des données mesurées par les capteurs lorsque les cartes correspondent à la solution recherchée (sans et avec ajout de bruit blanc) et aux résultats obtenus pour les deux initialisations (les sismogrammes sont maintenant représentés en niveaux de gris étant donné le nombre élevé de capteurs).

On remarque tout d'abord que l'algorithme arrive bien à convergence : pour les deux initialisations, les cartes obtenues sont similaires et le critère converge vers la même valeur. Les cartes obtenues montrent que la forme de l'objet diffractant est plutôt bien reconstruite. On remarque cependant que les contours sont plus marqués sur les cartes des $\sigma_s = \ln{\mathit{v}_s}$. Sur les différentes cartes, les deux régions (terre et béton) ne sont pas parfaitement homogènes mais les valeurs caractéristiques correspondantes sont proches de celles du milieu recherché. Comme pour le milieu de petite taille, la comparaison des données reconstruites montrent que les mesures obtenues sont semblables pour les deux initialisations et que l'on retrouve bien les données utilisées pour l'inversion.

Il a fallu davantage de temps (presque deux jours) par rapport au milieu de petite taille pour arriver à convergence mais plusieurs modifications pourraient encore être apportées afin de réduire le temps de calcul comme par exemple :
\begin{itemize}
\item pouvoir se ramener à des données non redondantes afin de réduire la quantité d'informations traitées (éliminer certaines fréquences ou certains capteurs) ;
\item modifier l'algorithme de minimisation en ayant recours au préconditionnement ou à des décompositions LU incomplètes afin d'arriver plus rapidement à convergence pour chaque groupe de fréquences considéré.
\end{itemize}

\chapter*{Conclusion et perspectives \markboth{Conclusion et perspectives}{}}
\addcontentsline{toc}{chapter}{Conclusion et perspectives}

Nous avons abordé l'imagerie de la subsurface en distinguant deux familles de méthodes : la première regroupe les méthodes faisant intervenir des variables auxiliaires et la seconde comprend les méthodes utilisant une formulation "primale" du problème direct.


Les méthodes faisant intervenir des variables auxiliaires ont l'avantage d'utiliser des formulations pour lesquelles le critère et le gradient sont rapides à calculer. Cependant, elles font intervenir un grand nombre de grandeurs à optimiser ce qui tend à augmenter le nombre d'itérations nécessaires et donc le temps de calcul global. Les méthodes s'appuyant sur une formulation primale permettent de minimiser le nombre de variables à optimiser et donc le nombre d'itérations nécessaire. Cependant, la formulation utilisée est complexe, ce qui implique un temps de calcul élevé à chaque itération. La question du compromis entre nombre de variables à manipuler, nombre total d'itérations et volume de calcul par itération est donc critique pour une résolution satisfaisante du problème.


Nous avons étudié différentes méthodes basées sur une formulation
bilinéaire du problème. Toutes font apparaître des formes algébriques
différentes et ont des qualités qui leur sont propres. Certaines
peuvent être appliquées sur une zone d'étude, limitant ainsi le nombre
d'inconnues, tandis que d'autres bénéficient d'estimateurs simples à
calculer. Nous ne sommes cependant parvenus à en faire converger
aucune, ce qui rend difficile toute comparaison quantitative. Nous
pouvons tout de même souligner que la méthode du gradient modifié
semble être la plus intéressante des formulations bilinéaires,
puisqu'elle permet d'obtenir les résultats plus proches de la solution
après un nombre d'itération donné. La méthode CFSI a un important
problème de conditionnement croisé de deux de ses estimateurs, tandis
que la méthode CSI a un coût de calcul par itération plus important
que la méthode GM. Finalement, la méthode sans contraste est la plus
gourmande en place mémoire et semble converger plus lentement que
toutes les autres. Dans tous les cas, le critère décroît très
lentement après quelques centaines d'itérations et aucune méthode
semble converger après plusieurs jours de calcul. La lenteur avec
laquelle le critère décroit peut être expliquée par le mauvais
conditionnement de la plupart des matrices normales. Ce
conditionnement se détériore au fur et à mesure des itérations ; il en
résulte une diminution du déplacement relatif de la solution, ce qui
entraîne une stagnation du critère. Les changements de variables
proposés ont eu des effets différents, mais aucun ne s'est avéré être
une solution au problème de convergence de la méthode à laquelle il a
été appliqué. Des tests supplémentaires ont été menés pour tenter de déterminer la raison des problèmes de convergence rencontrés. Les résultats de ces tests sont donnés en annexe de ce document et ne permettent pas de conclure avec certitude sur l'origine de ces problèmes de convergence.

En ce qui concerne la formulation primale, nous avons tout d'abord proposé une formulation optimisée de manière à diminuer à la fois le temps et l'espace mémoire nécessaires au calcul du critère et du gradient. Une première analyse des méthodes envisagées nous a amené à retenir celle du gradient conjugué non linéaire. Les premiers essais effectués sur le milieu de petite taille nous ont conduit à exprimer le critère en fonction d'autres variables que le contraste ($\ln{\mathit{v_p}}$ et $\ln{\mathit{v_s}}$) et à utiliser ces mêmes variables pour la régularisation.

Pour les deux familles de méthodes, l'introduction progressive des fréquences nous a permis de diminuer le temps de calcul. Dans le cas des méthodes bilinéaires, cette accélération permet seulement d'atteindre plus rapidement le palier de convergence lente du critère. Par contre, pour la formulation primale, la procédure d'inversion ainsi obtenue s'est révélée plutôt efficace et nous a permis de faire des tests sur le milieu de taille intermédiaire. Plusieurs pistes pourraient cependant être explorées pour accélerer davantage l'algorithme d'inversion.


Au vu des résultats obtenus pour les différentes méthodes abordées, l'utilisation de l'algorithme utilisant la formulation primale semble plus adaptée à notre problème. Cependant, des évolutions restent à prévoir. D'une part, l'algorithme doit être en mesure de traiter des problèmes de plus grande taille : les milieux traités dans la réalité seront de plus grande dimension et avec une résolution spatiale plus fine. D'autre part, il faut l'adapter pour qu'il prenne en compte la présence d'une surface libre (pour les problèmes traités jusqu'à maintenant, l'objet diffractant ainsi que la source et les capteurs étaient enfouis sous terre).

Enfin, certains points restent encore en suspens comme, par exemple, la détermination du milieu de référence qui est choisi par l'utilisateur et les fréquences nécessaires à une inversion de qualité.

\bigskip

Les travaux effectués jusqu'à présent nous ont donc permis de
développer une première méthode d'inversion 2D de type "cartographie"
n'incluant pas d'information \emph{a priori} concernant les valeurs caractéristiques du béton 
et la géométrie de l'objet recherché. Celle-ci s'est avérée performante sur des jeux de données
synthétiques de tailles petite et intermédiaire. Dans une certaine mesure, cette méthode
constitue donc une \og preuve de concept \fg de la faisabilité de
l'inversion de données sismiques 2D pour l'examen de fondations de
pylônes, et elle servira de base à la suite du projet. Pour
aboutir à une méthode satisfaisant les objectifs fixés au début de
l'étude, les étapes ultérieures seront les suivantes :
\begin{itemize}
\item \textbf{Poursuite de l'étude de la méthode développée
  jusqu'ici -- } L'objectif sera d'en caractériser plus précisément les
  performances, d'en améliorer le comportement numérique et d'élargir la gamme
  de données qu'elle est capable de traiter.
\item \textbf{Développement de modélisations plus précises du milieu
    -- } Il s'agira d'inclure davantage d'informations \emph{a priori}
  sur la nature de l'objet recherché, notamment la présence de deux
  zones différentes (la fondation et le sous-sol) dans le milieu à
  imager . Deux approches,
  réparties entre les institutions impliquées dans l'inversion, sont
  envisagées :  l'IRCCyN étudiera les techniques de type contour
  tandis que l'Ecole Polytechnique de Montréal abordera les méthodes
  de type région. Dans un cas comme dans l'autre, ceci devrait
  permettre de mieux identifier la forme de la fondation et donc
  contribuer à atteindre les objectifs du projet.
\end{itemize}
L'ensemble des méthodes sera testé sur les divers types de données, en
fonction de leur disponibilité : données synthétiques de plus grande
taille avec surface libre, données obtenues avec les maquettes simulées
par le LCPC, données réelles recueillies lors des campagnes de
mesures.


\bibliography{baseLION}
\bibliographystyle{IEEEtran}

\addcontentsline{toc}{chapter}{Annexe}
\setcounter{figure}{0}
\renewcommand*\thefigure{\Alph{figure}}
\chapter*{Annexe : Tests sur la convergence de la méthode du gradient modifié}

\section*{Introduction}

Suite aux résultats obtenus par les méthodes basées sur l'approche bilinéaire, un certain
nombre de questions ont été posées auxquelles nous donnons des
éléments de réponse dans cette annexe. Les trois questions soulevées concernaient :
\begin{itemize}
\item la possibilité de convergence de la méthode GM vers un minimum local ;
\item l'effet de la pénalisation quadratique du champ de vitesse sur le résultat de la reconstruction ;
\item la comparaison entre les variables auxiliaires réelles et estimées.
\end{itemize}
Pour obtenir des résultats rapidement nous avons effectué ces tests
sur le milieu de petite taille. Les tests ont été effectués avec la
représentation bilinéaire en utilisant la méthode qui donnait les meilleurs résultats, à savoir la méthode GM avec changement de variable en $log$.

\section*{Test de convergence vers un minimum local}

Pour essayer de déterminer si le critère que nous tentons de minimiser
possède des minima locaux, nous avons effectués deux tests. Le premier
test consiste à observer la variation du critère lorsque l'on fait
varier linéairement les valeurs du contraste entre la valeur
d'initialisation (on initialise à la terre) et l'estimée obtenue en
initialisant l'algorithme avec la solution. Le deuxième test est
identique dans la manière, mais utilise les valeurs du critère
obtenues par la méthode GM jusqu'à la valeur obtenue lors de l'arrêt
de la méthode après $8000$ itérations, puis les contrastes sont calculés
de façon linéaire entre cette estimée et l'estimée obtenue en
initialisant l'algorithme avec la solution. Dans les deux cas, les
variables auxiliaires sont déduites des contrastes en utilisant
l'estimateur direct issu de la formulation du GM donnée par les équations~\ref{eq_DefNotationsGM}. De même, l'expression du critère de la
méthode GM que nous calculons est donné par l'équation~\ref{eq_CritereGM}.

\begin{figure}[!htb]
\centering
 \subfloat[Critère partant de l'initialisation à la terre des contrastes]{\includegraphics[scale=0.35]{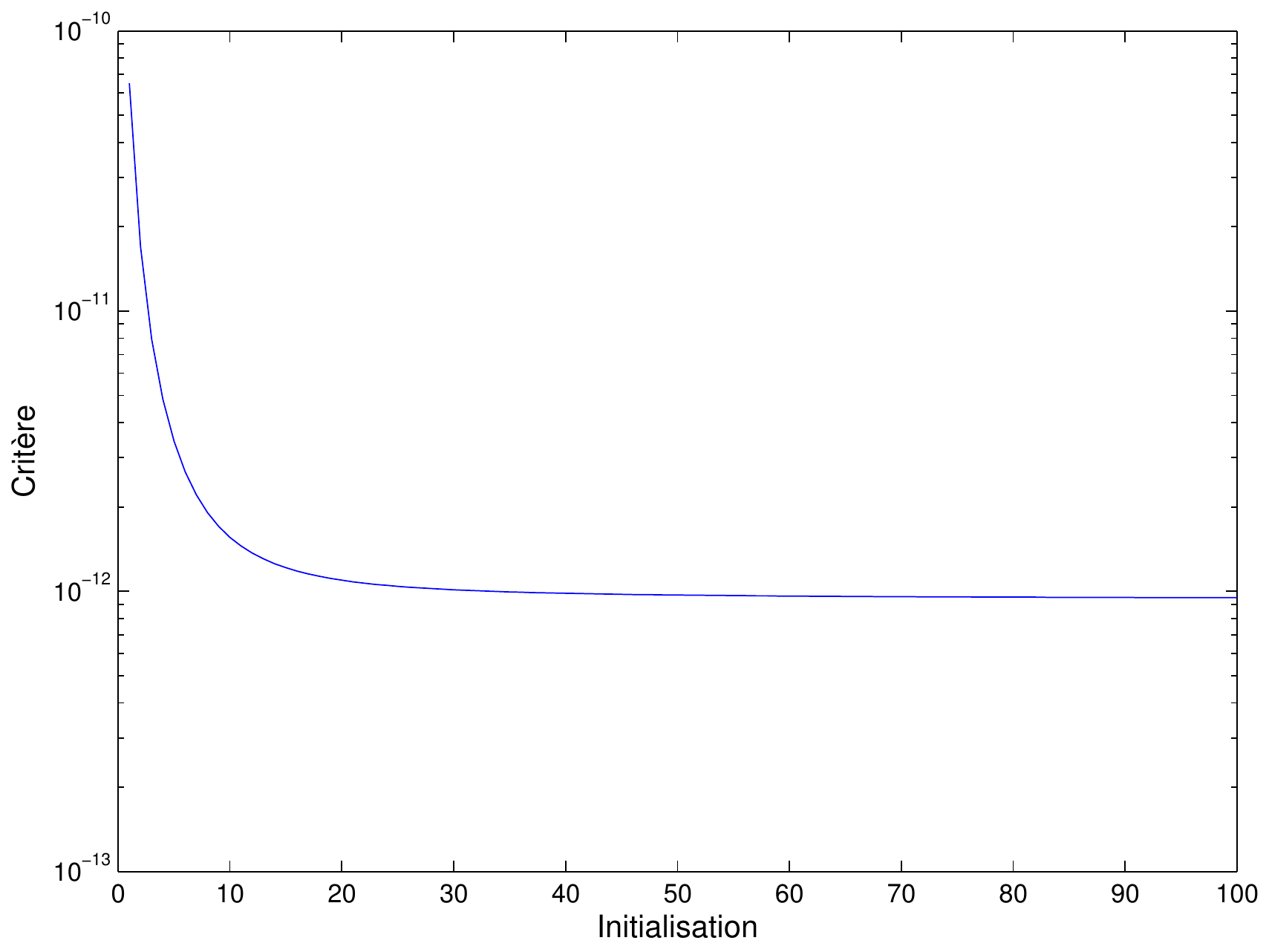}}
 \subfloat[Critère partant de l'estimée obtenue par la méthode GM]{\includegraphics[scale=0.35]{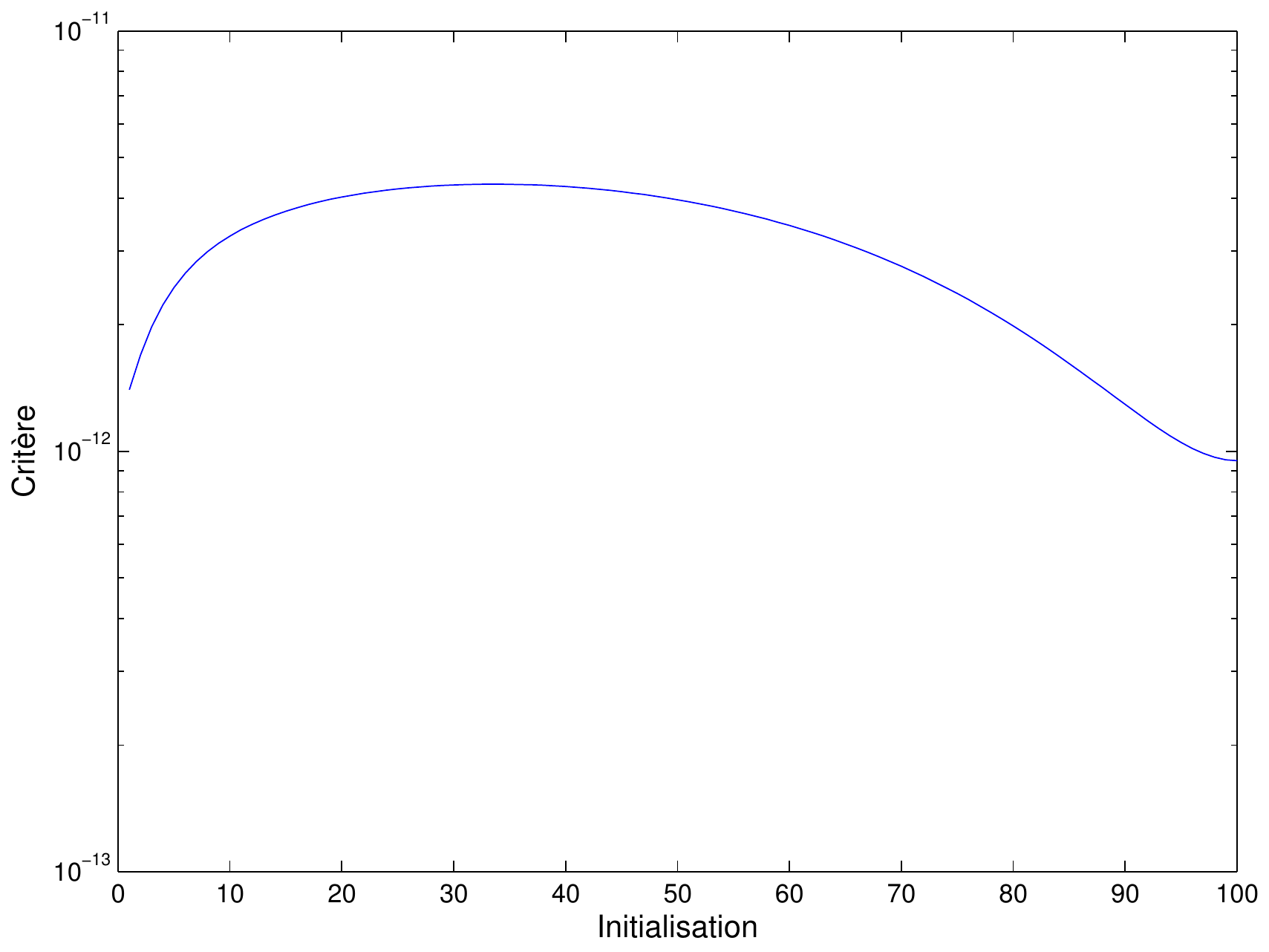}}\\
 \subfloat[Coût lié aux variables auxiliaires]{\includegraphics[scale=0.35]{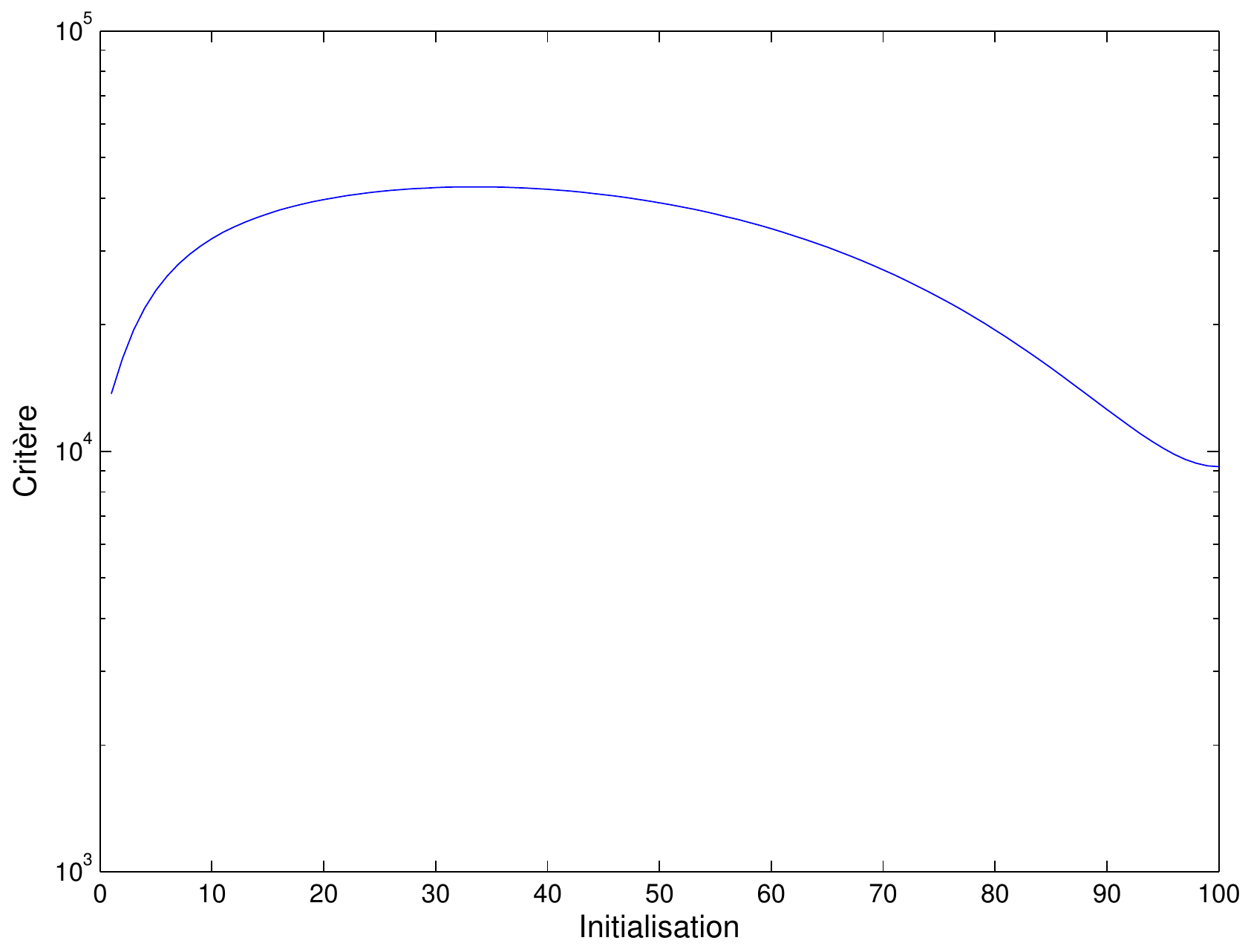}}\\
\caption{Variation de la fonction de coût de la méthode GM, en faisant varier le contraste linéairement à partir de l'initialisation à la terre (a) ou de la valeur estimée avec la méthode GM après 8000 itérations (b). La courbe (c) montre la partie du critère liée à l'estimation des variables auxiliaires.}
\label{fig:Test_min_local}
\end{figure}

Le résultat de ce test est présenté à la
figure~\ref{fig:Test_min_local}. Son interprétation est délicate. Lorsque l'on part de
valeurs du contraste nulles, ce qui est le cas si on initialise la
méthode avec les valeurs caractéristiques de la terre, alors le
critère décroit à mesure que le contraste se rapproche de la
solution. Par contre, si nous partons de l'estimée obtenue après
$8000$ itérations de la méthode GM, alors on peut voir que le critère
n'est pas monotone décroissant. Cependant, on ne peut pas conclure que
la méthode GM décroit obligatoirement vers un minimum
local. En effet, la forme du critère est très probablement due à un problème de conditionnement de l'estimateur des variables auxiliaires comme le laisse suggérer la courbe (c) de la figure~\ref{fig:Test_min_local} représentant le coût relié aux variables auxiliaires et qui présente la même allure que le critère partant de l'estimée de la méthode GM. 

\section*{Effet de la régularisation sur la variable auxiliaire}
Les méthodes d'inversion de type bilinéaires que nous proposons dans ce rapport utilisent une fonction de pénalisation quadratique sur les variables auxiliaires. Pour mesurer l'effet de cette régularisation sur le résultat, nous avons fait varier le paramètre de régularisation associé à cette fonction de pénalisation et nous avons observé les variations résultantes sur l'erreur quadratique moyenne du contraste et de la variable auxiliaire, ainsi que sur le temps de calcul. Dans tous les cas l'inversion s'arrêtait lorsque l'on atteignait un seuil de la norme du gradient de $2.10^{-18}$.

Nous avons limité le choix de l'hyperparamètre aux valeurs comprises dans l'intervalle $[10^{-18}; 10^{-6}]$ car en deçà de $10^{-18}$ l'effet de la pénalisation est négligeable, et au delà de $10^{-6}$ on observait une sur-régularisation marquée.

\begin{figure}[!htb]
\begin{center}
 \includegraphics[scale=0.35]{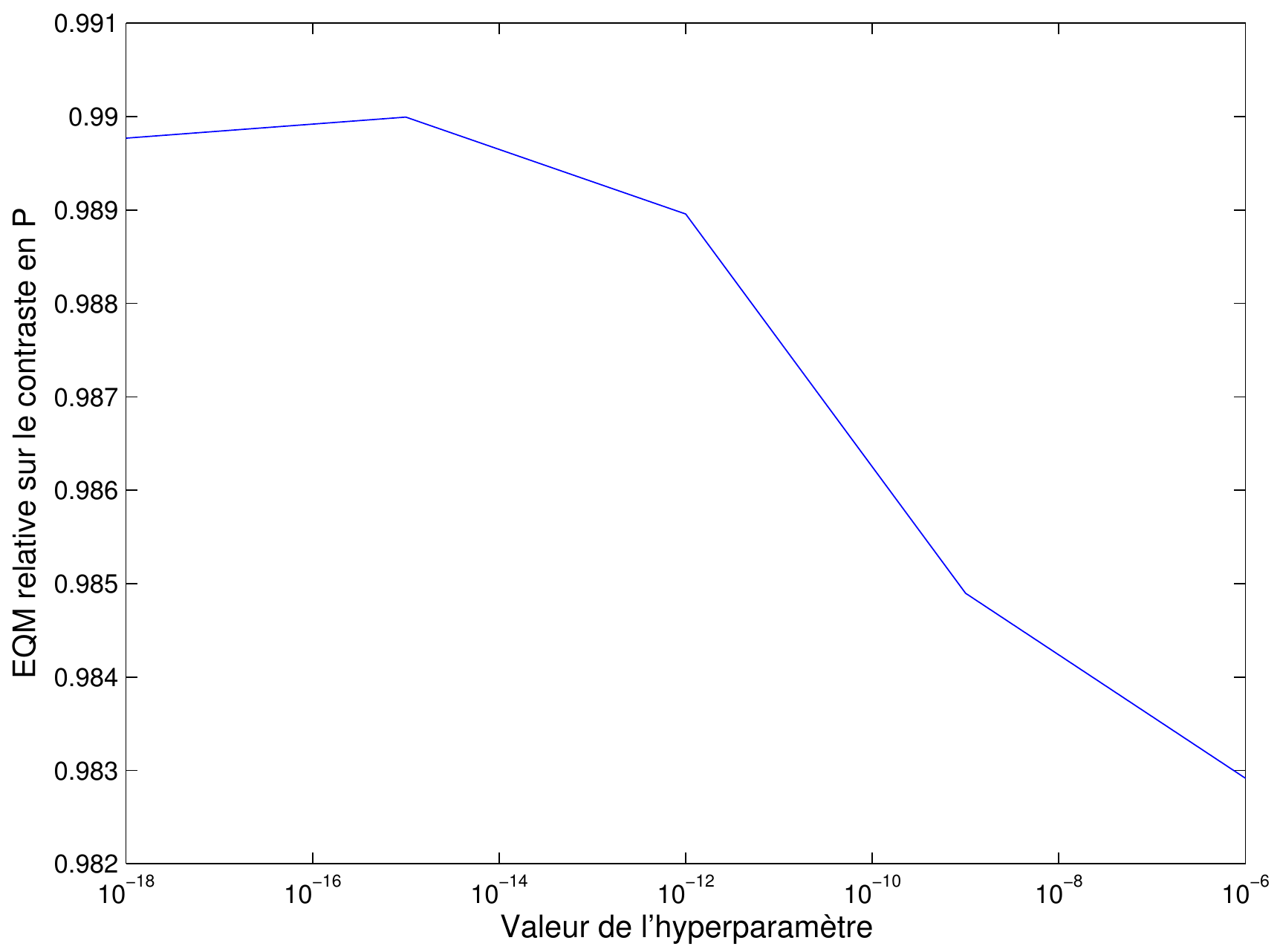}
 \includegraphics[scale=0.35]{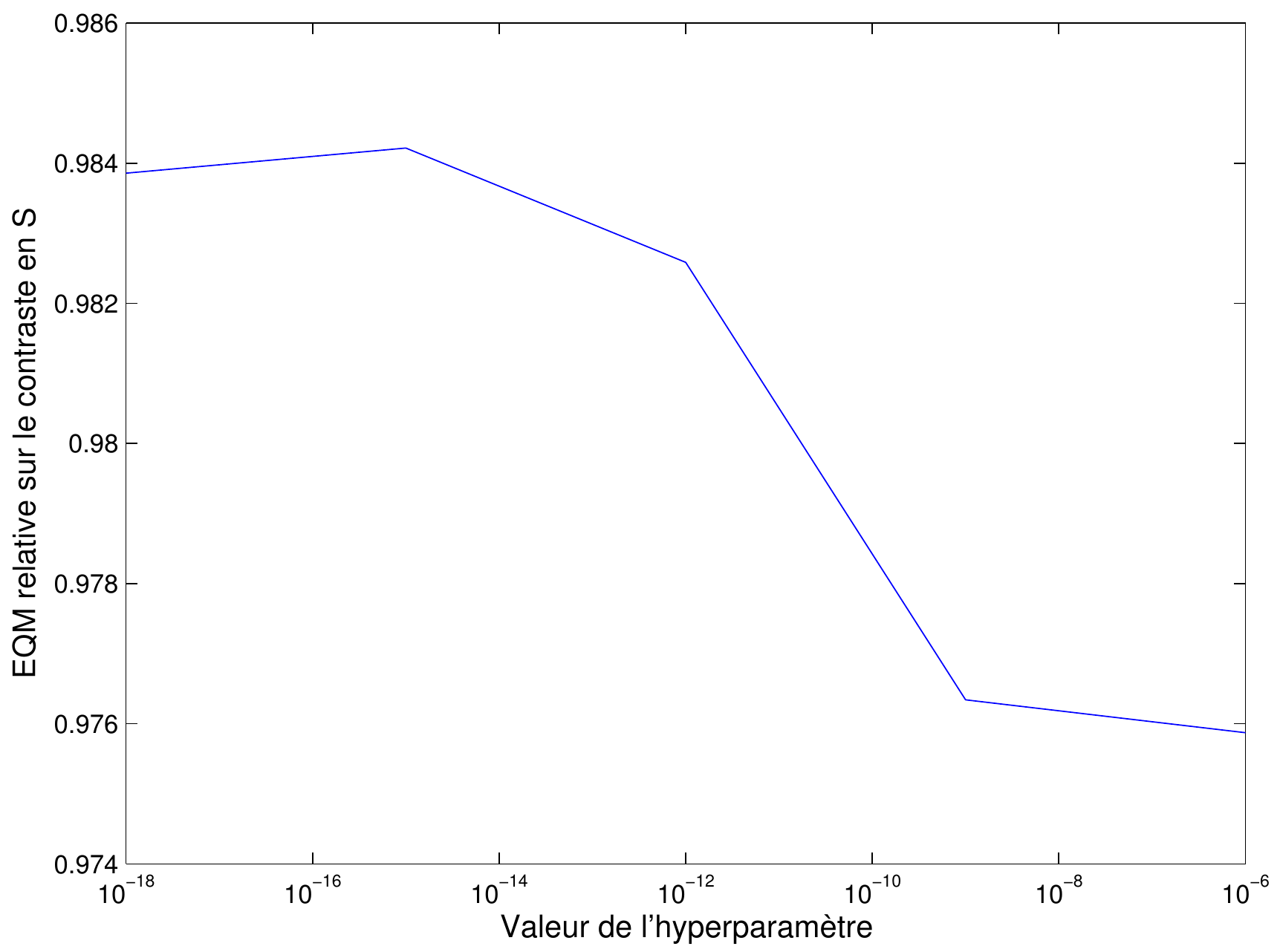}\\
\caption{Évolution de l'erreur quadratique relative sur les contrastes en P et S en fonction de la valeur de l'hyperparamètre}
\label{fig:Test_reg}
\end{center}
\end{figure}

\begin{figure}[!htb]
\begin{center}
 \includegraphics[scale=0.35]{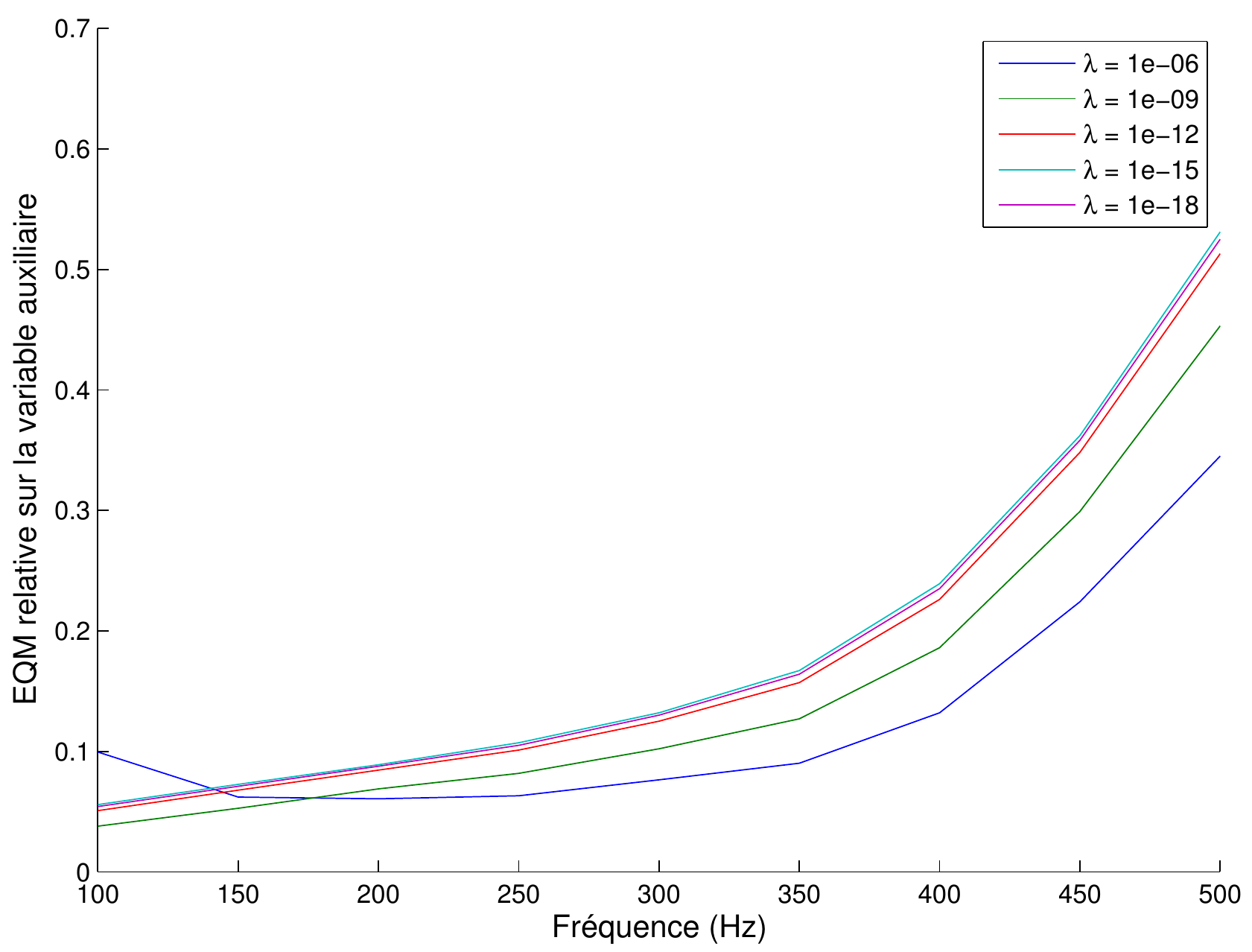}
\caption{Évolution de l'erreur quadratique relative sur la variable auxiliaire en fonction de la valeur de l'hyperparamètre}
\label{fig:Test_reg2}
\end{center}
\end{figure}

\begin{figure}[!htb]
\begin{center}
 \includegraphics[scale=0.35]{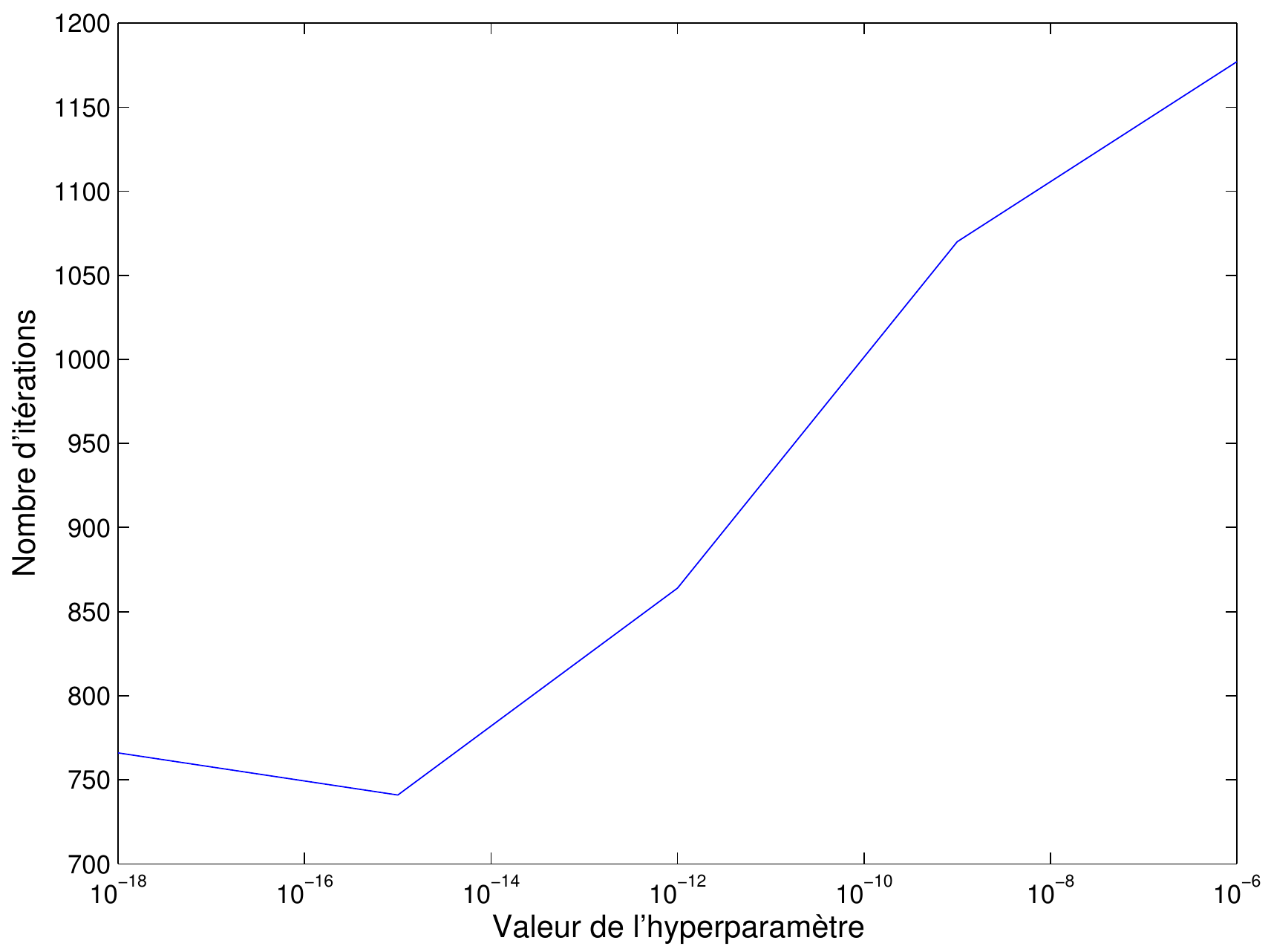}
\caption{Évolution du temps de calcul en fonction de la valeur de l'hyperparamètre}
\label{fig:Test_reg3}
\end{center}
\end{figure}

Ces résultats indiquent que la pénalisation quadratique des variables auxiliaires a un impact significatif sur le nombre d'itérations de la méthode, et donc le temps de calcul, ainsi que sur la qualité du résultat de l'estimation de toutes les variables : les contrastes en P et S et les variables auxiliaires. Le temps de calcul augmente lorsque la valeur de l'hyperparamètre augmente tandis que l'EQM des différentes variables estimées décroit. Il faut donc faire un compromis entre la qualité du résultat souhaitée et le temps de calcul.

\section*{Comparaison des variables auxiliaires réelles et estimées}
\begin{figure}
\centering
 \subfloat[Fréquence 100 Hz]{\includegraphics[scale=0.4]{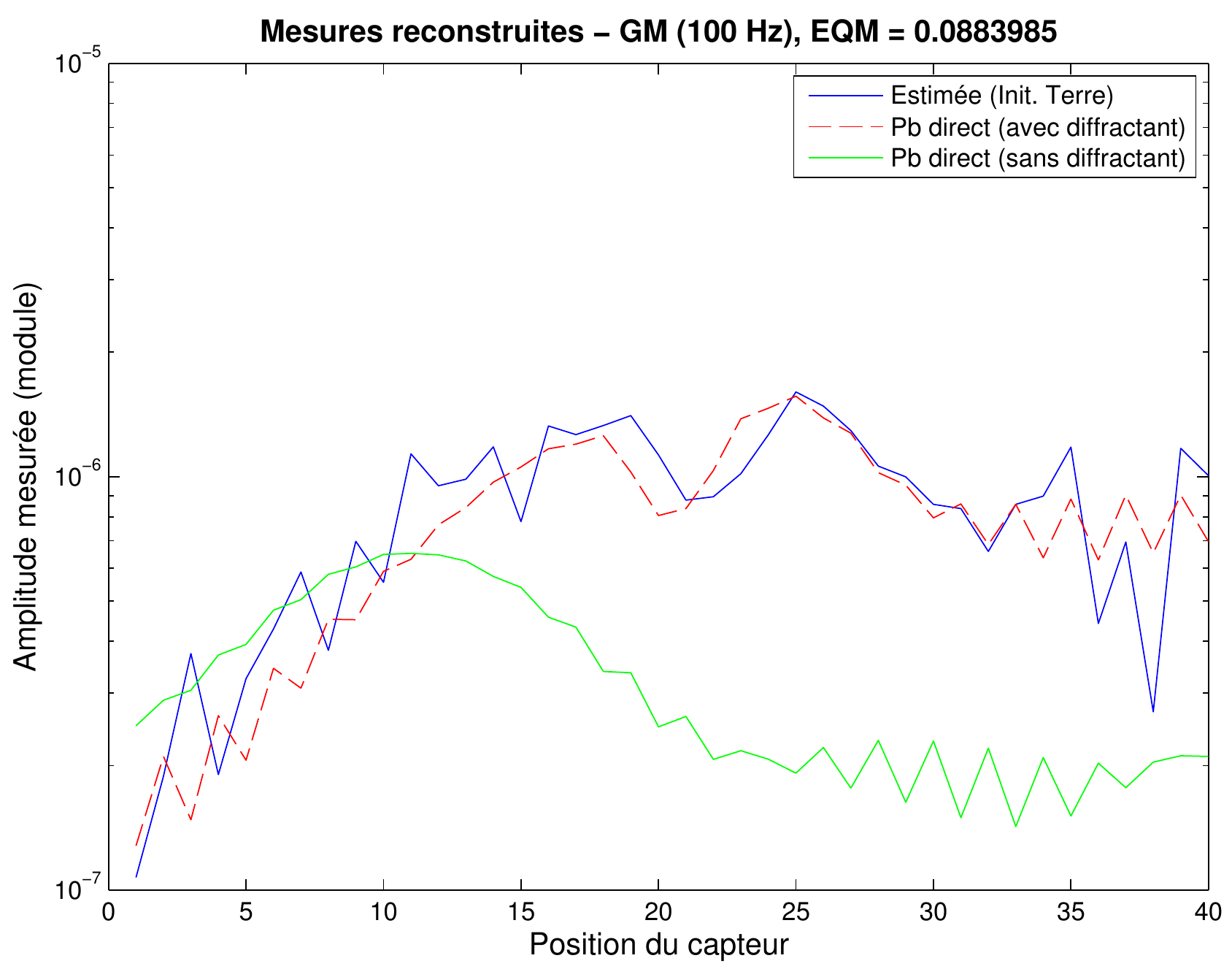}}
  \subfloat[Fréquence 250 Hz]{\includegraphics[scale=0.4]{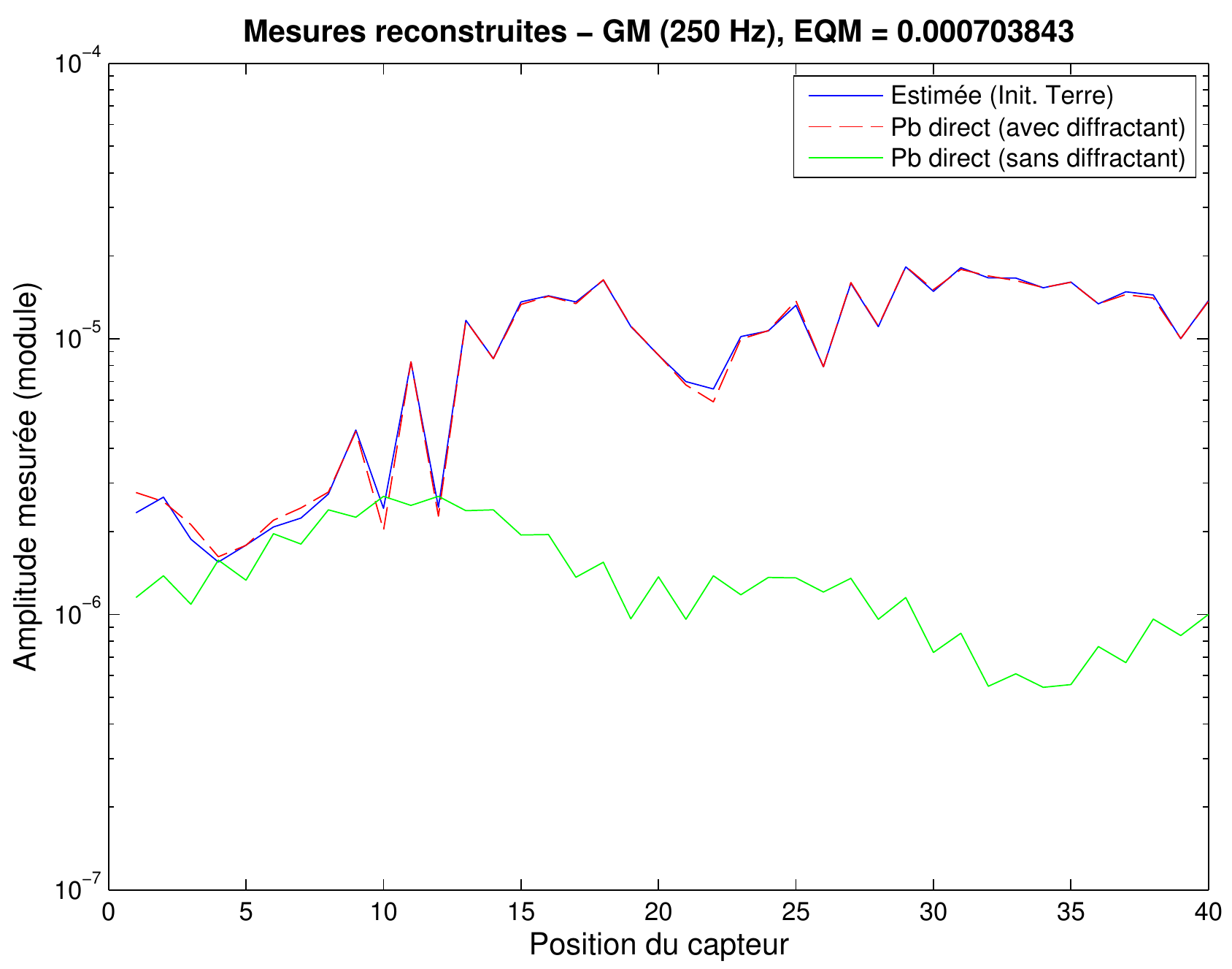}}\\
  \subfloat[Fréquence 400 Hz]{\includegraphics[scale=0.4]{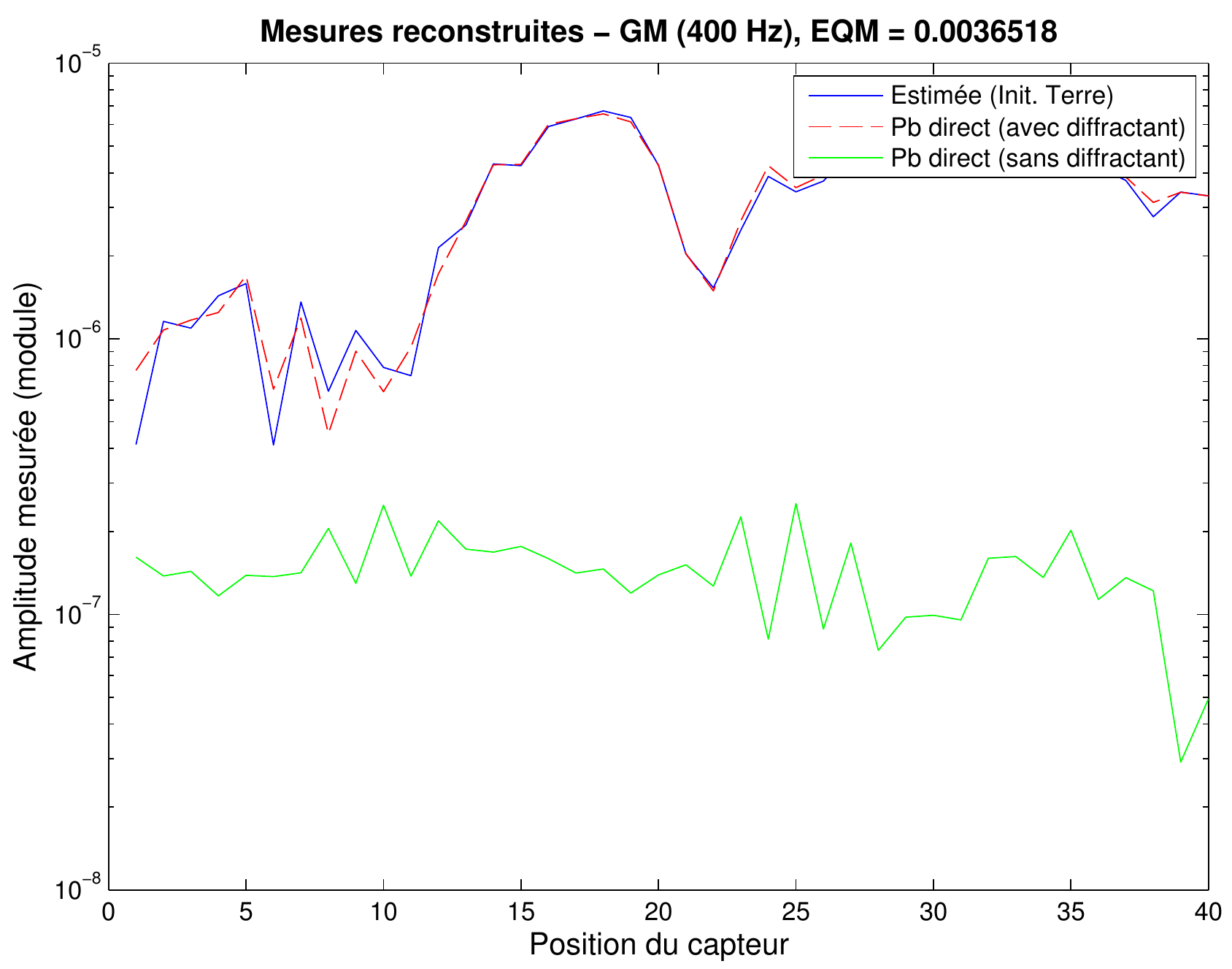}}
  \subfloat[Fréquence 550 Hz]{\includegraphics[scale=0.4]{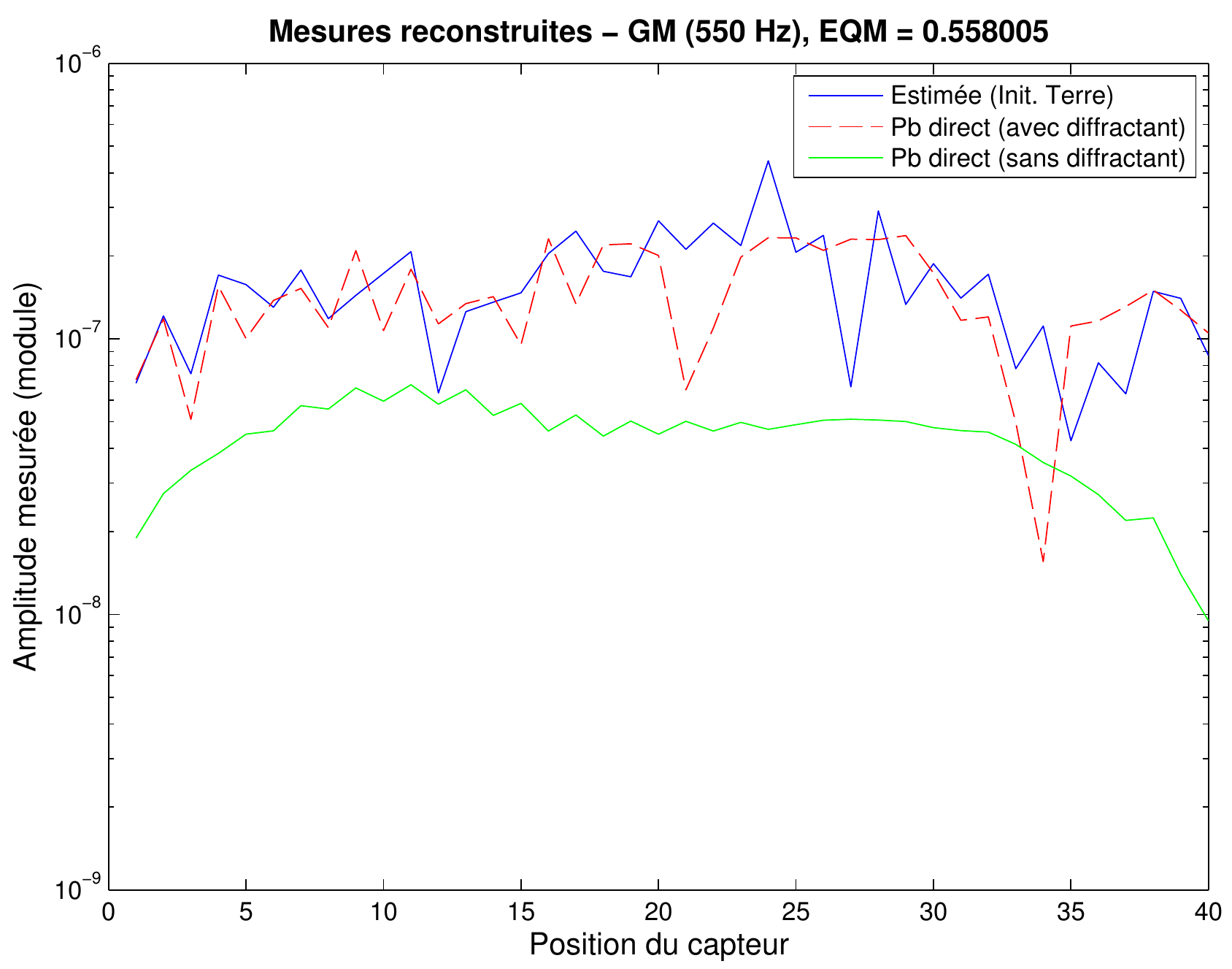}}\\
  \subfloat[Fréquence 700 Hz]{\includegraphics[scale=0.4]{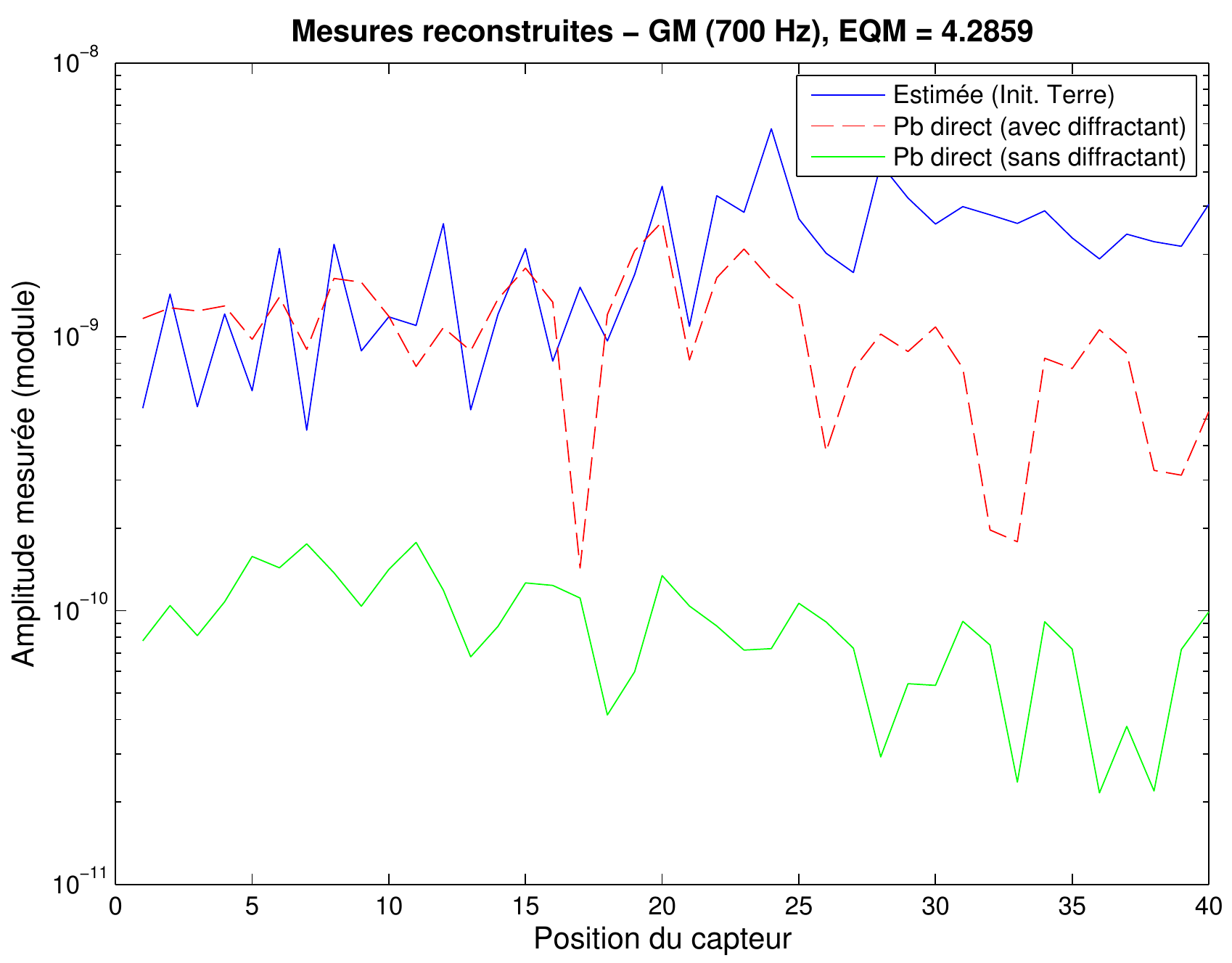}}
\caption{Comparaison entre les données générées par le problème direct et les données reconstruites par la méthode GM pour différentes fréquences}
\label{fig_comparaison1}
\end{figure}

Dans la figure~\ref{fig_comparaison1} nous présentons des courbes montrant les données du problème direct, les données initiales et les données reconstruites par la méthode GM pour différentes fréquences. La figure~\ref{fig_comparaison2} compare les cartes des champs de vitesse en x et y pour les mêmes fréquences obtenues par la méthode GM.

On constate que les champs de vitesses en x et y pour les basses fréquences sont bien reconstruits tandis que ceux correspondant aux hautes fréquences sont mal restitués.

\begin{figure}
\centering
  \subfloat[Fréquence 100 Hz]{\includegraphics[scale=0.35]{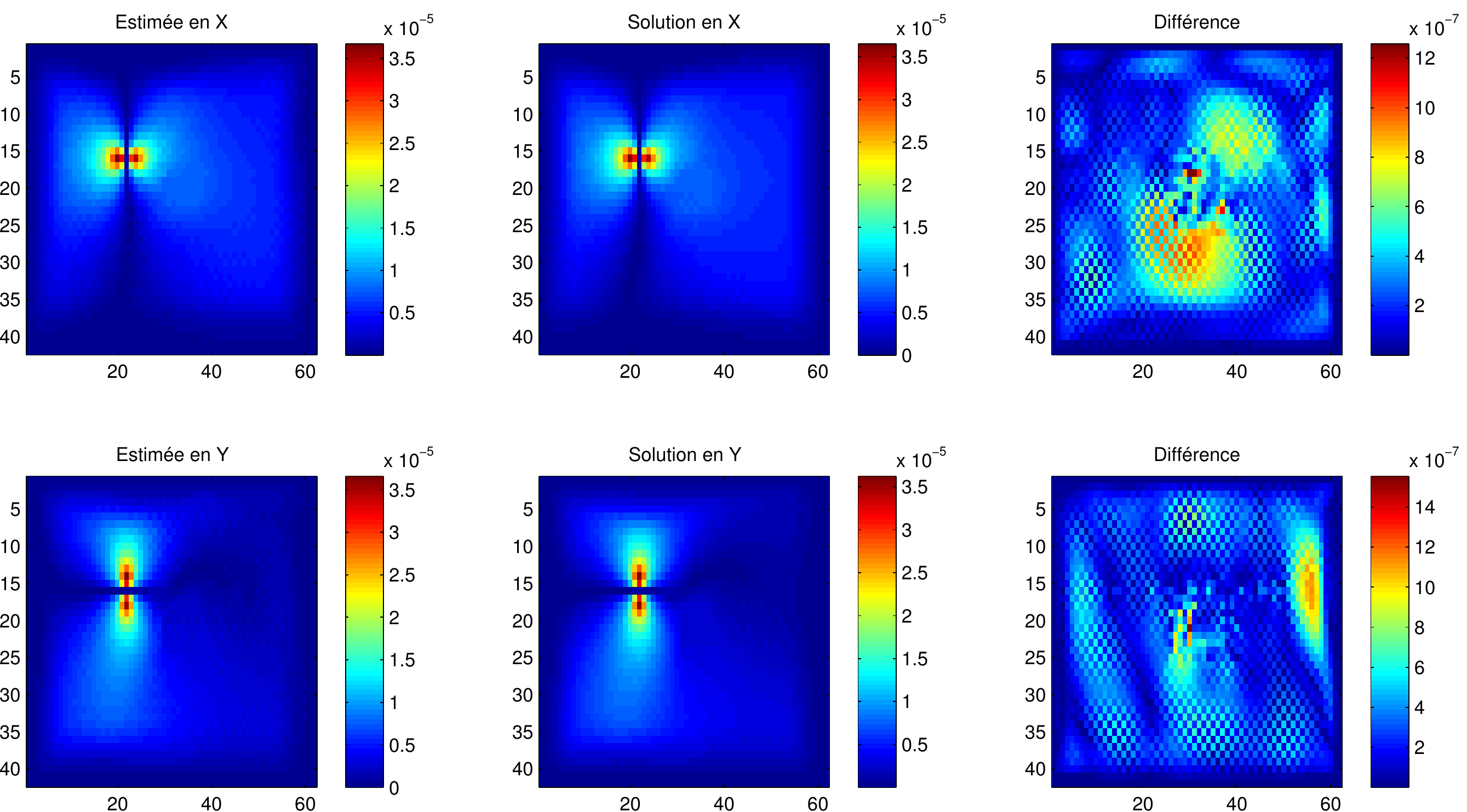}}\\
  \subfloat[Fréquence 250 Hz]{\includegraphics[scale=0.35]{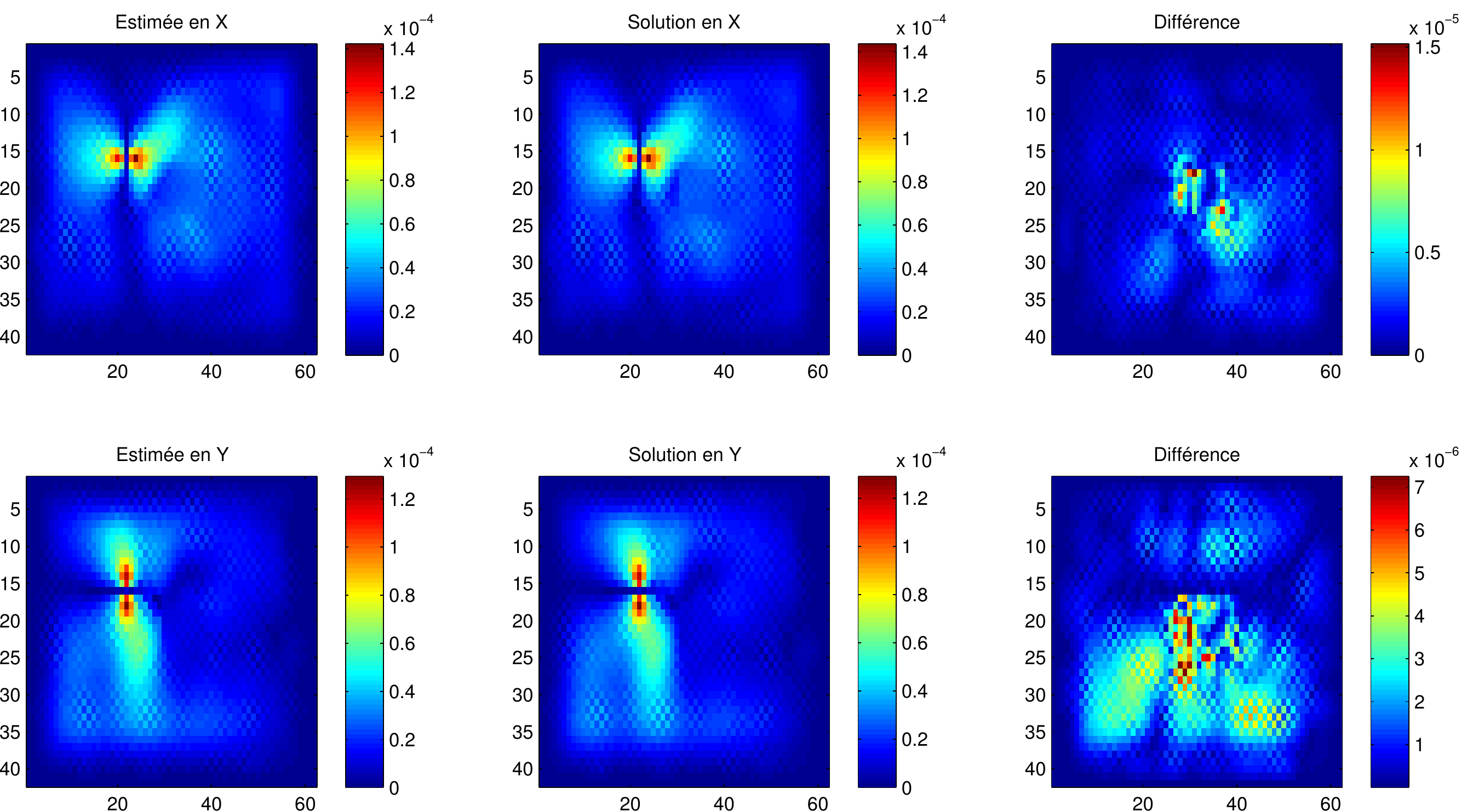}}\\
  \subfloat[Fréquence 400 Hz]{\includegraphics[scale=0.35]{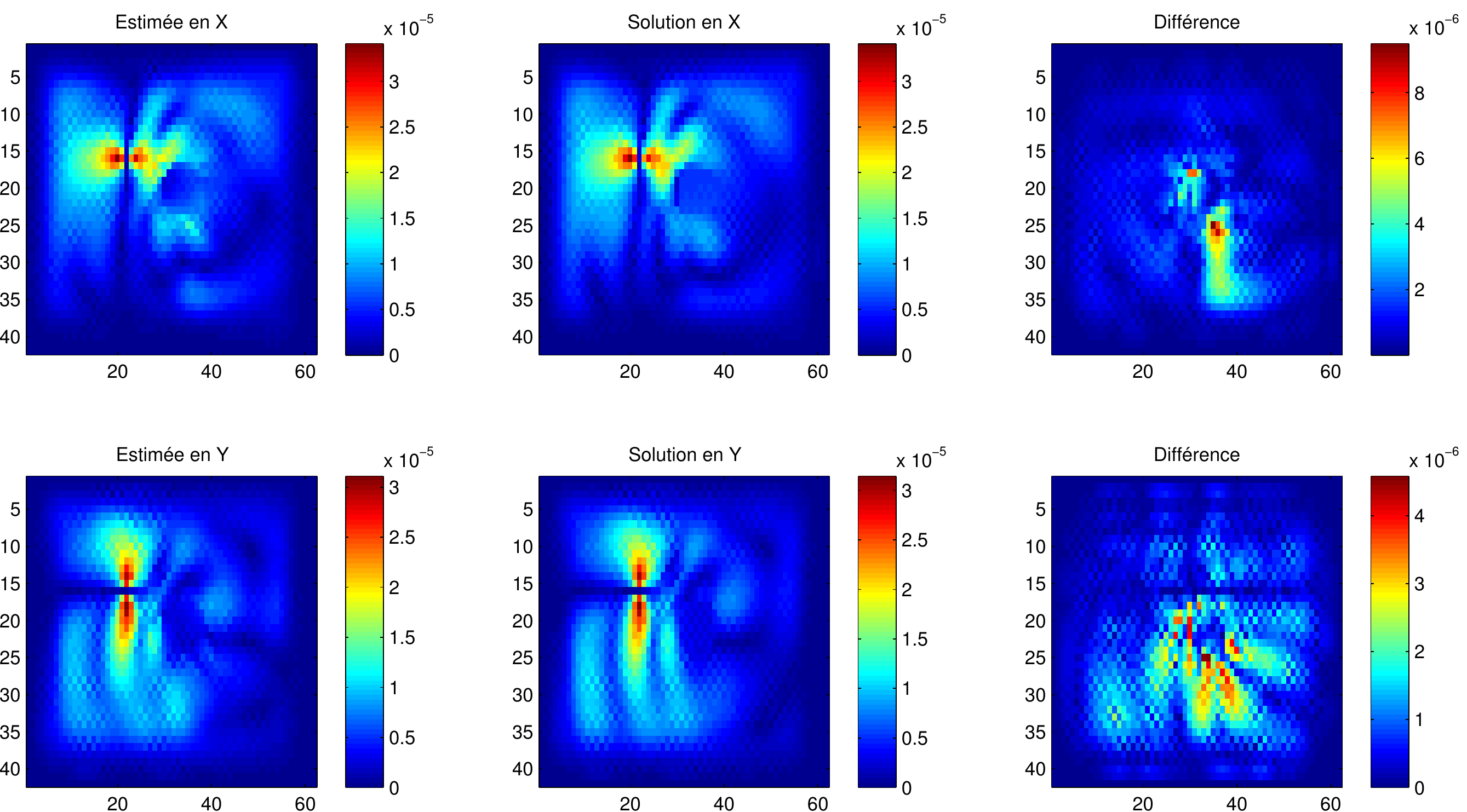}}\\
\caption{Comparaison entre les cartes du champ de vitesse données par le modèle direct et les cartes reconstruites par la méthode GM pour différentes fréquences}
\label{fig_comparaison2}
\end{figure}
\begin{figure}
\ContinuedFloat
\centering
  \subfloat[Fréquence 550 Hz]{\includegraphics[scale=0.35]{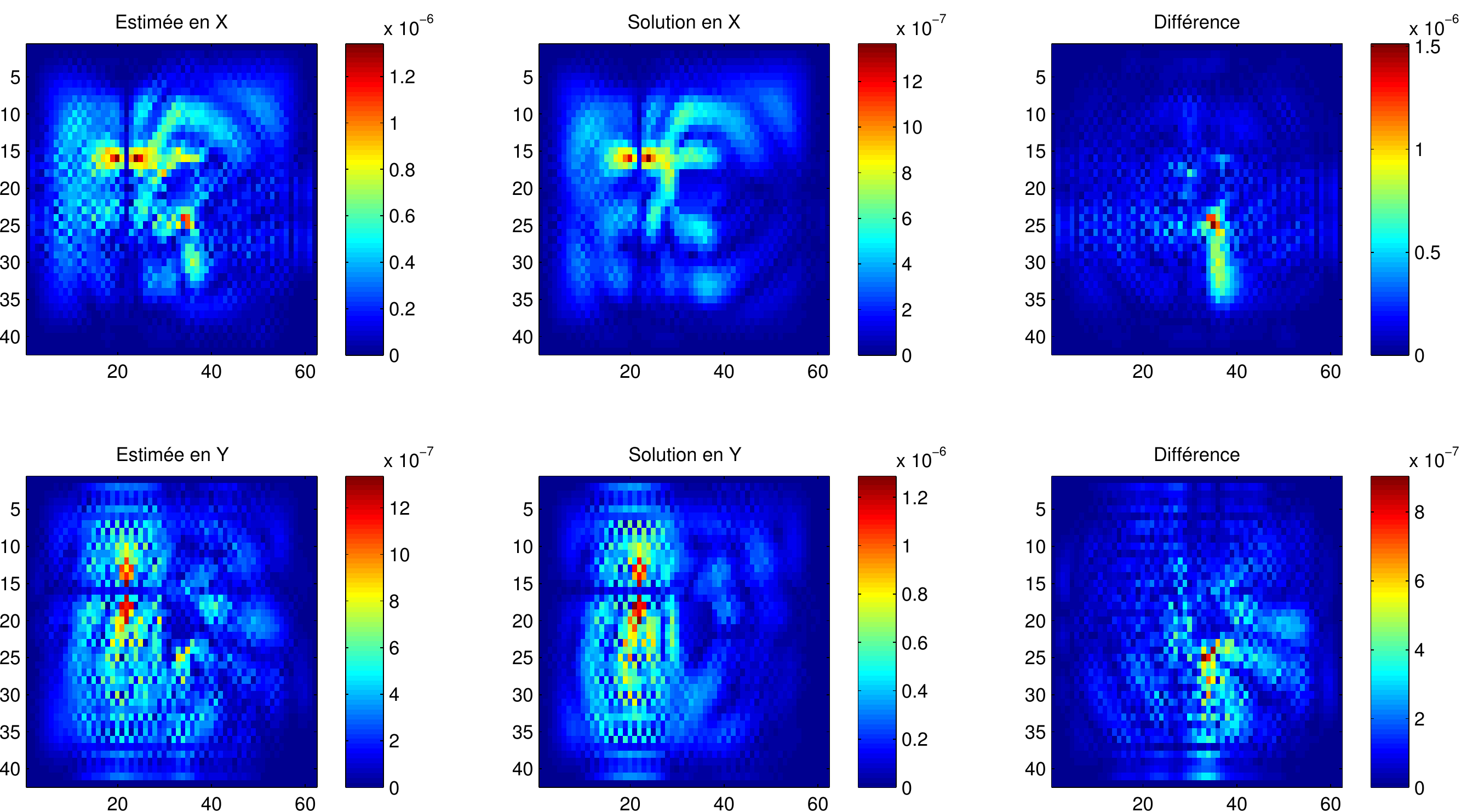}}\\
  \subfloat[Fréquence 700 Hz]{\includegraphics[scale=0.35]{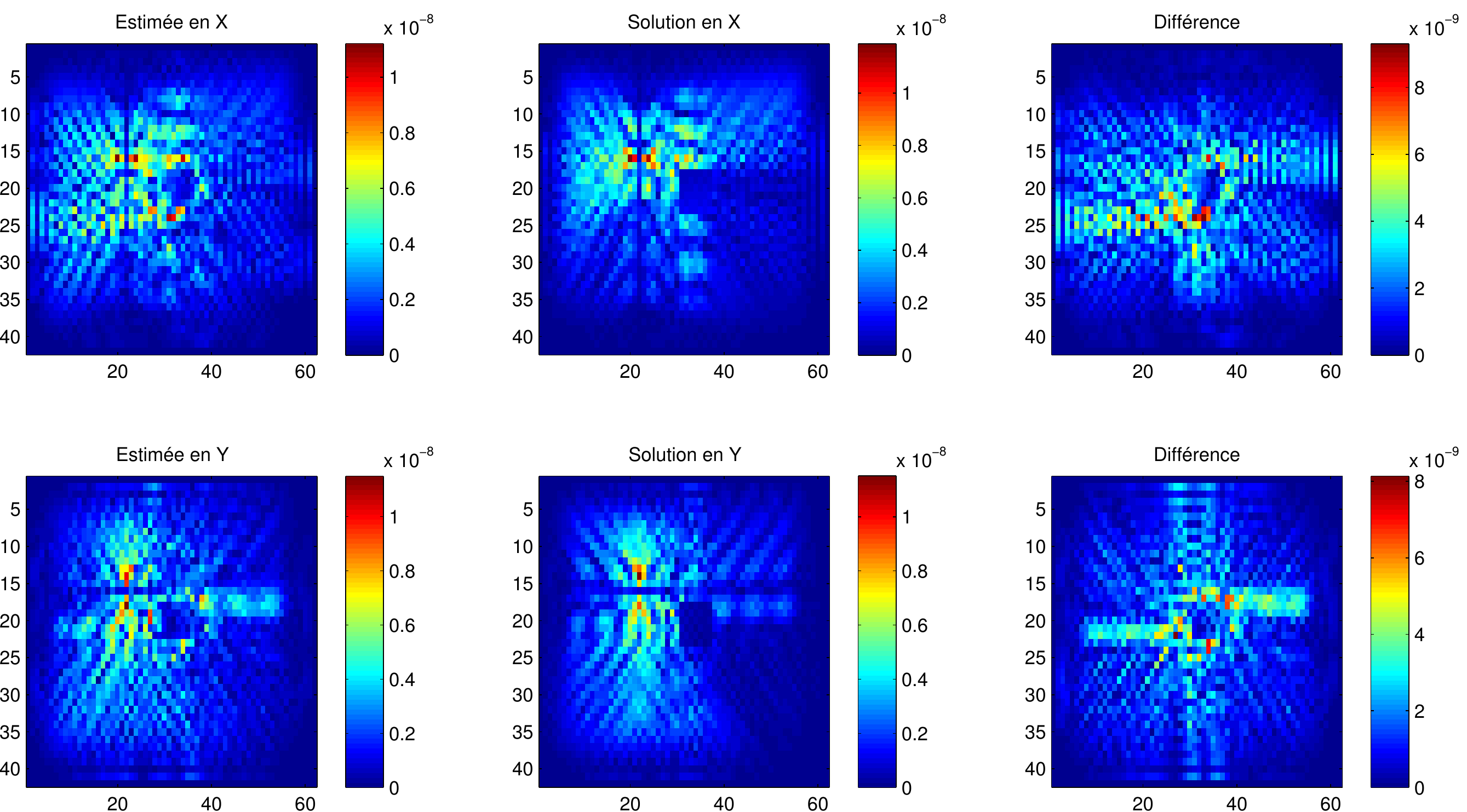}}
\caption{Comparaison entre les cartes du champ de vitesse données par le modèle direct et les cartes reconstruites par la méthode GM pour différentes fréquences (suite))}
\label{fig_comparaison2b}
\end{figure}

\end{document}